\documentclass[10pt]{article}
\usepackage[margin=1.2in]{geometry}
\usepackage[T1]{fontenc}
\usepackage{lmodern,microtype}
\usepackage{amsmath,amssymb,amsthm,mathtools}
\usepackage{enumitem}
\usepackage[dvipsnames]{xcolor}
\usepackage[colorlinks=true,linkcolor=BrickRed,citecolor=RoyalBlue,urlcolor=RoyalBlue]{hyperref}
\usepackage[nameinlink,capitalize,noabbrev]{cleveref}
\AddToHook{cmd/appendix/after}{\crefalias{section}{appendix}}
\usepackage{xltabular,booktabs}
\usepackage{aliascnt,needspace}
\usepackage{tikz,pdflscape,capt-of}
\usetikzlibrary{arrows.meta,calc}
\numberwithin{equation}{section}
\newtheorem{theorem}{Theorem}[section]
\newaliascnt{proposition}{theorem}\newtheorem{proposition}[proposition]{Proposition}\aliascntresetthe{proposition}
\newaliascnt{lemma}{theorem}\newtheorem{lemma}[lemma]{Lemma}\aliascntresetthe{lemma}
\newaliascnt{corollary}{theorem}\newtheorem{corollary}[corollary]{Corollary}\aliascntresetthe{corollary}
\newaliascnt{assumption}{theorem}\aliascntresetthe{assumption}
\newaliascnt{definition}{theorem}\newtheorem{definition}[definition]{Definition}\aliascntresetthe{definition}
\theoremstyle{remark}
\newaliascnt{remark}{theorem}\newtheorem{remark}[remark]{Remark}\aliascntresetthe{remark}
\newtheorem*{remark*}{Remark}
\newcommand{\Tr}{\operatorname{Tr}}
\newcommand{\supp}{\operatorname{supp}}
\newcommand{\dist}{\operatorname{dist}}
\newcommand{\ad}{\operatorname{ad}}
\newcommand{\id}{\operatorname{id}}
\newcommand{\idmat}{\mathbf{1}}
\newcommand{\E}{\mathbb E}
\newcommand{\R}{\mathbb R}
\newcommand{\N}{\mathbb N}
\newcommand{\C}{\mathbb C}
\newcommand{\dd}{\mathrm{d}}
\newcommand{\rhoB}{\rho_\beta}

\newcommand{\cL}{\mathcal L}
\newcommand{\cD}{\mathcal D}
\newcommand{\cU}{\mathcal U}
\newcommand{\cJ}{\mathcal J}
\newcommand{\cV}{\mathcal V}
\newcommand{\cB}{\mathcal B}
\newcommand{\cE}{\mathcal E}
\newcommand{\hL}{\widehat{\mathcal L}}
\newcommand{\hQ}{\widehat\Psi}
\newcommand{\LK}{\mathcal L_{\mathrm{KMS}}}
\newcommand{\HL}{H_{\mathrm{Lamb}}}
\newcommand{\EK}{\mathcal E_K}
\newcommand{\ED}{\mathcal E_{\mathrm{KMS}}}
\newcommand{\nr}[1]{\left\|#1\right\|_{\rhoB}}
\newcommand{\no}[1]{\left\|#1\right\|_{\mathrm{op}}}
\newcommand{\nm}[1]{\left\|#1\right\|_{\rhoB \to \rhoB}}
\newcommand{\ip}[2]{\left\langle #1,#2\right\rangle_{\rhoB}}
\newcommand{\mean}[1]{\Tr(\rhoB #1)}
\newcommand{\acr}{\widehat{\operatorname{ac}}}
\DeclareMathOperator{\csch}{csch}
\newcommand{\proofstep}[1]{\par\smallskip\noindent\textit{#1}\enspace}
\allowdisplaybreaks[1]

\newcommand{\red}[1]{{#1}}
\DeclareRobustCommand{\green}[1]{{#1}}
\newcommand{\pair}[2]{\langle #1,#2\rangle}
\newcommand{\norm}[1]{\lVert #1\rVert}
\newcommand{\Rea}{\operatorname{Re}}
\newcommand{\ac}{\operatorname{ac}}
\newcommand{\Q}{\mathcal Q_{\mathrm{Lamb}}}
\newcommand{\Gm}{G^{\mathrm{mod}}}

\usepackage{todonotes}

\title{Efficient thermalization of one-dimensional noncommuting Hamiltonians via system-bath interactions with $\Theta(n)$ ancilla qubits}

\author{%
  Zhiyan Ding\thanks{Department of Mathematics, University of Michigan,
    Ann Arbor, MI 48109, USA.
    Email: \href{mailto:zyding@umich.edu}{\texttt{zyding@umich.edu}}.}
  \and
  Bowen Li\thanks{Department of Mathematics, City University of Hong Kong,
    Hong Kong.
    Email: \href{mailto:bowen.li@cityu.edu.hk}{\texttt{bowen.li@cityu.edu.hk}}.}
  \and
  Ruizhe Zhang\thanks{Department of Computer Science, Purdue University,
    West Lafayette, IN 47907, USA.
    Email: \href{mailto:rzzhang@purdue.edu}{\texttt{rzzhang@purdue.edu}}.}
}
\date{}

\begin{document}
\pagenumbering{arabic}
\maketitle
\begin{abstract}
We prove rapid convergence of a system-bath interaction algorithm for preparing Gibbs states of one-dimensional noncommuting Hamiltonians at every fixed positive temperature. For a chain of $n$ qudits of local dimension $2^q$ with bounded nearest-neighbor interactions, the algorithm prepares the Gibbs state from any initial state to trace-norm error $0<\varepsilon\le1/2$ in expected total Hamiltonian evolution time $\widetilde{\mathcal O}(n/\varepsilon)$. This improves the previously proved $\widetilde{\mathcal O}(n^6/\varepsilon)$ bound for single-ancilla system-bath Gibbs samplers, up to logarithmic factors. The proposed algorithm couples $4^q-1$ resettable bath qubits to each site in parallel and randomizes the interaction time. Our analysis bounds both the one-round approximation error and the stationarity defect of an approximate stationary state in Gibbs-weighted $L^2$ norms, with only linear dependence on the system size. It also uses the dephasing induced by randomized evolution to control the extensive Lamb shift without treating it as a small perturbation of the dissipative dynamics.
\end{abstract}
\setcounter{tocdepth}{2}
\begingroup\small\tableofcontents\endgroup
\clearpage

\section{Introduction}\label{sec:intro}
Preparing the thermal state of a quantum many-body system, a task known as \emph{quantum Gibbs sampling}, is a central problem in quantum algorithms. Given a local Hamiltonian $H_n$ on $n$ sites and an inverse temperature $\beta>0$, the \emph{Gibbs state}
\[
  \rhoB=\frac{e^{-\beta H_n}}{\Tr(e^{-\beta H_n})}
\]
describes the system in thermal equilibrium and determines the finite-temperature properties of materials and molecules. Quantum Gibbs sampling is therefore a basic step in quantum simulation for condensed-matter physics and quantum chemistry~\cite{PoulinWocjan09,TOVPV11,Motta20,CKBG25}; it is also an important subroutine in many quantum algorithms for optimization and machine learning~\cite{BrandaoSvore17,KieferovaWiebe17,Amin18,vAGGdW20}.

A promising approach to quantum Gibbs sampling is to construct a Lindbladian whose dynamics drives every initial state toward $\rhoB$, and to simulate this dynamics on a quantum computer. The key innovation in recent constructions~\cite{CKBG25,DLL25} is to build the jump operators by filtering the Heisenberg evolution of local operators so that the Lindbladian satisfies the Kubo--Martin--Schwinger (KMS) detailed-balance condition, which makes $\rhoB$ an \emph{exact} stationary state of the dynamics while allowing efficient quantum simulation. The convergence of these dynamics is also well understood in several regimes. For example, rapid mixing, that is, a mixing time logarithmic or polylogarithmic in the system size, has been proved for local Hamiltonians at high temperature~\cite{RFA25,bakshi2026dobrushin} and for one-dimensional spin chains at every fixed temperature~\cite{chending2026}; see~\cref{subsec:intro-related} for more detailed discussions. However,
standard circuit implementations for these Lindbladian-based quantum Gibbs samplers require forward and backward Hamiltonian evolutions controlled by multiple ancilla registers~\cite{CKBG23,DLL25}, making these algorithms challenging to implement on current and near-term quantum hardware.

Recent system-bath interaction algorithms~\cite{ding2026simple,hahn2025towards,lloyd2026quantum} take a different route, modeled on how a physical system in contact with an environment relaxes toward thermal equilibrium. In each round, one or more ancilla qubits, serving as the bath, are prepared in a simple state. The system and the bath then evolve together under a chosen joint Hamiltonian for a finite time, after which the bath is discarded and reset. The bath state and the interaction are designed so that repeated rounds bring the system close to $\rhoB$. This framework requires only joint Hamiltonian evolution with a time-dependent coupling and resettable qubits~\cite{ding2026simple}.  Moreover, because the system-bath interaction is typically local, simulating a round of duration $t$ costs about as much as simulating the system Hamiltonian $e^{-iH_nt}$, making these algorithms well-suited to current and near-term quantum devices~\cite{farrell2026preparing}.

The simplicity of system-bath algorithms comes with a substantial gap in the available complexity guarantees. For one-dimensional noncommuting Hamiltonians with bounded, finite-range interactions, the spectral gap of~\cite{BergamaschiChen} implies that simulating the Lindbladian of~\cite{CKG} prepares the Gibbs state at any fixed temperature to accuracy $\varepsilon$ in total Hamiltonian evolution time $\widetilde{\mathcal{O}}(n^2)$, and the rapid mixing of~\cite{chending2026} improves this to $\widetilde{\mathcal{O}}(n)$.\footnote{Here and throughout, $\widetilde{\mathcal{O}}$ hides factors polylogarithmic in $n$ and $1/\varepsilon$.}
In contrast, the best previous guarantee for a system-bath algorithm on the same chains is $\widetilde{\mathcal{O}}(n^6/\varepsilon)$~\cite{chen2026overcoming}. Part of this gap is due to the structure of the system-bath algorithms. Each round of system-bath interaction can only fix $\rhoB$ \emph{approximately}, \emph{i.e.}, its stationary state is slightly biased. The existing end-to-end guarantees~\cite{ding2026simple,wang2025beyond,chen2026overcoming} control this bias and finite-interaction errors by tuning the coupling strength $\alpha$ or the pulse width $\sigma$ in the joint Hamiltonian, incurring a $\operatorname{poly}(1/\varepsilon)$ cost. On the other hand, it is unclear whether the drastic difference in the system-size dependence in the complexities is inherent. It is therefore natural to ask:
\begin{center}
    \emph{Can system-bath algorithms match the system-size dependence of Lindbladian-based algorithms while keeping their simpler implementation?}
\end{center}

One obstacle is that the convergence guarantees for KMS-detailed-balanced Lindbladians do not transfer directly to the system-bath interaction rounds. After the bath is traced out, the quantum channel $\Phi_{\alpha,T}$ applied to the system in a round of duration $2T$ can be approximated by:
\begin{align}\label{eq:intro_channel_approx}
  \Phi_{\alpha,T}\approx\cU(T)\circ e^{\alpha^2\cL}\circ\cU(T),\qquad \cU(t)(\rho)=e^{-itH_n}\rho\, e^{itH_n},
\end{align}
up to errors of higher order in the interaction strength $\alpha$.
 Here ${\cL}$ is an effective Lindbladian, which can be decomposed as
 \[
  \cL(\rho)=\cL_{\rm KMS}(\rho)-i[H_{\rm Lamb},\rho],\qquad \cL_{\rm KMS}(\rhoB)=0,
\]
in the constructions of the system-bath algorithms based on KMS detailed balance~\cite{chen2026overcoming,SlezakEtAl}.
The first term $\cL_{\rm KMS}$ is a KMS-detailed-balanced Lindbladian of the form in~\cite{CKG}. For one-dimensional chains, ${\cL_{\rm KMS}}$ has a system-size-independent spectral gap~\cite{BergamaschiChen} and a rapid-mixing guarantee~\cite{chending2026}. The second term is generated by a Hermitian operator $H_{\rm Lamb}$, called the \emph{Lamb shift}, which generally does not commute with $\rhoB$. The dissipative dynamics induced by $\cL_{\rm KMS}$ brings any two states closer together in a distance weighted by $\rhoB$, while the unitary evolution introduced by the Lamb shift can increase this distance, offsetting part of the contraction provided by dissipation. Hence, the convergence guarantees for $\cL_{\rm KMS}$ do not automatically extend to $\cL$.

In addition, the bath architecture imposes a further cost in the system size. The algorithms of~\cite{ding2026simple,wang2025beyond,chen2026overcoming} couple a single ancilla to one part of the system per round, so each site interacts with the bath only about once every $n$ rounds, although every round requires evolving the entire chain. By contrast, in the Lindbladian dynamics considered in~\cite{BergamaschiChen}, all $\Theta(n)$ jump operators act simultaneously. This motivates coupling bath qubits to all sites simultaneously using $\Theta(n)$ ancillas. Such parallel-bath architectures have been proposed in~\cite{hahn2025towards,lloyd2026quantum,ong2026rigorous} and experimentally implemented in~\cite{farrell2026preparing}. Their higher update rate, however, does not by itself reduce the end-to-end complexity. With parallel interactions, bounding the difference between the channel $\Phi_{\alpha,T}$ and its approximation in~\eqref{eq:intro_channel_approx} requires estimating the fourth-order terms in their expansions in $\alpha$; a naive analysis would give an error scaling as $n^2\alpha^4$ due to the contributions from every pair of sites. In addition, $H_{\rm Lamb}$ becomes a sum of $n$ quasi-local contributions, so global bounds on its effect in the $\rhoB$-weighted distance can grow linearly with $n$, whereas the spectral gap of ${\cL}_{\rm KMS}$ remains independent of $n$. Therefore, keeping the higher-order errors small may require a weaker coupling (smaller $\alpha$), and controlling the Lamb shift term may require a larger interaction time $T$. Either cost can potentially outweigh the benefit of the parallel architecture.

\subsection{Main results}\label{subsec:intro-results}

In this work, we show that for the one-dimensional quantum spin chain, our system-bath algorithm with a parallel-bath architecture can achieve end-to-end complexity $\widetilde{\mathcal O}(n/\varepsilon)$, matching the system-size dependence of Lindbladian-based algorithms \cite{BergamaschiChen,chending2026} up to polylogarithmic factors, thereby giving an affirmative answer to the above question.
Below, we first introduce the class of Hamiltonians and our system-bath interaction algorithm, and then present the main theorem (\cref{thm:intro-main}).

\subsubsection{Setup of the Hamiltonian}\label{sec:hamiltonian-setup}

We first introduce the local noncommuting Hamiltonian on a one-dimensional chain $[n]=\{1,\ldots,n\}$ equipped with the index distance $|x-y|$. For an index subset $S\subseteq[n]$, define its neighborhood of radius $r$ as
\[
 N_r(S)=\{y\in[n]:\dist(y,S)\le r\},
\]
and then $N_r(x)=[x-r,x+r]\cap[n]$ for $S = \{x\}$. As in~\cite{BergamaschiChen,chending2026}, each site $x\in[n]$ carries a qudit of local dimension $2^q$, which can be identified with $q$ qubits, $\C^{2^q}\cong(\C^2)^{\otimes q}$. Without loss of generality, we consider the nearest-neighbor Hamiltonian
\begin{equation}\label{eq:hamiltonian}
 H_n = \sum_{x=1}^{n-1} h_x,\quad h_x=h_x^\dagger,\quad
 \supp h_x\subseteq\{x,x+1\},\quad \no{h_x}\le 1.
\end{equation}
The general finite-range case can be dealt with by grouping consecutive qudits as in \cite[Section~I.A]{BergamaschiChen}.
 For an operator $O$, $\supp O\subseteq[n]$ is the smallest set of sites outside which $O$ acts as the identity; it need not be an interval. The norm $\|\cdot\|_{\mathrm{op}}$ denotes the operator norm.

The model~\eqref{eq:hamiltonian} is exactly the setting of~\cite[Definition~II.1]{BergamaschiChen}. For later use, we write $S_x=\{x,x+1\}$ for the assigned support interval of $h_x$, so that $\supp h_x\subseteq S_x$, and $H_I=\sum_{S_x\subseteq I}h_x$ for the Hamiltonian isolated to an interval $I$.

\subsubsection{The system-bath interaction algorithm}\label{sec:setup}

Our algorithm follows the system-bath interaction framework of~\cite{ding2026simple} with the randomized interaction time~\cite{chen2026overcoming}, and extends the bath architecture by coupling $4^q-1$ resettable bath qubits to every site of the chain in parallel. Given a fixed inverse temperature $\beta>0$ and starting from any initial state $\rho_0$, the system-bath algorithm prepares the target Gibbs state
\begin{equation*}
    \rhoB=\frac{e^{-\beta H_n}}{\Tr \left(e^{-\beta H_n}\right)}
\end{equation*}
by repeatedly applying the quantum channel
\begin{equation}\label{eq:intro-algorithm-channel}
 \begin{aligned}
 \rho_{k+1}=\Phi(\rho_k)= \E_T[\Phi_{\alpha,T}(\rho_k)],\qquad
 \Phi_{\alpha,T}(\rho)=\E_{\boldsymbol\omega}\Bigl[\Tr_E\bigl(U_{\alpha,\boldsymbol\omega,T}(\rho\otimes\tau_{E,\boldsymbol\omega})U_{\alpha,\boldsymbol\omega,T}^\dagger\bigr)\Bigr],
 \end{aligned}
\end{equation}
where $\tau_{E,\boldsymbol\omega}$ is the initial bath state (defined below), and the expectations average over the random bath frequencies $\boldsymbol\omega=(\omega_{x,a})_{x,a}$ and the random interaction time $T$. The unitary $U_{\alpha,\boldsymbol\omega,T}={\cal T}\exp(-i\int_{-T}^T H(t)\,\dd t)$, with ${\cal T}$ denoting time ordering, is the time-evolution of the following time-dependent Hamiltonian:
\begin{equation}\label{eq:intro-joint}
 H(t)=H_n+\sum_{x=1}^n\sum^{4^q-1}_{a=1}H_{E_{x,a}}(\omega_{x,a})
 +\alpha f_\sigma(t)\sum_{x=1}^n\sum^{4^q-1}_{a=1}\bigl(A_{x,a}\otimes B_{x,a}+A_{x,a}^\dagger\otimes B_{x,a}^\dagger\bigr),
\end{equation}
over $t\in[-T,T]$.
The bath is designed following the framework of~\cite{chen2026overcoming}:
\begin{itemize}
\item The bath consists of one qubit for each label $(x,a)\in [n]\times [4^q-1]$. Each bath qubit has Hamiltonian and bath coupling operator
\begin{align*}
    H_{E_{x,a}}(\omega_{x,a})=-\frac{\omega_{x,a}}{2}Z_{E_{x,a}},\qquad B_{x,a}=|1\rangle\langle0|,
\end{align*}
where the bath frequency $\omega_{x,a}$ is sampled, as in~\cite[Assumption~3.3]{chen2026overcoming}, from the following probability density:
\begin{equation}
g_{\beta,\sigma}(\omega) =\frac{\beta}{\sqrt{2\pi s_{\beta,\sigma}}} e^{-(\beta\omega+1)^2/(2s_{\beta,\sigma})}\,, \qquad s_{\beta,\sigma} = 2-\frac{\beta^2}{4\sigma^2}\,.
 \label{eq:frequency}
\end{equation}
\item At the beginning of each round, the $(4^q-1)n$ bath qubits are initialized in the following product state ${\tau_{E,\boldsymbol{\omega}}}$
\begin{equation*}
 \tau_{E,\boldsymbol\omega}=\bigotimes_{x,a}
 \frac{e^{-\widetilde\beta H_{E_{x,a}}(\omega_{x,a})}}
 {\Tr(e^{-\widetilde\beta H_{E_{x,a}}(\omega_{x,a})})},
 \qquad \widetilde\beta=\frac{2\beta}{2-\beta^2/(4\sigma^2)}.
\end{equation*}

\item The system-bath interaction time $T$, corresponding to a round duration $2T$, is sampled from a distribution with the following probability density function:
\begin{equation}\label{eqn:T_0}
 \mu_{T_0}(t)=\frac1{6T_0}\left(\frac t{T_0}-1\right)^3
             e^{-(t/T_0-1)}\mathbf1_{\{t\ge T_0\}},\qquad
 \widehat\mu_{T_0}(\omega)=e^{i\omega T_0}(1-i\omega T_0)^{-4},
\end{equation}
where $T_0$ is a parameter to be chosen later, and $\widehat{\mu}_{T_0}$ is the Fourier transform of $\mu_{T_0}$.
\end{itemize}

For convenience, we set the system coupling operators $A_{x,a}$ as $4^q-1$ nonidentity Pauli strings on the $q$ qubits  at each site $x\in[n]$:
\begin{equation}\label{eq:couplings}
 \{A_{x,a}\}_{a=1}^{4^q-1}=\{P_{x,a}\}_{a=1}^{4^q-1}
 :=\bigl\{\idmat,X,Y,Z\bigr\}^{\otimes q}_x\setminus\{\idmat\}.
\end{equation}
We write $Q:=4^q-1$. Each $A_{x,a}$ is Hermitian, has norm one, and acts only on site $x$.
We choose the Gaussian pulse $f_\sigma$ in~\eqref{eq:intro-joint} as in \cite[Assumption~3.3]{chen2026overcoming}: for $\sigma = \beta/2$,
\begin{align}
 f_\sigma(t)=\frac{e^{-t^2/(4\sigma^2)}}{\sqrt{\sigma\sqrt{2\pi}}}, \qquad \text{with normalization} \quad
\|f_\sigma\|_1^2 = 2\sqrt{2\pi}\,\sigma, \quad \|f_\sigma\|_2=1.
 \label{eq:pulse}
\end{align}
At this width $\sigma$, we have $s_{\beta,\sigma}=1$ and $\widetilde\beta=2\beta$, so the bath frequency law becomes
\begin{align*}
    g_{\beta,\sigma}(\omega)=\frac{\beta}{\sqrt{2\pi}} e^{-(\beta\omega+1)^2/2}.
\end{align*}

Each round of the algorithm requires implementing the quantum channel $\Phi$ in~\eqref{eq:intro-algorithm-channel}, which involves simulating $H(t)$ for an interaction duration $2T$. Because both the bath Hamiltonians and the system-bath couplings are local, we only measure the computational cost in terms of the \emph{total Hamiltonian evolution time}, which has expectation
\begin{equation}\label{eq:intro-total-time}
T_{\mathrm{total}}
:=\E\left[\sum_{j=1}^m 2T_j\right]
=2m\E[T]
=10mT_0\,,
\end{equation}
where $m$ is the number of rounds and
we use~\eqref{eqn:T_0} for the last equality.

\subsubsection{Main result}\label{sec:intro-main-result}
Our algorithm uses fresh randomness in each round, and the output of $m$ rounds is $\Phi^m(\rho_0)$, which is a mixed state over all the randomness. The following main theorem bounds the number of rounds that is sufficient to prepare the target Gibbs state $\rhoB$ within $\varepsilon$ error:

\begin{theorem}[Informal version of \Cref{cor:complexity}]\label{thm:intro-main}
Let $H_n$ be a one-dimensional noncommuting Hamiltonian of the form in~\eqref{eq:hamiltonian}, and let $\beta>0$ be fixed. For every $0<\varepsilon\le1/2$, run the algorithm~\eqref{eq:intro-algorithm-channel} with $\alpha^2=c\varepsilon/n$ for a sufficiently small constant $c>0$, and $T_0=\Theta(\sqrt{\log(n/\varepsilon)})$ for $m=\widetilde{\mathcal O}(n/\varepsilon)$ rounds. Then
\begin{align*}
    \|\Phi^m(\rho_0)-\rhoB\|_1 \le \varepsilon,
\end{align*}
for every initial state $\rho_0$. The expected total Hamiltonian evolution time~\eqref{eq:intro-total-time} is
\[
  T_{\mathrm{total}} = \widetilde{\mathcal O}(n/\varepsilon).
\]
\end{theorem}

\Cref{thm:intro-main} improves the best previous end-to-end complexity result for system-bath algorithms~\cite{chen2026overcoming} by a factor of $n^5$, and keeps the linear dependence on $1/\varepsilon$. Its dependence on $n$ matches the state-of-the-art result for Lindbladian-based algorithms that follows from~\cite{chending2026}. A comparison between our result and previous results is summarized in Table~\ref{tab:intro-comparison}.

The remaining difference between these two approaches to quantum Gibbs sampling lies in the dependence on the accuracy, which is $\operatorname{polylog}(1/\varepsilon)$ for Lindbladian-based algorithms and $1/\varepsilon$ in our $T_{\rm total}$. It comes from the bias of the stationary state: since one round of the system-bath interaction does not preserve $\rhoB$ exactly, the stationary state of $\Phi$ differs from $\rhoB$. Randomizing the interaction time cancels the leading effect of the Lamb shift on this bias~\cite{chen2026overcoming}, and we show that the remaining bias is $\mathcal O(n\alpha^2)$ even with parallel coupling, so our analysis takes $\alpha^2=c\varepsilon/n$.

\begin{table}[htbp]
\centering
\begin{tabular}{@{}lcc@{}}
\toprule
\textbf{References}                                             & \textbf{Bath architecture}                      & \textbf{\shortstack{Total Hamiltonian\\evolution time}}                \\ \midrule
\cite{ding2026simple} (analyzed by~\cite{SlezakEtAl})           & One qubit                                       & $\widetilde{\mathcal{O}}(n^{10}/\varepsilon^4)$         \\ \midrule
\cite{wang2025beyond}                                           & One qubit                                       & $\widetilde{\mathcal{O}}(n^7/\varepsilon^2)$            \\ \midrule
\cite{scandi2026thermalization} (analyzed by~\cite{SlezakEtAl}) & Macroscopic Gaussian bath                       & $\widetilde{\mathcal{O}}(n^7/\varepsilon^2)$            \\ \midrule
\cite{chen2026overcoming}                                       & \begin{tabular}[c]{@{}c@{}}One qubit, \\ \shortstack{randomized\\interaction time}\end{tabular}              & $\widetilde{\mathcal{O}}(n^6/\varepsilon)$              \\ \midrule
\textbf{This work}                                              & \textbf{\begin{tabular}[c]{@{}c@{}}$\mathcal{O}(n)$ qubits, \\ \shortstack{randomized\\interaction time}\end{tabular}} & $\boldsymbol{\widetilde{\mathcal O}(n/\varepsilon)}$ \\ \bottomrule
\end{tabular}
\caption{Comparisons of the system-bath interaction algorithms for preparing Gibbs states of one-dimensional noncommuting Hamiltonians with $n$ sites and accuracy $\varepsilon$. \cite{scandi2026thermalization} requires high temperature, while the other results apply to any fixed $\beta>0$. %
}\label{tab:intro-comparison}
\end{table}

\subsection{Related work}\label{subsec:intro-related}

\paragraph{Davies generators.}
Davies generators arise in the natural thermalization process and have the Gibbs state as an exact stationary state~\cite{Davies74,Davies76}. For commuting Hamiltonians, their convergence has been studied through spectral gaps and modified logarithmic Sobolev inequalities (MLSIs). Spectral-gap bounds follow from clustering properties of the Gibbs state~\cite{KastoryanoBrandao}, while MLSIs control relative-entropy decay and provide stronger mixing estimates~\cite{KastoryanoTemme13}. In particular, MLSI methods yield logarithmic or polylogarithmic mixing times for commuting Hamiltonians on one-dimensional chains at every fixed positive temperature, and on higher-dimensional nearest-neighbor lattices at sufficiently high temperature~\cite{BCGLPR23,KACR}. For general noncommuting Hamiltonians, however, simulating Davies generators can require resolving exponentially close Bohr frequencies, making efficient quantum implementation challenging~\cite{CKBG25}.

\paragraph{Engineered Lindbladian Gibbs samplers.}
Recent breakthroughs in energy-filtering Lindbladian constructions have led to efficiently implementable Gibbs samplers for noncommuting Hamiltonians~\cite{CKBG23,CKBG25}, including generators satisfying exact KMS detailed balance~\cite{CKG,DLL25} and quantum analogs of Glauber and Metropolis dynamics~\cite{GCDK24}. System-size-independent spectral gaps have been established in several perturbative regimes, including high-temperature local Hamiltonians~\cite{RFA24}, high-temperature all-to-all systems~\cite{bergamaschi2026fast}, and weakly interacting fermionic systems~\cite{TongZhan25,SMBB25}, mainly through perturbative analyses of the parent Hamiltonians associated with the Lindbladian dynamics. Beyond such perturbative regimes, Bergamaschi and Chen established a system-size-independent spectral gap for bounded, finite-range one-dimensional Hamiltonians at every fixed positive temperature~\cite{BergamaschiChen}. Stronger logarithmic mixing-time bounds have also been obtained for high-temperature lattice systems~\cite{RFA25,bakshi2026dobrushin}, including systems with arbitrary on-site external fields~\cite{bakshi2026rapid}, as well as for weakly interacting qudit systems~\cite{vsmid2025rapid}.

\paragraph{System-bath samplers.}
Gibbs-state preparation through repeated interactions with resettable baths dates back to~\cite{TD00} and is closely related to collision models~\cite{Ciccarello22}. Early analyses established thermalization guarantees under the eigenstate thermalization hypothesis~\cite{shtanko2021preparing,CB23}. More recent protocols instead design the system-bath interactions based on newly developed Lindbladian constructions~\cite{lloyd2026quantum,hahn2025towards,ding2026simple}. Ref.~\cite{ding2026simple} established the first end-to-end guarantees for free-fermion systems and high-temperature commuting Hamiltonians using a single reusable bath qubit. Ref.~\cite{SlezakEtAl} extended such guarantees to noncommuting systems, including one-dimensional chains, via spectral-gap comparisons for Gaussian-filtered Lindbladians. Subsequently, Ref.~\cite{wang2025beyond} established convergence guarantees for system-bath interaction beyond the Lindbladian approximation regime and improved the complexity bounds of Ref.~\cite{SlezakEtAl}. These bounds were further improved in Ref.~\cite{chen2026overcoming} by introducing randomized interaction durations, which enable control of the Lamb shift contribution and yield sharper end-to-end complexity guarantees. Related stationary-state bias bounds were independently obtained by Ong et al.~\cite{ong2026rigorous}. Our work builds most directly on the randomized-interaction framework of Ref.~\cite{chen2026overcoming}, while developing new techniques to obtain substantially improved complexity bounds in the many-body settings considered here.

\paragraph{Macroscopic baths and physical thermalization.}
Another system-bath interaction model couples the system continuously to a large environment, initially in thermal equilibrium, without resetting the environment during the evolution. Scandi--Alhambra derived a master equation satisfying KMS detailed balance for this setting, incorporating the Lamb shift into a renormalized system Hamiltonian~\cite{scandi2026thermalization}. Under suitable bath-correlation assumptions, Slezak et al.\ proved high-temperature thermalization in time $\widetilde{\mathcal{O}}(n^7\varepsilon^{-2})$~\cite{SlezakEtAl}. In our algorithm, the bath instead consists of explicitly controlled qubits that are prepared, coupled to the system for a finite time, and reset after each round. We establish guarantees at every fixed positive temperature in one dimension, and the extension to a continuously coupled reservoir remains open.

\paragraph{One-dimensional thermal estimates.}
Our analysis uses complex-time locality in one dimension, developed by Araki~\cite{Araki69} and quantified by P{\'e}rez-Garc{\'\i}a and P{\'e}rez-Hern{\'a}ndez~\cite{perez2023locality}. We combine these estimates with the auxiliary spectral gap of Bergamaschi--Chen~\cite{BergamaschiChen} and exponential clustering for gapped ground states~\cite{NachtergaeleSims} to control correlations in the purified Gibbs state. Appendix~\ref{app:1d} collects the resulting thermal estimates.

\paragraph*{Concurrent work.}
During the preparation of this paper, many concurrent works appeared that establish further results for one-dimensional Gibbs states at every fixed finite temperature~\cite{pilatowskycameo2026,kim2026rapidmixingquantumspin,cps26,rf26,chending2026,chen2026sizeindependentlogarithmicsobolevinequalities,gao2026completemodifiedlogarithmicsobolev}. Ref.~\cite{pilatowskycameo2026} proves an exact decomposition of the Gibbs state into a mixture of product states over contiguous blocks of bounded size, leading to a constant-depth local-unitary preparation algorithm with polynomial-time classical preprocessing. Refs.~\cite{kim2026rapidmixingquantumspin,rf26} establish an $\mathcal O(\log(n/\varepsilon))$ mixing time for their respective Gibbs samplers. Ref.~\cite{cps26} establishes an $\mathcal O(\log n\,\log(n/\varepsilon))$ mixing time for the heat-bath and regularized heat-bath dynamics. Because these algorithms have structures fundamentally different from Hamiltonian-simulation based dynamics, their complexities cannot be directly compared with ours using total Hamiltonian evolution time. The most closely related work is Ref.~\cite{chending2026}, which directly studies the CKG sampler~\cite{CKG} and proves rapid mixing through spectral analysis in the KMS $L^2$ space. In particular, the Lindbladian considered there is the same dynamics that our system-bath interaction algorithm aims to approximate. Our rapid-mixing proof in \cref{sec:complexity} combines their enhanced-dissipation estimate with the physical-channel analysis to obtain $\widetilde{\mathcal O}(n/\varepsilon)$ expected evolution time.

\paragraph{Organization.}
Section~\ref{subsec:intro-results} presents the Hamiltonian setup, the system-bath interaction algorithm, and the main complexity bound.
Sections~\ref{subsec:preliminaries} and~\ref{subsec:technical-overview} introduce the notation and outline the three steps of the proof: channel approximation, construction of an approximate stationary state, and contraction and mixing-time analysis.
Section~\ref{sec:generator} derives the second-order generator and its Lamb shift from the Dyson expansion and bath parity.
Section~\ref{sec:locality} establishes quasi-locality of the effective terms, which Section~\ref{sec:comparison} uses to prove the one-round comparison between the physical and ideal channels.
Section~\ref{sec:weighted} develops weighted field and moment estimates using spectral conditional expectations; these estimates support the approximate stationary-state construction in Section~\ref{sec:reference} and the contraction analysis in Section~\ref{sec:mixing}.
Section~\ref{sec:reference} constructs an approximate stationary state for the averaged physical channel, while Section~\ref{sec:mixing} proves contraction of the averaged ideal channel by controlling the Lamb shift through dissipation and dephasing, retaining a fixed fraction of the averaged Dirichlet form.
Finally, Section~\ref{sec:complexity} transfers contraction to the averaged physical channel and bounds its stationary-state bias, then combines enhanced dissipation with contraction toward the physical stationary state to prove the $\widetilde{\mathcal O}(n/\varepsilon)$ expected evolution-time bound in \cref{cor:complexity}.
Appendix~\ref{app:1d} proves the one-dimensional locality and thermal estimates used in the analysis.
Throughout, the inverse temperature $\beta$ and the local dimension $2^q$ are fixed independently of $n$ and $\varepsilon$, and the pulse width is $\sigma=\beta/2$.

\subsection*{Use of AI tools.}

From late June to late July 2026, the authors used ChatGPT 5.5 Pro and ChatGPT 5.6 Sol to explore possible proof strategies for achieving an $\widetilde O(n^2)$ system-size dependence for system--bath Gibbs samplers with $\Theta(n)$ parallel bath ancillas for one-dimensional noncommuting Hamiltonians, motivated by the fast mixing result \cite{BergamaschiChen}. The idea of employing a parallel bath architecture originated with the authors, and they wrote research notes analyzing the fixed-point approximation of parallel bath system-bath interaction channels for some simpler systems (with little help from ChatGPT 5.5 Pro). Later, with the authors' guided prompting using the notes, ChatGPT 5.6 Sol worked out a fixed-point proof for more general, noncommuting systems. However, the first version of the proof is very complicated and involves several system-dependent assumptions. The authors verified, digested, and simplified it using ChatGPT 6 Pro and Fable 5.1, and formulated the cleanest part (for 1D noncommuting systems) as Steps 1 and 2 in~\cref{subsec:technical-overview}.  The contraction estimate for the random-time evolution and the bound on the Lamb-shift term in~\cref{subsec:contraction-theorem}, under the assumptions developed in~\cref{sec:weighted}, were initially proposed by ChatGPT 5.6 Sol. The underlying idea was first investigated in the one-dimensional commuting case using ChatGPT 5.5 Pro and later extended to the noncommuting case using ChatGPT 5.6 Sol, yielding a complexity bound of $\widetilde O(n^2)$. The authors independently reorganized the proof, clarified the required conditions, and formulated the technical overview in Step 3 of~\cref{subsec:technical-overview}. During the preparation of this manuscript, new rapid-mixing results for one-dimensional noncommuting Gibbs samplers were obtained by one of the authors and by other research groups in late September and early October 2026. The authors incorporated the enhanced-dissipation estimate~\eqref{eq:enhancedestimate} from~\cite{chending2026} into their analysis, improving the system-size dependence to $\widetilde O(n)$ in~\cref{cor:complexity}.

The authors also used AI assistants, including ChatGPT 5.6 Sol, ChatGPT 6 Pro, ChatGPT 6 Astra, and Claude Opus 5.5, to assist in drafting and revising the exposition, plotting the figures, improving the presentation of proofs, checking calculations, and ensuring consistency of notation. All mathematical statements, proofs, and references were independently verified by the authors, who take full responsibility for the content of this manuscript.

\newpage
\section{Preliminaries and technical overview}
In this section, we provide a brief overview of the proof of \cref{thm:intro-main}, and the technical tools and concepts involved in the analysis.

\subsection[Preliminaries]{\texorpdfstring{\red{Preliminaries}}{Preliminaries}}\label{subsec:preliminaries}\label{sec:model}

We fix the notation and preliminary concepts used in the subsequent analysis. The Hamiltonian and algorithm are defined in Sections~\ref{sec:hamiltonian-setup} and~\ref{sec:setup}, respectively.

\paragraph{Operators, norms, and asymptotic notation.}
For $S\subseteq[n]$, write $\mathcal H_S=\bigotimes_{x\in S}\C^{2^q}$ and let $\mathcal B(\mathcal H_S)$ denote its full matrix algebra. An operator supported on $S$ acts as the identity on $S^c=[n]\setminus S$; tensor factors are always kept in their original order. We write $\idmat$ for the identity matrix on the relevant Hilbert space, $\idmat_S$ for the identity on $\mathcal H_S$, and $\id$ for the identity superoperator. Letters such as $I,J$ denote intervals of sites.
We use $\N=\{1,2,\ldots\}$ and $\N_0=\N\cup\{0\}$, and write $[k]=\{1,\ldots,k\}$. A state is a positive semidefinite matrix of trace one.

For a matrix $O$, its Hermitian adjoint is $O^\dagger$. Writing $s_j(O)$ for its singular values, the Schatten norms and Hilbert--Schmidt inner product are
\[
 \|O\|_p=\Big(\sum_j s_j(O)^p\Big)^{1/p},\quad 1\le p<\infty, \qquad
 \langle O,P\rangle_2=\Tr(O^\dagger P),
\]
where $\|O\|_\infty=\no O=\max_j s_j(O)$.
We use the trace norm to measure the preparation error throughout the paper. For scalar functions $f$, $\|f\|_p$ denotes its standard $L^p$-norm. For a superoperator $\mathcal T$, we use the induced norms
\[
 \|\mathcal T\|_{p\to p}=\sup_{O\ne0}\frac{\|\mathcal T(O)\|_p}{\|O\|_p},\qquad p=1,2,
\]
and its \green{Hilbert-Schmidt adjoint} is determined by
\begin{equation}\label{eq:trace-adjoint}
 \Tr\bigl(O^\dagger\mathcal T(P)\bigr)
 =\Tr\bigl((\mathcal T^\dagger(O))^\dagger P\bigr).
\end{equation}
We write $f=\Omega(g)$ when $g=\mathcal O(f)$ and $f=\Theta(g)$ when both bounds hold; $\widetilde{\mathcal O}$ suppresses polylogarithmic factors in $n$ and $1/\varepsilon$. Constants $C,c,\ldots$ may change from line to line and depend on the fixed local parameters $\beta,q$, but not on $n,\alpha,T$, or $T_0$.

Throughout the manuscript, the Gaussian pulse width is fixed to $\sigma=\beta/2$, for simplicity. \red{We retain $\sigma$ in the formulas as shorthand for this fixed value.}

\paragraph{The KMS inner product and weighted state differences.}
Since $\rhoB$ is strictly positive, it defines the Kubo--Martin--Schwinger (KMS) inner product, its norm, and the corresponding induced map norm:
\begin{equation}\label{eq:rho-space}
 \begin{aligned}
 \ip OP&=\Tr(\rhoB^{1/2}O^\dagger\rhoB^{1/2}P),\\
 \nr O&=\|\rhoB^{1/4}O\rhoB^{1/4}\|_2,\qquad
 \nm{\mathcal T}=\sup_{O\ne0}\frac{\nr{\mathcal T(O)}}{\nr O}.
 \end{aligned}
\end{equation}
Define the change of coordinates and its powers by
\begin{equation}\label{eq:Gamma}
 \Gamma^s(O)=\rhoB^{s/2}O\rhoB^{s/2}\quad(s\in\R),\qquad
 \Gamma(O)=\rhoB^{1/2}O\rhoB^{1/2},\qquad
 \Gamma^{-1}(\delta)=\rhoB^{-1/2}\delta\rhoB^{-1/2}.
\end{equation}
In particular, $\Gamma^{1/2}$ is an isometry from the KMS space to the Hilbert--Schmidt space. The normalization identities are
\begin{equation}\label{eq:normalization-identities}
 \Gamma(\idmat)=\rhoB,\qquad \nr{\idmat}=1,\qquad
 \ip{\idmat}{O}=\mean O,\qquad \Tr\Gamma(O)=\mean O.
\end{equation}
We call $O$ Gibbs-centered when its \green{Gibbs mean} satisfies $\mean O=0$; this is equivalent to $\Gamma(O)$ being traceless. For a state $\rho$, the matrix $\Gamma^{-1}(\rho)=\rhoB^{-1/2}\rho\rhoB^{-1/2}$ is the (KMS-symmetrized) relative density of $\rho$ with respect to $\rhoB$, and $\nr{\Gamma^{-1}(\rho-\rhoB)}^2$ is the corresponding $\chi^2$-divergence. A state difference $\delta$ is measured in the \green{weighted state-difference norm} $\nr{\Gamma^{-1}(\delta)}=\|\Gamma^{-1/2}(\delta)\|_2$. Schatten H\"older gives
\begin{equation}\label{eq:trace-conversion}
 \begin{aligned}
 \|\delta\|_1
 &=\|\rhoB^{1/4}\Gamma^{-1/2}(\delta)\rhoB^{1/4}\|_1
 \le\|\rhoB^{1/4}\|_4^2\|\Gamma^{-1/2}(\delta)\|_2
 =\nr{\Gamma^{-1}(\delta)},\qquad \nr O\le\no O.
 \end{aligned}
\end{equation}
Thus weighted state estimates imply the trace-norm error required by the algorithm.

For a physical state map $\mathcal T$, write $\widehat{\mathcal T}=\Gamma^{-1}\mathcal T\Gamma$; since $\Gamma^{-1}(\mathcal T(\rho))=\widehat{\mathcal T}(\Gamma^{-1}(\rho))$, the map $\widehat{\mathcal T}$ sends the relative density of $\rho$ to that of $\mathcal T(\rho)$, and we call it the action of $\mathcal T$ on relative densities. A star denotes the adjoint in the KMS inner product, so
\begin{equation}\label{eq:map-duality}
 \mathcal A^\star=\Gamma^{-1}\mathcal A^\dagger\Gamma,\qquad
 (\widehat{\mathcal T})^\star=\mathcal T^\dagger,\qquad
 \nm{\widehat{\mathcal T}}=\nm{\mathcal T^\dagger}
 =\|\Gamma^{-1/2}\mathcal T\Gamma^{1/2}\|_{2\to2}.
\end{equation}
These identities follow from \eqref{eq:trace-adjoint} and the isometry $\Gamma^{1/2}$. They let us use observable estimates to control differences of physical states.

\paragraph{Forward evolution, Bohr-frequency components, and Fourier transforms.}
For $z\in\C$ and $t\in\R$, our conventions are
\begin{equation}\label{eq:evolutions}
 O(z)=e^{izH_n}Oe^{-izH_n},\qquad
 \cU(t)(O)=e^{-itH_n}Oe^{itH_n}=O(-t),\qquad
 \ad_{H_n}(O)=[H_n,O].
\end{equation}
The commutator and anticommutator are $[B,C]=BC-CB$ and $\{B,C\}=BC+CB$; we also write $\ac_B(O)=\{B,O\}$. Since $[H_n,\rhoB]=0$,
\begin{equation}\label{eq:free-isometry}
 \Gamma^{-1}\cU(t)\Gamma=\cU(t),\qquad
 \nr{\cU(t)O}=\nr O,\qquad \cU(t)^{-1}=\cU(-t).
\end{equation}
If $H_n=\sum_E E\Pi_E$ is its spectral decomposition, then
\begin{equation}\label{eq:Bohr}
 O=\sum_\nu O_\nu,\qquad O_\nu=\sum_{E-E'=\nu}\Pi_EO\Pi_{E'},\qquad
 \ad_{H_n}(O_\nu)=\nu O_\nu,\qquad \cU(t)O_\nu=e^{-it\nu}O_\nu.
\end{equation}
Here $\nu$ ranges over the Bohr frequencies, or energy differences. Distinct components are orthogonal in both the Hilbert--Schmidt and KMS inner products. Throughout the manuscript, scalar Fourier transforms use the convention
\begin{equation}\label{eq:fourier-convention}
 \widehat h(\omega)=\int_\R h(t)e^{i\omega t}\,\dd t,\qquad
 h(t)=\frac1{2\pi}\int_\R\widehat h(\omega)e^{-i\omega t}\,\dd\omega.
\end{equation}
In particular, averaging the forward evolution of a Bohr component gives
\[
 \int_\R h(t)\cU(t)(O_\nu)\,\dd t=\widehat h(-\nu)O_\nu.
\]

\paragraph{Quantum channels and the interaction picture.}
A quantum channel is a completely positive, trace-preserving linear map; its \green{Hilbert-Schmidt adjoint} is unital, $\Phi^\dagger(\idmat)=\idmat$. The \green{one-round physical channel} from \eqref{eq:intro-algorithm-channel} is
\begin{equation}\label{eq:physical-channel}
 \Phi_{\alpha,T}(\rho)=\E_{\boldsymbol\omega}\Tr_E\bigl[
 U_{\alpha,\boldsymbol\omega,T}(\rho\otimes\tau_{E,\boldsymbol\omega})
 U_{\alpha,\boldsymbol\omega,T}^\dagger\bigr],\qquad
 \Phi=\E_T[\Phi_{\alpha,T}]=\Phi_\alpha^{\mu_{T_0}}.
\end{equation}
The propagator evolves from $-T$ to $T$, so one \green{round} has Hamiltonian evolution time $2T$ and expected Hamiltonian evolution time $10T_0$.

To remove the system and bath evolution, we define $H_0=H_n+\sum^n_{x=1}\sum^Q_{a=1}H_{E_{x,a}}(\omega_{x,a})$ from~\eqref{eq:intro-joint}. The interaction-picture coupling and propagator are
\begin{equation}\label{eq:W}
 \begin{aligned}
 W_{\boldsymbol\omega}(t)
 &=f_\sigma(t)\sum_{x,a}\bigl(e^{i\omega_{x,a}t}A_{x,a}(t)\otimes B_{x,a}
       +e^{-i\omega_{x,a}t}A_{x,a}(t)^\dagger\otimes B_{x,a}^\dagger\bigr),\\
 U^I_{\alpha,\boldsymbol\omega,T}
 &={\cal T}\exp\left(-i\alpha\int_{-T}^T W_{\boldsymbol\omega}(t)\,\dd t\right),
 \end{aligned}
\end{equation}
where ${\cal T}$ orders later times to the left. Define
\begin{equation}\label{eq:interaction-channel}
 \Psi_{\alpha,T}(\rho)=\E_{\boldsymbol\omega}\Tr_E\bigl[
 U^I_{\alpha,\boldsymbol\omega,T}(\rho\otimes\tau_{E,\boldsymbol\omega})
 (U^I_{\alpha,\boldsymbol\omega,T})^\dagger\bigr].
\end{equation}
The identity $U_{\alpha,\boldsymbol\omega,T}=e^{-iTH_0}U^I_{\alpha,\boldsymbol\omega,T}e^{-iTH_0}$, invariance of the bath state under its bath evolution, and the partial trace give
\begin{equation}\label{eq:free-frame}
 \Phi_{\alpha,T}=\cU(T)\Psi_{\alpha,T}\cU(T).
\end{equation}

\paragraph{\green{Lindbladians} and KMS detailed balance.}
A \green{Lindbladian} $\mathcal G$ generates the channels $e^{t\mathcal G}$ for $t\ge0$ and has the form:
\begin{align}\label{eq:def_lindbladian}
 \mathcal G(\rho)=-i[H_{\mathrm{coh}},\rho]+\sum_j\mathcal D_{V_j}(\rho),\qquad
 \mathcal D_V(\rho)=V\rho V^\dagger-\frac12\{V^\dagger V,\rho\},
 \qquad H_{\mathrm{coh}}=H_{\mathrm{coh}}^\dagger.
\end{align}
We call $\mathcal J(\rho)=\sum_jV_j\rho V_j^\dagger$ the transition map and $-\tfrac12\{M,\cdot\}$, with $M=\sum_jV_j^\dagger V_j$, the decay part. Sums may also include integrals with nonnegative weights. Trace preservation is equivalent to $\mathcal G^\dagger(\idmat)=0$.

A state map $\mathcal T$ satisfies KMS detailed balance with respect to $\rhoB$ when $\mathcal T^\dagger$ is self-adjoint in \eqref{eq:rho-space}, equivalently
\[
 \ip O{\mathcal T^\dagger(P)}=\ip{\mathcal T^\dagger(O)}P
 \quad\text{for all }O,P,
 \qquad \Gamma^{-1}\mathcal T\Gamma=\mathcal T^\dagger.
\]
For a channel this implies $\mathcal T(\rhoB)=\rhoB$, and for a \green{Lindbladian} it implies $\mathcal T(\rhoB)=0$, by applying the last identity to $\idmat$. Detailed balance of the transition map alone does not give stationarity of the full generator. When $\mathcal J$ satisfies detailed balance, the \green{KMS coherent correction}
\begin{equation} \label{eq:lambshift}
     G_D=\frac i2\sum_\nu\tanh(\beta\nu/4)M_\nu
\end{equation}
makes $\mathcal J-\tfrac12\{M,\cdot\}-i[G_D,\cdot]$ \green{KMS-detailed-balanced}~\cite{CKG,DLL25}. Section~\ref{sec:generator} verifies this construction for our transition map and identifies the remaining coherent term as $-i[\HL,\cdot]$.

\paragraph{Dirichlet forms, spectral gap, and mixing.}
For the \green{KMS-detailed-balanced} \green{Lindbladian} $\LK$, \green{its observable generator $\LK^\dagger$} is \green{KMS-self-adjoint} with nonpositive real spectrum. Its Dirichlet form and the Pauli energy are
\begin{equation}\label{eq:Pauli-energy}
 \ED(O)=-\ip{O-\mean O \idmat}{\LK^\dagger(O-\mean O \idmat)},\qquad
 \EK(O)=\sum_{x\in[n]}\sum_{a=1}^{Q}\nr{[P_{x,a},O]}^2.
\end{equation}
On \green{Gibbs-centered} observables, $\ED(O)=-\ip O{\LK^\dagger(O)}$. If the kernel of $\LK^\dagger$ consists only of scalar matrices, its spectral gap is the variational quantity
\[
 \inf_{\substack{O\ne0\\\mean O=0}}\frac{\ED(O)}{\nr O^2}.
\]
A positive lower bound $\lambda$ gives $\ED(O)\ge\lambda\nr O^2$ on \green{Gibbs-centered} observables and hence
\[
 \nr{\Gamma^{-1}e^{t\LK}(\delta)}\le e^{-\lambda t}\nr{\Gamma^{-1}\delta},\qquad \Tr\delta=0.
\]

For the averaged physical channel, the number of rounds required to prepare the target Gibbs state is
\[
 m_{\mathrm{mix}}(\varepsilon;\rhoB)
 =\min\left\{m\in\N_0:\sup_{\rho_0\text{ state}}\|\Phi^m(\rho_0)-\rhoB\|_1\le\varepsilon\right\}.
\]
If the set is empty, this quantity is infinite. The channel's stationary state, denoted $\green{\rho_\infty}$, may differ from $\rhoB$. We therefore control both convergence to that stationary state and its bias from Gibbs.

\paragraph{Standard identities and inequalities.}
We use the following \emph{Schatten H\"older inequality}: for $p,q,r\in[1,\infty]$ with $1/r=1/p+1/q$, and for any operators $A,B$,
\begin{align*}
\|AB\|_r\le \|A\|_p\|B\|_q.
\end{align*}
It implies that for $1\leq p\leq \infty$,
\begin{align}\label{eq:commutator_norm}
\|AXB\|_p\le \no A\,\|X\|_p\,\no B,
\qquad
\|[A,X]\|_p,\ \|\{A,X\}\|_p
\le 2\no A\,\|X\|_p.
\end{align}

We also use the following standard results on the induced norm:
\begin{align*}
\nm{\mathcal A\mathcal B}
\le\nm{\mathcal A}\nm{\mathcal B},\quad
\nm{\mathcal A^\star}=\nm{\mathcal A},\quad
\nm{\mathcal A^\star\mathcal A}
=\nm{\mathcal A}^{2}.
\end{align*}
\red{The definition of the KMS adjoint also gives the following identities} for any operator $O$:
\begin{align*}
\operatorname{Re}\ip{O}{\mathcal A(O)}
=
\ip{O}{\frac{\mathcal A+\mathcal A^\star}{2}(O)},\qquad
\nr{\mathcal A^\star(O)}^2-\nr{\mathcal A(O)}^2=\ip{O}{[\mathcal A,\mathcal A^\star](O)}.
\end{align*}

In addition, we use the following elementary \red{commutator} and anticommutator identities:
\begin{align*}
[A,BC]=[A,B]C+B[A,C], \qquad [\ad_A,\ad_B]=\ad_{[A,B]},
\qquad
[\ad_A,\ac_B]=\ac_{[A,B]},
\end{align*}
where $\ad_A(O):=[A,O]$ and $\ac_A(O):=\{A,O\}$.

\subsection{Proof overview and challenges}\label{subsec:technical-overview}
In this section, we provide a high-level overview of the proof of \cref{thm:intro-main}, highlighting the main steps and challenges involved in the analysis. The main roadmap of the proof is inspired by~\cite{chen2026overcoming} and
consists of three main steps: (1) Channel approximation; (2) Approximate stationary state construction; (3) Contraction and \green{mixing-time analysis}. We briefly describe each step below. In this section, we often use $\|\cdot\|$ to represent an unspecified norm for simplicity.

\paragraph{Step 1: Channel approximation (Sections~\ref{sec:generator},~\ref{sec:locality},~\ref{sec:comparison}).}
Recall the \red{one-round} physical channel $\Phi_{\alpha,T}$ in~\eqref{eq:free-frame},
\[
 \Phi_{\alpha,T}=\cU(T)\Psi_{\alpha,T}\cU(T)\,,
\]
where $\Psi_{\alpha,T}$ is the interaction-picture channel in~\eqref{eq:interaction-channel} and can be calculated explicitly using the Dyson series expansion in~\cref{subsec:dyson}. The first \red{observation is that the leading nontrivial term of $\Psi_{\alpha,T}$ is a \green{Lindbladian}} $\mathcal{L}_T$ with quasi-local jump operators:
\begin{equation}\label{eqn:Psi_Lindbladian_approximation}
 \Psi_{\alpha,T}=\id{}+\alpha^2\cL_T+\mathcal{O}(\alpha^4),\qquad \cL_T:=\sum_{x\in[n]}\cL_{x,T}\,.
\end{equation}
Here, each $\cL_{x,T}$ is \red{a \green{Lindbladian}} with quasi-local jump operators and coherent terms that are quasi-localized around the site $x$. This is demonstrated in~\cref{subsec:second-order,subsec:coefficients}. Because \red{a \green{Lindbladian} has a simpler mathematical structure than the Dyson series expansion}, we approximate $\Psi_{\alpha,T}$ by $\exp(\alpha^2\cL)$, where $\cL=\lim_{T\to\infty}\cL_T$ is the \green{infinite-window effective Lindbladian}, and define the \emph{ideal channel}:
\begin{equation}\label{eq:intro-ideal}
 \widetilde\Phi_{\alpha,T}:=\cU(T)e^{\alpha^2\cL}\cU(T)=\cU(T)(\mathrm{id}+\alpha^2\mathcal{L}+\mathcal{O}(\alpha^4))\cU(T)\,.
\end{equation}
After bounding the error between $\Psi_{\alpha,T}$ and $e^{\alpha^2\cL}$ as in Section~\ref{sec:comparison}, it suffices to analyze the ideal channel $\widetilde\Phi_{\alpha,T}$ and transfer the results to the physical channel $\Phi_{\alpha,T}$ using perturbation theory.

\red{Although a similar approximation has been studied previously}~\cite{ding2026simple,chen2026overcoming,wang2025beyond}, \red{those analyses consider interactions with a single bath qubit and directly bound} the $\mathcal{O}(\alpha^4)$ error term in~\eqref{eqn:Psi_Lindbladian_approximation} and~\eqref{eq:intro-ideal}. Because our model involves $\Theta(n)$ bath qubits, which induces $\Theta(n^2)$ quasi-local superoperators in the fourth-order expansion in~\eqref{eqn:Psi_Lindbladian_approximation} and~\eqref{eq:intro-ideal}, \red{directly bounding these terms gives an error of order $\mathcal{O}(n^2\alpha^4)$, and hence an unfavorable scaling of $\alpha$ and of the mixing time}. This gives us \textbf{the first main challenge} in the analysis, which is to prove the following theorem:

\begin{theorem}[Informal statement of Theorem~\ref{thm:comparison}]\label{chall:channel-approximation}
\red{For sufficiently large $T$ and sufficiently small $\alpha$,}
\[
 \left\|\Gamma^{-1}(\Phi_{\alpha,T}-\widetilde\Phi_{\alpha,T})\Gamma\right\|
 =\mathcal O(n\alpha^4),
\]
where $\|\cdot\|$ is a certain norm that will be specified later.
\end{theorem}

\begin{proof}[Informal proof of \cref{chall:channel-approximation}]
We show that the higher-order terms in the expansions of $\Psi_{\alpha,T}$ and $e^{\alpha^2\cL}$ can be matched term by term. Given a single Pauli operator $A_x$ supported at site $x$, we define $A_x(t)=e^{itH_n}A_xe^{-itH_n}$ as in~\eqref{eq:evolutions}. Informally, fix $k\geq 2$. The order-$\alpha^{2k}$ term in the expansion of $e^{\alpha^2\cL}$ is a sum of terms of the form:
\begin{equation}\label{eqn:expansion-form_Lindbladian}
\begin{aligned}
\int_{\mathbb{R}^{k_1}}\int_{\mathbb{R}^{k_2}}&K(\textbf{t},\textbf{s})\dd t_1\cdots \dd t_{k_1} \dd s_1\cdots \dd s_{k_2}A_{x_1}(t_1)\dots A_{x_{k_1}}(t_{k_1})\rho A_{z_1}(s_1)\cdots A_{z_{k_2}}(s_{k_2})
\end{aligned}
\end{equation}
where $K(\textbf{t},\textbf{s})$ is some function, depending on $x_1,\ldots,x_{k_1},z_1,\ldots,z_{k_2}$, of $\textbf{t}=(t_1,\ldots,t_{k_1})$ and $\textbf{s}=(s_1,\ldots,s_{k_2})$, and $k_1+k_2=2k$.

We first consider the terms whose site indices satisfy
\begin{equation}\label{eqn:subindex-condition}
\#\{i\in[k_1]:x_i=x\}
+\#\{j\in[k_2]:z_j=x\}
\in\{0,2\},
\qquad \forall x\in[n].
\end{equation}
The remaining cases can already be bounded by $\mathcal{O}(n^{k-1}\alpha^{2k})$, which is the desired error bound.

For $\Psi_{\alpha,T}$, when we restrict to the same subindex condition in~\eqref{eqn:subindex-condition}, the order $\alpha^{2k}$ term in the expansion of $\Psi_{\alpha,T}$ is \red{a sum of terms of the form}:
\begin{equation}\label{eqn:expansion-form_Dyson}
\begin{aligned}
&\int_{-T<t_{\pi(1)}<\dots<t_{\pi(k_1)}<T}\int_{-T<s_{\pi'(1)}<\dots<s_{\pi'(k_2)}<T}K(\textbf{t},\textbf{s})\dd t_{\pi(1)}\cdots \dd t_{\pi(k_1)} \dd s_{\pi'(1)}\cdots \dd s_{\pi'(k_2)}\\
&A_{x_{\pi(k_1)}}(t_{\pi(k_1)})
\cdots
A_{x_{\pi(1)}}(t_{\pi(1)}) \rho A_{z_{\pi'(1)}}(s_{\pi'(1)})\cdots A_{z_{\pi'(k_2)}}(s_{\pi'(k_2)})\,,
\end{aligned}
\end{equation}
where $\pi,\pi'$ are permutations of $[k_1]$ and $[k_2]$, respectively, and $K(\textbf{t},\textbf{s})$ is the same function as before. A weighted version of the above integrals can be found in~\cref{lem:two-forms}.

Comparing~\eqref{eqn:expansion-form_Lindbladian} and~\eqref{eqn:expansion-form_Dyson}, we can see that the only difference is the integration domain and the order of the operators $A_{x_i}(t_i)$ and $A_{z_j}(s_j)$. If $A_{x}(t)$ and $A_y(s)$ \red{commute, summing}~\eqref{eqn:expansion-form_Dyson} over all permutations $\pi$ and $\pi'$ will give exactly the same integral as in~\eqref{eqn:expansion-form_Lindbladian} up to a time truncation error. The key observation is that, under \eqref{eqn:subindex-condition},
\[
\text{\eqref{eqn:expansion-form_Lindbladian}}-\sum_{\pi,\pi'}\text{\eqref{eqn:expansion-form_Dyson}}=\text{Commutator error ($[A_x(t),A_y(s)]$)}+\text{Time truncation error}\,.
\]
The commutator error can be bounded in~\cref{lem:reorder-cost}
and the time truncation error can be bounded by the Gaussian decay of the pulse function $f_\sigma(t)$. \red{Summing the commutator errors over $k\ge2$ gives $\mathcal O(n\alpha^4)$. The Gaussian truncation error can also be bounded by $\mathcal O(n\alpha^4)$ when $T$ is sufficiently large.} The rigorous version can be found in~\cref{thm:comparison}.
\end{proof}

Earlier system-bath analyses~\cite{ding2026simple,chen2026overcoming,wang2025beyond} use the induced trace norm $\|\cdot\|_{1\to1}$ to control channel-approximation errors. Such estimates yield perturbation bounds for fixed points and mixing times; see, for example,~\cite[Theorem~S3]{ding2026simple}. However, although $\|\Phi_{\alpha,T} - \widetilde{\Phi}_{\alpha,T}\|_{\red{1\to1}}$ is compatible with $\|\rho_n-\rho_\beta\|_1$, the mixing time of \red{$\mathbb E_T\widetilde{\Phi}_{\alpha,T}$} is studied in the \green{weighted state-difference norm (the square root of the $\chi^2$-divergence)}, namely $\left\|\Gamma^{-1}(\rho_n-\rhoB)\right\|_{\rhoB}=\left\|\rhoB^{-1/4}(\rho_n-\rhoB)\rhoB^{-1/4}\right\|_{2}$, see \cref{chall:contraction} in Step 3 for details. The transition between two different norms introduces additional $n$ factors in the error bound. \red{To avoid losing a factor of $n$}, in~\cref{thm:comparison}, we bound the channel distance in the $\chi^2$-divergence norm, i.e.
\begin{equation}\label{eq:channel-norm-proof_idea}
\left\|\Gamma^{-1}\left(\Phi_{\alpha,T}-\widetilde{\Phi}_{\alpha,T}\right)\Gamma\right\|_{\red{\rho_\beta\to\rho_\beta}}=\left\|\Gamma^{-1/2}\left(\Phi_{\alpha,T}-\widetilde{\Phi}_{\alpha,T}\right)\Gamma^{1/2}\right\|_{2\rightarrow 2}\,.
\end{equation}
This approximation bound can then be combined with the contraction bound in~\cref{chall:contraction} to give the final mixing time bound in Step 3.

According to~\eqref{eqn:expansion-form_Lindbladian},~\eqref{eqn:expansion-form_Dyson}, this requires us to control
\[
 \red{\nm{\Gamma^{-1}\bigl(B\,\Gamma(\cdot)\,C\bigr)}}
 \le\no{B(-i\beta/4)}\,\no{C(i\beta/4)},
 \qquad B(z)=e^{izH_n}Be^{-izH_n}\,,
\]
where both $B,C$ are quasi-local operators \red{formed from products and commutators} of $A_{x_i}(t_i)$ and $A_{z_j}(s_j)$. This involves a technical challenge of controlling the imaginary-time translates of quasi-local operators, which one-dimensional complex-time locality~\cite{perez2023locality} bounds independently of $n$. Specifically, the \red{bound on these superoperators} is shown in~\cref{lem:multiplication-maps}, the preservation of quasi-locality under imaginary-time evolution is shown in~\cref{lem:Araki}, and the bound of the commutator of quasi-local operators is shown in~\cref{lem:reorder-cost}. A complete bound \red{on each term appearing} in the Lindbladian generator $\mathcal{L}$ is shown in~\cref{prop:quasilocal}.

\paragraph{Step 2: Approximate stationary state construction (Sections~\ref{sec:weighted},~\ref{sec:reference}).}
Recall the averaged physical channel $\Phi=\E_T[\Phi_{\alpha,T}]$ in~\eqref{eq:physical-channel}, and let $\widetilde{\Phi}:=\E_T\left[\widetilde{\Phi}_{\alpha,T}\right]$ be the averaged ideal channel. By~\cref{chall:channel-approximation}, the two channels differ by $\mathcal O(n\alpha^4)$. The first observation in this step is that the ideal channel does not fix the Gibbs state. By~\Cref{prop:KMS}, the generator splits into a \green{KMS-detailed-balanced} part and the Lamb shift,
\begin{equation}\label{eqn:overview-Lamb-split}
 \cL=\LK-i[\HL,\cdot\,],\qquad \LK(\rhoB)=0,\qquad \cL(\rhoB)=-i[\HL,\rhoB],
\end{equation}
where $\HL$ generally does not commute with $\rhoB$. Hence one round moves the Gibbs state already at first order, $\widetilde\Phi(\rhoB)-\rhoB=\alpha^2\E_T[\cU(T)\cL(\rhoB)]+\mathcal O(\alpha^4)$. Since the contraction of Step~3 is only $1-\Omega(\alpha^2)$ per round, leading to $\mathcal{O}(\alpha^{-2})$ rounds to reach stationarity, the first-order drift accumulates to a bias of $\mathcal{O}(1)$ according to the traditional perturbation theory. To overcome this difficulty,~\cite[Appendix~C]{chen2026overcoming} constructs an \green{approximate stationary state} $\rho_*$ such that
\begin{equation}\label{eqn:overview-corrected-state_1}
\left\|\widetilde{\Phi}(\rho_*)-\rho_*\right\|=\mathcal{O}(\alpha^4),\quad \left\|\rho_*-\rhoB\right\|=\mathcal{O}(\alpha^2),
\end{equation}
where we ignore the factor of $n$ here. Specifically, we find $\rho_*$ by solving the following equation:
\begin{equation}\label{eqn:overview-reference-state}
 \rho_*=\rhoB+\alpha^2\Gamma(X),\qquad
 (\E_T[\cU(2T)]-\id)\Gamma(X)+\E_T[\cU(T)\cL(\rhoB)]=0.
\end{equation}
See~\cref{subsec:reference-construction} for the detailed construction. \red{Following that construction, first replace each $B_x^{\mathrm{Lamb}}$ by $B_x^{\mathrm{Lamb}}-\Tr(\rhoB B_x^{\mathrm{Lamb}})\idmat$, retaining the same notation; these scalar shifts leave $\cL$, $K$, and $\cD$ unchanged. Define $Y_0=-i\int_{-\beta/2}^{\beta/2}\HL(-iu)\,\dd u$. The solution $X$ takes the form:}
\begin{equation}\label{eqn:overview-correction}
 X=-\int_\R\cU(t)(Y_0)\,\nu_{T_0}(t)\,\dd t,\qquad
 \|\nu_{T_0}\|_1\le\frac C{T_0},\qquad
 \int_\R|t|\,|\nu_{T_0}(t)|\,\dd t\le C,
\end{equation}
for a scalar kernel $\nu_{T_0}$ determined by the law~\eqref{eqn:T_0} of $T$. Because $H_{\rm Lamb}$ contains only $\Theta(n)$ quasi-local terms, it is straightforward to show $\|\Gamma(X)\|=\mathcal O(n)$ using the results of~\cref{sec:locality}. This implies $\|\rho_*-\rhoB\|=\mathcal O(n\alpha^2)$, which is the second part of~\eqref{eqn:overview-corrected-state_1}. \red{By the Taylor expansion~\eqref{eq:two-remainders}, controlling the defect reduces to bounding $\cL^2(\rhoB)$ and $\cL(\Gamma(X_T))$ in the \green{weighted state-difference norm}, where $X_T=\cU(T)(X)$.}

Because $\cL=\sum_x\cL_x$ contains $\Theta(n)$ quasi-local jump operators and coherent \red{terms}, we have \red{$\nm{\Gamma^{-1}\cL\Gamma}=\mathcal O(n)$} and a naive bound gives $\left\|\widetilde{\Phi}(\rho_*)-\rho_*\right\|=\mathcal{O}(n^2\alpha^4)$ in~\eqref{eqn:overview-corrected-state_1}. This again requires the unsatisfying $\alpha^2=\mathcal{O}(n^{-2})$ to control the final approximation error. This gives us \textbf{the second main challenge} in the analysis, which is to prove the following theorem:

\begin{theorem}[Informal statement of Theorem~\ref{thm:reference}]\label{chall:reference}
\red{Under the parameter conditions~\eqref{eq:reference-regime}, the matrix} $\rho_*$ in~\eqref{eqn:overview-reference-state} is a state, and
\[
 \nr{\Gamma^{-1}\bigl(\Phi(\rho_*)-\rho_*\bigr)}=\mathcal O(n\alpha^4).
\]
\end{theorem}

\begin{proof}[Informal proof of \cref{chall:reference}]
By~\cref{chall:channel-approximation}, replacing $\Phi$ by $\widetilde\Phi$ changes the \green{stationarity defect} by $\mathcal O(n\alpha^4)$, so it suffices to bound $\widetilde\Phi(\rho_*)-\rho_*$. Define $\hL=\Gamma^{-1}\cL\Gamma$. \red{Using $\Gamma^{-1}(\rhoB)=\idmat$, the KMS isometry of $\cU(T)$, and $\nm{e^{u\hL}}\le e^{Cnu}$ in~\eqref{eq:two-remainders} gives, for sufficiently small $n\alpha^2$,}
{
\[
\nr{\Gamma^{-1}\bigl(\widetilde\Phi(\rho_*)-\rho_*\bigr)}
\le C\alpha^4\left(\nr{\hL^2(\idmat)}+\E_T\nr{\hL(X_T)}\right).
\]
}
\red{It suffices to bound $\nr{\hL^2(\idmat)}$ and $\E_T\nr{\hL(X_T)}$ by $\mathcal O(n)$.} The key observation is that $\hL$ splits into a part that annihilates $\idmat$ and a symmetric multiplication by the drift (see~\eqref{eq:affine}):
\begin{equation}\label{eqn:overview-affine}
 \hL(O)=\cD(O)+\tfrac12\{K,O\},\qquad
 \cD(\idmat)=0,\qquad
 K=\hL(\idmat)=\Gamma^{-1}\cL(\rhoB).
\end{equation}
Accordingly,
\begin{equation}\label{eqn:overview-remainders}
 \hL^2(\idmat)=\cD(K)+K^2,\qquad
 \hL(X_T)=\cD(X_T)+\tfrac12(KX_T+X_TK).
\end{equation}

The terms with $\cD$ are controlled by locality. Each local term of $\cD$ annihilates $\idmat$, so it acts through commutators with operators near its site; for a dissipator, for example, $V^\dagger OV-\frac12\{V^\dagger V,O\}=\frac12\bigl(V^\dagger[O,V]+[V^\dagger,O]V\bigr)=0$ when $\mathrm{Supp}(O)\cap\mathrm{Supp}(V)=\emptyset$. \red{Summing the quasi-local commutator bounds over the separation of sites gives $O(n)$ at time zero and $O(n(1+|t|))$ after real-time evolution (\Cref{lem:D-action}):}
\begin{equation}\label{eqn:overview-D-action}
 \nr{\cD(K)}\le Cn,\qquad \nr{\cD(\cU(t)(Y_0))}\le Cn(1+|t|).
\end{equation}

\red{The same argument} does not apply to $K^2$ and $KX_T$. Two distant local terms do not multiply to zero, so locality gives nothing, and for $K=\sum_xK_x$ the triangle inequality only gives $\nr{K^2}\le\sum_{x,y}\nr{K_xK_y}=\mathcal O(n^2)$. What saves us is centering. Recall $X=-\int_\R\cU(t)(Y_0)\,\nu_{T_0}(t)\,\dd t$. Both $K=\sum_xK_x$ and $Y_0=\sum_xY_x$ are sums of Gibbs-centered local terms, $\mean{K_x}=\mean{Y_x}=0$, and for a sum of $n$ weakly dependent random variables with mean zero, the second moment is $\mathcal O(n)$ and the fourth moment is $\mathcal O(n^2)$. The key estimate of this work,~\cref{key:centered-sum} below, makes this heuristic rigorous for noncommuting Gibbs states and gives (\Cref{cor:moments})
\begin{equation}\label{eqn:overview-moments}
 \nr K+\nr{Y_0}\le C\sqrt n,\qquad
 \nr{K^2}\le Cn,\qquad
 \sup_{t\in\R}\Bigl(\nr{K\,\cU(t)(Y_0)}+\nr{\cU(t)(Y_0)\,K}\Bigr)\le Cn.
\end{equation}
Together with~\eqref{eqn:overview-correction} and~\eqref{eqn:overview-D-action}, \red{and using $\E T=5T_0$, these bounds give $\nr{\hL^2(\idmat)}+\E_T\nr{\hL(X_T)}=\mathcal O(n)$}, so the stationary error is $\mathcal O(n\alpha^4)$. They also give $\nr{\Gamma^{-1}(\rho_*-\rhoB)}=\alpha^2\nr X\le\alpha^2\|\nu_{T_0}\|_1\nr{Y_0}\le C\alpha^2\sqrt n/T_0$, while $\no X\le Cn/T_0$ ensures $\rho_*\succeq\rhoB/2$ when $n\alpha^2$ is small. The rigorous version can be found in~\cref{thm:reference}.
\end{proof}

The product terms are where the noncommutativity of the Gibbs state enters most directly, and controlling them is the main technical contribution of this work. The moment bounds~\eqref{eqn:overview-moments} are consequences of the following estimate on \green{Gibbs-centered sums of coefficient-quasi-local maps}, which is used again in Step~3.

\begin{theorem}[Informal statement of Theorem~\ref{thm:field}]\label{key:centered-sum}
For each site $x\in[n]$, let \red{$\mathcal A_x(O)=\Gamma^{-1/2}\bigl(B_x\Gamma^{1/2}(O)C_x\bigr)$, and assume that $(\mathcal A_x)_x$ is uniformly coefficient-quasi-local as in~\eqref{eq:coefficient-locality}, with constants independent of $n,x$}. If $\ip{\idmat}{\mathcal A_x(\idmat)}=0$, then for any operator $O$,
\begin{equation}\label{eqn:overview-centered-sum}
 \nr{\sum_x\mathcal A_x(O)}^2\le Cn\bigl(\EK(O)+|\mean O|^2\bigr),
\end{equation}
where $\EK(O)=\sum_{x,a}\nr{[P_{x,a},O]}^2$ is the Pauli energy~\eqref{eq:Pauli-energy}.
\end{theorem}

The bound has a single factor of $n$, whereas the triangle inequality gives $n^2$, and the input $O$ enters only through its local fluctuations and its \green{Gibbs mean}. To see how it gives~\eqref{eqn:overview-moments}, let $\mathcal A_x$ be the left multiplication by a \green{Gibbs-centered} local term $F_x$, and let $F=\sum_xF_x$. For $O=\idmat$, the Pauli energy vanishes, and~\eqref{eqn:overview-centered-sum} reads $\nr F^2\le Cn$, the second moment. For $O=G$, another sum of \green{Gibbs-centered} local terms, each commutator $[P_{y,a},G]$ involves only the terms of $G$ near $y$, so $\EK(G)=\mathcal O(n)$, and~\eqref{eqn:overview-centered-sum} gives $\nr{FG}^2\le Cn^2$, the fourth moment.

\begin{proof}[Informal proof of \cref{key:centered-sum}]
For simplicity, we consider, for a fixed $\ell$,
\[
 \mathcal A_x^\star=\mathcal A_x\in\mathfrak A(N_\ell(x)),
 \qquad \nm{\mathcal A_x}\le1,
 \qquad \ip{\idmat}{\mathcal A_x(\idmat)}=0.
\]
Here $\star$ is the KMS adjoint, and $\mathfrak A(S)$ denotes maps with coefficient support in $S$, as defined in~\cref{subsec:projections}. Let $p_I$ be the KMS-orthogonal projection onto operators that act as the identity on $I$. Write $\mathcal A=\sum_x\mathcal A_x$, and, for $R\ge0$, set
$p_x=p_{N_{\ell+R}(x)}$, $q_x=\id-p_x$, and $\mathcal R_x=p_x\mathcal A_xp_x$.
By~\eqref{eq:conditional} and~\eqref{eq:screen}, we have
\begin{equation} \label{axueq11}
\begin{aligned}
 \sum_x\nr{q_xO}^2
 &\le C\sum_x\sum_{y\in N_{\ell+R}(x),a}\nr{[P_{y,a},O]}^2
 \le C_\ell(R+1)\EK(O),\\
 \nm{\mathcal R_x}&\le C_\ell e^{-cR}.
 \end{aligned}
\end{equation}
Here and below, $C_\ell$ is independent of $n$ and $R$. Split each map $\mathcal A_x$ as
\[
 \mathcal A_x=\mathcal A_xq_x+q_x\mathcal A_xp_x+\mathcal R_x.
\]
We consider the second term in the expansion $\ip{\mathcal A(O)}{\mathcal A(O)} = \sum_x \ip{\mathcal A(O)}{(\mathcal A_xq_x+q_x\mathcal A_xp_x+\mathcal R_x)(O)}$:
\[
 \begin{aligned}
 \ip{\mathcal A(O)}{q_x\mathcal A_xp_x(O)}
 &=\ip{\mathcal A_x(q_xO)}{p_x\mathcal A(O)}
   +\ip{O}{[\mathcal A,q_x\mathcal A_xp_x](O)}.
 \end{aligned}
\]
The local approximants ~\eqref{eq:projection-locality}, together with commutation of maps with disjoint coefficient supports, give the estimate $\nm{[\mathcal A_y,q_x\mathcal A_xp_x]}\le C_\ell\min\{1,e^{-c(|x-y|-2\ell-R)}\}$ and hence
\begin{equation} \label{axueq22}
     \begin{aligned}
 \sum_x\nm{[\mathcal A,q_x\mathcal A_xp_x]}
 &\le\sum_{x,y}\nm{[\mathcal A_y,q_x\mathcal A_xp_x]}
 \le C_\ell n(R+1).
 \end{aligned}
\end{equation}
It follows from \eqref{axueq11}-\eqref{axueq22} that
\[
 \begin{aligned}
 \nr{\mathcal A(O)}^2
 &=\sum_x\ip{\mathcal A(O)}{\mathcal A_x(O)}\\
 &\le2\nr{\mathcal A(O)}\sum_x\nr{q_xO}
       +C_\ell n(R+1)\nr O^2
       +C_\ell ne^{-cR}\nr{\mathcal A(O)}\nr O.
 \end{aligned}
\]
Using $\sum_x\nr{q_xO}\le\sqrt{C_\ell n(R+1)\EK(O)}$ and Young's inequality, we arrive at
\[
 \nr{\mathcal A(O)}^2
 \le C_\ell n(R+1)\bigl(\EK(O)+\nr O^2\bigr)
       +C_\ell n^2e^{-2cR}\nr O^2.
\]
Choosing $R=\lceil\log(n+1)/(2c)\rceil$ then yields
\[
 \nr{\mathcal A(O)}^2
 \le C_\ell n\log(n+1)\bigl(\EK(O)+\nr O^2\bigr).
\]
The global conditional variance bound further gives $\nr O^2\le C\EK(O)+|\mean O|^2$, which yields~\eqref {eqn:overview-centered-sum} up to a logarithmic factor.

The logarithmic factor comes from choosing a single region radius $R=\mathcal O(\log n)$: each local fluctuation is then counted $\mathcal O(R)$ times. To remove this loss, we telescope over all radii and retain the exponential decay of the compressed maps. For self-adjoint maps, this gives an exact decomposition $\mathcal A_x=\mathcal T_x+\mathcal T_x^\star$ with $\mathcal T_x(\idmat)=0$ and exponentially decaying layers; see the proof of~\Cref{thm:field}.
The conditional variance bound then counts each local commutator with a total weight bounded by $\sum_{r\ge0}(r+1)e^{-cr}=\mathcal O(1)$. For $\mathcal T=\sum_x\mathcal T_x$, locality and the decay of the layers also give $\nm{[\mathcal T,\mathcal T^\star]}\le Cn$, which controls the adjoint contribution. This will remove the logarithmic factor; the full argument, including the extension to general coefficient-quasi-local maps, is given in Section~\ref{subsec:field}.
\end{proof}

\paragraph{Step 3: Contraction and \green{mixing-time analysis} (Sections~\ref{sec:mixing},~\ref{sec:complexity}).}

The last step is to derive an upper bound on the mixing time of the quantum channel $\mathbb{E}_T\left(\Phi_{\alpha,T}\right)$. Combining this \green{mixing-time analysis} with the approximate preservation of the approximate stationary state in~\cref{chall:reference} gives the final bound on the distance between the stationary state $\rho_\infty$ of the physical channel $\Phi$ and the Gibbs state $\rhoB$ and the final complexity of the algorithm.

Write $\Phi=\E_T\Phi_{\alpha,T}$ and $\widetilde\Phi=\E_T\widetilde\Phi_{\alpha,T}$. We first prove contraction of the averaged ideal channel $\widetilde\Phi$ in~\eqref{eq:intro-ideal}, then transfer it to the averaged physical channel $\Phi$ using~\cref{chall:channel-approximation}. Specifically, we prove the following informal theorem:
\begin{theorem}[Informal statement of Theorem~\ref{thm:ideal-contraction}]\label{chall:contraction}
If $n\alpha^2$ is sufficiently small and $T_0\ge1$, then for any states $\rho_1,\rho_2$, with $X=\rho_1-\rho_2$,
\begin{equation}\label{eqn:rho_1_rho_2_contraction}
 \nr{\Gamma^{-1}\widetilde\Phi(X)}^2
 \le \nr{\Gamma^{-1}X}^2-\frac{\alpha^2}{2}E_\mu(X)
 \le \Bigl(1-\tfrac12\lambda\alpha^2\Bigr)\nr{\Gamma^{-1}X}^2,
\end{equation}
where $E_\mu(X)=\E_T\ED(\Gamma^{-1}(\cU(T)(X)))$ and $\lambda>0$ is the gap lower bound in~\eqref{eq:gap}.
\end{theorem}
\begin{proof}[Informal proof of \cref{chall:contraction}]
Recall $\widetilde\Phi_{\alpha,T}:=\cU(T)e^{\alpha^2\cL}\cU(T)$, $\cL=\mathcal{Q}_{\rm Lamb}+\mathcal{L}_{\rm KMS}$. As proved in~\cite{BergamaschiChen}, $\mathcal{L}_{\rm KMS}$ already has a $\Omega(1)$ spectral gap (stated in~\cref{assump:gap} for completeness), implying
\begin{equation}\label{eqn:overview-KMS-contraction}
 \nr{\Gamma^{-1}\cU(T)e^{s\mathcal L_{\rm KMS}}\cU(T)(X)}
 \le e^{-\lambda s}\nr{\Gamma^{-1}\cU(T)(X)}
 =e^{-\lambda s}\nr{\Gamma^{-1}X},
\end{equation}
for every $T$, since $\cU(T)$ is a KMS isometry and preserves trace. In~\cite[Lemma D.3]{chen2026overcoming}, the contraction argument controls the Gibbs-commutation defect of $H_{\rm Lamb}$, which bounds the self-adjoint part of the \green{weighted Lamb shift generator}. Summing such bounds over $\Theta(n)$ bath qubits gives an extensive upper bound; at fixed $\sigma$, this estimate does not guarantee contraction uniformly in $n$. This leads to \textbf{the third main technical challenge}:

\emph{How to control the effect of the Lamb shift term in the contraction of
\begin{align*}
    \left\|\Gamma^{-1}\mathbb{E}_T\left(\mathcal{U}(T)e^{s\mathcal{L}}\mathcal{U}(T)\right)(\rho_1-\rho_2)\right\|_{\rhoB}
\end{align*}
when $\sigma=\Theta(1)$?}

In our work, we overcome this challenge by using the random forward evolution $\mathcal{U}(T)$. The dephasing produced by averaging over $T$ controls the \green{Lamb shift contribution} together with the \green{averaged Dirichlet form}, without requiring a small norm of the Lamb shift. To the best of our knowledge, this is the first time that the random forward evolution is used to suppress the effect of the coherent term in the contraction analysis of the quantum channel, which might be of independent interest. We now briefly explain the mechanism below:

Define $\ad_{H_n}(O):=[H_n,O]$. Write $H_n=\sum_e eP_e$, where $P_e$ is the spectral projection onto energy $e$.
For each Bohr frequency $\omega\in\{e-e':e,e'\in\operatorname{spec}(H_n)\}$, define
\[
 X_\omega:=\sum_{e-e'=\omega}P_eXP_{e'},\qquad
 X=\sum_\omega X_\omega,\qquad \ad_{H_n}(X_\omega)=\omega X_\omega.
\]
We have $\E_T[\cU(2T)(X_\omega)]=\hat\mu_{T_0}(-2\omega)X_\omega$, where $\hat\mu_{T_0}(\omega)=e^{i\omega T_0}(1-i\omega T_0)^{-4}$ {according to~\eqref{eqn:T_0}}. Thus averaging the \green{forward evolution} suppresses the \green{Bohr components} with $|\omega|>1/T_0$. This mechanism is rigorously incorporated into the analysis by defining
\begin{equation}\label{eq:intro-D}
 \begin{aligned}
 D_\mu(X)&=\|\Gamma^{-1}(X)\|_{\rhoB}^2
       -\|\Gamma^{-1}(\E_T\cU(2T)X)\|_{\rhoB}^2=\sum_\omega\bigl[1-(1+4T_0^2\omega^2)^{-4}\bigr]
                      \|\Gamma^{-1}(X_\omega)\|_{\rhoB}^2
 \end{aligned}
\end{equation}
{ For sufficiently small $n\alpha^2$}, the one-step expansion gives
\begin{equation}\label{eqn:single_step_contraction}
 \begin{aligned}
 \nr{\Gamma^{-1}({\widetilde{\Phi}}X)}^2
 \le{}&\nr{\Gamma^{-1}X}^2-{\frac{1}{2}D_\mu(X)-\alpha^2E_\mu(X)}+2\alpha^2\E_T\Rea\pair{\Gamma^{-1}(Y_T)}{\Gamma^{-1}(\Q Y_T)}_{\rhoB}
 \end{aligned}
\end{equation}
Here, $Y_T=\cU(T)X$. The rigorous bound is given in Lemma~\ref{lem:step}. The extra terms are absorded into the leading decaying using Young's inequality and $n\alpha^2<1$. It remains to control the Lamb shift term by $D_\mu(X)$ and $E_\mu(X)$, as in~\cref{lem:lamb}.

Thanks to the random forward evolution, it suffices to consider the low frequency of the Lamb shift term because high frequency can already be bounded by $D_\mu(X)$, meaning it will be suppressed by the dephasing mechanism. To control the low frequency term, the key observation is a special structure of the self-adjoint part $\mathcal S_{\rm Lamb}$ of $\Gamma^{-1/2}\Q\Gamma^{1/2}$. Set $F_T=\Gamma^{-1/2}(Y_T)=\rho^{-1/4}_{\beta}Y_T\rho^{-1/4}_{\beta}$. We have
\[
\Rea\left(\pair{\Gamma^{-1}(Y_T)}{\Gamma^{-1}(\Q Y_T)}_{\rhoB}\right)=\left\langle F_T,\mathcal{S}_{\rm Lamb}[F_T]\right\rangle_2\,,
\]
where $\mathcal{S}_{\rm Lamb}=\frac{1}{2}\left(\Gamma^{-1/2}\Q\Gamma^{1/2}+\Gamma^{1/2}\Q^\dagger\Gamma^{-1/2}\right)$. The crucial identity is
\begin{equation}\label{eq:intro-modular-commutator}
 \mathcal S_{\rm Lamb}=[\ad_{H_n},-i\ac_{\Gm}],\qquad
 \Gm=\frac12\int_{-\beta/4}^{\beta/4}e^{uH_n}\HL e^{-uH_n}\,\dd u,
\end{equation}
where $\ad_{H_n}=[H_n,\cdot]$ and $\ac_G(V)=\{G,V\}$. Let $\Pi$ project onto $|\omega|\le1/T_0$. Direct calculation gives
\[
 \|(\id-\Pi)F_T\|_2^2\le 2D_\mu(X),\qquad
 \|\ad_{H_n}\Pi F_T\|_2^2\le 2T_0^{-2}D_\mu(X)\,.
\]
The second bound controls the low-frequency factor $\ad_{H_n}\Pi F_T$ by $\sqrt{2}T_0^{-1}\sqrt{D_\mu(X)}$. Combining these bounds with the local field estimates gives Lemma~\ref{lem:lamb}:
\begin{equation}\label{eq:intro-lamb}
 \left|2\E_T\Re\langle\Gamma^{-1}(Y_T),
                         \Gamma^{-1}(\Q Y_T)\rangle_{\rhoB}\right|
 \le C\sqrt{nE_\mu(X)D_\mu(X)}+CnD_\mu(X).
\end{equation}
This holds for every $T_0\ge1$. Young's inequality absorbs the Lamb shift contribution and the Taylor remainders into the negative dephasing and dissipative terms when $ns$ is sufficiently small.
\end{proof}

Theorem~\ref{chall:contraction} therefore yields contraction by $1-\Theta(\alpha^2)$. The frequency split is used only in the proof and requires no lower bound on nonzero energy differences. Under the parameter conditions~\eqref{eq:reference-regime}, combining ideal contraction with the one-round approximation \eqref{eq:channel-norm-proof_idea} gives
\begin{equation}\label{eq:intro-contraction}
 \|\Gamma^{-1}{\Phi_{\alpha}}(\rho_1-\rho_2)\|_{\rho_\beta}\le(1-\kappa\alpha^2)
             \|\Gamma^{-1}(\rho_1-\rho_2)\|_{\rho_\beta},\quad \forall\
             \text{density operators}\ \rho_1,\rho_2\,,
\end{equation}
where $\kappa>0$ is independent of $n$ and $\alpha$ at fixed parameters (Theorem~\ref{thm:exact-mixing}). In particular, the physical channel has a unique stationary state $\rho_\infty$.

For arbitrary initial states,~\eqref{eq:intro-contraction} gives a gap-based bound of $O((n+\log(1/\varepsilon))/\alpha^2)$ rounds once the \green{stationary-state bias} is at most $\varepsilon/2$. The final improvement comes from~\cite{chending2026}, which establishes rapid mixing of $\mathcal L_{\rm KMS}$ under the same Hamiltonian assumptions. We combine its \green{enhanced-dissipation estimate} with~\cref{chall:contraction}. For every traceless $X$, the channel comparison and sufficiently small $n\alpha^2$ give
\[
 \begin{aligned}
 \nr{\Gamma^{-1}\Phi(X)}^2\le\nr{\Gamma^{-1}X}^2-\tfrac12\alpha^2E_\mu(X)
       +Cn\alpha^4\nr{\Gamma^{-1}X}^2\le\nr{\Gamma^{-1}X}^2-\tfrac14\alpha^2E_\mu(X),
 \end{aligned}
\]
where the last step uses $E_\mu(X)\ge\lambda\nr{\Gamma^{-1}X}^2$; see~\eqref{eq:simple-physical-dissipation}. Here $E_\mu$ is the \green{averaged Dirichlet form}. The estimate of~\cite{chending2026}, averaged over $T$, gives, for every state $\rho$,
\begin{equation} \label{eq:enhancedestimate}
     E_\mu(\rho-\rhoB)\geq \frac{C}{\log^3(en)} \bigl(1+\nr{\Gamma^{-1}(\rho-\rhoB)}^2\bigr) \log\bigl(1+\nr{\Gamma^{-1}(\rho-\rhoB)}^2\bigr),
\end{equation}
The logarithmic factor on the right-hand side improves the round bound to
\[
 m=\mathcal O_{\beta,q}\!\left(\frac{\log^4(en)+\log(1/\varepsilon)}{\alpha^2}\right).
\]
Choose $\alpha^2=c\varepsilon/n$ and $T_0=\Theta_{\beta,q}(\sqrt{\log(en/\varepsilon)})$ as in~\eqref{eq:parameters}. The latter choice controls the Gaussian cutoff error; ideal-channel contraction only requires $T_0\ge1$. Since $\E T=5T_0$ and each round lasts $2T$, the expected \green{total Hamiltonian evolution time} is $10mT_0=\widetilde{\mathcal O}_{\beta,q}(n/\varepsilon)$, achieving trace-norm error at most $\varepsilon$. The detailed statement and proof are given in Corollary~\ref{cor:complexity}.

\section{The second-order generator and the   Lamb shift}\label{sec:generator}

In this section, we use the Dyson expansion and bath parity to identify the second-order generator $\cL_T$ of the interaction-picture channel $\Psi_{\alpha,T}$ in~\eqref{eq:interaction-channel}. We then define the \green{infinite-window effective Lindbladian} $\cL$, separate its KMS-detailed-balanced part from the Lamb shift, and write it at the level of relative densities (\cref{subsec:second-order,subsec:kms-lamb,subsec:weighted-generator}). The Dyson expansion and generator are used in the channel comparison in~\cref {sec:comparison}, and the weighted form is used in the error estimates in~\cref {sec:weighted,sec:reference}.

\subsection{Dyson expansion and bath parity}\label{subsec:dyson}
We first recall the interaction-picture channel in~\eqref{eq:interaction-channel}
\begin{align*}
    \Psi_{\alpha,T}(\rho)=\E_{\boldsymbol\omega}\Tr_E\bigl[U^I_{\alpha,\boldsymbol\omega,T}(\rho\otimes\tau_{E,\boldsymbol\omega})(U^I_{\alpha,\boldsymbol\omega,T})^\dagger\bigr]
\end{align*}
with the coupling and propagator defined as in~\eqref{eq:W}:
\begin{align}
    W_{\boldsymbol\omega}(t)=&~ f_\sigma(t)\sum_{x,a}\bigl(e^{i\omega_{x,a}t}A_{x,a}(t)\otimes B_{x,a}+e^{-i\omega_{x,a}t}A_{x,a}(t)^\dagger\otimes B_{x,a}^\dagger\bigr), \label{def:Womega}\\
    U^I_{\alpha,\boldsymbol\omega,T}=&~ {\cal T}\exp\left(-i\alpha\int_{-T}^T W_{\boldsymbol\omega}(t)\,\dd t\right), \notag
\end{align}
We will expand it into the Dyson series and determine which terms survive after tracing out the bath. This yields the second-order generator, called the \green{finite-window effective Lindbladian}, in the next subsection. We also group the full series by site multiplicities, which will be used in~\cref{sec:comparison} to compare the channel with the effective \green{Lindbladian dynamics}.

For a Hermitian system coupling operator $A$ and a bath frequency $\omega$, we write the single bath qubit interaction-picture coupling as
\begin{equation}\label{eq:single-bath-coupling}
 \begin{aligned}
 W_{A,\omega}(t):=f_\sigma(t)A(t)\otimes b_\omega(t),\qquad
 b_\omega(t):=e^{i\omega t}B+e^{-i\omega t}B^\dagger,\qquad B=|1\rangle\langle0|.
 \end{aligned}
\end{equation}
This factorization uses $A=A^\dagger$, which holds for every coupling in~\eqref{eq:couplings}. The bath operator satisfies
\begin{align*}
    b_\omega(t)^\dagger=b_\omega(t),\qquad b_\omega(t)^2=\idmat,\qquad Zb_\omega(t)Z=-b_\omega(t).
\end{align*}
The initial state of this bath qubit is
\begin{align*}
 \tau_\omega:=\frac{e^{\widetilde\beta\omega Z/2}}{\Tr[e^{\widetilde\beta\omega Z/2}]},
\end{align*}
that is, the Gibbs state of the bath Hamiltonian $-\omega Z/2$ in~\cref{sec:setup} at inverse temperature $\widetilde\beta$.

For a bath label $\ell=(x,a)\in[n]\times[Q]$, let $W_\ell(t)$ be the single-bath coupling in~\eqref{eq:single-bath-coupling}, with $A=A_{x,a}$, $\omega=\omega_{x,a}$, and $B=B_{x,a}$ acting on the bath qubit $E_{x,a}$. Thus, we write $W_{\boldsymbol\omega}$ in \eqref{def:Womega} as
\begin{align*}
    W_{\boldsymbol\omega}(t)=\sum_{\ell\in [n]\times [Q]} W_\ell(t).
\end{align*}
Define the interaction insertion as the whole commutator with the joint state $\varrho$:
\begin{equation}\label{eq:sumexp}
\cV_\ell(t)(\varrho):=-i[W_\ell(t),\varrho],
 \qquad -i[\alpha W_{\boldsymbol\omega}(t),\varrho]
       =\alpha\sum_{\ell\in [n]\times [Q]}\cV_\ell(t)(\varrho).
\end{equation}
Define the time-ordered simplex:
\begin{align*}
    \Delta_p(T)=\{(t_1,\ldots,t_p):-T<t_1<\cdots<t_p<T\}
\end{align*}
and $\dd\boldsymbol t=\dd t_1\cdots\dd t_p$. Then, we have the following Dyson series expansion for the channel $\Psi_{\alpha, T}$:
\begin{equation}\label{eq:dyson}
\begin{aligned}
 \Psi_{\alpha,T}(\rho)
 &=\sum_{p=0}^\infty\alpha^p
   \sum_{\ell_1,\ldots,\ell_p}\int_{\Delta_p(T)}
   \E_{\boldsymbol\omega}\Tr_E\bigl[\cV_{\ell_p}(t_p)\cdots\cV_{\ell_1}(t_1)
     (\rho\otimes\tau_{E,\boldsymbol\omega})\bigr]\,\dd\boldsymbol t.
 \end{aligned}
\end{equation}
The term $p=0$ in the series is $\rho$. By~\cref{lem:bath-parity}~\textnormal{(i)} below, the series is absolutely and uniformly convergent in the bath frequencies. Each insertion carries one factor of $\alpha$, and later insertions act on the left.

\begin{definition}[Words and site multiplicities]
A word $\boldsymbol\ell=(\ell_1,\ldots,\ell_p)\in ([n]\times [Q])^{\times p}$ contains the bath label of each insertion. Its numbers of insertions at a bath label $\ell$ and at a site $x$ are
\begin{equation*}
 m_\ell(\boldsymbol\ell):=\#\{i:\ell_i=\ell\},
 \qquad m_x(\boldsymbol\ell):=\sum_{a=1}^Qm_{(x,a)}(\boldsymbol\ell).
\end{equation*}
\end{definition}
For $\boldsymbol{k}=(k_1,\dots,k_n)\in\N_0^n$, let $|\boldsymbol{k}|=\sum_xk_x$ and collect the words with exactly $2k_x$ insertions at every site, leaving out their common factor $\alpha^{2|\boldsymbol{k}|}$:
\begin{equation}\label{eq:Psi-k}
\begin{aligned}
 \Psi_T^{(\boldsymbol{k})}(\rho)
 &:=\sum_{\substack{\boldsymbol\ell=(\ell_1,\ldots,\ell_{2|\boldsymbol{k}|})\\
                   m_x(\boldsymbol\ell)=2k_x\ \forall x\in [n]}}
 \int_{\Delta_{2|\boldsymbol{k}|}(T)}
 \E_{\boldsymbol\omega}\Tr_E\bigl[\cV_{\ell_{2|\boldsymbol{k}|}}(t_{2|\boldsymbol{k}|})\cdots\cV_{\ell_1}(t_1)
 (\rho\otimes\tau_{E,\boldsymbol\omega})\bigr]\,\dd\boldsymbol t.
 \end{aligned}
\end{equation}
In particular, $\Psi_T^{(\boldsymbol{0})}=\id$. The next lemma shows that only even site multiplicities are needed.

\begin{lemma}[Dyson expansion and bath parity]\label{lem:bath-parity}
Let $T>0$ and $\alpha\in\R$.
\begin{enumerate}[label=(\roman*)]
 \item \emph{(Convergence.)} At order $p$ in~\eqref{eq:dyson}, the sum of the \green{induced trace norms} of the time-integrated words, including their common factor $|\alpha|^p$, is at most
 \begin{equation*}
  \frac{(4Qn\|f_\sigma\|_1|\alpha|)^p}{p!}.
 \end{equation*}
 This bound also holds before the frequency average, uniformly in $\boldsymbol\omega$. In particular, the Dyson series converges absolutely.
 \item \emph{(Label parity.)} For every word $\boldsymbol\ell$, every $\boldsymbol\omega$, and all insertion times,
 \begin{equation*}
  \begin{aligned}
  \Tr_E\bigl[\cV_{\ell_p}(t_p)\cdots\cV_{\ell_1}(t_1)
                 (\rho\otimes\tau_{E,\boldsymbol\omega})\bigr]=0\quad \text{if $m_\ell(\boldsymbol\ell)$ is odd for some bath label $\ell$.}
  \end{aligned}
 \end{equation*}
 \item \emph{(Grouping by site multiplicities.)} We have
 \begin{equation}\label{eq:sitewise-parity}
    \Psi_{\alpha,T}=\sum_{\boldsymbol{k}\in\N_0^n}\alpha^{2|\boldsymbol{k}|}\Psi_T^{(\boldsymbol{k})}=\sum_{m=0}^\infty\alpha^{2m}\sum_{|\boldsymbol{k}|=m}\Psi_T^{(\boldsymbol{k})}.
 \end{equation}
 Thus, all odd-order terms vanish. The maps $\Psi_T^{(\boldsymbol{k})}$ are contributions with specified site multiplicities, and their sum over $|\boldsymbol{k}|=m$ is the coefficient of $\alpha^{2m}$.
 \item If a word has exactly two insertions at a site $x$ and its bath trace is nonzero, those two insertions have the same bath label $(x,a)$.
\end{enumerate}
\end{lemma}

\begin{proof}
For Part~(i), since
\begin{align*}
    \no{W_\ell(t)}=\no{f_\sigma(t)A_{x,a}(t)\otimes e^{i\omega_{x,a}t}B_{x,a} + f_\sigma(t)A_{x,a}(t)\otimes e^{-i\omega_{x,a}t}B_{x,a}^\dagger}\leq 2f_\sigma(t),
\end{align*}
the commutator's trace norm inequality~\eqref{eq:commutator_norm} and~\eqref{eq:sumexp} give
\begin{equation*}
\|\cV_\ell(t)\|_{1\to1}\le2\no{W_\ell(t)}\le4f_\sigma(t),
 \qquad \sum_{\ell\in [n]\times [Q]}\|\cV_\ell(t)\|_{1\to1}\le4Qnf_\sigma(t).
\end{equation*}
Note that the embedding $\rho\mapsto\rho\otimes\tau_{E,\boldsymbol\omega}$ preserves the trace norm, while the partial trace and the frequency average do not increase it. Submultiplicativity therefore bounds the sum at order $p$ by
\begin{equation*}
\begin{aligned}
 |\alpha|^p\int_{\Delta_p(T)}
       \prod_{i=1}^p\sum_{\ell_i\in [n]\times [Q]}\|\cV_{\ell_i}(t_i)\|_{1\to1}\,\dd\boldsymbol t
 &\le\frac{(4Qn|\alpha|)^p}{p!}
          \left(\int_{-T}^Tf_\sigma(t)\,\dd t\right)^p\\
 &\le\frac{(4Qn\|f_\sigma\|_1|\alpha|)^p}{p!}.
 \end{aligned}
\end{equation*}
Here the factor $1/p!$ follows from the time ordering. Summing the bound over $p$ proves absolute convergence.

For Part~(ii), fix a bath label $\ell$ and define $\mathcal Z_\ell(\varrho)=Z_\ell\varrho Z_\ell$, where $Z_\ell$ is Pauli $Z$ on that bath qubit. Since $ZBZ=-B$ and $ZB^\dagger Z=-B^\dagger$, conjugation by $Z_\ell$ changes the sign of $W_\ell$ and leaves every $W_{\ell'}$ with $\ell'\ne\ell$ unchanged. Hence
\begin{equation*}
\mathcal Z_\ell\cV_{\ell'}(t)\mathcal Z_\ell=(-1)^{\mathbf1_{\{\ell'=\ell\}}}\cV_{\ell'}(t).
\end{equation*}
The initial bath state $\tau_\omega$ is diagonal, so $\mathcal Z_\ell(\rho\otimes\tau_{E,\boldsymbol\omega})=\rho\otimes\tau_{E,\boldsymbol\omega}$. Also $\Tr_E\circ\mathcal Z_\ell=\Tr_E$. Inserting $\mathcal Z_\ell^2=\id$ between successive factors therefore gives
\begin{equation}\label{eq:word-parity}
\begin{aligned}
\Tr_E\bigl[\cV_{\ell_p}(t_p)\cdots\cV_{\ell_1}(t_1)(\rho\otimes\tau_{E,\boldsymbol\omega})\bigr]=&~ \Tr_E\bigl[\mathcal Z_\ell\cV_{\ell_p}(t_p)\mathcal Z_\ell
          \cdots\mathcal Z_\ell\cV_{\ell_1}(t_1)\mathcal Z_\ell(\rho\otimes\tau_{E,\boldsymbol\omega})\bigr]\\
=&~ (-1)^{m_\ell(\boldsymbol\ell)}
          \Tr_E\bigl[\cV_{\ell_p}(t_p)\cdots\cV_{\ell_1}(t_1)(\rho\otimes\tau_{E,\boldsymbol\omega})\bigr].
 \end{aligned}
\end{equation}
If $m_\ell(\boldsymbol\ell)$ is odd, this quantity equals its negative and must vanish.

For Part~(iii), by (ii), every bath-label multiplicity of a nonzero word is even. Its site multiplicities, being sums of bath-label multiplicities, are therefore $m_x=2k_x$ for some $\boldsymbol{k}\in\N_0^n$. Such a word has length $2|\boldsymbol{k}|$ and scalar factor $\alpha^{2|\boldsymbol{k}|}$. Absolute convergence from (i) permits regrouping by $\boldsymbol{k}$, which proves~\eqref{eq:sitewise-parity}.

For Part~(iv), if the two insertions at $x$ had different bath labels, each of those labels would occur exactly once. Their bath trace would then vanish by (ii).
\end{proof}

The lemma now lets us identify the second-order generator: part~(iv) reduces each surviving term to \emph{one} bath qubit, while parts~(i) and~(iii) control the remainder. The higher-order grouping~\eqref{eq:sitewise-parity} will be used again in~\cref{sec:comparison}, together with the locality estimates of~\cref{sec:locality}.

\subsection{The second-order generator}\label{subsec:second-order}

By \Cref{lem:bath-parity}~(ii), both insertions of a surviving second-order term use the same bath label. Accordingly, for $x\in[n]$, define the site generator
\begin{equation}\label{eq:dyson-second}
 \cL_{x,T}(\rho):=-\sum_{a=1}^Q\int_\R g_{\beta,\sigma}(\omega)\int_{-T<s<t<T}
 \Tr_{E_{x,a}}\bigl[W_{A_{x,a},\omega}(t),\bigl[W_{A_{x,a},\omega}(s),\rho\otimes\tau_\omega\bigr]\bigr]
 \,\dd s\,\dd t\,\dd\omega,
\end{equation}
where $\Tr_{E_{x,a}}$ denotes the partial trace over the bath qubit labeled $(x,a)$.

Let $\boldsymbol{e}_x\in\N_0^n$ be all zeros except for the $x$-th entry, which is one. By~\Cref{lem:bath-parity}~(iv), the only surviving words in $\Psi_T^{(\boldsymbol{e}_x)}$ use a bath label $(x,a)$ twice. The product of their commutator maps is
\[
 \cV_{(x,a)}(t)\cV_{(x,a)}(s)(\varrho)
 =-[W_{(x,a)}(t),[W_{(x,a)}(s),\varrho]].
\]
All unused bath factors have trace one. In~\eqref{eq:Psi-k}, by taking the remaining bath trace and frequency average and summing over $a$, we have
\begin{align}\label{eq:psi_T_ex}
    \Psi_T^{(\boldsymbol{e}_x)}=\cL_{x,T}.
\end{align}
Since $\Psi_T^{(\boldsymbol{0})}=\id$ and the indices with $|\boldsymbol{k}|=1$ are exactly the $\boldsymbol{e}_x$, the grouped expansion~\eqref{eq:sitewise-parity} becomes
\[
 \Psi_{\alpha,T}
 =\id+\alpha^2\sum_x\Psi_T^{(\boldsymbol{e}_x)}
       +\sum_{|\boldsymbol{k}|\ge2}\alpha^{2|\boldsymbol{k}|}\Psi_T^{(\boldsymbol{k})}.
\]
Thus
\begin{equation}\label{eq:second-order-expansion}
 \Psi_{\alpha,T}=\id{}+\alpha^2\cL_T+\mathcal{O}(\alpha^4),\qquad \cL_T:=\sum_{x\in[n]}\cL_{x,T}.
\end{equation}
Every word in the remaining sum has length $2|\boldsymbol{k}|\ge4$. Applying \Cref{lem:bath-parity}~(i) to each order and summing over all orders gives the explicit remainder bound
\begin{equation}\label{eq:second-order-remainder}
 \bigl\|\Psi_{\alpha,T}-\id-\alpha^2\cL_T\bigr\|_{1\to1}\leq\sum_{p\geq4}\frac{(4Qn\|f_\sigma\|_1|\alpha|)^p}{p!}.
\end{equation}
We call $\cL_T$ the \green{\emph{finite-window effective Lindbladian}} of the \red{system-bath} sampler~\eqref{eq:physical-channel}.

We next derive the Lindblad form of $\cL_{x,T}$ in~\eqref{eq:dyson-second} using the filtered coupling and bath correlation defined below.
For a Hermitian system matrix $A$, define the \emph{filtered coupling} by
\begin{equation}\label{eq:filtered-operator}
 V_{A,T}(u):=\int_{-T}^T f_\sigma(t)A(t)e^{-iut}\,\dd t,\qquad
 V_{A,T}(u)^\dagger=V_{A,T}(-u),
\end{equation}
where the second identity holds because $f_\sigma$ is real and $A$ is Hermitian.
The \emph{bath correlation function} is the \green{Fourier transform} of the frequency law~\eqref{eq:frequency}:
\begin{equation}\label{eq:filters}
 c_{\beta,\sigma}(v):=\int_\R g_{\beta,\sigma}(u)e^{iuv}\,\dd u
 =e^{-s_{\beta,\sigma}v^2/(2\beta^2)-iv/\beta}.
\end{equation}
In particular, $c_{\beta,\sigma}(-v)=\overline{c_{\beta,\sigma}(v)}$ and $|c_{\beta,\sigma}(v)|\le1$.

\begin{proposition}[Second-order state dynamics]\label{prop:second}
For every $T>0$ and $x\in[n]$, define the site transition map:
\begin{align}\label{eq:site-jump}
    \cJ_{x,T}(\rho):=\sum_{a=1}^Q\int_\R g_{\beta,\sigma}(u)V_{A_{x,a},T}(u)\,\rho\,V_{A_{x,a},T}(u)^\dagger\,\dd u,
\end{align}
the ordered coefficient:
\begin{align}\label{eq:Q}
    Q_{x,T}:=\sum_{a=1}^Q\int_{-T<s<t<T}f_\sigma(t)f_\sigma(s)c_{\beta,\sigma}(t-s)
                         A_{x,a}(t) A_{x,a}(s)\,\dd s\,\dd t,
\end{align}
and the decay matrix and the coherent term:
\begin{align}\label{eq:MH}
M_{x,T}:=\sum_{a=1}^Q\int_\R g_{\beta,\sigma}(u)V_{A_{x,a},T}(u)^\dagger V_{A_{x,a},T}(u)\,\dd u,
 \qquad H_{x,T}:=\frac{Q_{x,T}-Q_{x,T}^\dagger}{2i}.
\end{align}
Then,
\begin{align*}
    M_{x,T}=Q_{x,T}+Q_{x,T}^\dagger,\qquad Q_{x,T}=\tfrac12M_{x,T}+iH_{x,T}.
\end{align*}
Moreover, the site generator $\cL_{x,T}$ in~\eqref{eq:dyson-second} satisfies
\begin{equation}\label{eq:GKLS}
\begin{aligned}
 \cL_{x,T}(\rho)=&~\cJ_{x,T}(\rho)-Q_{x,T}\rho-\rho Q_{x,T}^\dagger\\
 =&~ -i[H_{x,T},\rho] + \sum_{a=1}^Q\int_\R g_{\beta,\sigma}(u)\,\mathcal D_{V_{A_{x,a},T}(u)}(\rho)\,\dd u,
\end{aligned}
\end{equation}
where $\mathcal D_V$ is defined in~\eqref{eq:def_lindbladian}. In particular, $\cL_{x,T}$ is a trace-preserving \green{Lindbladian} with transition map $\cJ_{x,T}$, decay part $-\tfrac12\{M_{x,T},\cdot\}$, and coherent term $H_{x,T}$.
\end{proposition}

\begin{proof}
We compute one summand of $\cL_{x,T}(\rho)$ in~\eqref{eq:dyson-second}. Fix $a\in[Q]$ and write $A=A_{x,a}$, which is Hermitian by~\eqref{eq:couplings}. Let $E=E_{x,a}$ be the corresponding bath qubit, with frequency $\omega$, coupling operator $B=|1\rangle\langle0|$, and initial state $\tau_\omega$.
All interchanges of integrals below are justified because $f_\sigma\in L^1$ and $g_{\beta,\sigma}$ is a probability density.

\proofstep{The ordered coefficient and unitarity.}
At fixed $\omega$, define
\begin{align*}
 F:=\int_{-T}^T W_{A,\omega}(t)\,\dd t,\qquad
 Y:=\int_{-T<s<t<T}W_{A,\omega}(t)W_{A,\omega}(s)\,\dd s\,\dd t,
\end{align*}
and define the two system objects
\begin{align*}
 \mathcal J_\omega(\rho):=\Tr_E[F(\rho\otimes\tau_\omega)F],\qquad
 Q_\omega:=\Tr_E[Y(\idmat\otimes\tau_\omega)].
\end{align*}
Expanding the double commutator gives two ordered products and two cross terms. Exchanging $s$ and $t$ in one cross term combines the two triangles into the full square, and hence
\begin{equation}\label{eq:one-bath-split}
 \begin{aligned}
-\int_{-T<s<t<T}\Tr_E[W_{A,\omega}(t),[W_{A,\omega}(s),\rho\otimes\tau_\omega]]\,\dd s\,\dd t=\mathcal J_\omega(\rho)-Q_\omega\rho-\rho Q_\omega^\dagger.
 \end{aligned}
\end{equation}
Here the system matrix $\rho$ can be pulled outside the partial trace in each ordered term.  Hermiticity of $W_{A,\omega}(t)$ defined as in~\eqref{eq:single-bath-coupling} also gives $F^\dagger=F$ and
\begin{align*}
 Y+Y^\dagger & = \int_{-T<s<t<T}\bigl(W_{A,\omega}(t)W_{A,\omega}(s)+W_{A,\omega}(s)W_{A,\omega}(t)\bigr)\,\dd s\,\dd t \\
& = \iint_{[-T,T]^2}W_{A,\omega}(t)W_{A,\omega}(s)\,\dd s\,\dd t=F^2.
\end{align*}
Cyclicity of the full trace gives
\begin{align*}
 \Tr[\mathcal J_\omega(\rho)]=\Tr[F(\rho\otimes\tau_\omega)F]=\Tr[(\rho\otimes\tau_\omega)F^2]=\Tr\!\left[\rho\Tr_E\bigl(F^2(\idmat\otimes\tau_\omega)\bigr)\right],
\end{align*}
where the last step follows from moving the bath-only factor $\tau_\omega$ to the right by cyclicity and then taking the partial trace. Since this holds for every $\rho$, the definition of the \green{Hilbert-Schmidt adjoint} and $F^2=Y+Y^\dagger$ imply
\begin{align}\label{eq:J_to_Q_omega}
 \mathcal J_\omega^\dagger(\idmat)=\Tr_E[F^2(\idmat\otimes\tau_\omega)]=\Tr_E[(Y+Y^\dagger)(\idmat\otimes\tau_\omega)]=Q_\omega+Q_\omega^\dagger.
\end{align}

\proofstep{Combining the bath populations.}
The diagonal entries of $\tau_\omega=e^{\widetilde\beta\omega Z/2}/\Tr[e^{\widetilde\beta\omega Z/2}]$ are
\begin{align*}
 p_0(\omega)=\frac1{1+e^{-\widetilde\beta\omega}},\qquad
 p_1(\omega)=\frac1{1+e^{\widetilde\beta\omega}}.
\end{align*}
The Gaussian density defined in~\eqref{eq:frequency} and the choice $\widetilde\beta=2\beta/s_{\beta,\sigma}$ give
\begin{align*}
 \frac{g_{\beta,\sigma}(-u)}{g_{\beta,\sigma}(u)}
 =\exp\!\left(\frac{(\beta u+1)^2-(1-\beta u)^2}{2s_{\beta,\sigma}}\right)
 =e^{2\beta u/s_{\beta,\sigma}}=e^{\widetilde\beta u}.
\end{align*}
Therefore, for every bounded measurable matrix-valued function $\varphi$,
\begin{equation}\label{eq:matching}
\begin{aligned}
 \int_\R g_{\beta,\sigma}(\omega)
 (p_0(\omega)\varphi(-\omega)+p_1(\omega)\varphi(\omega))\,\dd\omega=&~ \int_\R g_{\beta,\sigma}(u)
 \left(e^{\widetilde\beta u}p_0(-u)+p_1(u)\right)\varphi(u)\,\dd u\\
 = &~ \int_\R g_{\beta,\sigma}(u)\varphi(u)\,\dd u,
 \end{aligned}
\end{equation}
where the first step substitutes $u=-\omega$ in the $p_0$ term, and the second step uses $p_0(-u)=p_1(u)$ and $(1+e^{\widetilde\beta u})p_1(u)=1$.

\proofstep{Identifying the transition, decay, and coherent terms.}
By~\eqref{eq:filtered-operator} and~\eqref{eq:single-bath-coupling},
\begin{align*}
 F=V_{A,T}(-\omega)\otimes B+V_{A,T}(\omega)\otimes B^\dagger.
\end{align*}
Since $B^2=(B^\dagger)^2=0$, $\Tr[B\tau_\omega B^\dagger]=p_0(\omega)$, and $\Tr[B^\dagger\tau_\omega B]=p_1(\omega)$, we obtain
\begin{align*}
 \mathcal J_\omega(\rho)=\Tr_E[F(\rho\otimes\tau_\omega)F]
 =p_0(\omega)V_{A,T}(-\omega)\rho V_{A,T}(-\omega)^\dagger
 +p_1(\omega)V_{A,T}(\omega)\rho V_{A,T}(\omega)^\dagger.
\end{align*}
Applying~\eqref{eq:matching} to $\varphi(u)=V_{A,T}(u)\rho V_{A,T}(u)^\dagger$ and summing over $a$ yields $\cJ_{x,T}$ in~\eqref{eq:site-jump}:
\begin{align*}
    \sum_{a=1}^Q\int_{\R} g_{\beta,\sigma}(\omega)\cJ_\omega(\rho)\, \dd\omega
    =&~ \sum_{a=1}^Q\int_{\R} g_{\beta,\sigma}(\omega) \left(p_0(\omega)V_{A,T}(-\omega)\rho V_{A,T}(-\omega)^\dagger
 +p_1(\omega)V_{A,T}(\omega)\rho V_{A,T}(\omega)^\dagger\right)\, \dd\omega\\
 = &~ \sum_{a=1}^Q\int_{\R} g_{\beta,\sigma}(u) V_{A,T}(u)\rho V_{A,T}(u)^\dagger\,\dd u=\cJ_{x,T}(\rho).
\end{align*}
For the ordered coefficient, direct multiplication of the bath factors gives
\begin{align*}
 \Tr[b_\omega(t)b_\omega(s)\tau_\omega]
 =p_1(\omega)e^{i\omega(t-s)}+p_0(\omega)e^{-i\omega(t-s)}.
\end{align*}
Hence,
\begin{equation}\label{eq:ordered-trace}
\begin{aligned}
 Q_\omega=&~ \int_{-T<s<t<T}\Tr_E[W_{A,\omega}(t)W_{A,\omega}(s)(\idmat\otimes\tau_\omega)]\,\dd s\,\dd t\\
 =&~ \int_{-T<s<t<T}f_\sigma(t)f_\sigma(s)\left(p_1(\omega)e^{i\omega(t-s)}+p_0(\omega)e^{-i\omega(t-s)}\right) A(t)A(s)\,\dd s\,\dd t.
\end{aligned}
\end{equation}
Applying~\eqref{eq:matching} to $\varphi(u)=e^{iu(t-s)}$ gives
\begin{align*}
    \int_\R g_{\beta,\sigma}(\omega) \left(p_1(\omega)e^{i\omega(t-s)}+p_0(\omega)e^{-i\omega(t-s)}\right)\,d\omega=\int_\R g_{\beta,\sigma}(u) e^{iu(t-s)}\,du=c_{\beta,\sigma}(t-s).
\end{align*}
Since $A(t)=A(t)^\dagger$, averaging and summing gives $Q_{x,T}$ in~\eqref{eq:Q}.

Thus~\eqref{eq:one-bath-split} gives the first equality in~\eqref{eq:GKLS}. Moreover, by averaging~\eqref{eq:J_to_Q_omega}, we have
\begin{align*}
 Q_{x,T}+Q_{x,T}^\dagger=\cJ_{x,T}^\dagger(\idmat)
 =\sum_{a=1}^Q\int_\R g_{\beta,\sigma}(u)V_{A_{x,a},T}(u)^\dagger V_{A_{x,a},T}(u)\,\dd u
 =M_{x,T}.
\end{align*}
With $H_{x,T}=(Q_{x,T}-Q_{x,T}^\dagger)/(2i)$, it follows that
\begin{align*}
 -Q_{x,T}\rho-\rho Q_{x,T}^\dagger
 =-\tfrac12\{M_{x,T},\rho\}-i[H_{x,T},\rho].
\end{align*}
This proves the Lindblad expression in~\eqref{eq:GKLS}. Its transition map is completely positive because $g_{\beta,\sigma}\ge0$, and $M_{x,T}=\cJ_{x,T}^\dagger(\idmat)$ implies $\cL_{x,T}^\dagger(\idmat)=0$.
\end{proof}

\paragraph{The infinite-window limit.}
Since $f_\sigma\in L^1$, dominated convergence shows that $V_{A,T}(u)$, $\cJ_{x,T}$, $M_{x,T}$, $Q_{x,T}$, $H_{x,T}$, and $\cL_{x,T}$ converge as $T\to\infty$ for every $n$. We denote the limits by $V_A$, $\cJ_x$, $M_x$, $Q_x$, $H_x$, and $\cL_x$. We also define the \green{infinite-window effective Lindbladian}:
\begin{equation}\label{eq:Llimit}
 \begin{gathered}
 \cL=\sum_x\cL_x,\qquad \cJ=\sum_x\cJ_x,\qquad
 M=\sum_xM_x,\qquad H_{\mathrm{coh}}=\sum_xH_x,\\
 \cL(\rho)=-i[H_{\mathrm{coh}},\rho]+\cJ(\rho)-\tfrac12\{M,\rho\}.
 \end{gathered}
\end{equation}

\subsection{KMS detailed balance and the Lamb shift}\label{subsec:kms-lamb}

By~\eqref{eq:GKLS} and~\eqref{eq:Llimit}, $\cL$ has transition map $\cJ$ and decay part $-\tfrac12\{M,\cdot\}$. In this subsection, we show that $\cJ$ is \green{KMS-detailed-balanced}. Then, the \green{KMS coherent correction} $G_D$ in~\eqref{eq:lambshift} makes the \green{KMS Lindbladian} exactly \green{KMS-detailed-balanced}, and the remaining coherent term is the Lamb shift.

For the infinite-window coupling $V_A$, by the \green{Bohr decomposition}~\eqref{eq:Bohr} and the Gaussian Fourier transform, we have
\begin{align*}
 V_A(u)=\int_{\R} f_\sigma(t)A(t)e^{-iut}\,\dd t=\|f_\sigma\|_1\sum_\nu e^{-\sigma^2(\nu-u)^2}A_\nu.
\end{align*}
Since $A_\nu(-i\beta/2)=e^{\beta\nu/2}A_\nu$, completing the square as
\begin{align*}
 -\sigma^2(\nu-u)^2+\frac{\beta\nu}{2}
 =-\sigma^2\left(\nu-u-\frac{\beta}{4\sigma^2}\right)^2+\frac{\beta u}{2}+\frac{\beta^2}{16\sigma^2}
\end{align*}
shows that
\begin{equation}\label{eq:filter-modular-shift}
 V_A(u)(-i\beta/2)
 =e^{\beta u/2+\beta^2/(16\sigma^2)}V_A(u+\beta/(4\sigma^2)).
\end{equation}
The frequency density transforms under the \red{change of variables} $r(u):=-u-\beta/(4\sigma^2)$ by
\begin{align*}
 \log\frac{g_{\beta,\sigma}(r(u))}{g_{\beta,\sigma}(u)}
 &=\frac{(\beta u+1)^2-(1-\beta u-\beta^2/(4\sigma^2))^2}{2s_{\beta,\sigma}}\\
 &=\frac{(2\beta u+\beta^2/(4\sigma^2))(2-\beta^2/(4\sigma^2))}{2s_{\beta,\sigma}}
 =\beta u+\frac{\beta^2}{8\sigma^2},
\end{align*}
where the last step follows from $s_{\beta,\sigma}=2-\beta^2/(4\sigma^2)$ in~\eqref{eq:frequency}. Define the density-weighted filtered coupling $\widetilde V_A(u):=\sqrt{g_{\beta,\sigma}(u)}V_A(u)$. Combining the preceding two equations with $V_A(v)^\dagger=V_A(-v)$ yields
\begin{equation}\label{eq:filter-modular-reflection}
 \widetilde V_A(u)(-i\beta/2)=\sqrt{g_{\beta,\sigma}(r(u))}\,V_A(u+\beta/(4\sigma^2))=\widetilde V_A(r(u))^\dagger.
\end{equation}
This is the key identity that ensures the \green{transition map} ${\cal J}$ satisfies the \green{KMS detailed-balance condition}, as shown in the proposition below. %

As in Section~\ref{subsec:preliminaries} and~\cite[Lemma~II.1]{CKG}, we split the \green{KMS coherent correction} over sites and define the \green{Lamb shift terms} by
\begin{equation}\label{eq:Lamb}
 \begin{aligned}
 G_{D,x}&:=\frac i2\sum_\nu\tanh(\beta\nu/4)(M_x)_\nu,\qquad & G_D&:=\sum_xG_{D,x},\\
 B_x^{\mathrm{Lamb}}&:=H_x-G_{D,x},\qquad & \HL&:=\sum_x B_x^{\mathrm{Lamb}}.
 \end{aligned}
\end{equation}
Each $G_{D,x}$ is Hermitian, because $(M_x)_\nu^\dagger=(M_x)_{-\nu}$ and $\tanh$ is odd. Hence, $B_x^{\mathrm{Lamb}}$ and $\HL$ are also Hermitian. By linearity of $O\mapsto O_\nu$, $G_D$ is the \green{KMS coherent correction}~\eqref{eq:lambshift} for this $M$. %

\Needspace{10\baselineskip}
\begin{proposition}[KMS decomposition]\label{prop:KMS}
Define the generator
\begin{equation}\label{eq:LK}
 \LK(\rho):=-i[G_D,\rho]+\sum_{x,a}\int g_{\beta,\sigma}(u)\mathcal D_{V_{A_{x,a}}(u)}(\rho)\,\dd u,
\end{equation}
and its site version
\begin{align*}
    \mathcal L_{\mathrm{KMS},x}(\rho) :=-i[G_{D,x},\rho] + \sum_a  \int g_{\beta,\sigma}(u)\mathcal D_{V_{A_{x,a}}(u)}(\rho)\,\dd u.
\end{align*}
Then $\Gamma^{-1}\LK\Gamma=\LK^\dagger$, and the \green{Hilbert-Schmidt adjoint} $\LK^\dagger$ is \green{KMS-self-adjoint} and nonpositive. The same holds for each $\mathcal L_{\mathrm{KMS},x}$. Moreover,
\begin{equation}\label{eq:thermal-split}
 \cL(\rho)=\LK(\rho)-i[\HL,\rho],\qquad
 \LK(\rhoB)=0,\qquad \cL(\rhoB)=-i[\HL,\rhoB].
\end{equation}
\end{proposition}
\begin{proof}
By~\eqref{eq:MH} and~\eqref{eq:Llimit},
\begin{align*}
 \sum_{x,a}\int g_{\beta,\sigma}(u)V_{A_{x,a}}(u)^\dagger V_{A_{x,a}}(u)\,\dd u=M.
\end{align*}
Thus~\eqref{eq:LK} gives that
\begin{align*}
    \LK(\rho)=-i[G_D,\rho]+\cJ(\rho)-\tfrac12\{M,\rho\}
\end{align*}
We deal with the transition map and the remaining decay and coherent parts separately.

\proofstep{The transition map.}
By~\eqref{eq:Gamma} and~\eqref{eq:evolutions}, for all matrices $X,Y,O$,
\begin{equation}\label{eq:Gamma-conjugation}
 \Gamma^{-1}\bigl(X\Gamma(O)Y\bigr)=X(-i\beta/2)\,O\,Y(i\beta/2).
\end{equation}
Writing $\cJ$ with the couplings $\widetilde V_A$ defined above, and using~\eqref{eq:filter-modular-reflection}, we obtain
\begin{align*}
 (\Gamma^{-1}\cJ\Gamma)(O)
 &=\sum_{x,a}\int_\R \widetilde V_{A_{x,a}}(u)(-i\beta/2)\,O\,
       \widetilde V_{A_{x,a}}(u)^\dagger(i\beta/2)\,\dd u\\
 &=\sum_{x,a}\int_\R \widetilde V_{A_{x,a}}(r(u))^\dagger\,O\,
       \widetilde V_{A_{x,a}}(r(u))\,\dd u\\
 &=\sum_{x,a}\int_\R \widetilde V_{A_{x,a}}(v)^\dagger\,O\,
       \widetilde V_{A_{x,a}}(v)\,\dd v
 =\cJ^\dagger(O).
\end{align*}
This proves $\Gamma^{-1}\cJ\Gamma=\cJ^\dagger$.

\proofstep{The decay and coherent parts.}
Let $R:=\tfrac12M+iG_D$. Since $G_D$ is Hermitian, \eqref{eq:Lamb} gives
\begin{align*}
 R_\nu=\tfrac12\bigl(1-\tanh(\beta\nu/4)\bigr)M_\nu,
 \qquad
 (R^\dagger)_\nu=\tfrac12\bigl(1+\tanh(\beta\nu/4)\bigr)M_\nu.
\end{align*}
By the identity $e^{2v}(1-\tanh v)=1+\tanh v$ with $v=\beta\nu/4$ and~\eqref{eq:Bohr}, we have
\begin{align*}
 \red{R(-i\beta/2)=\sum_\nu e^{\beta\nu/2}R_\nu=\frac12\sum_\nu\bigl(1+\tanh(\beta\nu/4)\bigr)M_\nu=R^\dagger,}
\end{align*}
and taking adjoints gives $R^\dagger(i\beta/2)=R$. Therefore,
\begin{align}\label{eq:R_transform_M_GD}
 \Gamma^{-1}\bigl(-R\Gamma(O)-\Gamma(O)R^\dagger\bigr)
 =&~{-R(-i\beta/2)O}-OR^\dagger(i\beta/2)\notag\\
 =&~{-R^\dagger O}-OR\notag\\
 =&~{-\tfrac12\{M,O\}}+i[G_D,O],
\end{align}
where the first step follows from~\eqref{eq:Gamma-conjugation}, the second step follows from $R(-i\beta/2)=R^\dagger$ and $R^\dagger(i\beta/2)=R$, and the last step follows from the definition of $R$. Since
\begin{align*}
    -R\rho-\rho R^\dagger=-\tfrac12\{M,\rho\}-i[G_D,\rho],
\end{align*}
the right-hand side of~\eqref{eq:R_transform_M_GD} is the \green{Hilbert-Schmidt adjoint} of the decay and coherent parts of $\LK$. Together with the identity for $\cJ$, this proves $\Gamma^{-1}\LK\Gamma=\LK^\dagger$.

\proofstep{Stationarity and sign.}
Since $\mathcal D_V^\dagger(\idmat)=0$ for every $V$ and $[G_D,\idmat]=0$, we have $\LK^\dagger(\idmat)=0$. Hence
\begin{align*}
 \LK(\rhoB)=\LK\Gamma(\idmat)=\Gamma\LK^\dagger(\idmat)=0,
\end{align*}
where the first step follows from $\Gamma(\idmat)=\rhoB$, and the second step follows from $\Gamma^{-1}\LK\Gamma=\LK^\dagger$. The map $\LK^\dagger$ is self-adjoint in the KMS inner product and generates a contractive semigroup, so its spectrum is nonpositive.
The same argument applies to each $\mathcal L_{\mathrm{KMS},x}$.

\proofstep{The Lamb shift term.}
Subtracting~\eqref{eq:LK} from~\eqref{eq:Llimit} and using $H_{\mathrm{coh}}-G_D=\HL$ gives the first identity in~\eqref{eq:thermal-split}. Applying it to $\rhoB$ and using $\LK(\rhoB)=0$ gives the last identity.
\end{proof}

\subsection[The generator on relative densities]{The generator on relative densities}\label{subsec:weighted-generator}

Noting $\Gamma^{-1}(\cL(\rho))=\Gamma^{-1}\cL\Gamma(\Gamma^{-1}(\rho))$ for a state $\rho$, the generator acts on relative densities as $\Gamma^{-1}\cL\Gamma$. By \Cref{prop:KMS}, $\rho_\beta$ need not be a stationary state of $\cL$. \red{Equivalently,}
\begin{align*}
\Gamma^{-1}\cL\Gamma(\idmat)=\Gamma^{-1}\bigl(\cL(\rhoB)\bigr)=\Gamma^{-1}\bigl(-i[\HL,\rhoB]\bigr)
\end{align*}
\red{need not vanish.}
In \Cref{sec:reference}, we construct an approximate stationary state and bound its \green{one-round stationarity defect}.
This requires bounding $\Gamma^{-1}\cL\Gamma$ applied to sums of $n$ quasi-local terms, such as $(\Gamma^{-1}\cL\Gamma)(\Gamma^{-1}\cL\Gamma(\idmat))$.
If we bound such terms by operator norms, we shall obtain a double sum over sites and an error of order $n^2$. To obtain an error of order $n$, we write $\Gamma^{-1}\cL\Gamma$ as the sum of two parts that are controlled by different arguments:
\begin{itemize}
 \item a map $\cD$ with $\cD(\idmat)=0$. It is a sum of local terms, each of which can be written using commutators, so its action on a sum of local terms is controlled by locality (\Cref{lem:D-action});
 \item the symmetric multiplication $O\mapsto\tfrac12\{\Gamma^{-1}\cL\Gamma(\idmat),O\}$. It produces products of two \green{Gibbs-centered} sums of local terms, which are controlled by the moment bounds of \Cref{cor:moments}.
\end{itemize}

By \Cref{prop:KMS}, $\cL=\LK+\Q$, where $\Q(\rho):=-i[\HL,\rho]$ is the \green{Lamb shift map}. Hence, we can write
\begin{align*}
 \Gamma^{-1}\cL\Gamma=\LK^\dagger+\Gamma^{-1}\mathcal {Q}_{\mathrm{Lamb}}\Gamma,
\end{align*}
\red{where, by}~\eqref{eq:Gamma-conjugation},
\begin{equation}\label{eq:relative-generator}
 \Gamma^{-1}\mathcal {Q}_{\mathrm{Lamb}}\Gamma(O)=-i\bigl(\HL(-i\beta/2)\,O-O\,\HL(i\beta/2)\bigr).
\end{equation}

To separate the commutator part of~\eqref{eq:relative-generator}, using $\HL(i\beta/2)=\HL(-i\beta/2)^\dagger$, we split $\HL(-i\beta/2)$ into its Hermitian and anti-Hermitian parts:
\begin{equation}\label{eq:Hs-K}
\begin{aligned}
\HL(-i\beta/2)&=\HL^{\mathrm s}+\frac{i}{2}K,\\
 \HL^{\mathrm s}:=\frac12\bigl(\HL(-i\beta/2)+\HL(i\beta/2)\bigr),&\qquad
 K:=-i\bigl(\HL(-i\beta/2)-\HL(i\beta/2)\bigr).
\end{aligned}
\end{equation}
Both $\HL^{\mathrm s}$ and $K$ are Hermitian. Substituting $\HL(\mp i\beta/2)=\HL^{\mathrm s}\pm\tfrac i2K$ into~\eqref{eq:relative-generator} gives
\begin{align*}
 \Gamma^{-1}\mathcal {Q}_{\mathrm{Lamb}}\Gamma(O)=-i[\HL^{\mathrm s},O]+\frac12\{K,O\}.
\end{align*}
Collecting the terms in $\Gamma^{-1}\cL\Gamma$ that annihilate $\idmat$, we can define the \green{unital generator} $\mathcal{D}$:
\begin{equation}\label{eq:affine}
 \cD(\cdot):=\LK^\dagger(\cdot)-i[\HL^{\mathrm s},\cdot\,] = \Gamma^{-1}\cL\Gamma (\cdot) - \tfrac12\{K, \cdot\},
\end{equation}
satisfying $\cD(\idmat)=0$, which gives
\begin{equation}\label{eq:drift-identity} K=\Gamma^{-1}\cL\Gamma(\idmat)=\Gamma^{-1}\bigl(\cL(\rhoB)\bigr).
\end{equation}
Thus $K$ is the \emph{drift} at the Gibbs state. It is produced entirely by the Lamb shift, and it vanishes when $[\HL,H_n]=0$. Moreover, $K$ is \green{Gibbs-centered}, i.e., $\Tr(\rhoB K)=\Tr(\cL(\rhoB))=0$ by~\eqref{eq:normalization-identities}.

We have now identified the channel expansion, its second-order generator $\cL_T$, and the drift $K$ at the Gibbs state. To estimate these objects uniformly in the system size, we next establish locality of their matrix coefficients. These estimates will then be applied to the Dyson series in Section~\ref{sec:comparison} and to sums of local terms in Section~\ref{sec:weighted}.

\section{Locality of the effective terms}\label{sec:locality}

This section establishes norm and locality bounds, uniform in the system size, for the site generators of~\cref{sec:generator} and the matrices that define them. In~\cref{subsec:complex-time}, we bound multiplication by a coupling operator at the level of relative densities, using one-dimensional locality estimates for imaginary-time evolution~\cite{perez2023locality}. In~\cref{subsec:coefficients}, we apply these bounds to $M_x$, $Q_x$, $H_x$, $G_{D,x}$, and $B_x^{\mathrm{Lamb}}$, and to the site generators. These estimates will be used in the channel comparison of~\cref{sec:comparison} and in the bounds on sums of local terms in~\cref{sec:weighted}.

\subsection{Multiplication maps and complex-time locality}\label{subsec:complex-time}
After the commutators in~\eqref{eq:sumexp} are expanded, each insertion in the Dyson series~\eqref{eq:dyson} multiplies the system state by a coupling $A_{x,a}(t)$ on the left or on the right, and the bath operators contribute only a scalar factor after tracing out the bath. In this subsection, we bound these system multiplications at the level of relative densities (\cref{subsec:weighted-generator}), uniformly in the system size. Their norms are controlled by the imaginary-time evolved couplings $A_{x,a}(t\pm i\beta/4)$, which we bound in~\Cref{cor:coupling-bounds}.

Since $\nr O=\|\Gamma^{1/2}(O)\|_2$, the isometry $\Gamma^{1/2}$ from~\eqref{eq:Gamma} lets us estimate these maps in the Hilbert--Schmidt space. For any operators $B,C$, define the weighted multiplication map
\begin{equation}\label{eq:weighted-multiplication}
 \cB_{B,C}(O):=\Gamma^{-1/2}\bigl(B\,\Gamma^{1/2}(O)\,C\bigr) =\rhoB^{-1/4}B\rhoB^{1/4}\,O\,\rhoB^{1/4}C\rhoB^{-1/4},
\end{equation}
which is the two-sided multiplication $V\mapsto BVC$ on the Hilbert--Schmidt space, transported to the KMS space by the isometry $\Gamma^{1/2}$.

\begin{lemma}[Weighted multiplication maps]\label{lem:multiplication-maps}
Let $\star$ denote the KMS adjoint as in~\eqref{eq:map-duality}. Then, for all operators $B,C,D,E$, the weighted multiplication maps satisfy the following properties:
\begin{equation}\label{eq:weighted-algebra}
 \begin{aligned}
 \nm{\cB_{B,C}}&\le\no B\no C,\\
 \cB_{B,C}\cB_{D,E}&=\cB_{BD,EC},\\
 \cB_{B,C}^\star&=\cB_{B^\dagger,C^\dagger}.
 \end{aligned}
\end{equation}
\end{lemma}
\begin{proof}

For the first inequality, using $\nr O=\|\Gamma^{1/2}(O)\|_2$, we have
\begin{align*}
    \nr{\cB_{B,C}(O)}=\|B\Gamma^{1/2}(O)C\|_2
 \le\no B\no C\|\Gamma^{1/2}(O)\|_2=\no B\no C\nr O,
\end{align*}
which implies that $\nm{\cB_{B,C}}\le\no B\no C$.
For the second one, applying \eqref{eq:weighted-multiplication} twice, we have
\begin{align*}
    \cB_{B,C}\cB_{D,E}(O)=&~\Gamma^{-1/2}\left(B\,\Gamma^{1/2}\bigl(\cB_{D,E}(O)\bigr)\,C\right)=\Gamma^{-1/2}\left(B\,\Gamma^{1/2}\left(\Gamma^{-1/2}\bigl(D\Gamma^{1/2}(O)E\bigr)\right)C\right)\\
    =&~ \Gamma^{-1/2}\bigl((BD)\Gamma^{1/2}(O)(EC)\bigr)=\cB_{BD,EC}(O).
\end{align*}
For the last one, by~\eqref{eq:map-duality}, $\cB_{B,C}^\star(O)=\Gamma^{-1}\cB_{B,C}^\dagger \Gamma(O)$, where the \green{Hilbert-Schmidt adjoint} $\cB_{B,C}^\dagger$ can be derived as follows. For every operator $X$,
\begin{align*}
    \Tr[X^\dagger \cB_{B,C}(O)]=&~ \Tr\left[X^\dagger\Gamma^{-1/2}(B\Gamma^{1/2}(O)C)\right]=\Tr\left[C\Gamma^{-1/2}(X^\dagger)B\Gamma^{1/2}(O)\right]\\
    =&~ \Tr\left[\Gamma^{1/2}\left(C\Gamma^{-1/2}(X^\dagger)B\right)O\right]=\Tr\left[\left(\Gamma^{1/2}\left(B^\dagger \Gamma^{-1/2}(X)C^\dagger\right)\right)^\dagger O\right],
\end{align*}
where the second and third steps use the cyclicity of the trace, and the last step uses $\Gamma^{s}(Y)^\dagger=\Gamma^{s}(Y^\dagger)$. This implies that  $\cB_{B,C}^\dagger(X)=\Gamma^{1/2}\bigl(B^\dagger\Gamma^{-1/2}(X)C^\dagger\bigr)$. Thus, we have
\begin{align*}
    \cB_{B,C}^\star(O)=\Gamma^{-1}\cB_{B,C}^\dagger \Gamma(O)=\Gamma^{-1}\Gamma^{1/2}\left(B^\dagger \Gamma^{-1/2}(\Gamma(O) )C^\dagger\right)=\Gamma^{-1/2}\left(B^\dagger \Gamma^{1/2}(O)C^\dagger\right)=\cB_{B^\dagger, C^\dagger}(O).
\end{align*}

\end{proof}

We now apply \Cref{lem:multiplication-maps} to the system multiplications. For any map $O\mapsto \Gamma^{-1}(B\Gamma(O)C)$, we have
\begin{equation}\label{eq:vertex-matrices}
 \Gamma^{-1}(B\Gamma(O)C)
 =\rhoB^{-1/2}B\rhoB^{1/2}O\rhoB^{1/2}C\rhoB^{-1/2}
 =B(-i\beta/2)OC(i\beta/2)
 =\cB_{B(-i\beta/4),C(i\beta/4)}(O),
\end{equation}
where the first step follows from~\eqref{eq:Gamma}, the second step follows from~\eqref{eq:Gamma-conjugation}, and the last step follows from~\eqref{eq:weighted-multiplication} and $\rhoB^{\mp1/4}X\rhoB^{\pm1/4}=X(\mp i\beta/4)$ (see \eqref{eq:evolutions}). Equation~\eqref{eq:weighted-algebra} therefore gives
\begin{equation*}
\left\|O\mapsto\Gamma^{-1}(B\Gamma(O)C)\right\|_{\rhoB\to\rhoB}
 \le\no{B(-i\beta/4)}\,\no{C(i\beta/4)}.
\end{equation*}
In particular, on relative densities, the system multiplications $\rho\mapsto A_{x,a}(t)\rho$ and $\rho\mapsto\rho A_{x,a}(t)$, involved in the effective Lindbladian, become $\cB_{A_{x,a}(t-i\beta/4),\idmat}$ and $\cB_{\idmat,A_{x,a}(t+i\beta/4)}$, respectively. Thus, it remains to bound the coefficients $A_{x,a}(t\pm i\beta/4)$ uniformly in $n$ and $t$. Although $A_{x,a}$ acts only on site $x$, these evolved couplings are in general supported on the whole chain, and the later estimates in~\cref{subsec:coefficients,sec:comparison} also need them to be well approximated by local operators supported near $x$.

For this, let $S=N_r(x)$ and take the normalized partial trace over $S^c$:
\[
 \cE_{x,r}(B)=2^{-q|S^c|}\Tr_{S^c}(B)\otimes \idmat_{S^c},
\]
which maps onto matrices supported on $S$ and fixes each such matrix. Equivalently, $\mathcal{E}_{x,r}$ can be expressed as the average of $(\red{\idmat_S} \otimes U)\,B\,(\red{\idmat_S} \otimes U^\dagger)$ over Haar-random unitaries $U$ on $S^c$, yielding
\begin{equation}\label{eq:local-transfer}
 \begin{aligned}
 \no{\cE_{x,r}(B)}&\le\int\no{(\idmat_S\otimes U)B(\idmat_S\otimes U^\dagger)}\,\dd U=\no B,\\
 \no{B-\cE_{x,r}(B)}
 &=\no{(\id-\cE_{x,r})(B-B^{(r)})}
 \le2\no{B-B^{(r)}}
 \quad\text{if }\supp B^{(r)}\subseteq S,
 \end{aligned}
\end{equation}
where the second line uses $\cE_{x,r}(B^{(r)})=B^{(r)}$. Thus the normalized partial trace gives an error at most twice the approximation error of any operator supported on $S$.

We call $B_x$ \emph{quasi-local} at $x$ if
\[
 \no{B_x}\le C,\qquad\no{(\id-\cE_{x,r})B_x}\le Ce^{-cr}\qquad(r\ge0).
\]
Here $C,c>0$ are independent of $n,x,r$, so the approximation by $\cE_{x,r}(B_x)$ has exponentially small error.  The next lemma shows that a strictly local matrix becomes quasi-local under complex-time evolution, and that real-time evolution preserves quasi-locality up to a light cone, which is the specialization of {\cite[Theorems~2.3--2.4]{perez2023locality}}. In~\Cref{app:araki}, we verify the hypotheses of~\cite{perez2023locality} for~\eqref{eq:hamiltonian} and propagate the resulting imaginary-time approximations in real time.

\begin{lemma}[Complex-time locality]\label{lem:Araki}
Let $x\in[n]$ be a site.
\begin{enumerate}[label=(\roman*)]
 \item \emph{(Strictly local matrices at complex times.)} Let $B_x$ be a local operator with $\supp B_x\subseteq N_d(x)$. Then, for all $t\in\R$, $r\ge0$, and $|u|\le u_*$,
 \begin{equation}\label{eq:complex-cone}
  \begin{aligned}
  \no{B_x(t-iu)}&\le C\no{B_x},\\
  \no{(\id-\cE_{x,r})B_x(t-iu)}
  & \le C\no{B_x}\min\{1,e^{v|t|-cr}\}.
  \end{aligned}
 \end{equation}
 The constants only depend on $d$ and $u_*$. At $t=0$, the second bound can be replaced by $Ce^{-cr\log(r+1)}\no{B_x}$.
 \item \emph{(Quasi-local matrices at real times.)} Suppose that $D_x$ is quasi-local at $x$, with constants $M,\mu>0$:
 \begin{equation}\label{eq:quasilocal-initial}
  \no{D_x}\le M,\qquad
  \no{(\id-\cE_{x,r})D_x}\le Me^{-\mu r}
  \qquad(r\ge0).
 \end{equation}
 Then, for the real-time evolution $D_x(t)=e^{itH_n}D_xe^{-itH_n}$, we have, for all $t\in\R$ and $r\ge0$,
 \begin{equation}\label{eq:quasilocal-real-cone}
  \begin{aligned}
  \no{D_x(t)}&=\no{D_x}\le M,\\
  \no{(\id-\cE_{x,r})D_x(t)}
  &\le C_\mu M\min\{1,e^{v_\mu|t|-c_\mu r}\}.
  \end{aligned}
 \end{equation}
The constants $C_\mu,c_\mu>0$ and $v_\mu\ge0$ only depend on $\mu$.
\end{enumerate}
\end{lemma}

Applying part~(i) to the coupling operators gives the bounds on the system multiplications.

\begin{corollary}[Bounds for the coupling operators]\label{cor:coupling-bounds}
There is a constant $C_\beta$, independent of $n$ and of the bath frequencies, such that for all $x\in[n]$, $a\in[Q]$, and $t\in\R$,
\begin{equation}\label{eq:vertex-bound}
 \no{A_{x,a}(t\pm i\beta/4)}\le C_\beta.
\end{equation}
Consequently, the left and right multiplications by $A_{x,a}(t)$, acting on relative densities, satisfy
\begin{equation}\label{eq:coupling-maps}
 \nm{\cB_{A_{x,a}(t-i\beta/4),\idmat}}\le C_\beta,
 \qquad
 \nm{\cB_{\idmat,A_{x,a}(t+i\beta/4)}}\le C_\beta.
\end{equation}
Moreover, for all $r\ge0$,
\begin{equation*}
 \no{(\id-\cE_{x,r})A_{x,a}(t\pm i\beta/4)}\le C_\beta\min\{1,e^{v|t|-cr}\}.
\end{equation*}
\end{corollary}
\begin{proof}
Each $A_{x,a}$ in~\eqref{eq:couplings} has norm one and is supported at $x$. Applying \Cref{lem:Araki}~(i) with $d=0$, $u_*=\beta/4$, and $u=\mp\beta/4$, we obtain~\eqref{eq:vertex-bound} and the tail bound. By~\eqref{eq:vertex-matrices}, the two maps in~\eqref{eq:coupling-maps} are the left and right multiplications by $A_{x,a}(t)$ acting on relative densities. The \red{first} property in~\eqref{eq:weighted-algebra}, together with~\eqref{eq:vertex-bound} and $\no{\idmat}=1$, gives~\eqref{eq:coupling-maps}.
\end{proof}

\subsection{Bounds on the site matrices}\label{subsec:coefficients}

We now apply the estimates of~\cref{subsec:complex-time} to the matrices $M_x$, $Q_x$, $H_x$, $G_{D,x}$, and $B_x^{\mathrm{Lamb}}$ for the effective Lindbladian from~\cref{sec:generator}. The first three are integrals of products $A_{x,a}(t)A_{x,a}(s)$ of evolved couplings; see~\eqref{eq:Q} and~\eqref{eq:MH}. To treat $G_{D,x}$ in the same way, we first write the \green{KMS coherent correction}~\eqref{eq:Lamb} as an integral over real time.

\begin{lemma}[Integral representation of the KMS coherent correction]\label{lem:tanh}
For every system operator $M$,
\begin{equation}\label{eq:tanh-integral}
 \frac i2\tanh\!\left(\frac\beta4\ad_{H_n}\right)(M)
 =i\int_\R\psi_\beta(t)[H_n,M](t)\,\dd t,\qquad
 \psi_\beta(t)=\int_{|t|}^{\infty}\frac1\beta\csch(2\pi u/\beta)\,\dd u.
\end{equation}
The kernel $\psi_\beta$ is nonnegative and integrable, with
\begin{equation}\label{eq:psi-tail}
 \|\psi_\beta\|_1=\frac\beta8,\qquad
 \int_{|t|>L}\psi_\beta(t)\,\dd t\le\frac\beta8 e^{-2\pi L/\beta}\quad(L\ge0).
\end{equation}
\end{lemma}
\begin{proof}
For each Bohr frequency $\nu$, the scalar Fourier identity in~\cite[Fact~E.3]{chen2026overcoming} gives
\[
 \frac i2\tanh(\beta\nu/4)
 =\int_0^\infty\frac1\beta\csch(2\pi u/\beta)(e^{iu\nu}-e^{-iu\nu})\,\dd u.
\]
Since $\ad_{H_n}(M_\nu)=\nu M_\nu$ by~\eqref{eq:Bohr}, each Bohr component $M_\nu$ is an eigenvector of $\ad_{H_n}$ with eigenvalue $\nu$. Hence,
\begin{align*}
    \sum_\nu \frac{i}{2}\tanh\left(\frac{\beta}{4}\nu\right)M_\nu=\frac i2\tanh\!\left(\frac\beta4\ad_{H_n}\right)(M).
\end{align*}
Since $\sum_\nu e^{\pm iu\nu}M_\nu=M(\pm u)$, we obtain the desired \eqref{eq:tanh-integral}:
\[
 \begin{aligned}
 \frac i2\tanh\!\left(\frac\beta4\ad_{H_n}\right)(M)=&~ \int_0^\infty\frac1\beta\csch(2\pi u/\beta)(M(u)-M(-u))\,\dd u\\
 =&~ i\int_0^\infty\frac1\beta\csch(2\pi u/\beta)\int_{-u}^u[H_n,M](t)\,\dd t\,\dd u\\
=&~i\int_\R\left[\int_{|t|}^\infty\frac1\beta\csch(2\pi u/\beta)\,\dd u\right]  [H_n,M](t)\,\dd t =  i\int_\R\psi_\beta(t)[H_n,M](t)\,\dd t,
 \end{aligned}
\]
where the second step follows from $\dd M(t)/\dd t=i[H_n, M](t)$, and the third step follows from changing the order of integration, which turns the region $u\ge0$, $|t|\le u$ into $t\in\R$, $u\ge|t|$.

To estimate the kernel $\psi_\beta$, we can explicitly calculate that integral:
\begin{align*}
    \psi_\beta(t)=\frac1{2\pi}\log\coth(\pi|t|/\beta)=\frac{1}{2\pi}\log \frac{1+z}{1-z},\qquad z:=e^{-2\pi|t|/\beta}.
\end{align*}
Using the power series for $\log((1+z)/(1-z))$, we have
\begin{align*}
    \psi_\beta(t)=\frac1\pi\sum_{j\ge0}\frac{e^{-(2j+1)2\pi|t|/\beta}}{2j+1}.
\end{align*}
Since all the summands are nonnegative, by Tonelli's theorem, we can integrate each term:
\[
 \begin{aligned}
 \int_{|t|>L}\psi_\beta(t)\,\dd t=&~ \frac1\pi\sum_{j\ge0}\int_{|t|>L}\frac{e^{-(2j+1)2\pi|t|/\beta}}{2j+1}\,\dd t=\frac\beta{\pi^2}\sum_{j\ge0}\frac{e^{-(2j+1)2\pi L/\beta}}{(2j+1)^2}\\
 \le&~ \frac\beta{\pi^2}e^{-2\pi L/\beta}\sum_{j\ge0}(2j+1)^{-2}
 =\frac\beta8e^{-2\pi L/\beta},
 \end{aligned}
\]
where the equality holds at $L=0$, implying $\|\psi_\beta\|_1=\beta/8$.
\end{proof}

We now bound all the \red{site matrices}. The following proposition also gives the uniform bound on the site generators needed in~\cref{sec:comparison}.
\begin{proposition}[Site-matrix estimates]\label{prop:quasilocal}
Let $\beta>0$ and $\sigma=\beta/2$. For each fixed $u_*>0$ and $m\in\N_0$, there are constants $C,c>0$ such that, for all $x\in[n]$, $|u|\le u_*$, $r\ge0$, and
\[
 X_x\in\{M_x,Q_x,H_x,G_{D,x},B_x^{\mathrm{Lamb}}\},
\]
it holds that
\[
 \no{(\ad_{H_n}^mX_x)(-iu)}\le C,\qquad
 \no{(\id-\cE_{x,r})\left((\ad_{H_n}^mX_x)(-iu)\right)}\le Ce^{-cr}.
\]
The constants depend only on $\beta$, the fixed local parameters, $u_*$, and $m$. Thus these coefficients and their fixed-order commutators remain quasi-local under bounded imaginary-time evolution.

Moreover,
\begin{equation}\label{eq:site-norm}
 \sup_{T\ge0}\nm{\Gamma^{-1}\cL_{x,T}\Gamma}+\nm{\Gamma^{-1}\cL_x\Gamma}\le C_\beta\beta,
\end{equation}
where $C_\beta$ depends only on $\beta$ and the fixed local parameters.
\end{proposition}
\begin{proof}
\proofstep{The two-time products.}
We first show how locality bounds transfer to products. For system operators $B,D$, set $B_r=\cE_{x,r}(B)$ and $D_r=\cE_{x,r}(D)$. Since $B_rD_r$ is supported in $N_r(x)$,~\eqref{eq:local-transfer} gives
\[
 \begin{aligned}
 \no{(\id-\cE_{x,r})(BD)}
 &\le2\no{BD-B_rD_r}\\
 &\le2\no{B-B_r}\no D+2\no B\no{D-D_r},
 \end{aligned}
\]
where the last step follows from the operator-norm contractivity of $\cE_{x,r}$.

We now bound the imaginary-time evolutions of $Q_x$, $M_x$, and $H_x$. By~\eqref{eq:Q}, $Q_x$ is a double time integral of the products $A_{x,a}(t) A_{x,a}(s)$, so we first bound these products and then integrate against the pulses. For $|u|\le u_*$, it holds that
\[
 (A_{x,a}(t) A_{x,a}(s))(-iu)
 =A_{x,a}(t-iu)A_{x,a}(s-iu).
\]
By applying~\eqref{eq:complex-cone} to each factor and using the product estimate above, we have
\[
 \begin{aligned}
 \no{A_{x,a}(t-iu)A_{x,a}(s-iu)}&\le C,\\
 \no{(\id-\cE_{x,r})\left(A_{x,a}(t-iu)A_{x,a}(s-iu)\right)}
 &\le Ce^{-cr}(e^{v|t|}+e^{v|s|}).
 \end{aligned}
\]
Note that the $e^{v|t|}$ and $e^{v|s|}$ factors can be suppressed by the Gaussian pulse after integration:
\[
 \begin{aligned}
 \int_\R f_\sigma(t)e^{v|t|}\,\dd t=\frac{2e^{v^2\sigma^2}}{\sqrt{\sigma\sqrt{2\pi}}}
       \int_0^\infty e^{-(t-2v\sigma^2)^2/(4\sigma^2)}\,\dd t\le2e^{v^2\sigma^2}\|f_\sigma\|_1.
 \end{aligned}
\]
Substituting these estimates into the infinite-window version of~\eqref{eq:Q}, and using $|c_{\beta,\sigma}|\le1$ from~\eqref{eq:filters}, gives
\begin{align*}
    \no{Q_x(-iu)}\leq &~\sum_{a=1}^Q\int_{s<t}f_\sigma(t)f_\sigma(s)|c_{\beta,\sigma}(t-s)|\no{\left(A_{x,a}(t) A_{x,a}(s)\right)(-iu)}\,\dd s\,\dd t\\
    \leq &~ C\int_{\R^2}f_\sigma(t)f_\sigma(s)\,\dd s\,\dd t= C\|f_\sigma\|_1^2,
\end{align*}
and
\begin{align*}
    \no{(\id-\cE_{x,r})Q_x(-iu)}\leq &~ \sum_{a=1}^Q\int_{s<t}f_\sigma(t)f_\sigma(s)|c_{\beta,\sigma}(t-s)|\no{(\id-\cE_{x,r})\left(A_{x,a}(t) A_{x,a}(s)\right)(-iu)}\,\dd s\,\dd t\\
    \leq &~ Ce^{-cr}\int_{\R^2} f_\sigma(t)f_\sigma(s)(e^{v|t|}+e^{v|s|})\,\dd s\,\dd t\leq Ce^{-cr+v^2\sigma^2}\|f_\sigma\|_1^2,
\end{align*}
where the sum over $a\in[Q]$ is absorbed into $C$.

To obtain the same bounds for $M_x$ and $H_x$, we use $M_x=Q_x+Q_x^\dagger$ from~\Cref{prop:second}, the definition of $H_x$ in~\eqref{eq:MH}, and the adjoint identity $(Q_x^\dagger)(-iu)=Q_x(iu)^\dagger$:
\[
 M_x(-iu)=Q_x(-iu)+Q_x(iu)^\dagger,\qquad
 H_x(-iu)=\frac{Q_x(-iu)-Q_x(iu)^\dagger}{2i}.
\]
Since $|{-u}|\le u_*$ and $\cE_{x,r}$ commutes with taking adjoints, the norm and truncation bounds for $Q_x$ apply to both terms.

\proofstep{Commutators with the Hamiltonian.}
By~\eqref{eq:hamiltonian}, $H_n$ is a sum of nearest-neighbor terms $h_y$ with $\no{h_y}\le1$. Starting from $\supp A_{x,a}=\{x\}$, each commutator with $H_n$ enlarges the support by at most one site on either side. More precisely, if $\supp(\ad_{H_n}^jA_{x,a})\subseteq N_j(x)$, then only the terms $h_y$ with $y\in[x-j-1,x+j]\cap[n-1]$ can contribute to the next commutator. There are at most $2(j+1)$ such terms, and each satisfies $\no{[h_y,X]}\le2\no X$. Therefore,
\[
 \supp(\ad_{H_n}^{j+1}A_{x,a})\subseteq N_{j+1}(x),\qquad
 \no{\ad_{H_n}^{j+1}A_{x,a}}\le4(j+1)\no{\ad_{H_n}^{j}A_{x,a}}.
\]
Iterating this bound and using $\no{A_{x,a}}=1$ gives
\[
 \no{\ad_{H_n}^{m}A_{x,a}}\le4^m m!.
\]
Moreover, for system operators $B,D$ and $z\in\C$,
\[
 (\ad_{H_n}^mB)(z)=\ad_{H_n}^m(B(z)),\qquad
 \ad_{H_n}^m(BD)=\sum_{j=0}^m\binom mj(\ad_{H_n}^jB)(\ad_{H_n}^{m-j}D),
\]
where the first identity holds because $e^{izH_n}$ commutes with $H_n$, and the second follows by iterating $[H_n,BD]=[H_n,B]D+B[H_n,D]$.

Applying these identities inside the two-time integral for $Q_x$, we obtain
\begin{align*}
    \ad_{H_n}^m Q_x=&~ \sum_{a=1}^Q\int_{s<t}f_\sigma(t)f_\sigma(s)c_{\beta,\sigma}(t-s)\sum_{j=0}^m \binom{m}{j}\ad_{H_n}^j\bigl(A_{x,a}(t)\bigr)\,\ad_{H_n}^{m-j}\bigl(A_{x,a}(s)\bigr)\,\dd s\,\dd t\\
    = &~ \sum_{a=1}^Q\sum_{j=0}^m \binom{m}{j}\int_{s<t}f_\sigma(t)f_\sigma(s)c_{\beta,\sigma}(t-s)(\ad_{H_n}^jA_{x,a})(t)(\ad_{H_n}^{m-j} A_{x,a})(s)\,\dd s\,\dd t.
\end{align*}
Thus, each coupling $A_{x,a}$ is replaced by a commutator $\ad_{H_n}^jA_{x,a}$ with $j\le m$, which is supported in $N_m(x)$ and has norm at most $4^mm!$. Hence,~\eqref{eq:complex-cone} applies with $d=m$, and the argument for $Q_x$ above gives the same norm and truncation bounds for $\ad_{H_n}^mQ_x$, with constants that also depend on $m$.

Since $\ad_{H_n}^m(Q_x^\dagger)=(-1)^m(\ad_{H_n}^mQ_x)^\dagger$, the bounds for $\ad_{H_n}^mM_x$ and $\ad_{H_n}^mH_x$ follow as before.

\proofstep{The KMS coherent correction.}
Let $D_{x,u}=([H_n,M_x])(-iu)$. The preceding steps establish uniform bounds on the norm and exponential truncation for $|u|\le u_*$; that is, it is quasi-local at $x$ with constants independent of $u$. Applying the real-time estimate~\eqref{eq:quasilocal-real-cone} to this family gives
\[
 \no{(\id-\cE_{x,r})D_{x,u}(t)}\le C\min\{1,e^{v|t|-cr}\}.
\]
By~\eqref{eq:Lamb},~\Cref{lem:tanh} with $M=M_x$ gives $G_{D,x}$. Since the real- and imaginary-time conjugations commute, we have
\[
 G_{D,x}(-iu)=i\int_\R\psi_\beta(t)D_{x,u}(t)\,\dd t,
 \qquad\no{G_{D,x}(-iu)}\leq \no{D_{x,u}}\|\psi_\beta\|_1\le C\|\psi_\beta\|_1.
\]

For the truncation error, choose $a>0$ such that $va\le c/2$ and split the integral at $|t|=ar$. On $|t|\le ar$, use $\no{(\id-\cE_{x,r})D_{x,u}(t)}\leq Ce^{v|t|-cr}$; on the remaining region, use $\no{(\id-\cE_{x,r})D_{x,u}(t)}\leq C$ and the tail of $\psi_\beta$ in~\eqref{eq:psi-tail}. Thus,
\[
 \begin{aligned}
 \no{(\id-\cE_{x,r})G_{D,x}(-iu)}
 &\le C\int_\R\psi_\beta(t)\min\{1,e^{v|t|-cr}\}\,\dd t\\
 &\le Ce^{-(c-va)r}\|\psi_\beta\|_1
      +C\int_{|t|>ar}\psi_\beta(t)\,\dd t\\
 &\le C\frac\beta8(e^{-cr/2}+e^{-2\pi ar/\beta})\le C'e^{-c'r},
 \end{aligned}
\]
where the third line uses~\eqref{eq:psi-tail}.

To bound $\ad_{H_n}^mG_{D,x}$, pass the commutator through the integral. This replaces $D_{x,u}$ by $(\ad_{H_n}^{m+1}M_x)(-iu)$, whose norm and locality bounds were proved above. The integral remains absolutely convergent because $\psi_\beta\in L^1$.

Finally, since $B_x^{\mathrm{Lamb}}=H_x-G_{D,x}$ in~\eqref{eq:Lamb}, its bounds follow by the triangle inequality.

\proofstep{The induced norm of the site generator.}
To prove~\eqref{eq:site-norm}, we keep the bath correlation in the time integrals. Substituting~\eqref{eq:filtered-operator} into~\eqref{eq:site-jump} and using~\eqref{eq:filters} gives
\[
 \cJ_{x,T}(\rho)=\sum_{a=1}^Q\iint_{[-T,T]^2}
 f_\sigma(t)f_\sigma(s)c_{\beta,\sigma}(s-t)
 A_{x,a}(t)\rho A_{x,a}(s)\,\dd s\,\dd t.
\]
For each multiplication map in this expression,~\eqref{eq:vertex-matrices},~\eqref{eq:weighted-algebra}, and~\eqref{eq:vertex-bound} give
\[
 \begin{aligned}\nm{\Gamma^{-1}\bigl(A_{x,a}(t)\Gamma(\cdot)A_{x,a}(s)\bigr)}\le\no{A_{x,a}(t-i\beta/4)}\no{A_{x,a}(s+i\beta/4)}
 \le C_\beta^2.
 \end{aligned}
\]
Let
\begin{align*}
    I_T:=\iint_{[-T,T]^2}f_\sigma(t)f_\sigma(s)|c_{\beta,\sigma}(t-s)|\,\dd s\,\dd t.
\end{align*}
Then, we have
\begin{align*}
    \nm{\Gamma^{-1}\cJ_{x,T}\Gamma}\leq QC_\beta^2 I_T.
\end{align*}
The ordered terms $Q_{x,T}\rho$ and $\rho Q_{x,T}^\dagger$ in~\eqref{eq:GKLS} satisfy the same bound on each integrand:
\begin{align*}
    \nm{\Gamma^{-1}\left(A_{x,a}(t)A_{x,a}(s)\Gamma(\cdot) \right)}\leq &~ \no{\left(A_{x,a}(t)A_{x,a}(s)\right)(-i\beta/4)}\\
    = &~ \no{A_{x,a}(t-i\beta/4)A_{x,a}(s-i\beta/4)}\leq C_\beta^2.
\end{align*}
Thus, by~\eqref{eq:Q},
\begin{align*}
    \nm{\Gamma^{-1}(Q_{x,T}\Gamma(\cdot))}\leq QC_\beta^2\int_{-T\leq s<t\leq T}f_{\sigma}(t)f_\sigma(s)|c_{\beta,\sigma}(t-s)|\,\dd s\,\dd t\leq QC_\beta^2\frac{I_T}{2},
\end{align*}
where the last step follows from $|c_{\beta,\sigma}(s-t)|=|c_{\beta,\sigma}(t-s)|$. Similarly,
\begin{align*}
    \nm{\Gamma^{-1}(\Gamma(\cdot)Q_{x,T}^\dagger)}\leq QC_\beta^2\frac{I_T}{2}.
\end{align*}
Therefore, by the triangle inequality, we get that
\[
 \nm{\Gamma^{-1}\cL_{x,T}\Gamma}
 \le Q C_\beta^2\left(I_T+\frac{I_T}{2}+\frac{I_T}{2}\right)
 =2Q C_\beta^2 I_T.
\]
It remains to bound $I_T$. Since the integrand of $I_T$ is nonnegative, $I_T\le I_\infty$, so it suffices to compute $I_\infty$.  Set $u=(t+s)/2$ and $v=t-s$, so $t=u+v/2$, $s=u-v/2$, and the Jacobian has absolute value one. By~\eqref{eq:pulse} and~\eqref{eq:frequency},
\[
 \frac{t^2+s^2}{4\sigma^2}=\frac{u^2}{2\sigma^2}+\frac{v^2}{8\sigma^2},
 \qquad \frac{s_{\beta,\sigma}}{2\beta^2}=\frac1{\beta^2}-\frac1{8\sigma^2}.
\]
Thus,
\begin{equation}\label{eq:pair-integral}
\begin{aligned}
    I_\infty=&~ \iint_{\R^2}f_\sigma(t)f_\sigma(s)|c_{\beta,\sigma}(t-s)|\,\dd s\,\dd t=\iint_{\R^2}f_\sigma(t)f_\sigma(s)e^{-s_{\beta,\sigma}v^2/(2\beta^2)}\,\dd s\,\dd t\\
    = &~ \frac1{\sigma\sqrt{2\pi}}\iint_{\R^2}e^{-u^2/(2\sigma^2)}e^{-v^2/\beta^2}\,\dd u\,\dd v=\sqrt{\pi}\beta.
\end{aligned}
\end{equation}
Hence, for every $T\ge0$,
\begin{align*}
    \nm{\Gamma^{-1}\cL_{x,T}\Gamma}\leq 2\sqrt{\pi}Q C_\beta^2\beta.
\end{align*}
The same argument with $T=\infty$ gives $\nm{\Gamma^{-1}\cL_{x}\Gamma}\leq 2\sqrt{\pi}Q C_\beta^2\beta$. This proves~\eqref{eq:site-norm} after absorbing the constant factors.
\end{proof}

\section{One-round comparison}\label{sec:comparison}
We now compare the \green{physical channel}~\eqref{eq:physical-channel} with the effective dynamics generated by $\cL$ in~\cref{sec:generator}. The resulting bound,~\Cref{thm:comparison}, will be used in~\cref{sec:reference,sec:complexity}.

Recall the ideal channel from~\eqref{eq:intro-ideal}. With the notation $\widehat{\mathcal T}=\Gamma^{-1}\mathcal T\Gamma$ from~\cref{subsec:preliminaries}, we write the channel and the generators for relative densities:
\begin{equation}\label{eq:ideal-channel}
 \widetilde\Phi_{\alpha,T}=\cU(T)e^{\alpha^2\cL}\cU(T),\qquad
 \hQ_{\alpha,T}=\Gamma^{-1}\Psi_{\alpha,T}\Gamma,\qquad
 \hL_{T}=\Gamma^{-1}\cL_{T}\Gamma.
\end{equation}
\red{Similarly, $\hL_{x,T}=\Gamma^{-1}\cL_{x,T}\Gamma$, $\hL_x:=\Gamma^{-1}\cL_x\Gamma$, and $\hL=\Gamma^{-1}\cL\Gamma$.}

By~\eqref{eq:second-order-expansion}, the channel $\hQ_{\alpha,T}$ and the exponential $e^{\alpha^2\hL_T}$ agree up to second order. At higher orders, both contain products that involve several sites. To compare these products, we group both expansions by site multiplicities. Recall $\Psi_T^{(\boldsymbol k)}$ from~\eqref{eq:Psi-k}, where $k_x$ counts the pairs of insertions at site $x$. For $\boldsymbol k\in\N_0^n$ with $m=|\boldsymbol k|$, define
\begin{equation}\label{eq:comparison-contributions}
\begin{aligned}
 \hQ_T^{(\boldsymbol k)}:=\Gamma^{-1}\Psi_T^{(\boldsymbol k)}\Gamma,\qquad
 \hL_T^{(\boldsymbol k)}:=\frac1{m!}
  \sum_{\substack{x_1,\ldots,x_m\in[n]\\
       \#\{j:x_j=x\}=k_x\ \forall x}}
       \hL_{x_1,T}\cdots\hL_{x_m,T}.
 \end{aligned}
\end{equation}
For $m=0$, the empty product is $\id$. $\hL_T^{(\boldsymbol k)}$ collects the ordered products in $\hL_T^m=(\sum_x\hL_{x,T})^m$, in which each site generator $\hL_{x,T}$ occurs exactly $k_x$ times. Thus, conjugating~\eqref{eq:sitewise-parity} by $\Gamma$ and expanding the exponential series give
\begin{equation}\label{eq:two-series}
\hQ_{\alpha,T}
   =\sum_{\boldsymbol k\in\N_0^n}\alpha^{2|\boldsymbol k|}\hQ_T^{(\boldsymbol k)},
 \qquad
 e^{\alpha^2\hL_T}
   =\sum_{\boldsymbol k\in\N_0^n}\alpha^{2|\boldsymbol k|}\hL_T^{(\boldsymbol k)}.
\end{equation}
By~\eqref{eq:psi_T_ex}, the terms with $|\boldsymbol k|\le1$ agree:
\[
 \hQ_T^{(\boldsymbol 0)}=\hL_T^{(\boldsymbol 0)}=\id,
 \qquad
 \hQ_T^{(\boldsymbol e_x)}=\hL_T^{(\boldsymbol e_x)}=\hL_{x,T}.
\]
We call $x$ an \emph{active site} of $\boldsymbol k$ if $k_x\ge1$. The remaining indices, with $|\boldsymbol k|\ge2$, fall into two classes: either $k_x\ge2$ for some $x$, or $\boldsymbol k\in\{0,1\}^n$. In the first class, some site carries at least four insertions, and a direct norm bound suffices (\cref{subsec:repeated}). In the second class, each active site carries exactly two insertions, so the two contributions differ only in the order of insertions at different sites, and we bound this difference by locality (\cref{subsec:reordering}).

\subsection{Contributions with a repeated site}\label{subsec:repeated}

We first bound both contributions in~\eqref{eq:comparison-contributions} by a product over sites. Summing this bound over the indices with $k_x\ge2$ at some site $x$ gives a fourth-order error with only one factor of $n$. The following lemma also shows that both series in~\eqref{eq:two-series} converge absolutely, which justifies regrouping their terms.

\begin{lemma}[Repeated-site contribution]\label{lem:majorant}
Let $C_\beta$ be the constant in~\Cref{cor:coupling-bounds}, enlarged so that $C_\beta\ge1$, and let
\[
 \eta:= 2\alpha^2Q^2C_\beta^2\|f_\sigma\|_1^2.
\]
For every $\boldsymbol k\in\N_0^n$ and $T\ge0$,
\begin{equation}\label{eq:majorant}
 \nm{\alpha^{2|\boldsymbol k|}\hQ_T^{(\boldsymbol k)}},\quad
 \nm{\alpha^{2|\boldsymbol k|}\hL_T^{(\boldsymbol k)}}
 \le\prod_{x=1}^n\frac{\eta^{k_x}}{k_x!}.
\end{equation}
Both series in~\eqref{eq:two-series} converge absolutely in the induced KMS norm, and
\begin{equation}\label{eq:repeated-site-bound}
 \sum_{\boldsymbol k:\,\exists x,\,k_x\ge2}
 \alpha^{2|\boldsymbol k|}\nm{\hQ_T^{(\boldsymbol k)}-\hL_T^{(\boldsymbol k)}}
 \le n\eta^2e^{n\eta}.
\end{equation}
\end{lemma}
\begin{proof}
\proofstep{The weighted cost of the insertions.}
By~\eqref{eq:single-bath-coupling}, the insertion with bath label $\ell=(x,a)$ uses
\begin{align*}
    W_\ell(t)=f_\sigma(t)A_{x,a}(t)\otimes b_\ell(t),\qquad b_{\ell}(t)=e^{i\omega_\ell t} B_\ell + e^{-i\omega_\ell t}B_\ell^\dagger,
\end{align*}
\red{Here $\omega_\ell=\omega_{x,a}$ and $B_\ell=B_{x,a}$, and} $\no{b_\ell(t)}=1$. Expanding the commutator in~\eqref{eq:sumexp} gives
\begin{equation}\label{eq:insertion-sides}
 \cV_\ell(t)(\varrho)=-if_\sigma(t)\bigl(A_{x,a}(t)\otimes b_\ell(t)\bigr)\varrho+if_\sigma(t)\,\varrho\bigl(A_{x,a}(t)\otimes b_\ell(t)\bigr),
\end{equation}
that is, a left multiplication and a right multiplication of the joint state $\varrho$ by $f_\sigma(t)A_{x,a}(t)\otimes b_\ell(t)$.

Now fix a word $\boldsymbol\ell=(\ell_1,\ldots,\ell_p)$, with $\ell_j=(x_j,a_j)$, and times $t_1<\cdots<t_p$. Expanding every commutator in $\cV_{\ell_p}(t_p)\cdots\cV_{\ell_1}(t_1)$ by~\eqref{eq:insertion-sides} produces $2^p$ terms, one for each choice of side for each insertion. More specifically, each term acts as
\[
 \varrho\mapsto\epsilon\Bigl(\prod_{j=1}^pf_\sigma(t_j)\Bigr)(L_S\otimes D_L)\,\varrho\,(R_S\otimes D_R),\qquad |\epsilon|=1,
\]
where $L_S$ and $R_S$ are the ordered products of the couplings $A_{x_j,a_j}(t_j)$ placed on the left and on the right, respectively, and $D_L$ and $D_R$ are the corresponding products of the bath operators $b_{\ell_j}(t_j)$. Its bath trace factorizes as
\[
 \Tr_E\bigl[(L_S\otimes D_L)(\rho\otimes\tau_{E,\boldsymbol\omega})(R_S\otimes D_R)\bigr]
 =\Tr_E(D_L\tau_{E,\boldsymbol\omega}D_R)\,L_S\rho R_S.
\]
Since $D_L$ and $D_R$ are products of norm-one bath operators, we have
\[
 |\Tr_E(D_L\tau_{E,\boldsymbol\omega}D_R)|
 \le\|D_L\tau_{E,\boldsymbol\omega}D_R\|_1
 \le\no{D_L}\|\tau_{E,\boldsymbol\omega}\|_1\no{D_R}
 \le1,
\]
uniformly in the frequencies. The system map $\rho\mapsto L_S\rho R_S$ is the composition, in time order, of the one-sided multiplications by $A_{x_j,a_j}(t_j)$. On relative densities, these become $\cB_{A_{x_j,a_j}(t_j-i\beta/4),\idmat}$ or $\cB_{\idmat,A_{x_j,a_j}(t_j+i\beta/4)}$ by~\eqref{eq:vertex-matrices}, and each of them has induced norm at most $C_\beta$ by~\eqref{eq:coupling-maps}. Hence, on relative densities, each term has induced KMS norm at most $\prod_{j=1}^pC_\beta f_\sigma(t_j)$. Since these bounds are uniform in $\boldsymbol\omega$, they also hold after the expectation $\E_{\boldsymbol\omega}$.

Summing over the $2^p$ choices of sides and over the $Q$ bath labels at each prescribed site, we obtain, for all sites $x_1,\ldots,x_p$ and times $t_1<\cdots<t_p$,
\begin{equation}\label{eq:insertion-bound}
 \sum_{a_1,\ldots,a_p\in[Q]}
 \nm{\Gamma^{-1}\E_{\boldsymbol\omega}\Tr_E\bigl[\cV_{\ell_p}(t_p)\cdots\cV_{\ell_1}(t_1)
     (\Gamma(\cdot)\otimes\tau_{E,\boldsymbol\omega})\bigr]}
 \le\prod_{j=1}^{p}2QC_\beta f_\sigma(t_j),
\end{equation}
where $\ell_j=(x_j,a_j)$.

\proofstep{The two contributions.}
Let $m=|\boldsymbol k|$. A word in~\eqref{eq:Psi-k} is determined by its sequence of sites $(x_1,\ldots,x_{2m})$, in which each site $x$ occurs $2k_x$ times, together with the bath labels $a_1,\ldots,a_{2m}\in[Q]$. There are $(2m)!/\prod_x(2k_x)!$ such site sequences. Applying~\eqref{eq:insertion-bound} to~\eqref{eq:Psi-k}, we obtain
\[
 \begin{aligned}
 \nm{\hQ_T^{(\boldsymbol k)}}
 \le&~ \frac{(2m)!}{\prod_x(2k_x)!}(2QC_\beta)^{2m}
       \int_{\Delta_{2m}(T)}\prod_{j=1}^{2m}f_\sigma(t_j)\,\dd\boldsymbol t\\
 = &~ \frac{(2QC_\beta)^{2m}}{\prod_x(2k_x)!}
       \left(\int_{-T}^T f_\sigma(t)\,\dd t\right)^{2m}\\
 \le &~ \prod_x\frac{(2QC_\beta\|f_\sigma\|_1)^{2k_x}}{(2k_x)!}
 \le\prod_x\frac{(2Q^2C_\beta^2\|f_\sigma\|_1^2)^{k_x}}{k_x!},
 \end{aligned}
\]
where the second step uses that the integrand is symmetric in $t_1,\ldots,t_{2m}$ and that the simplex $\Delta_{2m}(T)$ is one of $(2m)!$ parts of $[-T,T]^{2m}$. Multiplying by $\alpha^{2m}$ proves~\eqref{eq:majorant} for $\hQ_T^{(\boldsymbol k)}$. In particular, the case $\boldsymbol k=\boldsymbol e_x$ gives
\begin{align*}
    \nm{\hL_{x,T}}\le2Q^2C_\beta^2\|f_\sigma\|_1^2.
\end{align*}

For $\hL_T^{(\boldsymbol k)}$, the sum in~\eqref{eq:comparison-contributions} contains $m!/\prod_xk_x!$ ordered products. By the submultiplicativity of the induced KMS norm and the bound on $\nm{\hL_{x,T}}$,
\[
 \nm{\alpha^{2m}\hL_T^{(\boldsymbol k)}}
 \le\frac{\alpha^{2m}}{m!}\cdot\frac{m!}{\prod_xk_x!}
          \left(2Q^2C_\beta^2\|f_\sigma\|_1^2\right)^m
 =\prod_x\frac{\eta^{k_x}}{k_x!}.
\]
This proves the second bound in~\eqref{eq:majorant}. Summing the right-hand side of~\eqref{eq:majorant} over all $\boldsymbol k\in\N_0^n$ gives
\[
 \prod_{x=1}^n\sum_{j\ge0}\frac{\eta^j}{j!}=e^{n\eta}<\infty,
\]
so both series in~\eqref{eq:two-series} converge absolutely.

\proofstep{Indices with a repeated site.}
Every index in the sum in~\eqref{eq:repeated-site-bound} has at least one site $x$ with $k_x\ge2$. Hence, by the triangle inequality,
\begin{equation}\label{eq:repeated-union}
 \sum_{\boldsymbol k:\,\exists x,\,k_x\ge2}
   \alpha^{2|\boldsymbol k|}\nm{\hQ_T^{(\boldsymbol k)}-\hL_T^{(\boldsymbol k)}}
 \le\sum_{x=1}^n\sum_{\boldsymbol k:\,k_x\ge2}
   \alpha^{2|\boldsymbol k|}\left(\nm{\hQ_T^{(\boldsymbol k)}}+\nm{\hL_T^{(\boldsymbol k)}}\right).
\end{equation}
An index with several repeated sites appears on the right-hand side of~\eqref{eq:repeated-union} once for each of them. Since all terms are nonnegative, this overcounting only increases the bound. For a fixed site $x$,~\eqref{eq:majorant} gives
\[
 \sum_{\boldsymbol k:\,k_x\ge2}
   \alpha^{2|\boldsymbol k|}\left(\nm{\hQ_T^{(\boldsymbol k)}}+\nm{\hL_T^{(\boldsymbol k)}}\right)
 \le2\sum_{\boldsymbol k:\,k_x\ge2}\prod_{y=1}^n\frac{\eta^{k_y}}{k_y!}
 =2(e^\eta-1-\eta)e^{(n-1)\eta},
\]
where the last step sums over $k_x\ge2$ and over $k_y\ge0$ for each $y\neq x$ separately. Since
\[
 e^\eta-1-\eta=\int_0^\eta(\eta-u)e^u\,\dd u\le\frac12\eta^2e^\eta,
\]
the right-hand side of~\eqref{eq:repeated-union} is at most $n\eta^2e^{n\eta}$. This proves~\eqref{eq:repeated-site-bound}.
\end{proof}

\subsection{Reordering contributions with two insertions per site}\label{subsec:reordering}

In this subsection, we bound $\hQ_T^{(\boldsymbol k)}-\hL_T^{(\boldsymbol k)}$ for $\boldsymbol k\in\{0,1\}^n$ with $|\boldsymbol k|\ge2$. Let $S=\{x:k_x=1\}$ be the set of active sites and $p=|S|$. Since each active site occurs exactly once in~\eqref{eq:comparison-contributions}, the comparison target is an average of products of site generators:
\begin{equation}\label{eq:target-average}
 \hL_T^{(\boldsymbol k)}=\frac1{p!}\sum_{\pi}\hL_{\pi(1),T}\cdots\hL_{\pi(p),T},
\end{equation}
where $\pi$ runs over the $p!$ orderings $\pi(1),\ldots,\pi(p)$ of $S$. We compare $\hQ_T^{(\boldsymbol k)}$ with each product $\hL_{\pi(1),T}\cdots\hL_{\pi(p),T}$. \Cref{lem:two-forms} writes both as integrals of the same terms over the same domain, which differ only in the order of their factors, and~\Cref{lem:reorder-cost} bounds the cost of reordering by locality.

We first write each site generator in terms of one-sided multiplications. Let $C_\beta\ge1$ be the constant in~\Cref{lem:majorant}, and define the normalized maps
\[
 \mathcal V^L_{x,a}(t)=C_\beta^{-1}\cB_{A_{x,a}(t-i\beta/4),\idmat},
 \qquad
 \mathcal V^R_{x,a}(t)=C_\beta^{-1}\cB_{\idmat,A_{x,a}(t+i\beta/4)}.
\]
By~\eqref{eq:vertex-matrices}, these are the left and right multiplications by $A_{x,a}(t)$ acting on relative densities, divided by $C_\beta$, and by~\eqref{eq:coupling-maps}, both have induced norm at most one. The following lemma expands a site generator in terms of $\mathcal V^L_{x,a}$ and $\mathcal V^R_{x,a}$.

\begin{lemma}[Branch form of the site generators]\label{lem:branch-form}
For every $x\in[n]$ and $T\ge0$,
\begin{equation}\label{eq:pair-branch}
 \begin{aligned}
 \hL_{x,T}=C_\beta^2\sum_a\int_{-T<u<t<T}f_\sigma(t)f_\sigma(u)
 \Bigl[&c_{\beta,\sigma}(u-t)\bigl(\mathcal V^L_{x,a}(t)\mathcal V^R_{x,a}(u)
                 -\mathcal V^R_{x,a}(t)\mathcal V^R_{x,a}(u)\bigr)\\
       +&c_{\beta,\sigma}(t-u)\bigl(\mathcal V^R_{x,a}(t)\mathcal V^L_{x,a}(u)
                 -\mathcal V^L_{x,a}(t)\mathcal V^L_{x,a}(u)\bigr)
 \Bigr]\,\dd u\,\dd t.
 \end{aligned}
\end{equation}
\end{lemma}
\begin{proof}
By~\eqref{eq:GKLS}, $\cL_{x,T}(\rho)=\cJ_{x,T}(\rho)-Q_{x,T}\rho-\rho Q_{x,T}^\dagger$. As in the proof of~\eqref{eq:site-norm}, substituting~\eqref{eq:filtered-operator} into~\eqref{eq:site-jump} and using~\eqref{eq:filters} gives
\[
 \cJ_{x,T}(\rho)=\sum_a\iint_{[-T,T]^2}f_\sigma(t)f_\sigma(s)c_{\beta,\sigma}(s-t)A_{x,a}(t)\rho A_{x,a}(s)\,\dd s\,\dd t.
\]
We split the square into the triangles $s<t$ and $s>t$, and rename the times so that $u<t$ in both. Together with~\eqref{eq:Q} and $\overline{c_{\beta,\sigma}(t-u)}=c_{\beta,\sigma}(u-t)$, this gives
\[
 \begin{aligned}
 \cL_{x,T}(\rho)=\sum_a\int_{-T<u<t<T}f_\sigma(t)f_\sigma(u)
 \Bigl[&c_{\beta,\sigma}(u-t)\bigl(A(t)\rho A(u)-\rho A(u)A(t)\bigr)\\
       +&c_{\beta,\sigma}(t-u)\bigl(A(u)\rho A(t)-A(t)A(u)\rho\bigr)\Bigr]\,\dd u\,\dd t,
 \end{aligned}
\]
where $A=A_{x,a}$. By~\eqref{eq:vertex-matrices} and the composition rule in~\eqref{eq:weighted-algebra}, the four maps become, on relative densities,
\[
 \begin{aligned}
 \Gamma^{-1}\bigl(A(t)\Gamma(\cdot)A(u)\bigr)&=C_\beta^2\mathcal V^L_{x,a}(t)\mathcal V^R_{x,a}(u),&
 \Gamma^{-1}\bigl(\Gamma(\cdot)A(u)A(t)\bigr)&=C_\beta^2\mathcal V^R_{x,a}(t)\mathcal V^R_{x,a}(u),\\
 \Gamma^{-1}\bigl(A(u)\Gamma(\cdot)A(t)\bigr)&=C_\beta^2\mathcal V^R_{x,a}(t)\mathcal V^L_{x,a}(u),&
 \Gamma^{-1}\bigl(A(t)A(u)\Gamma(\cdot)\bigr)&=C_\beta^2\mathcal V^L_{x,a}(t)\mathcal V^L_{x,a}(u),
 \end{aligned}
\]
where the second identity uses the fact that right multiplications compose in reverse order. Substituting these identities proves~\eqref{eq:pair-branch}.
\end{proof}

We call the four products $\mathcal V^s_{x,a}(t)\mathcal V^{s'}_{x,a}(u)$, $s,s'\in\{L,R\}$, in~\eqref{eq:pair-branch} the \emph{branches} of site $x$, and we write $w^{ss'}(u,t)=\pm C_\beta^2f_\sigma(t)f_\sigma(u) \red{c_{\beta,\sigma}}(\pm (t-u))$ for the scalar weight of $\mathcal V^s_{x,a}(t)\mathcal V^{s'}_{x,a}(u)$ in~\eqref{eq:pair-branch}. Then, $|w^{ss'}(u,t)|\le C_\beta^2f_\sigma(t)f_\sigma(u)$ since $|c_{\beta,\sigma}|\le1$.
We now align the product of site generators and $\hQ_T^{(\boldsymbol k)}$ as integrals of the same branches over the same domain.

\begin{lemma}[Grouped and time-ordered forms]\label{lem:two-forms}
For every ordering $\pi$ of $S$,
\begin{align}
 \hL_{\pi(1),T}\cdots\hL_{\pi(p),T}
 &=\sum_{\boldsymbol a\in [Q]^S,\boldsymbol s,\boldsymbol s'\in \{L,R\}^S}\int\Bigl(\prod_{x\in S}w^{s_xs'_x}(u_x,t_x)\Bigr)
   \prod_{x\in S}^{\pi}\bigl(\mathcal V^{s_x}_{x,a_x}(t_x)\mathcal V^{s'_x}_{x,a_x}(u_x)\bigr)\,\dd\boldsymbol u\,\dd\boldsymbol t,
 \label{eq:grouped-form}\\
 \hQ_T^{(\boldsymbol k)}
 &=\sum_{\boldsymbol a\in [Q]^S,\boldsymbol s,\boldsymbol s'\in \{L,R\}^S}\int\Bigl(\prod_{x\in S}w^{s_xs'_x}(u_x,t_x)\Bigr)
   \,\mathcal T\Bigl[\prod_{x\in S}\mathcal V^{s_x}_{x,a_x}(t_x)\mathcal V^{s'_x}_{x,a_x}(u_x)\Bigr]\,\dd\boldsymbol u\,\dd\boldsymbol t.
 \label{eq:time-ordered-form}
\end{align}
Here the integrals run over $-T<u_x<t_x<T$ for $x\in S$. The product $\prod^{\pi}_{x\in S}$ is taken in the order $\pi(1),\ldots,\pi(p)$ from left to right, and $\mathcal T$ arranges the $2p$ maps in decreasing order of their times from left to right.
\end{lemma}
\begin{proof}
For~\eqref{eq:grouped-form}, we multiply~\eqref{eq:pair-branch} over $x\in S$ in the order $\pi$ and combine the integrals over the independent pairs $(u_x,t_x)$.

For~\eqref{eq:time-ordered-form}, we start from~\eqref{eq:Psi-k}, in which $\hQ_T^{(\boldsymbol k)}$ is a sum over words with two insertions at each site of $S$, integrated over ordered times. By~\Cref{lem:bath-parity}~(iv), only words in which the two insertions at each $x\in S$ carry the same bath label $(x,a_x)$ contribute, which gives the sum over $\boldsymbol a$. Denote the two insertion times at $x$ by $u_x<t_x$. The order of these $2p$ times is a \emph{shuffle}, that is, a total order of the times $\{u_x,t_x\}_{x\in S}$ that preserves $u_x<t_x$ for every $x$; see~\Cref{fig:p54-shuffles}. A word with its ordered times is thus the same as a shuffle together with times ordered according to it. Since distinct times have a unique order, the regions of the different shuffles partition the product domain $\prod_{x\in S}\{-T<u_x<t_x<T\}$ up to a null set. Hence, summing over shuffles turns the ordered integrals in~\eqref{eq:Psi-k} into integrals over this product domain, without a combinatorial factor.

\begin{figure}[!htbp]
\centering
\begingroup
\definecolor{p54blue}{RGB}{39,91,133}
\definecolor{p54orange}{RGB}{176,92,36}
\newcommand{\pfaxis}[2]{\draw[-{Latex[length=1.4mm]},black!40,line width=.5pt] (#1-.2,#2)--(#1+2.45,#2);}
\newcommand{\pfshuffle}[7]{%
 \pfaxis{#2}{#3}
 \node[circle,fill=p54blue,inner sep=1.6pt,label={[font=\scriptsize,text=p54blue,inner sep=1pt]below:$u_x$}] (#1ux) at (#2+.55*#4-.3,#3) {};
 \node[circle,fill=p54blue,inner sep=1.6pt,label={[font=\scriptsize,text=p54blue,inner sep=1pt]below:$t_x$}] (#1tx) at (#2+.55*#5-.3,#3) {};
 \node[circle,fill=p54orange,inner sep=1.6pt,label={[font=\scriptsize,text=p54orange,inner sep=1pt]below:$u_y$}] (#1uy) at (#2+.55*#6-.3,#3) {};
 \node[circle,fill=p54orange,inner sep=1.6pt,label={[font=\scriptsize,text=p54orange,inner sep=1pt]below:$t_y$}] (#1ty) at (#2+.55*#7-.3,#3) {};
 \draw[p54blue,line width=.8pt] (#1ux) to[out=75,in=105] (#1tx);
 \draw[p54orange,line width=.8pt] (#1uy) to[out=75,in=105] (#1ty);
}
\begin{tikzpicture}[x=1cm,y=1cm]
\node[font=\small,align=center] at (1.05,2.05) {one ordered pair\\at each site};
\draw[-{Latex[length=1.4mm]},black!40,line width=.5pt] (-.2,1.05)--(2.35,1.05);
\node[circle,fill=p54blue,inner sep=1.6pt,label={[font=\scriptsize,text=p54blue,inner sep=1pt]below:$u_x$}] (lux) at (.45,1.05) {};
\node[circle,fill=p54blue,inner sep=1.6pt,label={[font=\scriptsize,text=p54blue,inner sep=1pt]below:$t_x$}] (ltx) at (1.65,1.05) {};
\draw[p54blue,line width=.8pt] (lux) to[out=75,in=105] (ltx);
\draw[-{Latex[length=1.4mm]},black!40,line width=.5pt] (-.2,-.25)--(2.35,-.25);
\node[circle,fill=p54orange,inner sep=1.6pt,label={[font=\scriptsize,text=p54orange,inner sep=1pt]below:$u_y$}] (luy) at (.45,-.25) {};
\node[circle,fill=p54orange,inner sep=1.6pt,label={[font=\scriptsize,text=p54orange,inner sep=1pt]below:$t_y$}] (lty) at (1.65,-.25) {};
\draw[p54orange,line width=.8pt] (luy) to[out=75,in=105] (lty);
\node[font=\scriptsize,text=p54blue,anchor=east] at (-.25,1.05) {site $x$};
\node[font=\scriptsize,text=p54orange,anchor=east] at (-.25,-.25) {site $y$};
\draw[-{Latex[length=2mm]},line width=.7pt] (2.75,.45)--(3.85,.45);
\node[font=\small] at (5.35,2.05) {separated};
\node[font=\small] at (8.25,2.05) {crossing};
\node[font=\small] at (11.15,2.05) {nested};
\pfshuffle{a}{4.35}{1.05}{1}{2}{3}{4}
\pfshuffle{b}{7.25}{1.05}{1}{3}{2}{4}
\pfshuffle{c}{10.15}{1.05}{1}{4}{2}{3}
\pfshuffle{d}{4.35}{-.25}{3}{4}{1}{2}
\pfshuffle{e}{7.25}{-.25}{2}{4}{1}{3}
\pfshuffle{f}{10.15}{-.25}{2}{3}{1}{4}
\node[font=\scriptsize,black!60,anchor=west] at (12.7,1.05) {$u_x<u_y$};
\node[font=\scriptsize,black!60,anchor=west] at (12.7,-.25) {$u_y<u_x$};
\end{tikzpicture}
\endgroup
\caption{The shuffles for two active sites $x$ and $y$ with insertion times $u_x<t_x$ and $u_y<t_y$, where the arcs connect the two times of the same site. Every choice of four distinct times gives exactly one shuffle, so the regions of the six shuffles partition the product of the two pair domains up to a null set.}
\label{fig:p54-shuffles}
\end{figure}
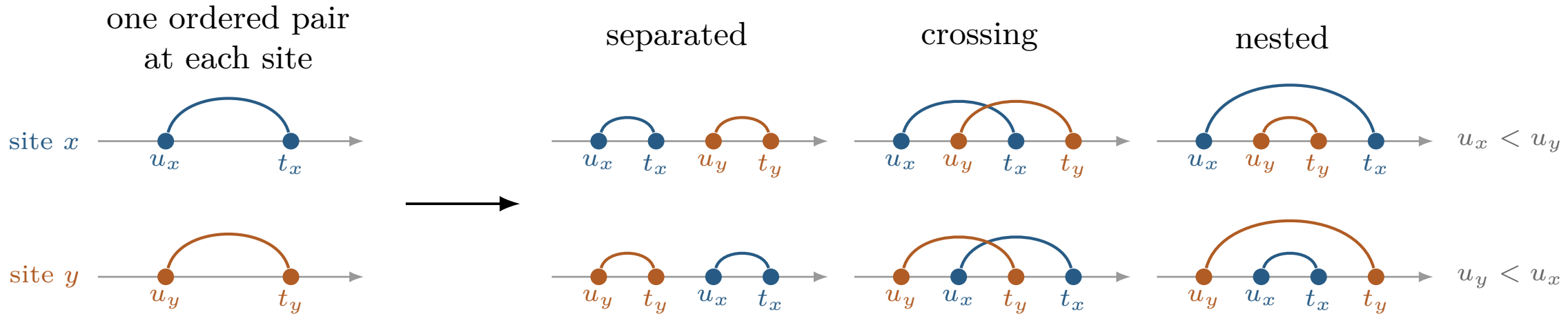

Next, we expand each commutator by~\eqref{eq:insertion-sides}, choosing the side $s_x$ at time $t_x$ and the side $s'_x$ at time $u_x$. By~\eqref{eq:vertex-matrices}, the system factors then form $C_\beta^{2p}$ times the time-ordered product on relative densities in~\eqref{eq:time-ordered-form}. They are multiplied by the pulse factors, the signs from~\eqref{eq:insertion-sides}, and the frequency-averaged bath trace. It remains to show that these scalars combine into $\prod_{x\in S}w^{s_xs'_x}(u_x,t_x)$.

The two bath operators of site $x$ act on the qubit $E_{x,a_x}$, and bath operators of different sites act on different qubits and hence commute. Since the initial bath state $\tau_{E,\boldsymbol\omega}$ is a product state and the frequencies are independent, the averaged bath trace is the product over $x\in S$ of single-site averages. After a cyclic permutation, the two bath operators $b_\ell(u_x)$ and $b_\ell(t_x)$ of site $x$, with $\ell=(x,a_x)$, appear in the order $b_\ell(u_x)b_\ell(t_x)$ if $s'_x=R$ and in the order $b_\ell(t_x)b_\ell(u_x)$ if $s'_x=L$. By the proof of~\Cref{prop:second}, $\E_\omega\Tr[b_\omega(t_1)b_\omega(t_2)\tau_\omega]=c_{\beta,\sigma}(t_1-t_2)$.

Then, by comparing with~\eqref{eq:pair-branch}, we can see that the scalar factor of site $x$ is exactly $w^{s_xs'_x}(u_x,t_x)$. This proves~\eqref{eq:time-ordered-form}.
\end{proof}

By~\Cref{lem:two-forms}, $\hL_{\pi(1),T}\cdots\hL_{\pi(p),T}$ and $\hQ_T^{(\boldsymbol k)}$ have the same weights and the same normalized maps, and they differ only in the order of the maps. The next lemma bounds the cost of reordering them.

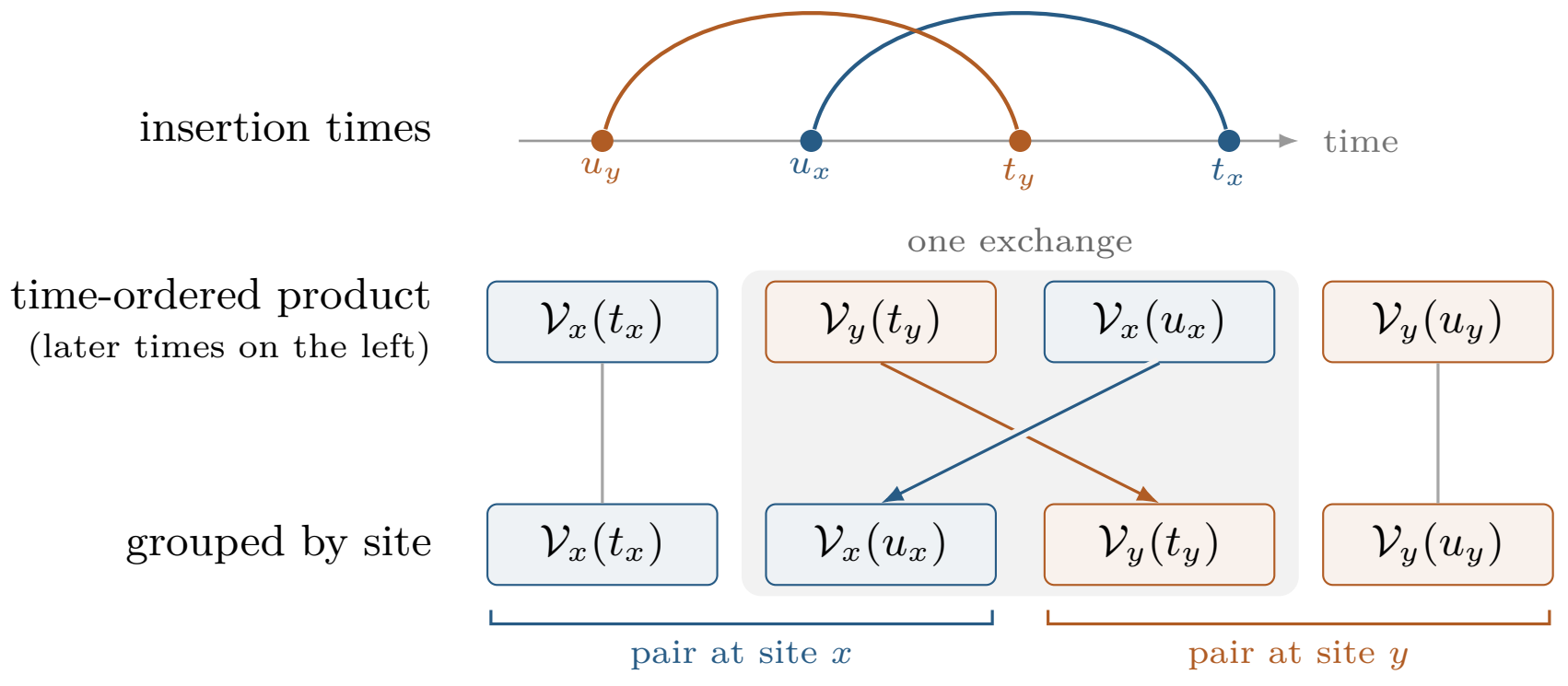
\begin{figure}[!htbp]
\centering
\begingroup
\definecolor{p54blue}{RGB}{39,91,133}
\definecolor{p54orange}{RGB}{176,92,36}
\begin{tikzpicture}[x=1cm,y=1cm,
 fx/.style={draw=p54blue,fill=p54blue!8,rounded corners=2pt,
   minimum width=1.65cm,minimum height=.58cm,inner sep=2pt,font=\small},
 fy/.style={draw=p54orange,fill=p54orange!8,rounded corners=2pt,
   minimum width=1.65cm,minimum height=.58cm,inner sep=2pt,font=\small},
 lbl/.style={font=\small,anchor=east,align=right},
 link/.style={line width=.6pt,black!35},
 arr/.style={-{Latex[length=1.6mm]},line width=.6pt}]
\draw[-{Latex[length=1.4mm]},black!40,line width=.5pt] (3.0,2.15)--(8.6,2.15);
\node[font=\scriptsize,black!55,anchor=west] at (8.65,2.15) {time};
\node[circle,fill=p54orange,inner sep=1.6pt,label={[font=\scriptsize,text=p54orange,inner sep=1pt]below:$u_y$}] (uy) at (3.6,2.15) {};
\node[circle,fill=p54blue,inner sep=1.6pt,label={[font=\scriptsize,text=p54blue,inner sep=1pt]below:$u_x$}] (ux) at (5.1,2.15) {};
\node[circle,fill=p54orange,inner sep=1.6pt,label={[font=\scriptsize,text=p54orange,inner sep=1pt]below:$t_y$}] (ty) at (6.6,2.15) {};
\node[circle,fill=p54blue,inner sep=1.6pt,label={[font=\scriptsize,text=p54blue,inner sep=1pt]below:$t_x$}] (tx) at (8.1,2.15) {};
\draw[p54blue,line width=.8pt] (ux) to[out=75,in=105] (tx);
\draw[p54orange,line width=.8pt] (uy) to[out=75,in=105] (ty);
\node[lbl] at (2.5,2.25) {insertion times};
\fill[black!5,rounded corners=4pt] (4.6,-1.12) rectangle (8.6,1.22);
\node[font=\scriptsize,black!60] at (6.6,1.42) {one exchange};
\node[lbl] at (2.5,.85) {time-ordered product\\[-1pt]{\scriptsize(later times on the left)}};
\node[lbl] at (2.5,-.75) {grouped by site};
\node[fx] (a1) at (3.6,.85) {$\mathcal V_x(t_x)$};
\node[fy] (a2) at (5.6,.85) {$\mathcal V_y(t_y)$};
\node[fx] (a3) at (7.6,.85) {$\mathcal V_x(u_x)$};
\node[fy] (a4) at (9.6,.85) {$\mathcal V_y(u_y)$};
\node[fx] (b1) at (3.6,-.75) {$\mathcal V_x(t_x)$};
\node[fx] (b2) at (5.6,-.75) {$\mathcal V_x(u_x)$};
\node[fy] (b3) at (7.6,-.75) {$\mathcal V_y(t_y)$};
\node[fy] (b4) at (9.6,-.75) {$\mathcal V_y(u_y)$};
\draw[link] (a1.south)--(b1.north);
\draw[link] (a4.south)--(b4.north);
\draw[arr,p54orange] (a2.south)--(b3.north);
\draw[arr,p54blue,preaction={draw=black!5,line width=2.6pt}] (a3.south)--(b2.north);
\draw[p54blue,line width=.6pt] (2.8,-1.22)--(2.8,-1.32)--(6.4,-1.32)--(6.4,-1.22);
\node[font=\scriptsize,text=p54blue] at (4.6,-1.55) {pair at site $x$};
\draw[p54orange,line width=.6pt] (6.8,-1.22)--(6.8,-1.32)--(10.4,-1.32)--(10.4,-1.22);
\node[font=\scriptsize,text=p54orange] at (8.6,-1.55) {pair at site $y$};
\end{tikzpicture}
\endgroup
\caption{Grouping the factors of one term by site, for the crossing shuffle $u_y<u_x<t_y<t_x$. Here, the exchange incurs an error $\mathcal V_x(t_x)\mathcal V_y(t_y)\mathcal V_x(u_x)\mathcal V_y(u_y)
 -\mathcal V_x(t_x)\mathcal V_x(u_x)\mathcal V_y(t_y)\mathcal V_y(u_y)
 =\mathcal V_x(t_x)\,[\mathcal V_y(t_y),\mathcal V_x(u_x)]\,\mathcal V_y(u_y)$, with a norm at most $\nm{[\mathcal V_y(t_y),\mathcal V_x(u_x)]}$.
}
\label{fig:p54-exchange}
\end{figure}

\begin{lemma}[The cost of reordering]\label{lem:reorder-cost}
\red{There are constants $C,c>0$ and $v\ge0$, depending only on the fixed local parameters, such that the following bounds hold.}
\begin{enumerate}[label=(\roman*)]
 \item Let $\mathcal V_x(t)$ and $\mathcal V_y(u)$ be normalized maps of the form $\mathcal V^L$ or $\mathcal V^R$, at sites $x\ne y$. Then
 \begin{equation}\label{eq:insertion-commutator}
  \nm{[\mathcal V_x(t),\mathcal V_y(u)]}
  \le C\min\{1,e^{v(|t|+|u|)-c|x-y|}\}.
 \end{equation}
 \item For every ordering $\pi$ of $S$, all $\boldsymbol a,\boldsymbol s,\boldsymbol s'$, and all times with $u_x<t_x$,
 \begin{equation}\label{eq:reorder-cost}
  \begin{aligned}
  &\nm{\mathcal T\Bigl[\prod_{x\in S}\mathcal V^{s_x}_{x,a_x}(t_x)\mathcal V^{s'_x}_{x,a_x}(u_x)\Bigr]
     -\prod_{x\in S}^{\pi}\bigl(\mathcal V^{s_x}_{x,a_x}(t_x)\mathcal V^{s'_x}_{x,a_x}(u_x)\bigr)}\\
  &\qquad\le4C\sum_{\{x,y\}\subseteq S}\min\{1,e^{v(|u_x|+|t_x|+|u_y|+|t_y|)-c|x-y|}\}.
  \end{aligned}
 \end{equation}
\end{enumerate}
\end{lemma}
\begin{proof}
\red{Let $v,c_1$ be the constants from~\Cref{cor:coupling-bounds}, and set $c=c_1/3$.} For Part~(i), if the two maps multiply on opposite sides, they commute by~\eqref{eq:weighted-algebra}. If they multiply on the same side, then~\eqref{eq:weighted-algebra} gives
\[
 [\cB_{B,\idmat},\cB_{D,\idmat}]=\cB_{[B,D],\idmat},
 \qquad
 [\cB_{\idmat,B},\cB_{\idmat,D}]=\cB_{\idmat,[D,B]},
\]
so the commutator has norm at most $\no{[B,D]}$, where $B$ and $D$ are the normalized coefficients $C_\beta^{-1}A_{x,a}(t\mp i\beta/4)$ and $C_\beta^{-1}A_{y,b}(u\mp i\beta/4)$, with the same sign. By~\eqref{eq:vertex-bound}, $\no B,\no D\le1$, and hence $\no{[B,D]}\le2$. For the decay in $|x-y|$, let $r=\lfloor(|x-y|-1)/3\rfloor$. Since $2r<|x-y|$, the operators $\cE_{x,r}(B)$ and $\cE_{y,r}(D)$ have disjoint supports and therefore commute. Thus,
\begin{equation*}
 \begin{aligned}
 \no{[B,D]}=&~ \no{[B,D]-[\cE_{x,r}(B),\cE_{y,r}(D)]}\\
    = &~ \no{[B-\cE_{x,r}(B),D]+[\cE_{x,r}(B),D-\cE_{y,r}(D)]}\\
    \le &~ 2\no{B-\cE_{x,r}(B)}\no{D}
       +2\no{\cE_{x,r}(B)}\no{D-\cE_{y,r}(D)}\\
 \le &~ \red{2e^{v|t|-c_1r}+2e^{v|u|-c_1r}}
 \red{\le Ce^{v(|t|+|u|)-c|x-y|},}
 \end{aligned}
\end{equation*}
where the fourth step uses the contractivity of $\cE_{x,r}$ and the tail bound in~\Cref{cor:coupling-bounds}, and the last step uses \red{$r\ge(|x-y|-4)/3$ and $c=c_1/3$}. Combining this bound with $\no{[B,D]}\le2$ proves~\eqref{eq:insertion-commutator}.

For Part~(ii), we pass from the time-ordered product to the grouped product by exchanging adjacent factors. Each exchange changes the product by a commutator:
\[
 UABV-UBAV=U[A,B]V,
 \qquad \nm{UABV-UBAV}\le\nm{[A,B]},
\]
since all factors in $U$ and $V$ have norm at most one. \Cref{fig:p54-exchange} shows one exchange. We reorder all factors in the first term of~\eqref{eq:reorder-cost} according to the positions of their sites in $\pi$. At each step, we exchange two adjacent factors whose sites appear in the wrong order with respect to $\pi$, as in bubble sort. Such an exchange puts one pair of factors into the correct relative order, and the relative order of two factors changes only when they are interchanged. Hence, each pair of factors at different sites is exchanged at most once, and the two factors at the same site are never exchanged. This procedure will end with the grouped product. Consequently, a pair of sites $\{x,y\}$ accounts for at most $2\cdot2=4$ exchanges, each involving two of the times $u_x,t_x,u_y,t_y$. Telescoping over the exchanges and applying Part~(i) proves~\eqref{eq:reorder-cost}.
\end{proof}

We now integrate~\eqref{eq:reorder-cost} against the weights. For distinct sites $x,y$, the integrated cost of the pair is
\begin{equation}\label{eq:swap-weight}
 \kappa_{xy}:=\iint_{u<t}\iint_{u'<t'}f_\sigma(t)f_\sigma(u)f_\sigma(t')f_\sigma(u')
 \min\{1,e^{v(|t|+|u|+|t'|+|u'|)-c|x-y|}\}\,\dd u\,\dd t\,\dd u'\,\dd t',
\end{equation}
with $c$ and $v$ as in~\eqref{eq:insertion-commutator}.

\begin{lemma}[Two insertions per active site]\label{lem:reordering}
Let $\boldsymbol k\in\{0,1\}^n$ with $p=|\boldsymbol k|\ge2$, and let $S$ be its set of active sites. Then
\begin{equation}\label{eq:reordering}
 \nm{\hQ_T^{(\boldsymbol k)}-\hL_T^{(\boldsymbol k)}}
 \le C\bigl(2QC_\beta^2\|f_\sigma\|_1^2\bigr)^{p-2}\sum_{\{x,y\}\subseteq S}\kappa_{xy},
\end{equation}
where $C$ depends only on $\beta$ and $q$.
\end{lemma}
\begin{proof}
Fix an ordering $\pi$ of $S$. We subtract~\eqref{eq:grouped-form} from~\eqref{eq:time-ordered-form}, and apply~\eqref{eq:reorder-cost} and $|w^{ss'}(u,t)|\le C_\beta^2f_\sigma(t)f_\sigma(u)$. Summing over the $Q$ bath labels and the four branches at each site, we obtain
\[
 \begin{aligned}
 &\nm{\hQ_T^{(\boldsymbol k)}-\hL_{\pi(1),T}\cdots\hL_{\pi(p),T}}\\
 &\qquad\le4C\sum_{\{x,y\}\subseteq S}\int\Bigl(\prod_{z\in S}4QC_\beta^2f_\sigma(t_z)f_\sigma(u_z)\Bigr)
 \min\{1,e^{v(|u_x|+|t_x|+|u_y|+|t_y|)-c|x-y|}\}\,\dd\boldsymbol u\,\dd\boldsymbol t,
 \end{aligned}
\]
where the integral runs over $-T<u_z<t_z<T$ for $z\in S$. For each pair $\{x,y\}$, the integral over the times of $x$ and $y$ is at most $(4QC_\beta^2)^2\kappa_{xy}$, and the integral over the times of each other site is at most $4QC_\beta^2\cdot\frac12\|f_\sigma\|_1^2$, since its integrand is symmetric in $u$ and $t$. Hence,
\[
 \nm{\hQ_T^{(\boldsymbol k)}-\hL_{\pi(1),T}\cdots\hL_{\pi(p),T}}
 \le4C(4QC_\beta^2)^2\bigl(2QC_\beta^2\|f_\sigma\|_1^2\bigr)^{p-2}\sum_{\{x,y\}\subseteq S}\kappa_{xy}.
\]
This bound is uniform in $\pi$. Averaging it over the $p!$ orderings, using~\eqref{eq:target-average} and the triangle inequality, proves~\eqref{eq:reordering}.
\end{proof}

\subsection{The comparison theorem}\label{subsec:comparison-theorem}

We now combine the two classes of contributions into a comparison at a fixed interaction window (\Cref{prop:finite-comparison}). We then replace $\cL_T$ by its infinite-window limit $\cL$ with the \green{forward evolution} $\cU(T)$, which gives the comparison of the physical and ideal channels (\Cref{thm:comparison}).

\begin{proposition}[Finite-window comparison]\label{prop:finite-comparison}
There is a constant $C>0$, depending only on $\beta$ and the number $q$ of qubits per site, such that, for $\sigma=\beta/2$, every $T\ge0$, and every $\alpha\in\R$,
\begin{equation}\label{eq:finite-comparison}
 \nm{\hQ_{\alpha,T}-e^{\alpha^2\hL_{T}}}
 \le Cn\alpha^4\sigma^2e^{Cn\alpha^2\sigma}.
\end{equation}
\end{proposition}
\begin{proof}
By~\Cref{lem:majorant}, both series in~\eqref{eq:two-series} converge absolutely. Hence, we have
\[
 \nm{\hQ_{\alpha,T}-e^{\alpha^2\hL_{T}}}
 \le\sum_{\boldsymbol k\in\N_0^n}\alpha^{2|\boldsymbol k|}\nm{\hQ_T^{(\boldsymbol k)}-\hL_T^{(\boldsymbol k)}}.
\]
The terms with $|\boldsymbol k|\le1$ vanish. We bound the remaining terms separately for $\boldsymbol k\in\{0,1\}^n$ and for the indices with a repeated site.

\proofstep{Indices without a repeated site.}
For $\boldsymbol k\in\{0,1\}^n$ with $|\boldsymbol k|\ge2$, we apply~\Cref{lem:reordering}, with $S$ the set of active sites of $\boldsymbol k$. Since $Q\ge1$, the parameter $\eta$ of~\Cref{lem:majorant} satisfies $2\alpha^2QC_\beta^2\|f_\sigma\|_1^2\le\eta$. Summing first over the pair $\{x,y\}$ and then over the remaining active sites $S'=S\setminus\{x,y\}$, we obtain
\[
 \begin{aligned}
 \sum_{\substack{\boldsymbol k\in\{0,1\}^n\\|\boldsymbol k|\ge2}}\alpha^{2|\boldsymbol k|}
       \nm{\hQ_T^{(\boldsymbol k)}-\hL_T^{(\boldsymbol k)}}
 &\le C\alpha^4\sum_{x<y}\kappa_{xy}
       \sum_{S'\subseteq[n]\setminus\{x,y\}}\eta^{|S'|}\\
 &=C\alpha^4(1+\eta)^{n-2}\sum_{x<y}\kappa_{xy}
 \le C\alpha^4e^{n\eta}\sum_{x<y}\kappa_{xy},
 \end{aligned}
\]
where each site outside $\{x,y\}$ is either inactive or contributes a factor at most $\eta$.

It remains to bound $\sum_{x<y}\kappa_{xy}$. Fix $x$ and $\tau\ge0$. The summand $\min\{1,e^{v\tau-c|x-y|}\}$ equals one when $|x-y|\le v\tau/c$ and decays exponentially beyond this distance. Since at most two sites of the chain lie at each distance $d\ge1$ from $x$, we have
\[
 \sum_{y:\,y\ne x}\min\{1,e^{v\tau-c|x-y|}\}
 \le2\sum_{1\le d<v\tau/c}1+2\sum_{d\ge v\tau/c}e^{-c(d-v\tau/c)}
 \le\frac{2v\tau}{c}+\frac{2}{1-e^{-c}}
 \le C(1+\tau).
\]
Substituting into~\eqref{eq:swap-weight} with $\tau=|t|+|u|+|t'|+|u'|$, we have
\[
 \begin{aligned}
 \sum_{x<y}\kappa_{xy}
 \le &~ \sum_{x=1}^n\sum_{y:\,y\ne x}\kappa_{xy}\\
 \le &~  Cn\iint_{u<t}\iint_{u'<t'}f_\sigma(t)f_\sigma(u)f_\sigma(t')f_\sigma(u')
       (1+|t|+|u|+|t'|+|u'|)\,\dd u\,\dd t\,\dd u'\,\dd t'\\
 = &~ Cn\left(\frac14\|f_\sigma\|_1^4+\|f_\sigma\|_1^3\int_\R|t|f_\sigma(t)\,\dd t\right)
 =Cn\|f_\sigma\|_1^4\left(\frac14+\frac{2\sigma}{\sqrt\pi}\right)
 \le Cn\sigma^2,
 \end{aligned}
\]
where the third step uses the symmetry of the integrand under $u\leftrightarrow t$ and $u'\leftrightarrow t'$, the fourth step uses $\int_\R|t|f_\sigma(t)\,\dd t=(2\sigma/\sqrt\pi)\|f_\sigma\|_1$ for the Gaussian pulse~\eqref{eq:pulse}, and the last step uses $\|f_\sigma\|_1^2=2\sqrt{2\pi}\,\sigma$ and $\sigma=\beta/2$. Therefore, the contributions with $\boldsymbol k\in\{0,1\}^n$ are at most $Cn\alpha^4\sigma^2e^{n\eta}$.

\proofstep{Indices with a repeated site.}
For these indices,~\Cref{lem:majorant} gives the bound $n\eta^2e^{n\eta}$. By~\eqref{eq:pulse},
\[
 \eta=2Q^2C_\beta^2\alpha^2\|f_\sigma\|_1^2
      =4\sqrt{2\pi}\,Q^2C_\beta^2\alpha^2\sigma,
\]
so this error is also at most $Cn\alpha^4\sigma^2e^{n\eta}$.

\smallskip

\noindent
Finally, adding the two bounds and using $\eta\le C\alpha^2\sigma$ proves~\eqref{eq:finite-comparison}.
\end{proof}

We are ready to compare the physical channel $\Phi_{\alpha,T}$ and the ideal channel $\widetilde\Phi_{\alpha,T}$ in~\eqref{eq:ideal-channel}.

\begin{theorem}[One-round approximation]\label{thm:comparison}
There are constants $c_0,C>0$, depending only on $\beta$ and $q$, such that, for $\sigma=\beta/2$, every $T\ge0$, and $n\alpha^2\sigma\le c_0$,
\begin{equation}\label{eq:comparison-main}
 \nm{\Gamma^{-1}(\Phi_{\alpha,T}-\widetilde\Phi_{\alpha,T})\Gamma}
 \le C\bigl(n\alpha^4\sigma^2+n\alpha^2\sigma e^{-T^2/(8\sigma^2)}\bigr).
\end{equation}
For any common law of the interaction time supported on $[T_0,\infty)$, the averaged channels satisfy the same bound with $T$ replaced by $T_0$ on the right-hand side.
\end{theorem}
\begin{proof}
\proofstep{Separating the two errors.}
We first note
\[
 \begin{aligned}
 \Gamma^{-1}(\Phi_{\alpha,T}-\widetilde\Phi_{\alpha,T})\Gamma= &~ \Gamma^{-1}\left(\cU(T)\Psi_{\alpha,T}\cU(T)-\cU(T)e^{\alpha^2\cL}\cU(T)\right)\Gamma\\
 = &~ \left(\Gamma^{-1}\cU(T)\Gamma\right)\Gamma^{-1}\left(\Psi_{\alpha,T}-e^{\alpha^2\cL}\right)\Gamma\left(\Gamma^{-1}\cU(T)\Gamma\right)\\
 = &~ \cU(T)\left(\hQ_{\alpha,T}-e^{\alpha^2\hL}\right)\cU(T),
 \end{aligned}
\]
where the first step uses~\eqref{eq:free-frame} for $\Phi_{\alpha,T}$ and~\eqref{eq:ideal-channel} for $\widetilde\Phi_{\alpha,T}$, and the last step uses $\Gamma^{-1}\cU(T)\Gamma=\cU(T)$ from~\eqref{eq:free-isometry}, together with $\hQ_{\alpha,T}=\Gamma^{-1}\Psi_{\alpha,T}\Gamma$ and $\Gamma^{-1}e^{\alpha^2\cL}\Gamma=e^{\alpha^2\hL}$. Note that $\nr{\cU(\pm T)O}=\nr O$ for any operator $O$ by~\eqref{eq:free-isometry}. Hence, for every map $\mathcal A$, we have
\[
 \nm{\cU(T)\mathcal A\,\cU(T)}\le\nm{\mathcal A}
 =\nm{\cU(-T)\bigl(\cU(T)\mathcal A\,\cU(T)\bigr)\cU(-T)}
 \le\nm{\cU(T)\mathcal A\,\cU(T)},
\]
by the submultiplicativity of the induced KMS norm. We then obtain
\begin{equation}\label{eq:comparison-error-split}
 \begin{aligned}
 \nm{\Gamma^{-1}(\Phi_{\alpha,T}-\widetilde\Phi_{\alpha,T})\Gamma}
 &=\nm{\hQ_{\alpha,T}-e^{\alpha^2\hL}}\\
 &\le\underbrace{\nm{\hQ_{\alpha,T}-e^{\alpha^2\hL_{T}}}}_{\text{finite window comparison}}
    +\underbrace{\nm{e^{\alpha^2\hL_{T}}-e^{\alpha^2\hL}}}_{\text{window truncation error}}.
 \end{aligned}
\end{equation}
By~\Cref{prop:finite-comparison} and $n\alpha^2\sigma\le c_0$, the first term is at most $\mathcal{O}(n\alpha^4\sigma^2)$.

\proofstep{Bounding the window truncation error.}
This part of the error comes from the pulses outside $[-T,T]$. We first bound the pulse tail. By~\eqref{eq:pulse}, we have
\[
 \int_{|t|>T}f_\sigma(t)\,\dd t = \int_{|t|>T} \frac{e^{-t^2/(4\sigma^2)}}{\sqrt{\sigma\sqrt{2\pi}}}\,\dd t
 \le\frac{e^{-T^2/(8\sigma^2)}}{\sqrt{\sigma\sqrt{2\pi}}}
                    \int_\R e^{-t^2/(8\sigma^2)}\,\dd t
 \le C\sqrt\sigma e^{-T^2/(8\sigma^2)}.
\]
Next, let $T\to\infty$ in~\eqref{eq:pair-branch}. By dominated convergence, it gives the same formula for $\hL_x$, with the integral over all $-\infty <u<t<\infty$. Hence, $\hL_{x,T}-\hL_x$ is minus the integral of the integrand in~\eqref{eq:pair-branch} over the pairs $u<t$ with $|t|>T$ or $|u|>T$. The normalized maps have norm at most one and $|w^{ss'}(u,t)|\le C_\beta^2f_\sigma(t)f_\sigma(u)$. It follows that
\begin{equation}\label{eq:bound_L_T_L}
 \begin{aligned}
 \nm{\hL_T-\hL}
 &\le4nQC_\beta^2\iint_{u<t}\mathbf1_{\{|t|>T\text{ or }|u|>T\}}f_\sigma(t)f_\sigma(u)\,\dd u\,\dd t\\
 &\le4nQC_\beta^2\|f_\sigma\|_1\int_{|t|>T}f_\sigma(t)\,\dd t
 \le Cn\sigma e^{-T^2/(8\sigma^2)},
 \end{aligned}
\end{equation}
where the second step uses the symmetry of the integrand under $u\leftrightarrow t$ and the union bound, and the last step uses the tail bound and $\|f_\sigma\|_1^2=2\sqrt{2\pi}\,\sigma$.

We also need the norms of the generators themselves. Summing~\eqref{eq:site-norm} over the sites and using $\beta=2\sigma$ gives
\begin{align}\label{eq:bound_L_T_plus_L}
 \nm{\hL_{T}}+\nm{\hL}\le Cn\sigma.
\end{align}

\proofstep{Comparing the exponentials.}
Differentiating $r\mapsto e^{(1-r)\alpha^2\hL_T}e^{r\alpha^2\hL}$ and integrating over $r\in[0,1]$ gives Duhamel's identity
\[
 e^{\alpha^2\hL_{T}}-e^{\alpha^2\hL} =\alpha^2\int_0^1 e^{(1-r)\alpha^2\hL_{T}} \left(\hL_{T}-\hL\right)e^{r\alpha^2\hL}\,\dd r.
\]
By submultiplicativity and $\nm{e^{s\mathcal A}}\le e^{s\nm{\mathcal A}}$ for $s\ge0$,
\[
 \begin{aligned}
 \nm{e^{\alpha^2\hL_{T}}-e^{\alpha^2\hL}}
 &\le\alpha^2\nm{\hL_{T}-\hL}
    \int_0^1e^{(1-r)\alpha^2\nm{\hL_{T}}+r\alpha^2\nm{\hL}}\,\dd r\\
 &\le Cn\alpha^2\sigma e^{-T^2/(8\sigma^2)}e^{Cn\alpha^2\sigma}
 \le C'n\alpha^2\sigma e^{-T^2/(8\sigma^2)},
 \end{aligned}
\]
where the second step follows from~\eqref{eq:bound_L_T_L} and~\eqref{eq:bound_L_T_plus_L}, and the last step uses $n\alpha^2\sigma\le c_0$. Substituting the two bounds into~\eqref{eq:comparison-error-split} proves~\eqref{eq:comparison-main}.

\proofstep{Averaging over the interaction time.}
Let $\mu$ be a common law of $T$ for the physical and ideal channels, supported on $[T_0,\infty)$. Taking the norm inside the integral and applying~\eqref{eq:comparison-main} for each $T$, we obtain
\[
 \begin{aligned}
 \nm{\Gamma^{-1}\E_{T\sim\mu}[\Phi_{\alpha,T}-\widetilde\Phi_{\alpha,T}]\Gamma}
 &\le\E_{T\sim\mu}\!\left[\nm{\Gamma^{-1}(\Phi_{\alpha,T}-\widetilde\Phi_{\alpha,T})\Gamma}\right]\\
 &\le Cn\alpha^4\sigma^2+Cn\alpha^2\sigma\,\E_{T\sim\mu}\bigl[e^{-T^2/(8\sigma^2)}\bigr]\\
 &\le Cn\alpha^4\sigma^2+Cn\alpha^2\sigma e^{-T_0^2/(8\sigma^2)},
 \end{aligned}
\]
where the last step uses $T\ge T_0$ almost surely.
The theorem is then proved.
\end{proof}

\section{Weighted estimates for sums of local terms}\label{sec:weighted}

In this section, we bound sums of $n$ local terms in the \green{KMS norm} $\nr{\cdot}$. If we apply the locality estimates of~\cref{sec:locality} to each term and use the triangle inequality, we bound such a sum by $\mathcal O(n)$, and hence its square by $\mathcal O(n^2)$, which would be too crude when the terms are \green{Gibbs-centered}. Indeed, for a sum $F=\sum_x F_x$ of Gibbs-centered terms, i.e., with $\mean{F_x}=0$, one may expect the sharper bound $\mathcal{O}(\sqrt{n})$, provided these local terms are only weakly correlated.

In~\cref{subsec:projections}, we introduce the spectral conditional expectations $p_I$ of~\cite{BergamaschiChen} and recall their elementary properties; the main result is \Cref{prop:thermal}. In~\cref{subsec:field}, we use these estimates to prove~\Cref{thm:field}, \red{which gives a squared-norm bound with a single factor of $n$ multiplying $\EK(O)+|\mean O|^2$ for each input $O$}. In~\cref{subsec:moments}, we apply it to bound the sum of quasi-local terms and their products (\Cref{cor:moments}) and bound the action of the \green{unital generator} $\cD$ of the effective Lindbladian in~\Cref{lem:D-action}. These results will be used for the approximate stationary state in~\cref{sec:reference} and for the contraction estimate in~\cref{sec:mixing}.

\subsection{Spectral conditional expectations and their estimates}\label{subsec:projections}

We use the \emph{spectral conditional expectation} introduced in~\cite[Definition III.2, Lemma III.1]{BergamaschiChen} to measure how strongly an operator depends on an interval $I$, namely, by its distance in the KMS norm from the operators that act as the identity on $I$.
\begin{definition}[Spectral conditional expectation~\cite{BergamaschiChen}]
For an interval $I\subseteq[n]$, the spectral conditional expectation $p_I$ is the orthogonal projection, with respect to the KMS inner product~\eqref{eq:rho-space}, onto the operators that act as the identity on $I$, that is, onto $\{\idmat_I\otimes B:B\in\mathcal B(\mathcal H_{I^c})\}$.
\end{definition}

\begin{remark} When the Gibbs state  $\rho_\beta = \rho_I\otimes \rho_{I^c}$ is a product state, $p_I$ is \red{the $\rho_I$-weighted partial trace over $I$, tensored with $\idmat_I$}. Specifically, if $O=O_I\otimes O_{I^c}$, then $p_I(O)=\Tr(\rho_I O_I)\idmat_I\otimes O_{I^c}$.
\end{remark}

Since $p_I$ is an orthogonal projection in the KMS space, it has the following elementary properties:
\begin{equation}\label{eq:projection-basic}
p_I^\star=p_I,\qquad p_I^2=p_I,\qquad \nm{p_I}\le1,\qquad
 \nr{(\id-p_I)O}=\inf_{B:\,\supp B\subseteq I^c}\nr{O-B},
\end{equation}
where $\star$ denotes the KMS adjoint defined in~\eqref{eq:map-duality}. Also note that $\idmat$  acts as the identity on every interval, so we have
\begin{equation}\label{eq:projection-identity}
p_I(\idmat)=\idmat,\qquad \forall I\subseteq [n].
\end{equation}
For the whole chain $I=[n]$, the range of $p_{[n]}$ consists only of the multiples of $\idmat$. By $\ip{\idmat}{O}=\mean O$ and $\nr{\idmat}=1$ from~\eqref{eq:normalization-identities}, we have
\begin{equation}\label{eq:projection-full}
p_{[n]}(O)=\frac{\ip{\idmat}{O}}{\nr{\idmat}^2}\,\idmat=\Tr(\rhoB O)\idmat.
\end{equation}
Furthermore, for any $I\subseteq J$, every operator that acts as the identity on $J$ also acts as the identity on $I$. Thus,~\cite[Lemma III.2]{BergamaschiChen} gives that
\begin{align}\label{eq:p_I_nesting}
    p_Ip_J=p_J p_I=p_J,\qquad \forall I\subseteq J\subseteq [n].
\end{align}
The spectral conditional expectations will be composed with weighted multiplication maps $\cB_{B,C}$ in \eqref{eq:weighted-multiplication}. We hence introduce the following notion of locality for maps.

\begin{definition}[Locality of maps]
For a set $S\subseteq[n]$, let $\mathfrak A(S)$ be the set of finite sums of multiplication maps $\cB_{B,C}$ defined as in~\eqref{eq:weighted-multiplication} with $B$ and $C$ supported on $S$. \green{We call these maps coefficient-local.} We say that a map $\mathcal A$ has coefficient support in $S$ if $\mathcal A\in\mathfrak A(S)$.
\green{We call a map $\mathcal A$ Gibbs-centered when $\ip{\idmat}{\mathcal A(\idmat)}=0$.} We call a family of maps $(\mathcal A_x)_{x\in[n]}$ uniformly coefficient-quasi-local, or simply coefficient-quasi-local, if
\begin{equation}\label{eq:coefficient-locality}
 \nm{\mathcal A_x}\le C,\qquad
 \inf_{\mathcal A_{x,r}\in\mathfrak A(N_r(x))}
          \nm{\mathcal A_x-\mathcal A_{x,r}}\le Ce^{-cr}\qquad(r\ge0),
\end{equation}
where $C,c>0$ are independent of $n$, $x$, and $r$.
\end{definition}

\begin{remark}\red{Having coefficient support in $S$ does not imply that} $\mathcal{A}(O)$ is supported in $S$ for every operator $O$ supported in $S$ because of the transformation $\Gamma(\cdot)=\rhoB^{1/2}\cdot\rhoB^{1/2}$. According to Lemma~\ref{lem:Araki}, we expect $\mathcal{A}(O)$ to be a quasi-local operator when $O$ is supported in $S$.
\end{remark}

By~\eqref{eq:weighted-algebra}, each $\mathfrak A(S)$ is closed under products and under the KMS adjoint $\star$, and the maps in $\mathfrak A(S_1)$ commute with those in $\mathfrak A(S_2)$ when $S_1\cap S_2=\varnothing$.

In addition, the coefficient-quasi-local property~\eqref{eq:coefficient-locality} is preserved under $\star$ and products: $\mathcal A_{x,r}^\star\in\mathfrak A(N_r(x))$ and $\nm{\mathcal A_x^\star-\mathcal A_{x,r}^\star}=\nm{\mathcal A_x-\mathcal A_{x,r}}$; and if $(\mathcal A_x)_{x\in[n]}$ and $(\mathcal B_x)_{x\in[n]}$ are coefficient-quasi-local, with \red{approximants} $\mathcal A_{x,r},\mathcal B_{x,r}\in\mathfrak A(N_r(x))$, then $\mathcal A_{x,r}\mathcal B_{x,r}\in\mathfrak A(N_r(x))$ and
\[
 \nm{\mathcal A_x\mathcal B_x-\mathcal A_{x,r}\mathcal B_{x,r}}
 \le\nm{\mathcal A_x-\mathcal A_{x,r}}\nm{\mathcal B_x}
   +\nm{\mathcal A_{x,r}}\nm{\mathcal B_x-\mathcal B_{x,r}}
 \le Ce^{-cr}.
\]

Beyond these basic properties, the proof of~\Cref{thm:field} also needs the following estimates on $p_I$.

\begin{proposition}[Estimates for spectral conditional expectations]\label{prop:thermal}
There are constants $C,c>0$, depending only on $\beta$ and $q$, such that the following hold for every $n$ and every interval $I\subseteq[n]$.
\begin{enumerate}[label=(\roman*)]
 \item \emph{(Conditional variance.)} For every $O$,
 \begin{equation}\label{eq:conditional}
  \nr{(\id-p_I)O}^2\le C\sum_{y\in I,a}\nr{[P_{y,a},O]}^2.
 \end{equation}
 \item \emph{(Local \red{approximants}.)} For every $r\ge0$, there is a map $p_I^r$ such that
 \begin{equation}\label{eq:projection-locality}
  \nm{p_I-p_I^r}\le Ce^{-cr},\qquad\nm{p_I^r}\le1,\qquad
  p_I^r\in\mathfrak A(N_r(I)).
 \end{equation}
 \item \emph{(Trivializing a \green{Gibbs-centered} island.)} If $\mathcal A^\star=\mathcal A\in\mathfrak A(S)$ for a set $S\subseteq I$ with $\dist(S,I^c)\ge R$, then
 \begin{equation}\label{eq:screen}
  \nm{p_I \mathcal A p_I -\ip{\idmat}{\mathcal A(\idmat)} p_I} \le C(1+|S|)e^{-cR}\nm{\mathcal A}.
 \end{equation}
 Consequently, if $(\mathcal A_x)_{x\in[n]}$ is uniformly coefficient-quasi-local, with $\mathcal A_x^\star=\mathcal A_x$ and $\ip{\idmat}{\mathcal A_x(\idmat)}=0$ for every $x$, then there are constants $C,c>0$, depending only on $\beta$, $q$, and the constants in~\eqref{eq:coefficient-locality}, such that for all $x\in[n]$ and $r\ge0$,
 \begin{equation}\label{eq:quasilocal-screen}
  \nm{p_{N_r(x)}\mathcal A_xp_{N_r(x)}}\le Ce^{-cr}.
 \end{equation}
\end{enumerate}
\end{proposition}

Part~\textnormal{(i)} follows from~\cite[Lemma~III.9 and Propositions~V.1--V.2]{BergamaschiChen}. Part~\textnormal{(ii)} extends~\cite[Lemma III.3]{BergamaschiChen} to give a locality estimate uniform in the interval length. Part~\text{(iii)} is a coefficient-local map version of~\cite[Lemma IV.1]{BergamaschiChen}. In particular, since $p_I$ projects onto the operators that act as the identity on $I$,~\eqref{eq:screen} is equivalent to \red{the following bound for all operators} $O_1$ and $O_2$ that act as the identity on $I$,
\[
 \bigl|\ip{O_2}{\mathcal A(O_1)}-\ip{\idmat}{\mathcal A(\idmat)}\ip{O_2}{O_1}\bigr|
 \le C(1+|S|)e^{-cR}\nm{\mathcal A}\nr{O_2}\nr{O_1}
\]
Thus, intuitively, seen from outside $I$, the map $\mathcal A$ acts as multiplication by the constant $\ip{\idmat}{\mathcal A(\idmat)}$.  We provide a complete proof of this proposition in Appendix~\ref{app:projections}.

\subsection[Bounding sums of Gibbs-centered coefficient-quasi-local maps]{\green{Bounding sums of Gibbs-centered coefficient-quasi-local maps}}\label{subsec:field}
In this subsection, we use~\Cref{prop:thermal} to prove that, for a \green{Gibbs-centered} coefficient-quasi-local family $(\mathcal A_x)_{x\in[n]}$ and every input $O$,
\[
 \nr{\sum_x\mathcal A_x(O)}^2\le Cn\bigl(\EK(O)+|\mean O|^2\bigr).
\]
We first bound maps that annihilate the identity (\Cref{lem:annihilator}). We then decompose each \green{Gibbs-centered} self-adjoint map into an identity-annihilating map and its adjoint; the general case follows by splitting into self-adjoint and skew-adjoint parts (\Cref{thm:field}).

\begin{lemma}[Maps that annihilate the identity]\label{lem:annihilator}
If $(\mathcal A_x)_{x\in[n]}$ is uniformly coefficient-quasi-local and $\mathcal A_x(\idmat)=0$ for every $x$, then, for all $x\in[n]$ and $O$,
\begin{equation}\label{eq:annihilator}
 \nr{\mathcal A_x(O)}^2
 \le C\sum_{y\in [n],a\in [Q]}e^{-c|x-y|}\nr{[P_{y,a},O]}^2,
 \qquad \sum_{x\in [n]}\nr{\mathcal A_x(O)}^2\le C\EK(O),
\end{equation}
where $\EK$ is the Pauli energy defined in~\eqref{eq:Pauli-energy}.
\end{lemma}

In the case when $\red{\mathcal{A}_x}$ has coefficient support in $N_r(x)$ and $\rho_\beta = \rho_{N_r(x)}\otimes \rho_{[n]\setminus N_r(x)}$, the condition $\mathcal A_x(\idmat)=0$ implies that $\mathcal A_x(O)=0$ if $O$ acts as the identity on $N_r(x)$. Thus,
\[
\mathcal{A}_x(O)=\mathcal{A}_xp_{N_r(x)}(O)+\mathcal{A}_x(\red{\id}-p_{N_r(x)})(O)=\mathcal{A}_x(\red{\id}-p_{N_r(x)})(O)\,.
\]
Using the \red{conditional} variance bound~\eqref{eq:conditional}, we can bound $\mathcal{A}_x(O)$ by the commutators of $O$ with the Pauli strings in $N_r(x)$. The proof of~\Cref{lem:annihilator} below generalizes this argument to the case when $\rhoB$ is not a product state and $\mathcal{A}_x$ is only coefficient-quasi-local with an exponentially decaying error to control the term $\mathcal{A}_xp_{N_r(x)}(O)$, using~\cref{prop:thermal}~\textnormal{(iii)}.

\begin{proof}
For simplicity, write $p_r=p_{N_r(x)}$. We first prove that $\nm{{\cal A}_xp_r}$ is small. Since coefficient-quasi-locality is preserved under products and $\star$, the map $\mathcal A_x^\star\mathcal A_x$ is self-adjoint and coefficient-quasi-local. Moreover, \red{it is \green{Gibbs-centered}, since}
\[
 \Tr[\rhoB{\cal A}_x^\star{\cal A}_x(\idmat)]=\ip{\idmat}{\mathcal A_x^\star\mathcal A_x(\idmat)}=\nr{\mathcal A_x(\idmat)}^2=0.
\]
Hence, by~\eqref{eq:quasilocal-screen} in~\cref{prop:thermal}~\textnormal{(iii)}, we have
\begin{equation}\label{eq:boundax}
 \nm{\mathcal A_xp_r}^2
 =\nm{(\mathcal A_xp_r)^\star(\mathcal A_xp_r)}
 =\nm{p_r\mathcal A_x^\star\mathcal A_xp_r}\le Ce^{-cr}.
\end{equation}

Since $N_n(x)\supseteq[n]$, we have $p_n=p_{[n]}$, and by~\eqref{eq:projection-full},
\begin{align*}
    \mathcal A_xp_n(O)=\mean O\,\mathcal A_x(\idmat)=0.
\end{align*}
By~\eqref{eq:p_I_nesting}, $p_r-p_{r+1}=p_r(\id-p_{r+1})$. Therefore,
\[
 \begin{aligned}
 \mathcal A_x=\mathcal A_x(\id-p_0)+\sum_{r=0}^{n-1}\mathcal A_x(p_r-p_{r+1})=\mathcal A_x(\id-p_0)+\sum_{r=0}^{n-1}\mathcal A_xp_r(\id-p_{r+1}),
 \end{aligned}
\]
and hence
\begin{align*}
 \nr{\mathcal A_x(O)}
 &\le C\nr{(\id-p_0)O}+C\sum_{r=0}^{n-1}e^{-cr/2}\nr{(\id-p_{r+1})O}\\
 &\le C'\sum_{r\ge0}e^{-c'r}
       \left(\sum_{y\in N_{r+1}(x),a}\nr{[P_{y,a},O]}^2\right)^{1/2},
\end{align*}
where the first inequality uses $\nm{\mathcal A_x}\le C$ and~\eqref{eq:boundax}, and the second uses~\eqref{eq:conditional} in~\cref{prop:thermal}~\textnormal{(i)}, with the first term absorbed into the term with $r=0$. Squaring and applying the Cauchy--Schwarz inequality with the weights $e^{-c'r}$ gives
\[
 \begin{aligned}
 \nr{\mathcal A_x(O)}^2
 &\le C\left(\sum_{r\ge0}e^{-c'r}\right)\left(\sum_{r\ge0}e^{-c'r}\sum_{y\in N_{r+1}(x),a}\nr{[P_{y,a},O]}^2\right)\\
 &\le C'\sum_{y,a}\left(\sum_{r\ge\max\{0,|x-y|-1\}}e^{-c'r}\right)
                      \nr{[P_{y,a},O]}^2\\
 &\le C''\sum_{y,a}e^{-c''|x-y|}\nr{[P_{y,a},O]}^2.
 \end{aligned}
\]
This proves the first bound in~\eqref{eq:annihilator}. Summing it over $x$ and using
\begin{align*}
    \sum_xe^{-c''|x-y|}\le1+2\sum_{d\ge1}e^{-c''d}
\end{align*}
for every $y$ proves the second bound.
\end{proof}

\begin{theorem}[\green{Gibbs-centered sums of coefficient-quasi-local maps}]\label{thm:field}
If $(\mathcal A_x)_{x\in[n]}$ is uniformly coefficient-quasi-local and $\ip{\idmat}{\mathcal A_x(\idmat)}=0$ for every $x$, then, for every $O$,
\begin{equation}\label{eq:field-bound}
 \nr{\sum_x\mathcal A_x(O)}^2
 \le Cn\bigl(\EK(O)+\nr O^2\bigr)
 \le C'n\bigl(\EK(O)+|\mean O|^2\bigr).
\end{equation}
The constants $C,C'$ depend only on $\beta$, $q$, and the constants in~\eqref{eq:coefficient-locality}.
\end{theorem}

\begin{proof}
We first assume that $\mathcal A_x^\star=\mathcal A_x$ for every $x$. Fix a site $x$, and let $p_{-1}=\id$ and $p_r=p_{N_r(x)}$ for $0\le r\le n$. The differences $d_r=p_{r-1}-p_r$ are mutually orthogonal projections in the KMS inner product, with ranges $\operatorname{ran}p_{r-1}\cap\ker p_r$. Since $p_n(O)=\mean O\,\idmat$, $O$ decomposes into:
\[
 O=\sum_{r=0}^n d_rO+\mean O\,\idmat.
\]
For each $r$, we have
\[
 p_{r-1}\mathcal A_xp_{r-1}-p_r\mathcal A_xp_r
 =p_{r-1}\mathcal A_xd_r+d_r\mathcal A_xp_r.
\]
Averaging this identity with its adjoint motivates the definition
\begin{equation}\label{eq:triangular-piece}
 \mathcal T_{x,r}=\frac12(p_{r-1}+p_r)\mathcal A_x\,(p_{r-1}-p_r) = p_r\mathcal A_xd_r+\frac12d_r\mathcal A_xd_r,
 \qquad\mathcal T_x=\sum_{r=0}^n\mathcal T_{x,r}\,,
\end{equation}
which satisfies, by $\mathcal A_x^\star=\mathcal A_x$,
\begin{equation}\label{eq:triangular-telescope}
\begin{aligned}
\mathcal T_{x,r}+\mathcal T_{x,r}^\star
 = &~ p_{r-1}\mathcal A_xp_{r-1}-p_r\mathcal A_xp_r,
\end{aligned}
\end{equation}
and $\mathcal T_x(\idmat)=0$, since $d_r(\idmat)=0$ for every $r$. Here the right factor $d_r$ in the definition \eqref{eq:triangular-piece} of $\mathcal T_{x,r}$ restricts the input to the $r$-th layer.
Summing~\eqref{eq:triangular-telescope} over $0\le r\le n$ gives
\begin{align*}
    \mathcal T_x+\mathcal T_x^\star=p_{-1}\mathcal A_xp_{-1} - p_n\mathcal A_xp_n = \mathcal A_x-p_n\mathcal A_xp_n,
\end{align*}
since $p_{-1}=\id$. Finally, $p_n=p_{[n]}$, so~\eqref{eq:projection-full} gives
\[
 p_n\mathcal A_xp_n(O)=\mean O\,\ip{\idmat}{\mathcal A_x(\idmat)}\idmat=0.
\]
Thus, we obtain the following decomposition of ${\cal A}_x$:
\begin{equation}\label{eq:triangular-decomposition}
 \mathcal A_x=\mathcal T_x+\mathcal T_x^\star,\qquad\mathcal T_x(\idmat)=0.
\end{equation}

We next show that the norm of ${\cal T}_{x,r}$ decays exponentially in the radius $r$.
By~\eqref{eq:projection-basic}, $p_r^2=p_r$. And by~\eqref{eq:p_I_nesting}, $p_{r-1}p_r=p_rp_{r-1}=p_r$. Hence, for $r\ge1$,
\begin{align}\label{eq:bound_T_x_r_tail}
 \red{\nm{\mathcal T_{x,r}}} &\red{{}=\frac12\nm{(p_{r-1}+p_r)\mathcal A_x(p_{r-1}-p_r)}}\notag\\
 &\red{{}=\frac12\nm{(p_{r-1}+p_r)\,(p_{r-1}\mathcal A_xp_{r-1})\,(p_{r-1}-p_r)}}\notag\\
 &\red{{}\le\nm{p_{r-1}\mathcal A_xp_{r-1}}\le Ce^{-c(r-1)},}
\end{align}
where
the last step follows from~\eqref{eq:quasilocal-screen}. The term with $r=0$ is bounded by $\nm{\mathcal A_x}$.

Let $\mathcal T=\sum_x\mathcal T_x$. Since $p_{r-1}p_r=p_r^2=p_r$, we have $(p_{r-1}-p_r)p_r=0$, and hence $\mathcal T_{x,r}=\mathcal T_{x,r}(\id-p_r)$. By \eqref{eq:bound_T_x_r_tail}, $\nr{\mathcal T_{x,r}(O)}\le Ce^{-cr}\nr{(\id-p_r)O}$ for $0\le r\le n$ and uniform constants $C,c$. By the triangle inequality, the Cauchy--Schwarz inequality with the weights $e^{-cr}$, and~\eqref{eq:conditional} with $I=N_r(x)$,
\[
 \begin{aligned}
 \nr{\mathcal T_x(O)}^2
 &\le\Bigl(C\sum_{r=0}^ne^{-cr}\nr{(\id-p_r)O}\Bigr)^2
 \le C'\sum_{r=0}^ne^{-cr}\nr{(\id-p_r)O}^2\\
 &\le C''\sum_{r\ge0}e^{-cr}\sum_{y\in N_r(x),a}\nr{[P_{y,a},O]}^2.
 \end{aligned}
\]
In the sum over $x$, each commutator $[P_{y,a},O]$ is charged with total weight $\sum_x\sum_{r\ge|x-y|}e^{-cr}\le\sum_{r\ge0}(2r+1)e^{-cr}$, which implies $\sum_x\nr{\mathcal T_x(O)}^2\le C\EK(O)$ and
\[
 \nr{\mathcal T(O)}^2
 \le n\sum_x\nr{\mathcal T_x(O)}^2\le Cn\EK(O).
\]

To bound $\mathcal T^\star$, we use the commutator $[\mathcal T,\mathcal T^\star]$. By~\eqref{eq:triangular-decomposition}, $\mathcal T+\mathcal T^\star=\mathcal A$ with $\mathcal A:=\sum_y\mathcal A_y$, so $[\mathcal T,\mathcal T^\star]=[\mathcal T,\mathcal A]$, and it suffices to consider $[\mathcal T_{x,r},\mathcal A_y]$. First, we note $$\nm{[\mathcal T_{x,r},\mathcal A_y]}\le2\nm{\mathcal T_{x,r}}\nm{\mathcal A_y}\le Ce^{-cr}.$$ Next, if $|x-y|\ge r+1$, let $s=\lfloor(|x-y|-r-1)/2\rfloor$, so that $N_{r+s}(x)\cap N_s(y)=\varnothing$. By~\eqref{eq:coefficient-locality}, there are $\mathcal A_{x,s}\in\mathfrak A(N_s(x))$ and $\mathcal A_{y,s}\in\mathfrak A(N_s(y))$ with $\nm{\mathcal A_x-\mathcal A_{x,s}}\le Ce^{-cs}$ and $\nm{\mathcal A_y-\mathcal A_{y,s}}\le Ce^{-cs}$; in particular, $\nm{\mathcal A_{x,s}},\nm{\mathcal A_{y,s}}\le2C$. For $r'\in\{r-1,r\}$ with $r'\ge0$, applying~\eqref{eq:projection-locality} with $I=N_{r'}(x)$ and radius $s$ gives a map $\widetilde p_{r'}\in\mathfrak A(N_s(N_{r'}(x)))=\mathfrak A(N_{r'+s}(x))$ with $\nm{p_{r'}-\widetilde p_{r'}}\le Ce^{-cs}$ and $\nm{\widetilde p_{r'}}\le1$; if $r=0$, we set $\widetilde p_{-1}=\id=\cB_{\idmat,\idmat}$. Since $\mathfrak A(S)$ increases with $S$ and $\mathfrak A(N_{r+s}(x))$ is closed under sums and products, replacing the factors in~\eqref{eq:triangular-piece} one at a time gives
\[
 \widetilde{\mathcal T}_{x,r}:=\frac12(\widetilde p_{r-1}+\widetilde p_r)\,\mathcal A_{x,s}\,(\widetilde p_{r-1}-\widetilde p_r)\in\mathfrak A(N_{r+s}(x)),
 \qquad
 \nm{\mathcal T_{x,r}-\widetilde{\mathcal T}_{x,r}}\le Ce^{-cs}.
\]
The maps $\widetilde{\mathcal T}_{x,r}$ and $\mathcal A_{y,s}$ have disjoint coefficient supports, and hence commute. Then,
\[
 \nm{[\mathcal T_{x,r},\mathcal A_y]}
 \le\nm{[\mathcal T_{x,r},\mathcal A_y-\mathcal A_{y,s}]}
   +\nm{[\mathcal T_{x,r}-\widetilde{\mathcal T}_{x,r},\mathcal A_{y,s}]}
 \le Ce^{-cs}.
\]
Since $2s\ge|x-y|-r-2$, we have $e^{-cs}\le e^{c}e^{-c(|x-y|-r)/2}$. As the minimum below equals $e^{-cr}$ when $|x-y|\le r$, the two bounds give, for some $C$,
\[
 \nm{[\mathcal T_{x,r},\mathcal A_y]}\le C\min\bigl\{e^{-cr},\,e^{-c(|x-y|-r)}\bigr\},\qquad x,y\in[n],\ 0\le r\le n.
\]
Splitting the sum over $y$ at $|x-y|=2r$ gives $\sum_y\nm{[\mathcal T_{x,r},\mathcal A_y]}\le C(r+1)e^{-cr}$. Summing over $x$ and $r$, we obtain
\[
 \nm{[\mathcal T,\mathcal T^\star]}
 =\nm{[\mathcal T,\mathcal A]}
 \le\sum_x\sum_{r=0}^n\sum_y\nm{[\mathcal T_{x,r},\mathcal A_y]}
 \le C\sum_x\sum_{r\ge0}(r+1)e^{-cr}\le C'n.
\]
It follows that
\[
 \begin{aligned}
 \nr{\mathcal T^\star(O)}^2
 &=\ip O{\mathcal T\mathcal T^\star(O)}
 =\ip O{\mathcal T^\star\mathcal T(O)}
       +\ip O{[\mathcal T,\mathcal T^\star](O)}\\
 &\le\nr{\mathcal T(O)}^2+\nm{[\mathcal T,\mathcal T^\star]}\nr O^2
 \le Cn\bigl(\EK(O)+\nr O^2\bigr).
 \end{aligned}
\]
Combining these bounds with~\eqref{eq:triangular-decomposition} gives
\[
 \nr{\sum_x\mathcal A_x(O)}^2
 =\nr{\mathcal T(O)+\mathcal T^\star(O)}^2
 \le2\nr{\mathcal T(O)}^2+2\nr{\mathcal T^\star(O)}^2
 \le Cn\bigl(\EK(O)+\nr O^2\bigr).
\]

Finally, we bound $\nr O$:
\[
 \nr O^2=\nr{O-\mean O\,\idmat}^2+|\mean O|^2=\nr{(\id-p_{[n]})O}^2+|\mean O|^2
 \le C\EK(O)+|\mean O|^2,
\]
where the second step follows from~\eqref{eq:projection-full}, and the third step follows from~\eqref{eq:conditional} in~\cref{prop:thermal}~\textnormal{(i)}. This proves~\eqref{eq:field-bound} for self-adjoint maps.

For general maps, we set $\mathcal A_x^{\mathrm{Re}}=(\mathcal A_x+\mathcal A_x^\star)/2$ and $\mathcal A_x^{\mathrm{Im}}=(\mathcal A_x-\mathcal A_x^\star)/(2i)$. Both families are self-adjoint and coefficient-quasi-local, and both are \green{Gibbs-centered}, since $\ip{\idmat}{\mathcal A_x^\star(\idmat)}=\overline{\ip{\idmat}{\mathcal A_x(\idmat)}}=0$. Applying the self-adjoint case to the two sums and using $\nr{X+iY}^2\le2\nr X^2+2\nr Y^2$ completes the proof.
\end{proof}

\subsection[Moments and the action of the unital generator]{Moments and the action of the \green{unital generator}}\label{subsec:moments}

In this subsection, we apply~\Cref{thm:field} to \green{Gibbs-centered} sums of quasi-local terms. We bound their second moments and the moments of their products in~\Cref{cor:moments}. %
We also bound the action of the \green{unital generator} $\cD$ on a time-evolved sum in~\Cref{lem:D-action}. These estimates will be used to approximate the stationary state in~\cref{sec:reference}.

\begin{corollary}[Second and fourth moments]\label{cor:moments}
Let $F=\sum_xF_x$ and $G=\sum_xG_x$, where each $F_x$ and $G_x$ is Gibbs-centered, i.e., $\mean{F_x}=\mean{G_x}=0$, and the imaginary-time evolutions $F_x(-iu)$ and $G_x(-iu)$ for $|u|\le\beta/4$ are quasi-local at $x$, i.e.,
\begin{align}\label{eq:F_G_quasi_local}
    \no{B(-iu)}\leq C,\qquad \no{(\id-{\cal E}_{x,r})B(-iu)}\leq Ce^{-cr}\qquad \forall r\geq 0,~B=F_x,G_x,
\end{align}
where $c,C$ are constants independent of $n$, $x$, and $u$. Then
\begin{equation}\label{eq:moments}
 \nr F+\nr G\le C\sqrt n,\qquad \nr{FG}+\nr{GF}\le Cn.
\end{equation}
The same bounds hold for adjoints, and
\begin{equation}\label{eq:invariant-product}
 \sup_{t\in\R}(\nr{FG(t)}+\nr{G(t)F})\le Cn.
\end{equation}
\end{corollary}
\begin{proof}
\proofstep{Second moments.}
Since $\rhoB^{-1/4}F_x(i\beta/4)\rhoB^{1/4}=F_x$, \eqref{eq:weighted-multiplication} shows that left multiplication by $F_x$ is the map $\cB_{F_x(i\beta/4),\idmat}$, which is coefficient-quasi-local, since $\cB_{\cE_{x,r}(F_x(i\beta/4)),\idmat}$ \red{lies} in $\mathfrak A(N_r(x))$, and
\[
 \nm{\cB_{F_x(i\beta/4),\idmat}-\cB_{\cE_{x,r}(F_x(i\beta/4)),\idmat}}
 \le\no{(\id-\cE_{x,r})F_x(i\beta/4)}\le Ce^{-cr},
\]
where the first step uses the norm bound in~\eqref{eq:weighted-algebra}, and the second step follows from~\eqref{eq:F_G_quasi_local}. It is also centered: $\ip{\idmat}{F_x\idmat}=\Tr(\rhoB F_x)=0$. Applying~\Cref{thm:field} with $O=\idmat$ gives
\[
 \nr F^2\le Cn\bigl(\EK(\idmat)+|\mean{\idmat}|^2\bigr)=Cn,
\]
since $[P_{x,a},\idmat]=0$ and $\mean{\idmat}=1$. The same argument gives $\nr G^2\le Cn$.

\proofstep{Products.} We shall view the product $FG$ as a left multiplication by $F$ with the input $G$ and then apply~\Cref{thm:field}.
We first bound the Pauli energy of $G$. For fixed $y$ and $a$, and $x\ne y$, the matrix $P_{y,a}$ commutes with $\cE_{x,r}(G_x)$ for $r=|x-y|-1$. Hence,
\[
 \nr{[P_{y,a},G_x]}\le\no{[P_{y,a},G_x]}\le2\no{(\id-\cE_{x,r})G_x}\le Ce^{-c|x-y|},
\]
where the first step uses $\nr B\le\no B$ from~\eqref{eq:trace-conversion}, the second step is by $[P_{y,a},\cE_{x,r}(G_x)]=0$, and the third step follows from~\eqref{eq:F_G_quasi_local}. For $x=y$, the same bound follows from $\no{[P_{y,a},G_y]}\le2\no{G_y}\leq C$. Therefore,
\[
 \EK(G)
 =\sum_{y,a}\nr{\sum_x[P_{y,a},G_x]}^2
 \le C\sum_{y,a}\left(\sum_xe^{-c|x-y|}\right)^2\le C'n,
\]
where the first inequality is the triangle inequality inside each norm, and the second uses the geometric sum over $x$ and the $Q$ Pauli strings per site. Applying~\Cref{thm:field} to the left multiplications by $F_x$ and to the Gibbs-centered input $G$ gives
\[
 \nr{FG}^2\le Cn\bigl(\EK(G)+|\mean G|^2\bigr)\le C'n^2.
\]
Interchanging $F$ and $G$, and replacing either of them by its adjoint, proves the remaining product bounds. The hypotheses are preserved under adjoints, since $\mean{F_x^\dagger}=\overline{\mean{F_x}}$ and $F_x^\dagger(-iu)=F_x(iu)^\dagger$.

\proofstep{Uniformity in time.}
Expanding the Schatten fourth powers and cycling the trace gives
\[
 \begin{aligned}
 \left\|\rhoB^{1/4}F\right\|_4^4
 &=\Tr(\rhoB^{1/2}FF^\dagger\rhoB^{1/2}FF^\dagger)=\nr{FF^\dagger}^2,\\
 \left\|G\rhoB^{1/4}\right\|_4^4
 &=\Tr(\rhoB^{1/2}G^\dagger G\rhoB^{1/2}G^\dagger G)=\nr{G^\dagger G}^2.
 \end{aligned}
\]
Moreover, $G(t)\rhoB^{1/4}=e^{itH_n}(G\rhoB^{1/4})e^{-itH_n}$, so its Schatten norms do not depend on $t$. Consequently,
\[
 \nr{FG(t)}
 =\left\|\rhoB^{1/4}F G(t)\rhoB^{1/4}\right\|_2
 \le \left\|\rhoB^{1/4}F\right\|_4\left\|G(t)\rhoB^{1/4}\right\|_4
 =\nr{FF^\dagger}^{1/2}\nr{G^\dagger G}^{1/2}\le Cn,
\]
where the second step is the Schatten H\"older inequality, the third uses the two identities above and the independence of $t$, and the last step is the product bound for the pairs $(F,F^\dagger)$ and $(G^\dagger,G)$.

For the reversed product, we use $\|\rhoB^{1/4}G(t)\|_4\|F\rhoB^{1/4}\|_4$ instead, and obtain the bound $\nr{GG^\dagger}^{1/2}\nr{F^\dagger F}^{1/2}\le Cn$.
\end{proof}

The last estimate concerns the \green{unital generator} $\cD$ in~\eqref{eq:affine}. \red{Since $\cD(\idmat)=0$, the semigroup $e^{t\cD}$ is unital.} Unlike the time-independent bound for the products above, the following lemma bounds it linearly in real time, because a time-evolved term spreads over a growing region.
\begin{lemma}[Action of the \green{unital generator}]\label{lem:D-action}
Let $F=\sum_yF_y$, where each $F_y$ is quasi-local at $y$. Then, for fixed $\beta$, $\sigma=\beta/2$, and all $t\in\R$,
\begin{equation}\label{eq:D-action}
 \nr{\cD(F(t))}\le Cn(1+|t|).
\end{equation}
\end{lemma}
\begin{proof}
\proofstep{Commutator form of $\cD$.}
Let $B_x^{\mathrm s}=(B_x^{\mathrm{Lamb}}(-i\beta/2)+B_x^{\mathrm{Lamb}}(i\beta/2))/2$. By~\eqref{eq:Lamb} and~\eqref{eq:Hs-K}, $\HL^{\mathrm s}=\sum_xB_x^{\mathrm s}$. We use the following identity:
\[
 V^\dagger OV-\frac12\{V^\dagger V,O\}
 =\frac12\bigl(V^\dagger[O,V]+[V^\dagger,O]V\bigr).
\]
The left-hand side is the adjoint of the dissipator $\mathcal D_V$ in~\eqref{eq:def_lindbladian} applied to $O$. Substituting this identity into the adjoint of~\eqref{eq:LK}, and using the definition $\cD=\LK^\dagger-i[\HL^{\mathrm s},\cdot\,]$ in~\eqref{eq:affine}, we obtain $\cD=\sum_x\cD_x$ with
\[
 \cD_x(O)
 =\frac12\sum_a\int_\R g_{\beta,\sigma}(u)
 \bigl(V_{A_{x,a}}(u)^\dagger[O,V_{A_{x,a}}(u)]
       +[V_{A_{x,a}}(u)^\dagger,O]V_{A_{x,a}}(u)\bigr)\,\dd u
 +i[G_{D,x}-B_x^{\mathrm s},O].
\]

\proofstep{Bounds on the coefficients.}
The filtered couplings $V_{A_{x,a}}\red{(u)}=\int_{-\infty}^\infty f_\sigma(t)A_{x,a}(t)e^{-iut}\dd t$ \red{satisfy the following bounds uniformly in $u$:}
\[
 \begin{aligned}
 \no{V_{A_{x,a}}(u)}&\le\int_\R f_\sigma(t)\no{A_{x,a}(t)}\,\dd t\le\|f_\sigma\|_1,\\
 \no{(\id-\cE_{x,r})V_{A_{x,a}}(u)}
 &\le Ce^{-cr}\int_\R f_\sigma(t)e^{v|t|}\,\dd t\le C_\sigma e^{-cr},
 \end{aligned}
\]
where the first bound uses $\no{A_{x,a}(t)}=\no{A_{x,a}}=1$, and the second integrates the real-time case of~\eqref{eq:complex-cone} against the pulse. The remaining coefficients $G_{D,x}$ and $B_x^{\mathrm s}$ obey bounds of the same type by~\Cref{prop:quasilocal}.

\proofstep{Commutators with distant terms.}
Let $B_x$ be any of these coefficients. For $x\ne y$, let $r=\lfloor(|x-y|-1)/3\rfloor$, so that $\cE_{x,r}(B_x)$ and $\cE_{y,r}(F_y(t))$ have disjoint supports and commute. Hence,
\[
 \begin{aligned}
 \no{[B_x,F_y(t)]}
 &\le2\no{B_x-\cE_{x,r}(B_x)}\no{F_y(t)}
     +2\no{\cE_{x,r}(B_x)}\no{F_y(t)-\cE_{y,r}(F_y(t))}\\
 &\le Ce^{-cr}+C\min\{1,e^{v|t|-cr}\}
 \le C'\min\{1,e^{v'|t|-c'|x-y|}\},
 \end{aligned}
\]
where the first step subtracts $[\cE_{x,r}(B_x),\cE_{y,r}(F_y(t))]=0$, the second step uses the coefficient bounds above and the real-time estimate~\eqref{eq:quasilocal-real-cone} for the quasi-local $F_y$. For $x=y$, we have $\no{[B_x,F_x(t)]}\le2\no{B_x}\no{F_x}\leq C'$ after enlarging $C'$ if necessary. Inserting these bounds into $\cD_x$, and using $\int g_{\beta,\sigma}=1$, we get that
\[
 \no{\cD_x(F_y(t))}\le C\min\{1,e^{v|t|-c|x-y|}\}.
\]

\proofstep{Summing over the sites.}
As in the proof of~\Cref{prop:finite-comparison}, the summand equals one for $|x-y|\le v|t|/c$ and decays geometrically beyond this distance, and at most two sites lie at each distance $d\ge1$ from $y$. Hence,
\[
 \sum_x\min\{1,e^{v|t|-c|x-y|}\}
 \le1+\frac{2v|t|}{c}+\frac{2}{1-e^{-c}}
 \le C(1+|t|),
\]
where the first term is the site $x=y$. Therefore,
\[
 \nr{\cD(F(t))}
 \le\no{\cD(F(t))}
 \le\sum_{x,y}\no{\cD_x(F_y(t))}
 \le Cn(1+|t|).
\]
This proves~\eqref{eq:D-action}.
\end{proof}

\section{The approximate stationary state}\label{sec:reference}

In this section, we construct a state $\rho_*$ close to $\rhoB$ that is nearly fixed by the averaged physical channel. A correction to $\rhoB$ is needed because of the exact identity $\cL(\rhoB)=-i[\HL,\rhoB]$ in~\eqref{eq:thermal-split}: the second-order generator need not preserve the Gibbs state, and its drift at the Gibbs state is the matrix $K=\Gamma^{-1}(\cL(\rhoB))$ of~\eqref{eq:drift-identity}. In~\cref{subsec:reference-construction}, we introduce the random interaction time, derive the cancellation equation that a first-order correction must satisfy after time averaging, construct $\rho_*$, and state and prove the main result, \cref{thm:reference}. In~\cref{subsec:kernel}, we prove that the correction kernel is integrable (\cref{lem:nu}) and that the correction solves the cancellation equation with the required bounds (\cref{lem:correction}). In~\cref{subsec:remainders}, we prove the estimates that control the two Taylor remainders (\cref{lem:remainders}). In~\cref{sec:complexity}, we combine \Cref{thm:reference} with the contraction of the averaged ideal channel proved in~\cref{sec:mixing} to bound the distance from the fixed point of the averaged physical channel to $\rhoB$. Throughout this section, $\beta$ and the local model parameters are fixed independently of $n$, and $\sigma=\beta/2$.

\subsection{Construction and main theorem}\label{subsec:reference-construction}
In this subsection, we fix the law of the interaction time, write the drift as a commutator $K=[H_n,Y_0]$, use $Y_0$ to construct the \green{candidate matrix} $\rho_*$, and prove \Cref{thm:reference}.

The interaction time $T$ is drawn from the shifted-gamma density
\begin{equation}\label{eq:time-law}
 \mu_{T_0}(t)=T_0^{-1}\mu_0(t/T_0),\qquad
 \mu_0(u)=\tfrac16(u-1)^3e^{-(u-1)}\mathbf1_{\{u\ge1\}},\qquad T_0>0.
\end{equation}
Its support is $[T_0,\infty)$, so $T_0$ is the minimum interaction time; its mean and \green{Fourier transform} are
\begin{equation}\label{eq:time-transform}
 \begin{aligned}
 \E[T]&=\frac{T_0}{6}\int_0^\infty(1+u)u^3e^{-u}\,\dd u
       =\frac{T_0}{6}(3!+4!)=5T_0,\\
 \widehat\mu_{T_0}(\omega)
 &=\frac{e^{i\omega T_0}}6\int_0^\infty u^3e^{-(1-i\omega T_0)u}\,\dd u
 =e^{i\omega T_0}(1-i\omega T_0)^{-4}.
 \end{aligned}
\end{equation}
Both identities follow from the substitution $t=T_0(1+u)$. The last integral equals $3!/(1-i\omega T_0)^4$ by three integrations by parts, whose boundary terms vanish because the coefficient $1-i\omega T_0$ in the exponent has real part one. Here and below, Fourier transforms follow the convention~\eqref{eq:fourier-convention}. With $\widetilde\Phi_{\alpha,T}$ from~\eqref{eq:ideal-channel}, we write the averaged physical and ideal channels as
\[
 \Phi_{\alpha}^{\mu_{T_0}}=\E_T[\Phi_{\alpha,T}],\qquad
 \widetilde\Phi_{\alpha}^{\mu_{T_0}}=\E_T[\widetilde\Phi_{\alpha,T}].
\]

Let $\xi$ be a Hermitian traceless correction, to be chosen below. To identify the equation that $\xi$ must satisfy, we replace $\alpha^2$ by $s$ in the averaged ideal channel and differentiate at $s=0$ its \green{one-round stationarity defect} at the input $\rhoB+s\xi$:
\[
 \begin{aligned}
 &\left.\frac{\dd}{\dd s}\left(\E_T[\cU(T)e^{s\cL}\cU(T)(\rhoB+s\xi)]
                          -\rhoB-s\xi\right)\right|_{s=0}\\
 &\qquad=\E_T[\cU(T)\cL(\rhoB)]
                   +(\E_T[\cU(2T)]-\id)\xi.
 \end{aligned}
\]
Here we used that the derivative of $e^{s\cL}$ at $s=0$ is $\cL$, that $\cU(T)\rhoB=\rhoB$, and that the term $s\xi$ contributes $\cU(T)\cU(T)\xi=\cU(2T)\xi$. Differentiation under $\E_T$ is justified because each $\cU(T)$ is an isometry by~\eqref{eq:free-isometry} and the dimension is finite. This computation motivates solving
\begin{equation}\label{eq:cancellation-target}
 (\E_T[\cU(2T)]-\id)\xi+\E_T[\cU(T)\cL(\rhoB)]=0.
\end{equation}
Dividing each Bohr component of~\eqref{eq:cancellation-target} by the corresponding energy difference does not by itself control the size of $\xi$, since nonzero Bohr frequencies may be arbitrarily small.

Instead, we write the drift itself as a commutator with $H_n$. We first replace each $B_x^{\mathrm{Lamb}}$ by its Gibbs-centered version $B_x^{\mathrm{Lamb}}-\Tr(\rhoB B_x^{\mathrm{Lamb}})\idmat$. This shifts $\HL$ and $\HL^{\mathrm s}$ by scalars and leaves $\cL$, $K$, and $\cD$ unchanged. Define
\begin{equation}\label{eq:drift-fields}
 Y_0=-i\int_{-\beta/2}^{\beta/2}\HL(-iu)\,\dd u.
\end{equation}
Then
\begin{equation}\label{eq:commutator-identity}
 \begin{aligned}
 [H_n,Y_0]
 &=-i\int_{-\beta/2}^{\beta/2}[H_n,\HL(-iu)]\,\dd u\\
 &=-i\int_{-\beta/2}^{\beta/2}\frac{\dd}{\dd u}\HL(-iu)\,\dd u
 =-i(\HL(-i\beta/2)-\HL(i\beta/2))=K,\\
 [H_n,\Gamma(Y_0)]&=\Gamma([H_n,Y_0])=\Gamma(K)=\cL(\rhoB).
 \end{aligned}
\end{equation}
The second equality uses $\frac{\dd}{\dd u}\HL(-iu)=[H_n,\HL(-iu)]$, obtained by differentiating $e^{uH_n}\HL e^{-uH_n}$. The last line uses that $\Gamma$ commutes with $\ad_{H_n}$, since $[H_n,\rhoB]=0$, together with~\eqref{eq:drift-identity}.

The Hermiticity of $\HL$ and the invariance of its \green{Gibbs mean} under imaginary-time translation give
\[
 \begin{aligned}
 Y_0^\dagger
 &=i\int_{-\beta/2}^{\beta/2}\HL(iu)\,\dd u
 =i\int_{-\beta/2}^{\beta/2}\HL(-iu)\,\dd u=-Y_0,\\
 \Tr(\rhoB Y_0)
 &=-i\int_{-\beta/2}^{\beta/2}\Tr(\rhoB\HL(-iu))\,\dd u
 =-i\beta\Tr(\rhoB\HL)=0.
 \end{aligned}
\]
The first line uses $\HL^\dagger=\HL$ and the substitution $u\mapsto-u$; the second cycles the conjugating exponentials through the trace and uses the centering of $\HL$.

We next record bounds on $Y_0$ and $K$. Write $Y_0=\sum_xY_x$ and $K=\sum_xK_x$, where
\[
 Y_x=-i\int_{-\beta/2}^{\beta/2}B_x^{\mathrm{Lamb}}(-iu)\,\dd u,
 \qquad K_x=-i(B_x^{\mathrm{Lamb}}(-i\beta/2)-B_x^{\mathrm{Lamb}}(i\beta/2)).
\]
Both $Y_x$ and $K_x$ are Gibbs-centered, since the \green{Gibbs mean} is invariant under imaginary-time translation and $B_x^{\mathrm{Lamb}}$ is \green{Gibbs-centered}. For $|u'|\le\beta/4$, the translates $Y_x(-iu')$ and $K_x(-iu')$ involve $B_x^{\mathrm{Lamb}}(-iw)$ only for $|w|\le3\beta/4$. Since the scalar shift $\Tr(\rhoB B_x^{\mathrm{Lamb}})\idmat$ is bounded and $\cE_{x,r}$ fixes scalar matrices, Proposition~\ref{prop:quasilocal} applies equally to the \green{Gibbs-centered} $B_x^{\mathrm{Lamb}}$. With $u_*=3\beta/4$ and $m=0$, it shows that these translates are quasi-local, uniformly in $x$ and in $u'$ with $|u'|\le\beta/4$. Thus $Y_x$ and $K_x$ satisfy the hypotheses of Corollary~\ref{cor:moments}, which together with Proposition~\ref{prop:quasilocal} gives
\begin{equation}\label{eq:drift-moments}
 \begin{aligned}
 \no{Y_0}&\le Cn,\qquad \nr{Y_0}+\nr K\le C\sqrt n,\\
 \nr{K^2}+\nr{Y_0^\dagger Y_0}&\le Cn,\qquad
 \sup_t(\nr{KY_0(t)}+\nr{Y_0(t)K})\le Cn.
 \end{aligned}
\end{equation}
The operator-norm bound follows from the triangle inequality, since each $Y_x$ is uniformly bounded. The weighted bounds are \eqref{eq:moments}, together with its version for adjoints, and \eqref{eq:invariant-product}, applied with $F,G\in\{Y_0,K\}$. They rest on the field bound of Theorem~\ref{thm:field}.

For an observable correction $X$ with $\xi=\Gamma(X)$, the $\omega$-component of $\Gamma^{-1}$ applied to~\eqref{eq:cancellation-target} is
\[
 (\widehat\mu_{T_0}(-2\omega)-1)X_\omega
       +\omega\widehat\mu_{T_0}(-\omega)Y_{0,\omega}=0.
\]
This follows from $\Gamma^{-1}\cU(t)\Gamma=\cU(t)$, $K_\omega=\omega Y_{0,\omega}$, and $\cU(t)Y_{0,\omega}=e^{-i\omega t}Y_{0,\omega}$. For $\omega\ne0$, it prescribes
\[
 X_\omega=\frac{\omega\widehat\mu_{T_0}(-\omega)}
                   {1-\widehat\mu_{T_0}(-2\omega)}Y_{0,\omega}.
\]
Both the numerator and the denominator vanish linearly at $\omega=0$, so the multiplier extends continuously to $\omega=0$. We define the corresponding time kernel now; its integrability is proved in \Cref{lem:nu}. Let
\begin{equation}\label{eq:nu-definition}
 h(\omega)=\frac{\omega\widehat\mu_0(\omega)}{1-\widehat\mu_0(2\omega)},
 \quad h(0)=i/10,\quad
 \nu_0(t)=\frac1{2\pi}\int_\R e^{-i\omega t}h(\omega)\,\dd\omega,
 \quad \nu_{T_0}(t)=T_0^{-2}\nu_0(t/T_0).
\end{equation}

We now define the correction and the \green{candidate matrix} $\rho_*$:
\begin{equation}\label{eq:reference-definition}
 \begin{aligned}
 X&=-\int_\R\cU(t)(Y_0)\nu_{T_0}(t)\,\dd t,
 &\xi&=\Gamma(X),&\rho_*&=\rhoB+\alpha^2\xi,\\
 X_T&=\cU(T)X,&\xi_T&=\Gamma(X_T)=\cU(T)\xi.
 \end{aligned}
\end{equation}
The integral converges in operator norm, since
$\int |\nu_{T_0}(t)|\no{\cU(t)Y_0}\,\dd t\le\|\nu_{T_0}\|_1\no{Y_0}<\infty$.
Positivity of $\rho_*$ is part of the following theorem, under a smallness condition on $\alpha$.

\begin{theorem}[Approximate stationary state]\label{thm:reference}
For sufficiently small $c_0>0$, suppose
\begin{equation}\label{eq:reference-regime}
 0<\alpha^2\sigma<1,\qquad n\alpha^2\sigma\le c_0,\qquad
 T_0\ge\max\left\{1,\,2\sigma\sqrt{2\log\frac1{\alpha^2\sigma}}\right\}.
\end{equation}
Then the matrix $\rho_*$ in \eqref{eq:reference-definition} is a state and satisfies
\begin{equation}\label{eq:reference-conclusion}
 \rho_*\succeq\rhoB/2,\qquad
 \nr{\Gamma^{-1}(\rho_*-\rhoB)}\le\frac{C\alpha^2\sqrt n}{T_0},\qquad
 \nr{\Gamma^{-1}(\Phi_{\alpha}^{\mu_{T_0}}(\rho_*)-\rho_*)}\le Cn\alpha^4.
\end{equation}
The same upper bounds hold in trace norm for the two differences.
\end{theorem}
The proof combines \Cref{lem:correction,lem:remainders}, proved in \cref{subsec:kernel,subsec:remainders}, with the channel comparison of \Cref{thm:comparison}.
\begin{proof}
\proofstep{Normalization, positivity, and bias.}
Lemma~\ref{lem:correction} gives Hermiticity and trace one. Also
\[
 \alpha^2\no X\le C\alpha^2n/T_0
 \le Cc_0/(\sigma T_0)\le Cc_0/\sigma\le1/2
\]
when $c_0$ is sufficiently small; here $c_0$ may depend on $\beta$, since $\sigma=\beta/2$. Thus
\[
 \rho_*=\rhoB^{1/2}(\idmat+\alpha^2X)\rhoB^{1/2}\succeq\rhoB/2,
 \qquad \nr{\Gamma^{-1}(\rho_*-\rhoB)}=\alpha^2\nr X
 \le C\alpha^2\sqrt n/T_0.
\]

\proofstep{Separate the channel error from the \green{ideal-channel stationarity defect}.}
We split the \green{one-round stationarity defect} as
\begin{equation}\label{eq:reference-error-split}
 \Phi_{\alpha}^{\mu_{T_0}}(\rho_*)-\rho_*
 =(\Phi_{\alpha}^{\mu_{T_0}}-\widetilde\Phi_{\alpha}^{\mu_{T_0}})(\rho_*)
   +\widetilde\Phi_{\alpha}^{\mu_{T_0}}(\rho_*)-\rho_*.
\end{equation}
The lower bound on $T_0$ in~\eqref{eq:reference-regime} implies $e^{-T_0^2/(8\sigma^2)}\le\alpha^2\sigma$. Moreover,
\[
 \nr{\Gamma^{-1}\rho_*}\le1+\alpha^2\nr X
 \le1+\frac{C c_0}{\sigma T_0\sqrt n}\le2.
\]
Consequently Theorem~\ref{thm:comparison} gives
\[
 \begin{aligned}
 &\nr{\Gamma^{-1}[(\Phi_{\alpha}^{\mu_{T_0}}-\widetilde\Phi_{\alpha}^{\mu_{T_0}})(\rho_*)]}\\
 &\quad\le\nm{\Gamma^{-1}(\Phi_{\alpha}^{\mu_{T_0}}-\widetilde\Phi_{\alpha}^{\mu_{T_0}})\Gamma}
             \nr{\Gamma^{-1}\rho_*}\\
 &\quad\le2C(n\alpha^4\sigma^2+n\alpha^2\sigma e^{-T_0^2/(8\sigma^2)})
 \le C'n\alpha^4\sigma^2\le C''n\alpha^4.
 \end{aligned}
\]
The first step is the definition of the induced norm, the second applies the channel comparison, the third uses $e^{-T_0^2/(8\sigma^2)}\le\alpha^2\sigma$, and the fourth uses $\sigma=\beta/2$. This bounds the first term of \eqref{eq:reference-error-split}.

\proofstep{Cancel the first-order ideal drift.}
Since $\cU(T)\rho_*=\rhoB+\alpha^2\xi_T$, we expand $e^{\alpha^2\cL}$ on the two terms separately by Taylor's formula with integral remainder:
\[
 \begin{aligned}
 e^{\alpha^2\cL}(\rhoB)
 &=\rhoB+\alpha^2\cL(\rhoB)
   +\int_0^{\alpha^2}(\alpha^2-u)e^{u\cL}\cL^2(\rhoB)\,\dd u,\\
 e^{\alpha^2\cL}(\xi_T)
 &=\xi_T+\int_0^{\alpha^2}e^{u\cL}\cL(\xi_T)\,\dd u.
 \end{aligned}
\]
The first formula integrates the second derivative of $u\mapsto e^{u\cL}(\rhoB)$ twice, and the second integrates the first derivative of $u\mapsto e^{u\cL}(\xi_T)$ once. Thus
\begin{equation}\label{eq:two-remainders}
 \begin{aligned}
 \widetilde\Phi_{\alpha}^{\mu_{T_0}}(\rho_*)-\rho_*
 ={}&\alpha^2\underbrace{\bigl[(\E_T[\cU(2T)]-\id)\xi
                      +\E_T[\cU(T)\cL(\rhoB)]\bigr]}_{=0\text{ by }\eqref{eq:cohomology}}\\
 &+\E_T\!\left[\cU(T)\int_0^{\alpha^2}(\alpha^2-u)e^{u\cL}\cL^2(\rhoB)\,\dd u\right]\\
 &+\alpha^2\E_T\!\left[\cU(T)\int_0^{\alpha^2}e^{u\cL}\cL(\xi_T)\,\dd u\right].
 \end{aligned}
\end{equation}
This identity follows by adding the first Taylor formula to $\alpha^2$ times the second, applying the outer forward evolution $\cU(T)$, using $\cU(T)\xi_T=\cU(2T)\xi$, and averaging over $T$. By~\eqref{eq:cohomology}, only the two integral remainders remain.

\proofstep{Bound the two remainders.}
For every matrix $\delta$, the exponential series gives
\[
 \Gamma^{-1}e^{u\cL}(\delta)=e^{u\hL}\Gamma^{-1}(\delta),
 \qquad \nm{e^{u\hL}}\le e^{u\nm{\hL}}\le e^{Cnu}.
\]
The last inequality sums~\eqref{eq:site-norm} of Proposition~\ref{prop:quasilocal} over the $n$ sites and uses $\sigma=\beta/2$. For $0\le u\le\alpha^2$, this is at most $e^{Cc_0/\sigma}$. The norm of the second line of \eqref{eq:two-remainders}, after $\Gamma^{-1}$, is at most
\[
 \begin{aligned}
 \int_0^{\alpha^2}(\alpha^2-u)e^{Cnu}\nr{\Gamma^{-1}\cL^2(\rhoB)}\,\dd u
 &\le Cn e^{Cn\alpha^2}\int_0^{\alpha^2}(\alpha^2-u)\,\dd u\\
 &=\tfrac12 Cn\alpha^4e^{Cn\alpha^2}\le C'n\alpha^4.
 \end{aligned}
\]
Here the isometry property~\eqref{eq:free-isometry} of the forward evolution and the triangle inequality for the integral remove $\cU(T)$ and the expectation, and the bound on $\nr{\Gamma^{-1}\cL^2(\rhoB)}$ is the first estimate in~\eqref{eq:remainder-bounds}. For the third line, the corresponding bound is
\[
 \begin{aligned}
 \alpha^2\int_0^{\alpha^2}e^{Cnu}\E_T\!\left[\nr{\Gamma^{-1}\cL(\xi_T)}\right]\,\dd u
 &\le Cn\alpha^2e^{Cn\alpha^2}\int_0^{\alpha^2}\dd u\\
 &=Cn\alpha^4e^{Cn\alpha^2}\le C'n\alpha^4.
 \end{aligned}
\]
Here we exchanged $\E_T$ with the $u$-integral and used the second estimate in~\eqref{eq:remainder-bounds}. Adding both bounds controls the \green{ideal-channel stationarity defect}. Together with the bound on the channel error in \eqref{eq:reference-error-split}, this proves \eqref{eq:reference-conclusion}. Equation~\eqref{eq:trace-conversion} then gives both trace-norm conclusions.
\end{proof}

\subsection{The correction kernel and the exact cancellation}\label{subsec:kernel}
In this subsection, we show that the kernel $\nu_{T_0}$ is purely imaginary and integrable with a finite first moment (\Cref{lem:nu}). We then show that the correction $X$ cancels the averaged first-order drift exactly and satisfies the bounds used in the proof of \Cref{thm:reference} (\Cref{lem:correction}).

\begin{lemma}[An integrable correction kernel]\label{lem:nu}
The function $\nu_0$ defined in \eqref{eq:nu-definition} is purely imaginary and exponentially decaying. In particular,
\begin{equation}\label{eq:nu-moments}
 \widehat\nu_{T_0}(\omega)
 =\frac{\omega\widehat\mu_{T_0}(\omega)}{1-\widehat\mu_{T_0}(2\omega)},\qquad
 \|\nu_{T_0}\|_1\le C/T_0,\qquad
 \int_\R|t||\nu_{T_0}(t)|\,\dd t\le C.
\end{equation}
At $\omega=0$, the quotient is understood as its continuous extension, with value $i/(10T_0)$.
\end{lemma}
\begin{proof}
\proofstep{Analyticity on a strip.}
We first show that $h$ is analytic on a strip about the real axis. By \eqref{eq:time-transform},
\[
 \widehat\mu_0(z)=e^{iz}(1-iz)^{-4},\qquad
 |\widehat\mu_0(2\omega)|=(1+4\omega^2)^{-2}<1\quad(\omega\ne0).
\]
At zero,
\[
 \widehat\mu_0(0)=1,\quad \widehat\mu_0'(0)=5i,\quad
 1-\widehat\mu_0(2z)=-10iz+\mathcal{O}(z^2),\quad
 z\widehat\mu_0(z)=z+\mathcal{O}(z^2).
\]
Thus $h$ has a removable singularity at zero, with value $1/(-10i)=i/10$. For $z=x+iy$ and $|y|\le1/4$,
\[
 |\widehat\mu_0(2z)|
 =\frac{e^{-2y}}{((1+2y)^2+4x^2)^2}
 \le\frac{e^{1/2}}{(1/4+4x^2)^2}.
\]
Choose $R$ such that this is at most $1/2$ for $|x|\ge R$. On the remaining compact rectangle, the analytic denominator has only finitely many zeros and no real zero except zero. Choose $a>0$, $a<1/4$, smaller than the distance of its nonreal zeros to the real axis. Then $h$ is holomorphic for $|\operatorname{Im}z|<a$. Outside the rectangle,
\[
 |h(x+iy)|
 \le2|x+iy|\frac{e^{-y}}{((1+y)^2+x^2)^2}
 \le C|x|^{-3}
 \qquad(|y|\le a).
\]
On a smaller closed strip, $h$ is also bounded on the remaining compact part. Thus the horizontal integrals of $|h|$ are uniformly finite.

\proofstep{Exponential decay.}
For $0<a'<a$, shift the inverse Fourier contour to $\operatorname{Im}z=-a'$ for $t>0$ and to $+a'$ for $t<0$. The vertical integrals vanish because $h(x+iy)=\mathcal{O}(|x|^{-3})$ uniformly in the strip. The exponential on the shifted contour has modulus $e^{-a'|t|}$, giving
\[
 |\nu_0(t)|
 \le\frac{e^{-a'|t|}}{2\pi}\sup_{|y|\le a'}\int_\R|h(x+iy)|\,\dd x
 \le C_{a'}e^{-a'|t|}.
\]
In particular, $\nu_0$ is integrable and has finite absolute moments of every order, and Fourier inversion gives $\widehat\nu_0=h$.

\proofstep{Imaginary values and scaling.}
For real $\omega$ and $t$,
\[
 \begin{aligned}
 h(-\omega)&=-\overline{h(\omega)},\\
 \overline{\nu_0(t)}
 &=\frac1{2\pi}\int_\R e^{i\omega t}\overline{h(\omega)}\,\dd\omega
 =\frac1{2\pi}\int_\R e^{-i\omega t}\overline{h(-\omega)}\,\dd\omega
 =-\nu_0(t).
 \end{aligned}
\]
The first identity follows from $\widehat\mu_0(-\omega)=\overline{\widehat\mu_0(\omega)}$; the second changes $\omega$ to $-\omega$. This proves that $\nu_0$ is purely imaginary.

Finally, the scaling $t=T_0u$ gives
\[
 \begin{aligned}
 \widehat\nu_{T_0}(\omega)&=T_0^{-1}h(T_0\omega)
 =\frac{\omega\widehat\mu_{T_0}(\omega)}{1-\widehat\mu_{T_0}(2\omega)},\\
 \int_\R |t|^p|\nu_{T_0}(t)|\,\dd t
 &=T_0^{p-1}\int_\R|u|^p|\nu_0(u)|\,\dd u\le C_pT_0^{p-1}.
 \end{aligned}
\]
The cases $p=0,1$ give the bounds in~\eqref{eq:nu-moments}, and the value at zero is $T_0^{-1}h(0)=i/(10T_0)$.
\end{proof}

\begin{lemma}[Cancellation and size of the correction]\label{lem:correction}
The matrices $X,\xi,\rho_*$ are Hermitian, $\Tr(\rhoB X)=\Tr\xi=0$, and $\Tr\rho_*=1$. They satisfy
\begin{equation}\label{eq:cohomology}
 (\E_T[\cU(2T)]-\id)\xi+\E_T[\cU(T)\cL(\rhoB)]=0,
\end{equation}
and
\begin{equation}\label{eq:correction-bounds}
 \nr X=\nr{X_T}\le C\sqrt n/T_0,\qquad
 \sup_{T\in\R}(\nr{KX_T}+\nr{X_TK})\le Cn/T_0,
 \qquad\no X\le Cn/T_0.
\end{equation}
\end{lemma}
\begin{proof}
\proofstep{Hermiticity and normalization.}
Since $Y_0$ is anti-Hermitian and $\nu_{T_0}$ is purely imaginary, $X$ is Hermitian:
\[
 X^\dagger
 =-\int\cU(t)(Y_0^\dagger)\overline{\nu_{T_0}(t)}\,\dd t
 =-\int\cU(t)(-Y_0)(-\nu_{T_0}(t))\,\dd t=X.
\]
The first step takes the adjoint through the norm-convergent integral; the second uses $Y_0^\dagger=-Y_0$ and Lemma~\ref{lem:nu}. Similarly,
\[
 \begin{aligned}
 \Tr(\rhoB X)
 &=-\int\nu_{T_0}(t)\Tr(\rhoB\cU(t)Y_0)\,\dd t
 =-\Tr(\rhoB Y_0)\int\nu_{T_0}(t)\,\dd t=0,\\
 \xi^\dagger&=\Gamma(X^\dagger)=\xi,\qquad
 \Tr\xi=\Tr(\rhoB X)=0,\qquad \Tr\rho_*=1.
 \end{aligned}
\]
The second equality uses $\Tr(\rhoB\cU(t)Y_0)=\Tr(\rhoB Y_0)$, which holds because $\rhoB$ is invariant under the forward evolution.

\proofstep{The cancellation equation.}
For each Bohr component of $Y_0$, equations \eqref{eq:commutator-identity} and \eqref{eq:reference-definition} give
\[
 K_\omega=\omega Y_{0,\omega},\qquad
 X_\omega=-\widehat\nu_{T_0}(-\omega)Y_{0,\omega}.
\]
For $\omega\ne0$, the $\omega$-component of $\Gamma^{-1}$ applied to the left-hand side of~\eqref{eq:cohomology} is
\[
 \begin{aligned}
 &(\widehat\mu_{T_0}(-2\omega)-1)X_\omega
        +\widehat\mu_{T_0}(-\omega)K_\omega\\
 &\quad=\left[(\widehat\mu_{T_0}(-2\omega)-1)
       \frac{\omega\widehat\mu_{T_0}(-\omega)}{1-\widehat\mu_{T_0}(-2\omega)}
       +\omega\widehat\mu_{T_0}(-\omega)\right]Y_{0,\omega}=0.
 \end{aligned}
\]
The equality substitutes the multiplier $\widehat\nu_{T_0}(-\omega)$ from \eqref{eq:nu-moments}. At $\omega=0$, both contributions vanish because $\widehat\mu_{T_0}(0)=1$ and $K_0=0$. Summing over frequencies and applying $\Gamma$, which commutes with $\cU(t)$, proves \eqref{eq:cohomology}.

\proofstep{Norm bounds.}
By~\eqref{eq:reference-definition},
\[
 \begin{aligned}
 \nr X
 &\le\int|\nu_{T_0}(t)|\nr{\cU(t)Y_0}\,\dd t
 =\|\nu_{T_0}\|_1\nr{Y_0}\le C\sqrt n/T_0,\\
 \no X&\le\|\nu_{T_0}\|_1\no{Y_0}\le Cn/T_0,\qquad
 \nr{X_T}=\nr X,\\
 \nr{KX_T}+\nr{X_TK}
 &\le\int|\nu_{T_0}(t)|
       (\nr{KY_0(-(T+t))}+\nr{Y_0(-(T+t))K})\,\dd t\\
 &\le Cn\|\nu_{T_0}\|_1\le C'n/T_0.
 \end{aligned}
\]
The first line uses the triangle inequality, the isometry property~\eqref{eq:free-isometry} of the forward evolution, and \eqref{eq:drift-moments}. The second uses the unitary invariance of the operator norm, the operator-norm bound in~\eqref{eq:drift-moments}, and again~\eqref{eq:free-isometry}. The third substitutes $X_T$ and uses $\cU(T)\cU(t)=\cU(T+t)$; the fourth uses the time-uniform mixed moment in \eqref{eq:drift-moments}. Each line also uses $\|\nu_{T_0}\|_1\le C/T_0$ from~\eqref{eq:nu-moments}. This proves~\eqref{eq:correction-bounds}.
\end{proof}

\subsection{The two remainders}\label{subsec:remainders}
In this subsection, we prove the two estimates used in the proof of \Cref{thm:reference} to bound the integral remainders in~\eqref{eq:two-remainders}, which apply $\cL$ twice to $\rhoB$ and once to $\xi_T$. Using \Cref{cor:moments,lem:D-action}, we bound the KMS norm of $\Gamma^{-1}\cL^2(\rhoB)$ and the time-averaged KMS norm of $\Gamma^{-1}\cL(\xi_T)$ by $Cn$, uniformly in $T_0\ge1$.

\begin{lemma}[Remainder estimates]\label{lem:remainders}
For $T_0\ge1$,
\begin{equation}\label{eq:remainder-bounds}
 \nr{\Gamma^{-1}\cL^2(\rhoB)}\le Cn,\qquad
 \E_T\!\left[\nr{\Gamma^{-1}\cL(\xi_T)}\right]\le Cn.
\end{equation}
\end{lemma}
\begin{proof}
Since $\cL\Gamma=\Gamma\hL$, we have
\[
 \begin{aligned}
 \Gamma^{-1}\cL^2(\rhoB)
 &=\hL^2(\idmat)=\hL(K)=\cD(K)+K^2,\\
 \nr{\Gamma^{-1}\cL^2(\rhoB)}
 &\le\nr{\cD(K)}+\nr{K^2}\le Cn.
 \end{aligned}
\]
The first line uses $\Gamma(\idmat)=\rhoB$, $\hL(\idmat)=K$, and \eqref{eq:affine}. The final inequality uses Lemma~\ref{lem:D-action} on the quasi-local field $K$ at time zero and \eqref{eq:drift-moments}.

For the correction, the same transformation gives
\[
 \Gamma^{-1}\cL(\xi_T)
 =\hL(X_T)=\cD(X_T)+\tfrac12(KX_T+X_TK).
\]
The anticommutator has norm at most $Cn/T_0$ by \eqref{eq:correction-bounds}. For the other part,
\[
 \begin{aligned}
 \nr{\cD(X_T)}
 &\le\int|\nu_{T_0}(t)|\nr{\cD(Y_0(-(T+t)))}\,\dd t\\
 &\le Cn\int(1+|T+t|)|\nu_{T_0}(t)|\,\dd t\\
 &\le Cn\left((1+T)\|\nu_{T_0}\|_1
                     +\int|t||\nu_{T_0}(t)|\,\dd t\right)\\
 &\le Cn\left(1+\frac{1+T}{T_0}\right)\qquad(T\ge T_0).
 \end{aligned}
\]
The first step substitutes $X_T$, the second applies Lemma~\ref{lem:D-action}, the third uses $|T+t|\le T+|t|$, and the fourth uses \eqref{eq:nu-moments}. Averaging therefore gives
\[
 \begin{aligned}
 \E_T\!\left[\nr{\Gamma^{-1}\cL(\xi_T)}\right]
 &\le Cn\left(1+\frac{1+\E[T]}{T_0}\right)+Cn/T_0\\
 &=Cn\left(6+\frac1{T_0}\right)+Cn/T_0\le C'n.
 \end{aligned}
\]
The equality uses $\E[T]=5T_0$, and the final step uses $T_0\ge1$.
\end{proof}
\section{\texorpdfstring{Contraction of the averaged ideal channel $\widetilde\Phi_{\alpha}^{\mu_{T_0}}$}{Contraction of the averaged ideal channel}}\label{sec:mixing}

The \green{approximate stationary state} constructed in \Cref{sec:reference} is close to $\rhoB$ and, by \cref{thm:reference}~\eqref{eq:reference-conclusion}, is approximately preserved by the averaged \red{physical channel $\Phi_{\alpha}^{\mu_{T_0}}$} up to an error of order $\mathcal{O}(n\alpha^4)$. The next step is to control the mixing time of the averaged channel, as stated in \cref{thm:exact-mixing}. Once such a mixing estimate is established, we can show that the fixed point of the averaged channel is close to the \green{approximate stationary state}, with error of order $\mathcal{O}(n\alpha^2)$; see \eqref{eq:fix-reference} in \cref{thm:exact-mixing}. Combined with the approximation of the \green{approximate stationary state} to $\rhoB$, this yields the desired control of the iterates toward $\rhoB$, as stated in \cref{thm:exact-mixing}~\eqref{eq:mixing-iterate}.

In this section, we establish contraction of the averaged ideal channel on \red{traceless matrices}, which provides the key ingredient for controlling its mixing time. The main result is summarized in \cref{thm:ideal-contraction}. The proof proceeds in two steps. We first derive a first-order expansion that captures both dissipative and dephasing effects. We then analyze the resulting \green{Lamb shift form} through a decomposition into low- and high-frequency contributions.

We begin by introducing the setup and the two propositions (Propositions~\ref{assump:local_bounded} and~\ref{assump:gap}) used in the contraction analysis. We then state the main contraction theorem and the two intermediate results underlying its proof. After proving these results in sequence, we establish the local estimates on which they rely.

\subsection{The setting and the contraction statement}\label{subsec:contraction-theorem}

\paragraph{\green{The model and the weighted state-difference norm.}}
We recall the nearest-neighbor Hamiltonian~\eqref{eq:hamiltonian} on the open chain of $n$ qudits, and the definition of neighborhoods
\begin{equation}\label{eq:mixing-neighborhood}
 N_r(S)=\{y\in[n]:\dist(y,S)\le r\},\qquad
 |N_r(x)|\le b(r):=2\lfloor r\rfloor+1.
\end{equation}
Recall the KMS norm and the change of coordinates from \eqref{eq:rho-space}--\eqref{eq:Gamma}:
\[
 \nr O=\|\Gamma^{1/2}(O)\|_2,\qquad
 \Gamma(O)=\rhoB^{1/2}O\rhoB^{1/2}.
\]
Here $\pair{V}{W}_2=\Tr(V^\dagger W)$ is the Hilbert--Schmidt inner product. A state difference $Y$ is often measured by $\nr{\Gamma^{-1}(Y)}=\|\Gamma^{-1/2}(Y)\|_2$, which bounds $\|Y\|_1$ by \eqref{eq:trace-conversion}. The adjoint identity \eqref{eq:map-duality} gives
\[
 \ip{\Gamma^{-1}(X)}{\Gamma^{-1}(\mathcal M Y)}
 =\ip{\mathcal M^\dagger(\Gamma^{-1}(X))}{\Gamma^{-1}(Y)}.
\]
Our goal is to show contraction in this \green{weighted state-difference norm} $\nr{\Gamma^{-1}(Y)}$ for \red{every traceless matrix $Y$}.

\paragraph{Local boundedness of the generator and the Lamb shift.}
Write the \green{effective Lindbladian} as
\begin{equation}\label{eq:mixing-generator}
 \cL=\LK+\Q,\qquad \Q=-i[\HL,\cdot\,],\qquad
 \HL=\sum_x b_x,\qquad b_x:=B_x^{\mathrm{Lamb}}.
\end{equation}
By \Cref{prop:KMS}, $\LK(\rhoB)=0$ and
$\LK^\dagger=\Gamma^{-1}\LK\Gamma$ is self-adjoint and nonpositive in the KMS inner product. Each $b_x$ is Hermitian. We first record the following boundedness and quasi-locality proposition; it follows from \cref{prop:KMS} and the estimates of Section~\ref{sec:locality}, as shown after the statement. This will be used to control the Taylor remainder of $e^{s\cL}$.

\begin{proposition}[Local boundedness and quasi-locality of $\cL$]\label{assump:local_bounded}
There are $C,c>0$, independent of $n,x,R$, such that
\begin{equation}\label{eq:effective-locality}
 \begin{aligned}
 &\nm{\Gamma^{-1}\LK\Gamma}\le Cn,\\
 &\sup_{|u|\le\beta}\no{e^{uH_n}b_xe^{-uH_n}}\le C,\\
 &\sup_{|u|\le\beta}\no{e^{uH_n}b_xe^{-uH_n}
       -\cE_{x,R}(e^{uH_n}b_xe^{-uH_n})}\le Ce^{-cR}.
 \end{aligned}
\end{equation}
Here $\cE_{x,R}$ is the normalized partial trace onto $N_R(x)$, embedded back into the full operator algebra:
\[
 \cE_{x,R}(O)
 =\frac{\Tr_{[n]\setminus N_R(x)}(O)}{2^{q|[n]\setminus N_R(x)|}}
        \otimes \idmat_{[n]\setminus N_R(x)},
\]
with tensor factors in their original order.
\end{proposition}
The second and third inequalities in~\eqref{eq:effective-locality} follow from the quasi-locality of the local terms $b_x$ on imaginary-time strips in~\cref{prop:quasilocal}. For the first, write $\LK=\sum_x(\cL_x+i[b_x,\cdot\,])$ by~\cref{prop:KMS}. The identity $\Gamma^{-1}([b_x,\Gamma(O)])=\cB_{b_x^+,\idmat}(O)-\cB_{\idmat,b_x^-}(O)$, with $b_x^{\pm}=e^{\pm\beta H_n/4}b_xe^{\mp\beta H_n/4}$, and~\eqref{eq:weighted-algebra} give
\[
 \nm{\Gamma^{-1}\LK\Gamma}
 \le\sum_x\Bigl(\nm{\hL_x}+\no{b_x^+}+\no{b_x^-}\Bigr)\le Cn,
\]
by~\eqref{eq:site-norm} and the second inequality in~\eqref{eq:effective-locality}.

\paragraph{The global gap and comparison estimates.}\label{subsec:comparison-assumption}
Recall the Pauli strings $P_{x,a}=A_{x,a}$, $a\in[Q]$, from~\eqref{eq:couplings}.
The following proposition combines a comparison with the Pauli energy and a global dissipative gap.
\begin{proposition}[Global gap and comparison]\label{assump:gap} When $\sigma=\beta/2$, there are constants $C_\Gamma,\lambda>0$, independent of $n$, such that every observable $O$ satisfies
\begin{equation}\label{eq:gap}
 \begin{aligned}
 \ED(O)&:=-\ip{O-\Tr(\rhoB O)\idmat}
          {\LK^\dagger(O-\Tr(\rhoB O)\idmat)}\\
 &\ge C_\Gamma\sum_{x,a}\nr{[P_{x,a},O]}^2
 \ge\lambda\nr{O-\Tr(\rhoB O)\idmat}^2.
 \end{aligned}
\end{equation}
\end{proposition}

\begin{proof}
At $\sigma=\beta/2$, $\LK$ is a constant multiple of the Lindbladian of~\cite[Eq.~(2.19)]{BergamaschiChen} with the single-site Pauli jumps and the Gaussian weight. To avoid clashes with our notation, write $\mathcal L^{\mathrm{BC}}$ for that Lindbladian, $\mathcal K^{\mathrm{BC}}$ for their auxiliary generator~\cite[Eq.~(3.6)]{BergamaschiChen}, and $\sigma_{\mathrm{BC}}$ for their Gaussian width. Take $\sigma_{\mathrm{BC}}=\beta^{-1}$. By~\eqref{eq:pulse},
\[
 f_{\beta/2}(t)=\Bigl(\frac{\sqrt{2/\pi}}{\beta}\Bigr)^{1/2}e^{-t^2/\beta^2},
\]
which is exactly the filter of~\cite[Eq. (2.16)]{BergamaschiChen} at width $\sigma_{\mathrm{BC}}=\beta^{-1}$. Hence $V_A(u)=\sqrt{2\pi}\,\widehat A(u)$, where $\widehat A$ is their operator Fourier transform ~\cite[Eq.~(2.15)]{BergamaschiChen}. At this width, their Gaussian weight~\cite[Eq.~(2.20)]{BergamaschiChen} is $\gamma_G(\omega)=e^{-(\beta\omega+1)^2/2}$, so $g_{\beta,\sigma}=\beta\gamma_G/\sqrt{2\pi}$. Therefore the transition and decay parts of $\LK$ in~\eqref{eq:LK} equal $c_\beta:=\sqrt{2\pi}\,\beta$ times those of $\mathcal L^{\mathrm{BC}}$ with jumps $\{P_{x,a}\}$; the identity string, if included in their jump set, contributes nothing. Both coherent parts are the \green{KMS coherent correction} of~\cite[Lemma~II.1]{CKG} (see~\eqref{eq:Lamb} and~\cite[Eq.~(2.21)]{BergamaschiChen}), which is linear in the transition part. Hence $\LK=c_\beta\mathcal L^{\mathrm{BC}}$.

Both observable generators annihilate $\idmat$ and are \green{KMS-self-adjoint}. Hence
\[
 \ED(O)=c_\beta\ip{O}{-(\mathcal L^{\mathrm{BC}})^\dagger(O)}.
\]
By~\cite[Theorem~VI.1]{BergamaschiChen}, there is $\alpha_{\mathrm{BC}}\in(0,\infty)$, depending only on $\beta$ and $q$, such that
\[
 \alpha_{\mathrm{BC}}\ip{O}{-(\mathcal L^{\mathrm{BC}})^\dagger(O)}
 \ge\ip{O}{-\mathcal K^{\mathrm{BC}}(O)}
 =\sum_{x,a}\nr{[P_{x,a},O]}^2 .
\]
This gives the first inequality in~\eqref{eq:gap} with $C_\Gamma=c_\beta/\alpha_{\mathrm{BC}}$. The second follows from \Cref{lem:aux-gap} with $I=[n]$, for which $H_I=H_n$ and $\rho_I=\rhoB$, with $\lambda=C_\Gamma\gamma$.
\end{proof}

\red{\Cref{assump:gap}} implies that the Dirichlet form associated with $\mathcal{L}_{\rm KMS}$ \red{dominates the Pauli energy up to a constant factor} and provides a global dissipative gap.

\paragraph{Results from Section~\ref{sec:weighted}.}
The local structure of the Gibbs state enters only through the spectral conditional expectations $p_I$ and the algebras $\mathfrak A(S)$ of \Cref{subsec:projections}, and through \Cref{prop:thermal} and \Cref{lem:aux-gap}, which are proved in Appendix~\ref{app:1d} for every chain considered here. In particular, \Cref{lem:annihilator} and \Cref{thm:field} are available in this section without further assumptions. We say that a map $\mathcal M$ has coefficient support in $S$ if $\mathcal M\in\mathfrak A(S)$; by \eqref{eq:weighted-algebra}, its Hilbert--Schmidt form $V\mapsto\Gamma^{1/2}\mathcal M\Gamma^{-1/2}(V)$ is then a finite sum of maps $V\mapsto BVC$ with $B,C$ supported in $S$.

\paragraph{The ideal channel and the contraction statement.}
Let $\cU(t)X=e^{-itH_n}Xe^{itH_n}$. For $s\ge0$, define
\begin{equation}\label{eq:theta}
 \Theta_s=\E_T[\cU(T)e^{s\cL}\cU(T)],\qquad
 M_\mu:=\Theta_0=\E_T\cU(2T),
\end{equation}
where the expectation \red{over} $T$ is taken with respect to \eqref{eq:time-law}:
\[
 \mu_{T_0}(t)=T_0^{-1}\mu_0(t/T_0),\qquad
 \mu_0(u)=\tfrac16(u-1)^3e^{-(u-1)}\mathbf1_{\{u\ge1\}},\qquad T_0>0\,.
\]
We note $\widetilde\Phi_{\alpha}^{\mu_{T_0}}=\E_T[\widetilde\Phi_{\alpha,T}]=\E_T[\cU(T)e^{\alpha^2\cL}\cU(T)]=\Theta_{\alpha^2}$, with $\widetilde\Phi_{\alpha,T}$ as in~\eqref{eq:ideal-channel}.

For \red{any traceless matrix $X$}, set
\begin{equation}\label{eq:DE}
 \begin{aligned}
 D_\mu(X)&:=\nr{\Gamma^{-1}(X)}^2-\nr{\Gamma^{-1}(M_\mu X)}^2=\E_T\nr{\Gamma^{-1}(\cU(2T)X-M_\mu X)}^2,\\
 E_\mu(X)&:=\E_T\ED(\Gamma^{-1}(\cU(T)X)).
 \end{aligned}
\end{equation}
In particular, according to \eqref{eq:gap} and $\Tr(\rhoB\Gamma^{-1}(\cU(T)X))=\Tr X=0$,
\begin{equation}\label{eq:averaged-gap}
 D_\mu(X)\ge0,\qquad E_\mu(X)\ge\lambda\nr{\Gamma^{-1}(X)}^2.
\end{equation}

Then, we state the main contraction result:
\begin{theorem}[Contraction of $\Theta_s$]\label{thm:ideal-contraction}
In the setting of Propositions~\ref{assump:local_bounded} and~\ref{assump:gap}, there is $c_*>0$, independent of $n,s,T_0$, such that
\[
 0<ns\le c_*,\qquad T_0\ge1
\]
imply, for every traceless $X$,
\begin{equation}\label{eq:ideal-contraction}
\begin{aligned}
 \nr{\Gamma^{-1}(\Theta_sX)}^2
 \le\nr{\Gamma^{-1}(X)}^2-\tfrac12D_\mu(X)-\tfrac12sE_\mu(X)\le(1-\tfrac12\lambda s)\nr{\Gamma^{-1}(X)}^2.
 \end{aligned}
\end{equation}
\end{theorem}

The proof rests on the following two estimates.
\paragraph{Step 1: first-order expansion.} Separate the random forward evolution from the generator step, and retain the negative dephasing and dissipative terms.
  \begin{lemma}[One-step estimate]\label{lem:step}
In the setting of Propositions~\ref{assump:local_bounded} and~\ref{assump:gap}, there is a constant $C$, independent of $n,s,T_0$, such that for $0\le s\le C^{-1}/n$ and traceless $X$, with $Y_T=\cU(T)X$,
\begin{align}
 \norm{\Gamma^{-1}(\Theta_sX)}_{\rhoB}^{2}
 \le{}&\norm{\Gamma^{-1}(X)}_{\rhoB}^{2}-D_\mu(X)-2sE_\mu(X)
 +2s\E_T\Rea\pair{\Gamma^{-1}(Y_T)}{\Gamma^{-1}(\Q Y_T)}_{\rhoB}\notag\\
 &+Cs\sqrt{nE_\mu(X)D_\mu(X)}+Cs^2nE_\mu(X).
 \label{eq:one-step}
\end{align}
\end{lemma}

\paragraph{Step 2: control of the \green{Lamb shift form}.} Split into low and high Bohr frequencies. The low-frequency estimate uses the field bound of \Cref{thm:field}, through \Cref{prop:ac_Gm}.
\begin{lemma}[The same-time \green{Lamb shift form}]\label{lem:lamb}
In the setting of Propositions~\ref{assump:local_bounded} and~\ref{assump:gap}, let $X$ be traceless and $Y_T=\cU(T)X$. For every $T_0\ge1$,
\begin{equation}\label{eq:lamb}
 \left|2\E_T\Rea\pair{\Gamma^{-1}(Y_T)}{\Gamma^{-1}(\Q Y_T)}_{\rhoB}\right|
 \le C\sqrt{nE_\mu(X)D_\mu(X)}+CnD_\mu(X).
\end{equation}
\end{lemma}
The proofs of the two lemmas are given in Sections~\ref{sec:step} and~\ref{sec:field}, respectively. We first combine these estimates to prove the theorem.
\begin{proof}[Proof of Theorem~\ref{thm:ideal-contraction}]
Since $T_0\ge1$, Lemma~\ref{lem:step} and Lemma~\ref{lem:lamb} yield
\[
 \begin{aligned}
 \nr{\Gamma^{-1}(\Theta_sX)}^2
 &\le\nr{\Gamma^{-1}(X)}^2-D_\mu(X)-2sE_\mu(X)\\
 &\quad+C_1s\sqrt{nE_\mu(X)D_\mu(X)}
       +C_2nsD_\mu(X)+C_3ns^2E_\mu(X).
 \end{aligned}
\]
where $C_1,C_2,C_3$ are independent of $n,s,T_0$. Young's inequality gives
\[
 C_1s\sqrt{nE_\mu(X)D_\mu(X)}\le\tfrac12sE_\mu(X)+C_4nsD_\mu(X).
\]
Consequently,
\[
 \nr{\Gamma^{-1}(\Theta_sX)}^2
 \le\nr{\Gamma^{-1}(X)}^2
  -\bigl(1-(C_2+C_4)ns\bigr)D_\mu(X)
  -\bigl(3/2-C_3ns\bigr)sE_\mu(X).
\]
Choose $c_*$ within the range of Lemma~\ref{lem:step}, with
$(C_2+C_4)c_*\le1/2$ and $C_3c_*\le1$. This gives the first inequality in \eqref{eq:ideal-contraction}; the second follows from \eqref{eq:averaged-gap}. Decrease $c_*$ if necessary so that $\lambda s/2\le1$ throughout this regime.
\end{proof}

\subsection{The one-step estimate: proof of Lemma~\ref{lem:step}}\label{sec:step}

We first separate the random forward evolution from the Lindbladian evolution, then expand the Lindbladian evolution to first order in $s$.

\begin{proof}[Proof of Lemma~\ref{lem:step}]

The first decomposition isolates $M_\mu X$. Define $L_T=\cU(T)(e^{s\cL}-\id)\cU(T)X$. We have
\begin{equation}\label{eqn:Gamma_theta_s}
\begin{aligned}
 \nr{\Gamma^{-1}(\Theta_sX)}^2
 &=\nr{\Gamma^{-1}(M_\mu X+\E_TL_T)}^2\\
 &=\nr{\Gamma^{-1}(M_\mu X)}^2
   +2\Rea\ip{\Gamma^{-1}(M_\mu X)}{\Gamma^{-1}(\E_TL_T)}
   +\nr{\Gamma^{-1}(\E_TL_T)}^2\\
 &=\nr{\Gamma^{-1}(X)}^2-D_\mu(X)
   +2\Rea\ip{\Gamma^{-1}(M_\mu X)}{\Gamma^{-1}(\E_TL_T)}
   +\nr{\Gamma^{-1}(\E_TL_T)}^2.
 \end{aligned}
\end{equation}

We next control the last two terms in~\eqref{eqn:Gamma_theta_s}. Recall that $Y_T=\cU(T)X$.

\paragraph{The squared increment.}

To bound the last term in~\eqref{eqn:Gamma_theta_s}, note that for every $T$,
\begin{equation}\label{eq:second_term_control}
\begin{aligned}
 \nr{\Gamma^{-1}(L_T)}^2
 &=\nr{\Gamma^{-1}((e^{s\cL}-\id)Y_T)}^2\\
 &=\left\|\int_0^s(\Gamma^{-1}e^{t\cL}\Gamma)
                    \Gamma^{-1}(\cL Y_T)\,\dd t\right\|_{\rhoB}^2.
 \end{aligned}
\end{equation}
Because $\norm{\Gamma^{-1}\cL\Gamma}_{\rhoB\to\rhoB}\leq C_0n$ according to Proposition~\ref{prop:Ln_bounded}~\eqref{eqn:L_n_bounded}, we have $\|\Gamma^{-1}e^{t\cL}\Gamma\|_{\rhoB\to\rhoB}
=\|e^{t\Gamma^{-1}\cL\Gamma}\|_{\rhoB\to\rhoB}
\le e^{C_0nt}$. Substituting this bound and using $s=\mathcal{O}(n^{-1})$ gives
\[
 \norm{\Gamma^{-1}(L_T)}_{\rhoB}^2=\norm{\Gamma^{-1}\left(\left(e^{s\cL}-\id\right)Y_T\right)}_{\rhoB}^2=\mathcal{O}\left(s^2\left\|\Gamma^{-1}\left(\cL Y_T\right)\right\|^2_{\rhoB}\right).
\]
By Proposition~\ref{prop:Ln_bounded}~\eqref{eq:AS} and Jensen's inequality, we have
\begin{equation}\label{eq:Csize}
\begin{aligned}
 \nr{\Gamma^{-1}(\E_TL_T)}^2
 \le\E_T\nr{\Gamma^{-1}(L_T)}^2\le Cs^2\E_T\nr{\Gamma^{-1}(\cL Y_T)}^2
 \le Cs^2nE_\mu(X).
 \end{aligned}
\end{equation}
This gives the last term in~\eqref{eq:one-step}.

\paragraph{The cross term.}

Insert and subtract the same realization $\cU(2T)X$ in the cross term:
\[
\begin{aligned}
 &\Rea\pair{\Gamma^{-1}(M_\mu X)}{\Gamma^{-1}(\E_T L_T)}_{\rhoB}\\
 =&\E_T\Rea\pair{\Gamma^{-1}(M_\mu X-\cU(2T)X)}{\Gamma^{-1}(L_T)}_{\rhoB}+\E_T\Rea\pair{\Gamma^{-1}(Y_T)}{\Gamma^{-1}((e^{s\cL}-\id)Y_T)}_{\rhoB}.
\end{aligned}
\]
By the definition in~\eqref{eq:DE}, Cauchy--Schwarz, and the estimate used in~\eqref{eq:Csize}, we have
\[
\begin{aligned}
\left|\E_T\Rea\pair{\Gamma^{-1}(M_\mu X-\cU(2T)X)}{\Gamma^{-1}(L_T)}_{\rhoB}\right|\le\sqrt{D_\mu(X)\,\E_T\norm{\Gamma^{-1}(L_T)}_{\rhoB}^2}
 \le Cs\sqrt{nE_\mu(X)D_\mu(X)}.
\end{aligned}
\]
This gives the mixed error term in~\eqref{eq:one-step}. It remains to control the same-time term
\[
\E_T\Rea\pair{\Gamma^{-1}(Y_T)}{\Gamma^{-1}((e^{s\cL}-\id)Y_T)}_{\rhoB}\,.
\]
We now separate the first-order term from the Taylor remainder.

\paragraph{The first-order term and its remainder.} For the same-time term,
\begin{equation}\label{eqn:separate_first_order}
\begin{aligned}
 &\E_T\Rea\pair{\Gamma^{-1}(Y_T)}{\Gamma^{-1}((e^{s\cL}-\id)Y_T)}_{\rhoB}\\
 =&\E_T\Rea\pair{\Gamma^{-1}(Y_T)}{\Gamma^{-1}((e^{s\cL}-\id-s\cL)(Y_T))}_{\rhoB}
   +s\E_T\Rea\pair{\Gamma^{-1}(Y_T)}{\Gamma^{-1}(\cL Y_T)}_{\rhoB}\\
 \le &-sE_\mu(X)+s\E_T\Rea\pair{\Gamma^{-1}(Y_T)}{\Gamma^{-1}(\Q Y_T)}_{\rhoB}+\left|\E_T\Rea\pair{\Gamma^{-1}(Y_T)}{\Gamma^{-1}((e^{s\cL}-\id-s\cL)(Y_T))}_{\rhoB}\right|.
\end{aligned}
\end{equation}
To bound the last term in~\eqref{eqn:separate_first_order},
\begin{align*}
 &\left|\mathbb{E}_T\pair{\Gamma^{-1}(Y_T)}{\Gamma^{-1}((e^{s\cL}-\id-s\cL)(Y_T))}_{\rhoB}\right|\\
 =&\left|\mathbb{E}_T\int_0^s(s-t)\pair{\Gamma^{-1}(Y_T)}{\Gamma^{-1}(e^{t\cL}\cL^2Y_T)}_{\rhoB}\,\dd t\right|=\left|\mathbb{E}_T\int_0^s(s-t)\pair{\cL^\dagger(\Gamma^{-1}(Y_T))}{\Gamma^{-1}(e^{t\cL}\cL Y_T)}_{\rhoB}\,\dd t\right|\\
 \leq &Cs^2\mathbb{E}_T\left[\norm{\cL^\dagger(\Gamma^{-1}(Y_T))}_{\rhoB}^2+\norm{\Gamma^{-1}(\cL Y_T)}_{\rhoB}^2\right]\\
 \leq &Cs^2nE_\mu(X).
\end{align*}
The first inequality follows from Cauchy--Schwarz, $ab\leq(a^2+b^2)/2$, and the previously established bound
\[
 \norm{\Gamma^{-1}e^{t\cL}\Gamma}_{\rhoB\to\rhoB}\leq e^{C_0nt}\leq C,\qquad 0\leq t\leq s,
\]
since $ns=\mathcal{O}(1)$ and $\int_0^s(s-t)\,\dd t=s^2/2$. The second inequality follows from Proposition~\ref{prop:Ln_bounded}~\eqref{eq:AS}. Combining these estimates in~\eqref{eqn:Gamma_theta_s} proves~\eqref{eq:one-step}.
\end{proof}

\subsection[The Lamb shift form: proof of Lemma~\ref{lem:lamb}]{The \green{Lamb shift form}: proof of Lemma~\ref{lem:lamb}}\label{sec:field}
We begin with the Bohr decomposition and the two frequency estimates supplied by dephasing. Define $\ad_{H_n}(O):=[H_n,O]$. For each Bohr frequency $\omega\in\{E-E':E,E'\in\operatorname{spec}(H_n)\}$, let $X_\omega$ be the Bohr component of $X$ from~\eqref{eq:Bohr}, with $\nu$ renamed $\omega$, so that
\[
 X=\sum_\omega X_\omega,\qquad \ad_{H_n}(X_\omega)=\omega X_\omega.
\]
Since $[\rhoB,H_n]=0$, distinct Bohr components are orthogonal after applying $\Gamma^{-1}$ in $\pair{\cdot}{\cdot}_{\rhoB}$, and
\[
 \cU(t)X_\omega=e^{-it\omega}X_\omega,\qquad
 M_\mu X_\omega=\widehat\mu_{T_0}(-2\omega)X_\omega.
\]
Thus \eqref{eq:DE} and \eqref{eq:time-transform} give
\[
\begin{aligned}
 &D_\mu(X)
 =\norm{\Gamma^{-1}(X)}_{\rhoB}^2-\norm{\Gamma^{-1}(M_\mu X)}_{\rhoB}^2=\sum_\omega\left(1-\left|\widehat\mu_{T_0}(-2\omega)\right|^2\right)\norm{\Gamma^{-1}(X_\omega)}_{\rhoB}^2\\
=&\sum_\omega\left(1-\frac{1}{(1+4T_0^2\omega^2)^4}\right)
             \norm{\Gamma^{-1}(X_\omega)}_{\rhoB}^2.
\end{aligned}
\]
To separate the components efficiently dephased by the random interaction time from those near zero frequency, define
\begin{equation}\label{eqn:Pi_def}
\Pi:=\mathbf1_{[-1/T_0,1/T_0]}(\ad_{H_n}),\qquad
\Pi(O)=\sum_{\substack{|\omega|\le 1/T_0}}O_{\omega}\,.
\end{equation}
The scalar bounds
$u^2\mathbf1_{\{|u|\le1\}}\le2[1-(1+4u^2)^{-4}]$ and
$\mathbf1_{\{|u|>1\}}\le2[1-(1+4u^2)^{-4}]$ give
\begin{equation}\label{eqn:high_freq}
 \left\|\Gamma^{-1}\left(\ad_{H_n}\left(\Pi X\right)\right)\right\|^2_{\rhoB}\leq \frac{2D_\mu(X)}{T^2_0} ,\quad \left\|\Gamma^{-1}((\id-\Pi) X)\right\|_{\rhoB}^2\leq 2D_\mu(X)\,.
\end{equation}
We will use the second inequality to control the high-frequency part of $Y_T$, which gives the second term in~\eqref{eq:lamb}.

For low-frequency components, $D_\mu(X)$ does not directly control $\norm{\Gamma^{-1}(\Pi X)}_{\rhoB}^2$ because the random forward evolution weakly dephases these components. The \green{symmetric Lamb shift map}, however, is a commutator with $\ad_{H_n}$. This supplies a factor of the Bohr frequency and gives the first term in~\eqref{eq:lamb}.

\begin{proof}[Proof of Lemma~\ref{lem:lamb}]
First observe that
\[
\pair{\Gamma^{-1}(Y_T)}{\Gamma^{-1}(\Q Y_T)}_{\rhoB}=\pair{\rho^{-1/4}_{\beta}Y_T\rho^{-1/4}_{\beta}}{\rho^{-1/4}_{\beta}\Q\left(\rho^{1/4}_\beta\left[\rho^{-1/4}_\beta Y_T\rho^{-1/4}_{\beta}\right]\rho^{1/4}_\beta\right)\rho^{-1/4}_\beta}_2
\]
Thus, we define $F_T=\Gamma^{-1/2}(Y_T)=\rho^{-1/4}_{\beta}Y_T\rho^{-1/4}_{\beta}$.  Taking the real part gives
\[
\Rea\left(\pair{\Gamma^{-1}(Y_T)}{\Gamma^{-1}(\Q Y_T)}_{\rhoB}\right)=\left\langle F_T,\mathcal{S}_{\rm Lamb}[F_T]\right\rangle_2\,,
\]
where $\mathcal{S}_{\rm Lamb}=\frac{1}{2}\left(\Gamma^{-1/2}\Q\Gamma^{1/2}+\Gamma^{1/2}\Q^\dagger\Gamma^{-1/2}\right)$. We use two representations of $\mathcal{S}_{\rm Lamb}$:
\begin{itemize}
\item $\mathcal{S}_{\rm Lamb}[F_T]=\{K^{\rm mod},F_T\}$, where
\[
K^{\rm mod}=\frac{1}{2i}\left(\rhoB^{-1/4}\HL\rhoB^{1/4}-\rhoB^{1/4}\HL\rhoB^{-1/4}\right).
\]
By the local bounds on the summands of $\HL$ in Proposition~\ref{assump:local_bounded}, we have $\|K^{\rm mod}\|_{\rm op}\leq Cn$ and $\|\mathcal{S}_{\rm Lamb}\|_{2\to 2}\leq Cn$.

\item $\mathcal{S}_{\rm Lamb}=[\ad_{H_n},-i\ac_{\Gm}]$, where
\[
\ad_{H_n}(O)=[H_n,O],\qquad \ac_{\Gm}(O)=\{\Gm,O\}.
\]
Explicitly,
\[
\mathcal{S}_{\rm Lamb}[F_T]=-i[H_n,\{\Gm,F_T\}]+i\{\Gm,[H_n,F_T]\},
\]
where
\[
\Gm=\frac12\int_{-\beta/4}^{\beta/4}
e^{uH_n}\HL e^{-uH_n}\,\dd u.
\]
Indeed, differentiation under the integral gives
\[
 -i[H_n,\Gm]
 =\frac1{2i}\left(\rhoB^{-1/4}\HL\rhoB^{1/4}
                  -\rhoB^{1/4}\HL\rhoB^{-1/4}\right)=K^{\rm mod},
\]
and $[\ad_{H_n},\ac_{\Gm}]=\ac_{[H_n,\Gm]}$. Proposition~\ref{assump:local_bounded} also gives $\|\Gm\|_{\rm op}\le Cn$.

\end{itemize}
To prove~\eqref{eq:lamb}, we separate the low-frequency part and high-frequency part of $F_T$ using $\Pi$, defined in~\eqref{eqn:Pi_def}:
\begin{equation}\label{eq:split}
\begin{aligned}
 \pair{F_T}{\mathcal S_{\rm Lamb}F_T}_2
 ={}&\pair{\Pi F_T}{\mathcal S_{\rm Lamb}\Pi F_T}_2\\
 &-\pair{(\id-\Pi)F_T}{\mathcal S_{\rm Lamb}(\id-\Pi)F_T}_2\\
 &+2\Rea\pair{(\id-\Pi)F_T}{\mathcal S_{\rm Lamb}F_T}_2.
 \end{aligned}
\end{equation}
For the second term, we use the fact that $\|\mathcal{S}_{\rm Lamb}\|_{2\to 2}\leq Cn$ and
\[
 \|(\id-\Pi)F_T\|^2_2=\left\|\Gamma^{-1}((\id-\Pi)Y_T)\right\|^2_{\rhoB}\leq
\left\|\Gamma^{-1}((\id-\Pi) X)\right\|_{\rhoB}^2\leq 2D_\mu(X)
\]
by~\eqref{eqn:high_freq}. Thus its contribution is at most $CnD_\mu(X)$. For the mixed term, Cauchy--Schwarz gives
\[
 2\left|\Rea\pair{(\id-\Pi)F_T}{\mathcal S_{\rm Lamb}F_T}_2\right|
 \le2\sqrt{2D_\mu(X)}\,\|\mathcal S_{\rm Lamb}F_T\|_2.
\]
To bound the remaining factor, use
\[
\begin{aligned}
 \mathcal{S}_{\rm Lamb}F_T=\frac{1}{2}\left(\Gamma^{-1/2}\Q\Gamma^{1/2}
   +\Gamma^{1/2}\Q^\dagger\Gamma^{-1/2}\right)F_T=\frac{1}{2}\Gamma^{-1/2}\left(\Q Y_T+\Gamma\Q^\dagger\Gamma^{-1}(Y_T)\right).
\end{aligned}
\]
Here $\Gamma^{1/2}F_T=Y_T$.
Because $\norm{\Gamma^{-1/2}(Y)}_2=\norm{\Gamma^{-1}(Y)}_{\rhoB}$, we have
\[
\|\mathcal{S}_{\rm Lamb}F_T\|_2\leq \frac{1}{2}\left(\norm{\Gamma^{-1}(\Q Y_T)}_{\rhoB}+\norm{\Q^\dagger(\Gamma^{-1}(Y_T))}_{\rhoB}\right).
\]
Applying Proposition~\ref{prop:Ln_bounded}~\eqref{eq:AS} gives
\[
\|\mathcal{S}_{\rm Lamb}F_T\|_2^2
\leq \frac{1}{2}\left(\norm{\Gamma^{-1}(\Q Y_T)}_{\rhoB}^{2}+\norm{\Q^\dagger(\Gamma^{-1}(Y_T))}_{\rhoB}^{2}\right)
\leq Cn\ED(\Gamma^{-1}(Y_T)).
\]
This implies
\[
2\mathbb{E}_T\left|\Rea\left\langle (\id-\Pi)F_T,\mathcal{S}_{\rm Lamb}F_T\right\rangle_2\right|\leq C\sqrt{nD_\mu(X)E_\mu(X)}.
\]

It remains to control the low-frequency part $\left\langle \Pi F_T,\mathcal{S}_{\rm Lamb}\Pi F_T\right\rangle_2$. Since $\ad_{H_n}$ is Hilbert--Schmidt self-adjoint and the scalar part of $\ac_{\Gm}$ commutes with it, the second representation gives
\[
\begin{aligned}
\left|\left\langle \Pi F_T,\mathcal{S}_{\rm Lamb}\Pi F_T\right\rangle_2\right|
&=2\left|\Rea\pair{\ad_{H_n}\Pi F_T}{-i(\ac_{\Gm}-2\Tr(\rho_\beta G^{\rm mod})\id)\Pi F_T}_2\right|\\
 &\leq 2\left\|\ad_{H_n}\Pi F_T\right\|_2\left\|(\ac_{\Gm}-2\Tr(\rho_\beta G^{\rm mod})\id)\Pi F_T\right\|_2.
\end{aligned}
\]
The two factors are bounded separately. First, \eqref{eqn:high_freq} gives
\[
 \|\ad_{H_n}\Pi F_T\|_2
 =\nr{\Gamma^{-1}(\ad_{H_n}\Pi X)}
 \le\frac{\sqrt{2D_\mu(X)}}{T_0}.
\]
For the second factor, set
$\mathcal A_G=\ac_{\Gm}-2\Tr(\rhoB\Gm)\id$. Adding and subtracting the full $F_T$ gives
\[
 \begin{aligned}
 \|\mathcal A_G\Pi F_T\|_2
 &\le\|\mathcal A_G(\id-\Pi)F_T\|_2+\|\mathcal A_GF_T\|_2\\
 &\le Cn\sqrt{D_\mu(X)}
       +C\sqrt{n\ED(\Gamma^{-1}(Y_T))}.
 \end{aligned}
\]
The first term uses $\|\mathcal A_G\|_{2\to2}\le Cn$ and \eqref{eqn:high_freq}. For the second, apply Proposition~\ref{prop:ac_Gm} with $V=F_T$. Its scalar term vanishes because
$\pair{\rhoB^{1/2}}{F_T}_2=\Tr Y_T=0$.
Averaging the product of these two bounds gives
\[
 \E_T\left|\pair{\Pi F_T}{\mathcal S_{\rm Lamb}\Pi F_T}_2\right|
 \le \frac{C}{T_0}\sqrt{nE_\mu(X)D_\mu(X)}
       +\frac{Cn}{T_0}D_\mu(X).
\]
Together with the high-frequency and mixed terms in \eqref{eq:split}, this proves \eqref{eq:lamb}, since $T_0\ge1$. The frequency split is an estimate only; it does not require a lower bound on nonzero Bohr frequencies or a modification of the physical channel.
\end{proof}

\subsection{Local estimates}\label{sec:local-estimates}

We now establish the generator and \green{Lamb shift estimates} used in the preceding two proofs.

The first proposition bounds the generator and its adjoint in terms of the \green{Dirichlet form}.
\begin{proposition}[Norm and energy-action bounds]\label{prop:Ln_bounded}
In the setting of Propositions~\ref{assump:local_bounded} and~\ref{assump:gap}, there is $C>0$, independent of $n$, such that
\begin{equation}\label{eqn:L_n_bounded}
 \nm{\Gamma^{-1}\cL\Gamma}\le Cn.
\end{equation}
For every traceless matrix $Y$,
\begin{equation}\label{eq:AS}
 \begin{aligned}
 &\nr{\Gamma^{-1}(\cL Y)}^2+\nr{\cL^\dagger(\Gamma^{-1}(Y))}^2\\
 &\quad+\nr{\Gamma^{-1}(\Q Y)}^2+\nr{\Q^\dagger(\Gamma^{-1}(Y))}^2
 \le Cn\ED(\Gamma^{-1}(Y)).
 \end{aligned}
\end{equation}
\end{proposition}
The proof of Proposition~\ref{prop:Ln_bounded} is given in Section~\ref{sec:AS}.

Second, we recall the definition of $G^{\rm mod}$ in the proof of Lemma~\ref{lem:lamb}:
\begin{equation}\label{eqn:G_mod_def}
G^{\rm mod}=\frac12\int_{-\beta/4}^{\beta/4}
e^{uH_n}\HL e^{-uH_n}\,\dd u.
\end{equation}
The following proposition bounds the deviation of $\ac_{\Gm}$ from a scalar multiple of the identity in terms of the Dirichlet form $\ED$:
\begin{proposition}[The \green{Gibbs-centered} Lamb shift term]\label{prop:ac_Gm}
In the setting of Propositions~\ref{assump:local_bounded} and~\ref{assump:gap}, there exists a constant $C>0$, independent of $n$, such that for every $V\in\mathcal B(\mathcal H_{[n]})$,
\begin{equation}\label{eq:PF}
 \norm{(\ac_{\Gm}-2\Tr(\rho_\beta G^{\rm mod})\id)V}_2^2
 \le Cn\left(\ED(\Gamma^{-1/2}(V))+|\pair{\rhoB^{1/2}}{V}_2|^2\right).
\end{equation}
Here $\Gamma^{-1/2}(V)=\rhoB^{-1/4}V\rhoB^{-1/4}$ and
$\pair{\rhoB^{1/2}}{V}_2=\Tr(\rhoB\Gamma^{-1/2}(V))$.
\end{proposition}
It is a direct consequence of \Cref{thm:field} and the comparison inequality in \eqref{eq:gap}. The proof of Proposition~\ref{prop:ac_Gm} is given in Section~\ref{sec:pf_ac_Gm}.

\subsubsection{Proof of Proposition~\ref{prop:Ln_bounded}}\label{sec:AS}

\begin{proof}
To prove~\eqref{eqn:L_n_bounded}, we first note that
\[
\left\|\Gamma^{-1}\cL\Gamma\right\|_{\rhoB\to\rhoB}=\left\|\Gamma^{-1/2}\cL\Gamma^{1/2}\right\|_{2\to 2}\leq \left\|\Gamma^{-1/2}\Q\Gamma^{1/2}\right\|_{2\to 2}+\left\|\Gamma^{-1/2}\mathcal{L}_{\mathrm{KMS}}\Gamma^{1/2}\right\|_{2\to 2}.
\]
According to Proposition~\ref{assump:local_bounded}, the second term is bounded by $Cn$.  \red{For the first term, use $\Q=-i\sum_x[b_x,\cdot]$ and Proposition~\ref{assump:local_bounded} to obtain}
\[
 \begin{aligned}
 \norm{\Gamma^{-1}[b_x,\cdot]\Gamma}_{\rhoB\to\rhoB}
 &\le\no{\rhoB^{-1/4}b_x\rhoB^{1/4}}
       +\no{\rhoB^{1/4}b_x\rhoB^{-1/4}}\le C.
 \end{aligned}
\]
Summing over $x$ proves~\eqref{eqn:L_n_bounded}.

Next, to prove~\eqref{eq:AS}, let $\Tr Y=0$. Since $\cL=\Q+\LK$ and $\norm{A+B}_{\rhoB}^2\le 2\norm{A}_{\rhoB}^2+2\norm{B}_{\rhoB}^2$, we have
\begin{equation}\label{eqn:Gamma_L_n_Y_bound}
\begin{aligned}
 &\norm{\Gamma^{-1}(\cL Y)}_{\rhoB}^{2}+\norm{\cL^\dagger(\Gamma^{-1}(Y))}_{\rhoB}^{2}\\
 &\le 2\norm{\Gamma^{-1}(\Q Y)}_{\rhoB}^{2}+2\norm{\Q^\dagger(\Gamma^{-1}(Y))}_{\rhoB}^{2}+2\norm{\Gamma^{-1}(\LK Y)}_{\rhoB}^{2}+2\norm{\LK^\dagger(\Gamma^{-1}(Y))}_{\rhoB}^{2}.
\end{aligned}
\end{equation}
KMS detailed balance gives $\Gamma^{-1}\LK\Gamma=\LK^\dagger$, and Proposition~\ref{assump:local_bounded} implies
\[
 \norm{\LK^\dagger}_{\rhoB\to\rhoB}
 =\norm{\Gamma^{-1/2}\LK\Gamma^{1/2}}_{2\to2}\le Cn.
\]
Because $-\LK^\dagger$ is self-adjoint and nonnegative in the KMS inner product, we have $\left(\LK^\dagger\right)^2\le Cn\left(-\LK^\dagger\right)$ as an operator inequality on the KMS Hilbert space. Consequently,
\begin{equation}\label{eqn:Gamma_L_k_Y_bound}
\begin{aligned}
 \norm{\Gamma^{-1}(\LK Y)}_{\rhoB}^2
 &=\norm{\LK^\dagger(\Gamma^{-1}(Y))}_{\rhoB}^2=\pair{\Gamma^{-1}(Y)}{(\LK^\dagger)^2\Gamma^{-1}(Y)}_{\rhoB}\\
 &\le Cn\bigl(-\pair{\Gamma^{-1}(Y)}{\LK^\dagger(\Gamma^{-1}(Y))}_{\rhoB}\bigr)
 =Cn\ED(\Gamma^{-1}(Y)).
\end{aligned}
\end{equation}

It remains to bound the two \green{Lamb shift terms} in~\eqref{eq:AS}. We first deal with the adjoint term:
\begin{equation}\label{eqn:Gamma_Q_Y_bound_adjoint}
\begin{aligned}
 \norm{\Q^\dagger(\Gamma^{-1}(Y))}_{\rhoB}^{2}
=&\norm{[\HL,\Gamma^{-1}(Y)]}_{\rhoB}^2\le n\sum_x\norm{[b_x,\Gamma^{-1}(Y)]}_{\rhoB}^2\\
 &\le Cn\sum_{y,a}\norm{[P_{y,a},\Gamma^{-1}(Y)]}_{\rhoB}^2\le Cn\ED(\Gamma^{-1}(Y)).
\end{aligned}
\end{equation}
The first inequality uses Cauchy--Schwarz. The second follows from Lemma~\ref{lem:local_b_x_bound}~\eqref{eqn:local_b_x_bound}; the last uses the comparison inequality in~\eqref{eq:gap} (Proposition~\ref{assump:gap}). Next, we bound the Lamb shift term itself. In the observable KMS inner product, the adjoint of $\Gamma^{-1}\Q\Gamma$ is $\Q^\dagger$, since $\Gamma^{-1}(\Gamma^{-1}\Q\Gamma)^\dagger\Gamma=\Q^\dagger$. Therefore,
\begin{equation}\label{eq:commutator}
 \norm{\Gamma^{-1}(\Q Y)}_{\rhoB}^2=\norm{\Q^\dagger(\Gamma^{-1}(Y))}_{\rhoB}^2+\pair{\Gamma^{-1}(Y)}
 {[\Q^\dagger,\Gamma^{-1}\Q\Gamma](\Gamma^{-1}(Y))}_{\rhoB}.
\end{equation}
To bound the commutator, we note $\HL=\sum_xb_x$, which gives
\[
 [\Q^\dagger,\Gamma^{-1}\Q\Gamma]
 =\sum_{x,y}\bigl[[b_x,\,\cdot\,],
       \Gamma^{-1}[b_y,\,\cdot\,]\Gamma\bigr].
\]
Expanding the two commutators gives, for every observable $O$,
\[
 \begin{aligned}
 &\bigl[[b_x,\cdot],\Gamma^{-1}[b_y,\cdot]\Gamma\bigr](O)\\
 &\qquad=[b_x,\rhoB^{-1/2}b_y\rhoB^{1/2}]O
              -O[b_x,\rhoB^{1/2}b_y\rhoB^{-1/2}].
 \end{aligned}
\]
Set $O=\Gamma^{-1/2}(V)$, so that $\|O\|_{\rhoB}=\|V\|_2$. Then
\begin{equation}\label{eqn:rho_B_transformation}
\left\|\bigl[[b_x,\,\cdot\,],\Gamma^{-1}[b_y,\,\cdot\,]\Gamma\bigr](O)\right\|_{\rhoB}=\left\|\Gamma^{1/2}\bigl[[b_x,\,\cdot\,],\Gamma^{-1}[b_y,\,\cdot\,]\Gamma\bigr](\Gamma^{-1/2}(V))\right\|_{2}\,,
\end{equation}
and
\[
\begin{aligned}
 \Gamma^{1/2}\Bigl(\bigl[[b_x,\cdot],\Gamma^{-1}[b_y,\cdot]\Gamma\bigr]
       (\Gamma^{-1/2}(V))\Bigr)=[\rhoB^{1/4}b_x\rhoB^{-1/4},\rhoB^{-1/4}b_y\rhoB^{1/4}]V-V[\rhoB^{-1/4}b_x\rhoB^{1/4},\rhoB^{1/4}b_y\rhoB^{-1/4}].
\end{aligned}
\]
Thus, we have
\begin{equation}\label{eqn:rho_B_transformation_end}
\begin{aligned}
 \norm{\bigl[[b_x,\cdot],\Gamma^{-1}[b_y,\cdot]\Gamma\bigr]}_{\rhoB\to\rhoB}\leq\no{[\rhoB^{1/4}b_x\rhoB^{-1/4},\rhoB^{-1/4}b_y\rhoB^{1/4}]}+\no{[\rhoB^{-1/4}b_x\rhoB^{1/4},\rhoB^{1/4}b_y\rhoB^{-1/4}]}.
\end{aligned}
\end{equation}
Without loss of generality, we bound the first term for $x\ne y$; the second is treated in the same way. Our goal is to use the quasi-locality of $b_x$ and $b_y$ to show that the commutator is exponentially small in $|x-y|$.

Take $R=|x-y|/3$. The sets $N_R(x)$ and $N_R(y)$ are disjoint, so
\[
 \bigl[\cE_{x,R}(\rhoB^{1/4}b_x\rhoB^{-1/4}),
       \cE_{y,R}(\rhoB^{-1/4}b_y\rhoB^{1/4})\bigr]=0.
\]
\red{Because $\cE_{x,R}$ and $\cE_{y,R}$ are contractions in operator norm and Proposition~\ref{assump:local_bounded} bounds both $\no{\rhoB^{1/4}b_x\rhoB^{-1/4}}$ and $\no{\rhoB^{-1/4}b_y\rhoB^{1/4}}$ by $C$, we have}
\[
\begin{aligned}
 &\no{[\rhoB^{1/4}b_x\rhoB^{-1/4},\rhoB^{-1/4}b_y\rhoB^{1/4}]}\\
 \leq &C\no{\rhoB^{1/4}b_x\rhoB^{-1/4}
       -\cE_{x,R}(\rhoB^{1/4}b_x\rhoB^{-1/4})}+C\no{\rhoB^{-1/4}b_y\rhoB^{1/4}
       -\cE_{y,R}(\rhoB^{-1/4}b_y\rhoB^{1/4})}\\
 \leq &Ce^{-cR}.
\end{aligned}
\]
Here the last inequality is the third bound of~\eqref{eq:effective-locality} in Proposition~\ref{assump:local_bounded}. For $|x-y|\leq 6$, $\no{\rhoB^{1/4}b_x\rhoB^{-1/4}}\leq C$, $\no{\rhoB^{-1/4}b_y\rhoB^{1/4}}\leq C$ give a constant bound directly. Thus, after adjusting $C,c>0$,
\[
 \norm{\bigl[[b_x,\cdot],\Gamma^{-1}[b_y,\cdot]\Gamma\bigr]}_{\rhoB\to\rhoB}
 \leq Ce^{-c|x-y|}.
\]
Summing over the one-dimensional chain now yields
\[
\begin{aligned}
 \norm{[\Q^\dagger,\Gamma^{-1}\Q\Gamma]}_{\rhoB\to\rhoB}
 \leq C\sum_{x,y\in[n]}e^{-c|x-y|}\leq Cn\left(1+2\sum_{r=1}^{\infty}e^{-cr}\right)\leq Cn,
\end{aligned}
\]
where $C$ is a constant independent of $n$. Substituting this into~\eqref{eq:commutator} and using Proposition~\ref{assump:gap} and the gap~\eqref{eq:gap} gives
\begin{equation}\label{eqn:Gamma_Q_Y_bound}
 \norm{\Gamma^{-1}(\Q Y)}_{\rhoB}^2
 \le\norm{\Q^\dagger(\Gamma^{-1}(Y))}_{\rhoB}^2
       +Cn\norm{\Gamma^{-1}(Y)}_{\rhoB}^2\le Cn\ED(\Gamma^{-1}(Y)),
\end{equation}
where we also use $\Tr(\rhoB\Gamma^{-1}(Y))=\Tr Y=0$. Substituting~\eqref{eqn:Gamma_L_k_Y_bound},~\eqref{eqn:Gamma_Q_Y_bound_adjoint}, and~\eqref{eqn:Gamma_Q_Y_bound} into~\eqref{eqn:Gamma_L_n_Y_bound} gives~\eqref{eq:AS}.

\end{proof}

To complete the proof of Proposition~\ref{prop:Ln_bounded}, we establish the local commutator estimate used above.
\begin{lemma}[Quasi-local commutator bound]\label{lem:local_b_x_bound}
In the setting of Proposition~\ref{assump:local_bounded}, for the \green{Lamb shift summands} satisfying~\eqref{eq:effective-locality}, there are $C,c>0$, independent of $x,n$, such that every observable $O$ satisfies
\begin{equation}\label{eqn:local_b_x_bound}
 \norm{[b_x,O]}_{\rhoB}^2\le C\sum_{y,a}e^{-c|x-y|}\norm{[P_{y,a},O]}_{\rhoB}^2.
\end{equation}
Consequently $\sum_x\norm{[b_x,O]}_{\rhoB}^2\le C\sum_{y,a}\norm{[P_{y,a},O]}_{\rhoB}^2$.
\end{lemma}

\begin{proof}[Proof of Lemma~\ref{lem:local_b_x_bound}]
This is \Cref{lem:annihilator} for $\mathcal A_x(O)=[b_x,O]$, which satisfies $\mathcal A_x(\idmat)=0$. With $b_x^{\pm}$ as in the justification of Proposition~\ref{assump:local_bounded}, so that $b_x^-=\rhoB^{1/4}b_x\rhoB^{-1/4}$ and $b_x^+=(b_x^-)^\dagger=\rhoB^{-1/4}b_x\rhoB^{1/4}$, \eqref{eq:weighted-multiplication} gives $\mathcal A_x=\cB_{b_x^-,\idmat}-\cB_{\idmat,b_x^+}$. Proposition~\ref{assump:local_bounded} bounds $\no{b_x^-}$ and $\no{b_x^+}$ uniformly, and, by \eqref{eq:weighted-algebra} and the third bound in \eqref{eq:effective-locality}, the map $\cB_{\cE_{x,r}(b_x^-),\idmat}-\cB_{\idmat,\cE_{x,r}(b_x^+)}\in\mathfrak A(N_r(x))$ approximates $\mathcal A_x$ within
\[
 \no{b_x^--\cE_{x,r}(b_x^-)}+\no{b_x^+-\cE_{x,r}(b_x^+)}\le Ce^{-cr}.
\]
Hence $(\mathcal A_x)_x$ is uniformly coefficient-quasi-local in the sense of \eqref{eq:coefficient-locality}, and \eqref{eq:annihilator} is \eqref{eqn:local_b_x_bound}.
\end{proof}

\subsubsection{Proof of Proposition~\ref{prop:ac_Gm}}\label{sec:pf_ac_Gm}
\begin{proof}
By the definition of $\Gm$ in~\eqref{eqn:G_mod_def},
\[
 \Gm=\sum_xg_x,\qquad
 g_x:=\frac12\int_{-\beta/4}^{\beta/4}e^{uH_n}b_xe^{-uH_n}\,\dd u,
 \qquad g_x^\dagger=g_x,\quad\no{g_x}\le C.
\]
Here Hermiticity follows by changing $u$ to $-u$ after taking the adjoint. Define the KMS maps
\begin{equation}\label{eq:insertion}
\begin{aligned}
 \mathcal M_x(O)&:=\Gamma^{-1/2}\ac_{g_x}\Gamma^{1/2}(O)
                 =J_xO+OJ_x^\dagger,\\
 \Gamma^{-1/2}\ac_{\Gm}\Gamma^{1/2}&=\acr_{\Gm}=\sum_x\mathcal M_x,\qquad
 J_x=\rhoB^{-1/4}g_x\rhoB^{1/4}.
\end{aligned}
\end{equation}
We have
\[
 \Gamma^{1/2}(\mathcal M_x(O))
 =g_x\Gamma^{1/2}(O)+\Gamma^{1/2}(O)g_x,
 \qquad \norm{\Gamma^{1/2}(O)}_2=\norm O_{\rhoB}.
\]
Since $g_x$ is Hermitian, we have $\mathcal M_x^\dagger\Gamma=\Gamma\mathcal M_x$, {\hypersetup{linkcolor=black}so $\mathcal M_x^\star=\mathcal M_x$ by~\eqref{eq:map-duality}}. In addition, we have
\[
\norm{\mathcal M_x}_{\rhoB\to\rhoB}
 =\norm{\ac_{g_x}}_{2\to2}\le2\no{g_x}\le C.
\]

We claim that the family $(\mathcal M_x)_x$ is uniformly coefficient-quasi-local in the sense of \eqref{eq:coefficient-locality}. More explicitly, we take
\[
\begin{aligned}
 \mathcal M_{x,r}(O):=\Gamma^{-1/2}\Bigl(\cE_{x,r}(g_x)\Gamma^{1/2}(O)+\Gamma^{1/2}(O)\cE_{x,r}(g_x)\Bigr).
\end{aligned}
\]
These maps are KMS-self-adjoint, have coefficient support in $N_r(x)$, and satisfy
\[
 \norm{\mathcal M_{x,r}}_{\rhoB\to\rhoB}\le C,
 \qquad \norm{\mathcal M_x-\mathcal M_{x,r}}_{\rhoB\to\rhoB}\le Ce^{-cr}.
\]
We verify this claim at the end of the proof.

Set
\[
 c_x=\pair{\idmat}{\mathcal M_x(\idmat)}_{\rhoB}
 =\Tr\bigl(\rhoB(J_x+J_x^\dagger)\bigr)
 =2\Tr(\rhoB g_x),
 \qquad
 \widetilde{\mathcal M}_x=\mathcal M_x-c_x\id.
\]
The family $(\widetilde{\mathcal M}_x)_x$ satisfies the hypotheses of \Cref{thm:field}. Since $g_x$ is Hermitian, $c_x$ is real and $\widetilde{\mathcal M}_x$ is KMS-self-adjoint; moreover, $\nm{\widetilde{\mathcal M}_x}\le\nm{\mathcal M_x}+|c_x|\le4\no{g_x}\le C$. In the notation of \eqref{eq:weighted-multiplication}, $\mathcal M_{x,r}=\cB_{\cE_{x,r}(g_x),\idmat}+\cB_{\idmat,\cE_{x,r}(g_x)}$ and $\id=\cB_{\idmat,\idmat}$, so $\mathcal M_{x,r}-c_x\id\in\mathfrak A(N_r(x))$. By the claim,
\[
 \nm{\widetilde{\mathcal M}_x-(\mathcal M_{x,r}-c_x\id)}
 =\nm{\mathcal M_x-\mathcal M_{x,r}}\le Ce^{-cr},
\]
so the family is uniformly coefficient-quasi-local in the sense of \eqref{eq:coefficient-locality}. Finally, $\pair{\idmat}{\widetilde{\mathcal M}_x(\idmat)}_{\rhoB}=c_x-c_x=0$. \Cref{thm:field} and the comparison inequality in \eqref{eq:gap} therefore give, for every observable $O$,
\begin{equation}\label{eq:field-observable}
 \norm{\Bigl(\sum_x\widetilde{\mathcal M}_x\Bigr)(O)}_{\rhoB}^2
 \le Cn\bigl(\EK(O)+|\Tr(\rhoB O)|^2\bigr)
 \le Cn\bigl(C_\Gamma^{-1}\ED(O)+|\Tr(\rhoB O)|^2\bigr).
\end{equation}

Finally, the explicit formula for $c_x$ gives
\[
 \sum_xc_x=2\sum_x\Tr(\rhoB g_x)=2\Tr(\rhoB\Gm).
\]
For any matrix $V$, take $O=\Gamma^{-1/2}(V)$. Then
\[
 \Gamma^{1/2}\Bigl(\sum_x\widetilde{\mathcal M}_x\Bigr)(O)
 =(\ac_{\Gm}-2\Tr(\rhoB\Gm)\id)V,
 \qquad \Tr(\rhoB O)=\pair{\rhoB^{1/2}}{V}_2.
\]
Using $\norm{\Gamma^{1/2}(O)}_2=\norm{O}_{\rhoB}$ in~\eqref{eq:field-observable} proves~\eqref{eq:PF}.

It remains to verify the claim, with
\[
\begin{aligned}
 \mathcal M_{x,r}(O):=\Gamma^{-1/2}\Bigl(\cE_{x,r}(g_x)\Gamma^{1/2}(O)+\Gamma^{1/2}(O)\cE_{x,r}(g_x)\Bigr).
\end{aligned}
\]
First, by the boundedness of $g_x$ according to Proposition~\ref{assump:local_bounded},
\[
 \norm{\mathcal M_{x,r}}_{\rhoB\to\rhoB}
 \le2\no{\cE_{x,r}(g_x)}
 \le2\no{g_x}\le C.
\]

By linearity of the normalized partial trace and the definition of $g_x$,
\[
 g_x-\cE_{x,r}(g_x)=\frac12\int_{-\beta/4}^{\beta/4}
 \Bigl(e^{uH_n}b_xe^{-uH_n}
       -\cE_{x,r}(e^{uH_n}b_xe^{-uH_n})\Bigr)\,\dd u.
\]
The triangle inequality and Proposition~\ref{assump:local_bounded} give
\[
 \no{g_x-\cE_{x,r}(g_x)}
 \le\frac12\int_{-\beta/4}^{\beta/4}
 \no{e^{uH_n}b_xe^{-uH_n}
       -\cE_{x,r}(e^{uH_n}b_xe^{-uH_n})}\,\dd u\le Ce^{-cr}.
\]
To convert the operator-norm approximation into the induced KMS norm, we use the same change of coordinates as in~\eqref{eqn:rho_B_transformation}--\eqref{eqn:rho_B_transformation_end}. Set $O=\Gamma^{-1/2}(V)$, so that $\norm O_{\rhoB}=\norm V_2$. Direct substitution gives
\[
\begin{aligned}
\Gamma^{1/2}(\mathcal M_x-\mathcal M_{x,r})(\Gamma^{-1/2}(V))=(g_x-\cE_{x,r}(g_x))V
   +V(g_x-\cE_{x,r}(g_x)).
\end{aligned}
\]
This implies
\[
\begin{aligned}
 \norm{\mathcal M_x-\mathcal M_{x,r}}_{\rhoB\to\rhoB}=\sup_{V\ne0}
 \frac{\norm{\Gamma^{1/2}(\mathcal M_x-\mathcal M_{x,r})(\Gamma^{-1/2}(V))}_2}{\norm V_2}\le2\no{g_x-\cE_{x,r}(g_x)}\le Ce^{-cr}.
\end{aligned}
\]
This proves the claim.
\end{proof}

\section{Rapid mixing of the physical channel and preparation complexity}\label{sec:complexity}

Throughout this section, fix $\sigma=\beta/2$, independently of $n$ and $\varepsilon$. The comparison estimates above now include this width, so they apply together with Proposition~\ref{assump:gap}.

Now, we are ready to present our \green{mixing-time result} and \green{stationary-state bias} bound:
\begin{theorem}[\green{Physical-channel mixing and stationary-state bias}]\label{thm:exact-mixing}
In the setting of Propositions~\ref{assump:local_bounded} and~\ref{assump:gap}, suppose that \eqref{eq:reference-regime} holds with sufficiently small $c_0>0$. The \green{averaged physical channel} $\Phi=\Phi_{\alpha}^{\mu_{T_0}}$ has a unique stationary state $\green{\rho_\infty}$. There is $\kappa>0$, independent of $n,\alpha,T_0$, such that
\begin{align}
 \nr{\Gamma^{-1}\Phi(\delta)}
 &\le(1-\kappa\alpha^2)\nr{\Gamma^{-1}\delta},\ \forall \delta \text{ with } \Tr\delta=0\quad (\text{\red{one-round contraction}}),\label{eq:exact-contraction}\\
 \nr{\Gamma^{-1}(\green{\rho_\infty}-\rho_*)}&\le Cn\alpha^2,\label{eq:fix-reference}\\
 \|\green{\rho_\infty}-\rhoB\|_1
 &\le\nr{\Gamma^{-1}(\green{\rho_\infty}-\rhoB)}
 \le Cn\alpha^2+\frac{C\alpha^2\sqrt n}{T_0}\quad (\text{\red{\green{stationary-state bias}}})\red{.}
 \label{eq:stationary-bias}
\end{align}
For every initial state $\rho_0$ and integer $m\ge0$,
\begin{equation}\label{eq:mixing-iterate}
 \|\Phi^m(\rho_0)-\rhoB\|_1
 \le e^{-\kappa\alpha^2m}\nr{\Gamma^{-1}(\rho_0-\green{\rho_\infty})}
            +Cn\alpha^2+\frac{C\alpha^2\sqrt n}{T_0}\quad (\text{\red{\green{end-to-end preparation-error bound}}})\red{.}
\end{equation}
\end{theorem}
\begin{proof}
\proofstep{Transfer contraction from the ideal channel.}
Theorem~\ref{thm:comparison} and~\cref{eq:reference-regime} give, with $\sigma=\beta/2$,
\[
 \begin{aligned}
 \nm{\Gamma^{-1}(\Phi-\Theta_{\alpha^2})\Gamma}\le C(n\alpha^4\sigma^2+n\alpha^2\sigma e^{-T_0^2/(8\sigma^2)})\le C'n\alpha^4\sigma^2\le C''n\alpha^4.
 \end{aligned}
\]
The second inequality uses $e^{-T_0^2/(8\sigma^2)}\le\alpha^2\sigma$ from \eqref{eq:reference-regime}. Since $T_0\ge1$ by \eqref{eq:reference-regime}, Theorem~\ref{thm:ideal-contraction} applies. For a traceless $\delta$, the triangle inequality and ideal contraction therefore give
\[
 \begin{aligned}
 \nr{\Gamma^{-1}\Phi(\delta)}
 &\le\nr{\Gamma^{-1}\Theta_{\alpha^2}(\delta)}
             +\nr{\Gamma^{-1}(\Phi-\Theta_{\alpha^2})(\delta)}\le\left(\sqrt{1-\lambda\alpha^2/2}+Cn\alpha^4\right)
                                \nr{\Gamma^{-1}\delta}\\
 &\le\left(1-\lambda\alpha^2/4+Cn\alpha^4\right)
                                \nr{\Gamma^{-1}\delta}\le(1-\kappa\alpha^2)\nr{\Gamma^{-1}\delta}.
 \end{aligned}
\]
The third step uses $\sqrt{1-z}\le1-z/2$ for $0\le z\le1$. Choose $c_0$ so that $n\alpha^2\le c_0/\sigma\le c_*$ and $Cn\alpha^2\le\lambda/8$, and take $\kappa=\lambda/8$. This proves \eqref{eq:exact-contraction}. Trace preservation ensures that this subspace is invariant:
$\Tr\Phi(\delta)=\Tr\delta=0$.
The set of density matrices is complete in the \green{weighted state-difference norm}, and $\Phi$ maps
it into itself. Thus \eqref{eq:exact-contraction} and the Banach fixed-point
theorem give a unique \green{stationary state} $\green{\rho_\infty}$.

\proofstep{\green{Convert the bias of $\rho_*$ into stationary-state bias.}}
Both $\green{\rho_\infty}-\rho_*$ and $\Phi(\rho_*)-\rho_*$ have zero trace. Linearity and stationarity give
\[
 \green{\rho_\infty}-\rho_*
   =\Phi(\green{\rho_\infty}-\rho_*)+\Phi(\rho_*)-\rho_*.
\]
Applying \eqref{eq:exact-contraction} to the first term and Theorem~\ref{thm:reference} to the second gives
\[
 \begin{aligned}
 \nr{\Gamma^{-1}(\green{\rho_\infty}-\rho_*)}
 &\le(1-\kappa\alpha^2)\nr{\Gamma^{-1}(\green{\rho_\infty}-\rho_*)}+Cn\alpha^4,\\
 \kappa\alpha^2\nr{\Gamma^{-1}(\green{\rho_\infty}-\rho_*)}
 &\le Cn\alpha^4,\\
 \nr{\Gamma^{-1}(\green{\rho_\infty}-\rho_*)}&\le(C/\kappa)n\alpha^2.
 \end{aligned}
\]
The second line subtracts the contraction term from both sides; the third divides by the positive contraction deficit. Adding the \green{bias of $\rho_*$} and using \eqref{eq:trace-conversion} yields
\[
 \begin{aligned}
 \|\green{\rho_\infty}-\rhoB\|_1
 &\le\nr{\Gamma^{-1}(\green{\rho_\infty}-\rhoB)}
 \le\nr{\Gamma^{-1}(\green{\rho_\infty}-\rho_*)}
                  +\nr{\Gamma^{-1}(\rho_*-\rhoB)}\\
 &\le Cn\alpha^2+C\alpha^2\sqrt n/T_0.
 \end{aligned}
\]

\proofstep{Iterate and add the bias.}
Because each successive difference remains traceless,
\[
 \begin{aligned}
 \|\Phi^m(\rho_0)-\rhoB\|_1
 &\le\|\Phi^m(\rho_0)-\green{\rho_\infty}\|_1+\|\green{\rho_\infty}-\rhoB\|_1\\
 &\le\nr{\Gamma^{-1}\Phi^m(\rho_0-\green{\rho_\infty})}
                    +Cn\alpha^2+C\alpha^2\sqrt n/T_0\\
 &\le(1-\kappa\alpha^2)^m\nr{\Gamma^{-1}(\rho_0-\green{\rho_\infty})}
                    +Cn\alpha^2+C\alpha^2\sqrt n/T_0\\
 &\le e^{-\kappa\alpha^2m}\nr{\Gamma^{-1}(\rho_0-\green{\rho_\infty})}
                    +Cn\alpha^2+C\alpha^2\sqrt n/T_0.
 \end{aligned}
\]
The second step uses stationarity and trace conversion, the third applies contraction $m$ times, and the last uses $1-z\le e^{-z}$. This proves \eqref{eq:mixing-iterate}.
\end{proof}

\begin{corollary}[Rapid mixing and preparation complexity]\label{cor:complexity}
In the setting of Propositions~\ref{assump:local_bounded} and~\ref{assump:gap}, for $0<\varepsilon\le1/2$ choose
\begin{equation}\label{eq:parameters}
 \alpha^2=c\varepsilon/n,\qquad T_0=\max\left\{1,2\sigma\sqrt{2\log\frac1{\alpha^2\sigma}}\right\},
\end{equation}
where $c>0$ is a sufficiently small constant. Then,
\begin{equation}\label{eq:round-complexity}
 \begin{aligned}
 m_{\mathrm{mix}}(\varepsilon;\rhoB):=\min\left\{m\in\N_0:\sup_{\rho_0\text{ state}}
       \|\Phi^m(\rho_0)-\rhoB\|_1\le\varepsilon\right\}\le C_{\mathrm{mix}}\frac n\varepsilon
       \left(\log^4(en)+\log\frac1\varepsilon\right),
 \end{aligned}
\end{equation}
where $C_{\mathrm{mix}}>0$ is independent of $n$ and $\varepsilon$.
For $m=m_{\mathrm{mix}}(\varepsilon;\rhoB)$, the expected \green{total Hamiltonian evolution time} satisfies
\[
T_{\rm total}=\mathcal{O}_{\beta,q}\!\left(
 \frac n\varepsilon\left(\log^4(en)+\log\frac1\varepsilon\right)\sqrt{\log\frac{en}{\varepsilon}}\right).
\]
\end{corollary}
\begin{proof}
According to Theorems~\ref{thm:reference} and~\ref{thm:exact-mixing}, specifically~\eqref{eq:reference-conclusion} and~\eqref{eq:fix-reference},
\[
 \nr{\Gamma^{-1}(\green{\rho_\infty}-\rhoB)}
 \le Cn\alpha^2+C\alpha^2\sqrt n/T_0
 \le2Cc\varepsilon\le\varepsilon/2
\]
for sufficiently small $c$.
Decrease $c$ also so that $n\alpha^2\sigma\le c_0$, $\alpha^2\sigma<1$,
and the small-step conditions below hold, including $\alpha^2\le1$.
Then \eqref{eq:parameters} ensures $T_0\ge1$ and
$e^{-T_0^2/(8\sigma^2)}\le\alpha^2\sigma$, so \eqref{eq:reference-regime} holds.

\proofstep{Retain dissipation in the \green{physical channel}.}
The weighted channel comparison of Theorem~\ref{thm:comparison}, together with~\eqref{eq:reference-regime}, gives
\[
 \nm{\Gamma^{-1}(\Phi-\Theta_{\alpha^2})\Gamma}\le Cn\alpha^4.
\]
Using Theorem~\ref{thm:ideal-contraction}, specifically the first inequality of
\eqref{eq:ideal-contraction}, we obtain, for every traceless $X$,
\begin{equation}\label{eq:simple-physical-dissipation}
 \begin{aligned}
 \nr{\Gamma^{-1}\Phi(X)}^2
 &\le\nr{\Gamma^{-1}X}^2-\tfrac12\alpha^2E_\mu(X)
          +Cn\alpha^4\nr{\Gamma^{-1}X}^2\\
 &\le\nr{\Gamma^{-1}X}^2-\tfrac14\alpha^2E_\mu(X).
 \end{aligned}
\end{equation}
The last step uses $E_\mu(X)\ge\lambda\nr{\Gamma^{-1}X}^2$ from~\eqref{eq:averaged-gap}
and sufficiently small $n\alpha^2$.
Moreover, \eqref{eq:exact-contraction} and \eqref{eq:reference-conclusion} imply
\begin{equation}\label{eq:simple-gibbs-drift}
 \begin{aligned}
 \nr{\Gamma^{-1}(\Phi(\rhoB)-\rhoB)}
 &\le2\nr{\Gamma^{-1}(\rho_*-\rhoB)}
       +\nr{\Gamma^{-1}(\Phi(\rho_*)-\rho_*)}\\
 &\le C\alpha^2\sqrt n/T_0+Cn\alpha^4\le C\alpha^2\sqrt n.
 \end{aligned}
\end{equation}

\proofstep{Dissipation iteration.}
For any state $\rho$, the \green{enhanced-dissipation estimate} of
\cite[Theorem~4]{chending2026} and \eqref{eq:gap} give, after averaging as in~\eqref{eq:DE},
\begin{equation}\label{eq:simple-enhanced}
 E_\mu(\rho-\rhoB)\ge
 \frac{c_1}{\log^3(en)}
 \bigl(1+\nr{\Gamma^{-1}(\rho-\rhoB)}^2\bigr)
 \log\bigl(1+\nr{\Gamma^{-1}(\rho-\rhoB)}^2\bigr),
\end{equation}
for sufficiently large $n$, with the threshold depending only on the fixed local parameters. We note that \cite[Theorem~4]{chending2026} applies to $\Gamma^{-1}(\cU(T)\rho-\rhoB)$:
this observable is Hermitian, has zero Gibbs mean by~\eqref{eq:normalization-identities}, and satisfies
\[
 \idmat+\Gamma^{-1}(\cU(T)\rho-\rhoB)
 =\rhoB^{-1/2}\cU(T)\rho\rhoB^{-1/2}\succeq0.
\]
Its KMS norm equals $\nr{\Gamma^{-1}(\rho-\rhoB)}$, since $\cU(T)$ commutes with $\Gamma$ and preserves the KMS norm by~\eqref{eq:free-isometry}.

\red{Let $n_0\ge3$ be a threshold above which the \green{enhanced-dissipation estimate} holds, depending only on the fixed local parameters. For $n<n_0$, the bound $\lambda_{\min}(\rhoB)\ge2^{-qn}e^{-2\beta(n-1)}$ and the weighted-norm identities give}
\[
 \red{\nr{\Gamma^{-1}(\rho_0-\green{\rho_\infty})}
 \le2\lambda_{\min}(\rhoB)^{-1/2}\le C.}
\]
\red{Together with the \green{stationary-state bias} bound $\varepsilon/2$ established above, \eqref{eq:mixing-iterate} therefore gives error at most $\varepsilon$ after $\mathcal O(\alpha^{-2}\log(1/\varepsilon))$ rounds. Substituting $\alpha^2=c\varepsilon/n$ proves \eqref{eq:round-complexity} for these chain lengths after enlarging $C_{\mathrm{mix}}$. We henceforth assume $n\ge n_0$.}

Writing
$\Phi(\rho)-\rhoB=\Phi(\rho-\rhoB)+\Phi(\rhoB)-\rhoB$,
and using \eqref{eq:simple-physical-dissipation}--\eqref{eq:simple-enhanced},
gives
\begin{equation}\label{eq:simple-rapid-recursion}
 \begin{aligned}
 \nr{\Gamma^{-1}(\Phi(\rho)-\rhoB)}^2
 \le{}&\nr{\Gamma^{-1}(\rho-\rhoB)}^2\\
 &-\frac{c_2\alpha^2}{\log^3(en)}
       \bigl(1+\nr{\Gamma^{-1}(\rho-\rhoB)}^2\bigr)
       \log\bigl(1+\nr{\Gamma^{-1}(\rho-\rhoB)}^2\bigr)\\
 &+C\alpha^2\sqrt n\,\nr{\Gamma^{-1}(\rho-\rhoB)}+Cn\alpha^4.
 \end{aligned}
\end{equation}
The cross term is bounded using \eqref{eq:simple-gibbs-drift} and
$\nr{\Gamma^{-1}\Phi(\rho-\rhoB)}
 \le\nr{\Gamma^{-1}(\rho-\rhoB)}$,
which follows from \eqref{eq:simple-physical-dissipation}.

\proofstep{Stage 1: enhanced convergence to a polynomial threshold.}
Fix a sufficiently large constant $C_{\mathrm{th}}$.
Whenever $\nr{\Gamma^{-1}(\rho-\rhoB)}^2\ge C_{\mathrm{th}}n\log^4(en)$ and $n\ge3$,
\[
 \frac{\sqrt n}{\nr{\Gamma^{-1}(\rho-\rhoB)}}
 \le\frac1{\sqrt{C_{\mathrm{th}}}\log^2(en)},
 \qquad
 \log\bigl(\nr{\Gamma^{-1}(\rho-\rhoB)}^2\bigr)
 \ge\log n\ge\tfrac12\log(en).
\]
Thus both error terms in \eqref{eq:simple-rapid-recursion} are absorbed into
half of its dissipative term, giving
\[
 \log\bigl(\nr{\Gamma^{-1}(\Phi(\rho)-\rhoB)}^2\bigr)
 \le\left(1-\frac{c_3\alpha^2}{\log^3(en)}\right)
       \log\bigl(\nr{\Gamma^{-1}(\rho-\rhoB)}^2\bigr).
\]
Since $\|H_n\|_{\rm op}\le n-1$ by~\eqref{eq:hamiltonian},
$\lambda_{\min}(\rhoB)\ge 2^{-qn}e^{-2\beta(n-1)}$ and the weighted-norm identities~\eqref{eq:rho-space}--\eqref{eq:Gamma} give
\[
 \nr{\Gamma^{-1}(\rho_0-\rhoB)}^2
 \le4\lambda_{\min}(\rhoB)^{-1}\le e^{Cn},
\]
iteration reaches
$\nr{\Gamma^{-1}(\rho-\rhoB)}^2\le C_{\mathrm{th}}n\log^4(en)$
within $\mathcal O(\log^4(en)/\alpha^2)$ rounds.

\proofstep{Stage 2: contraction toward the physical stationary state.}
Any state $\rho$ reached at the end of Stage 1 satisfies, by the triangle inequality and the \green{stationary-state bias} bound~\eqref{eq:stationary-bias},
\[
 \nr{\Gamma^{-1}(\rho-\green{\rho_\infty})}
 \le\sqrt{C_{\mathrm{th}}n}\log^2(en)+\varepsilon/2
 \le2\sqrt{C_{\mathrm{th}}n}\log^2(en).
\]
By \eqref{eq:exact-contraction} and trace conversion \eqref{eq:trace-conversion}, after $j$ further rounds,
\[
 \|\Phi^j(\rho)-\rhoB\|_1
 \le2e^{-\kappa\alpha^2j}\sqrt{C_{\mathrm{th}}n}\log^2(en)+\varepsilon/2.
\]
Taking $j=\mathcal O((\log(en)+\log(1/\varepsilon))/\alpha^2)$
makes this bound at most $\varepsilon$. Substituting $\alpha^2=c\varepsilon/n$ from~\eqref{eq:parameters} proves \eqref{eq:round-complexity}.
Finally, each round has expected duration $2\E[T]=10T_0$ by~\eqref{eq:time-transform}, and
$T_0=\mathcal O(\sqrt{\log(en/\varepsilon)})$ by~\eqref{eq:parameters}, which gives the stated \green{total Hamiltonian evolution time} via~\eqref{eq:intro-total-time}.

\end{proof}

\appendix

\section{One-dimensional estimates}\label{app:1d}

\subsection{Complex-time locality: proof of \texorpdfstring{\Cref{lem:Araki}}{Lemma \getrefnumber{lem:Araki}}}\label{app:araki}

Before proving \Cref{lem:Araki}, we first unify the notation and introduce relevant definitions and results from \cite{perez2023locality}.

For a spin chain on $\mathbb Z$, let $(h_Z)$ be an interaction of self-adjoint operators supported on finite sets $Z$, with
\[
 \Omega_k:=\sup_{j\in\mathbb{Z}}\sum_{\substack{j\in Z\\\operatorname{diam}Z\ge k}}\no{h_Z},
 \qquad k\in\N_0,\qquad \Omega_0<\infty.
\]
For finite $\Lambda\subset\mathbb Z$, write
\[
 H_\Lambda:=\sum_{Z\subseteq\Lambda}h_Z,
 \qquad \tau_\Lambda^s(A):=e^{isH_\Lambda}Ae^{-isH_\Lambda}.
\]
For $x\ge0$, define the nonnegative coefficients
\[
 \Omega_0^*(x):=1,\qquad
 \Omega_k^*(x):=\sum_{p=1}^k\frac{x^p}{p!}
   \sum_{\substack{\alpha_1,\ldots,\alpha_p\ge1\\
                    \alpha_1+\cdots+\alpha_p=k}}
       \prod_{j=1}^p\Omega_{\alpha_j}\quad(k\ge1).
\]
Equivalently, as a formal power series,
\begin{equation}\label{eq:pgph-generating}
 \sum_{k\ge0}\Omega_k^*(x)z^k
 =\exp\!\left(x\sum_{k\ge1}\Omega_k z^k\right).
\end{equation}

\begin{theorem}[Complex-time locality, {\cite[Theorem~2.3]{perez2023locality}}]\label{thm:pgph-complex}
With the interaction and coefficients above, take an observable $A$ supported in an interval $I_0=[a,b]\subset\mathbb Z$, and set $I_k=[a-k,b+k]$.
For integers $0\le\ell\le L$ and $s\in\C$, the following bounds hold:
\begin{equation}\label{eq:pgph-complex}
 \begin{aligned}
 \no{\tau_{I_L}^s(A)}
 &\le\no{A}e^{2|s|\Omega_0|I_0|}
       \sum_{k=0}^L e^{4|s|\Omega_0 k}\Omega_k^*(4|s|),\\
 \no{\tau_{I_L}^s(A)-\tau_{I_\ell}^s(A)}
 &\le\no{A}e^{2|s|\Omega_0|I_0|}
       \sum_{k=\ell+1}^L e^{4|s|\Omega_0 k}\Omega_k^*(4|s|).
 \end{aligned}
\end{equation}
\end{theorem}

\begin{theorem}[Real-time locality, {\cite[Theorem~2.4]{perez2023locality}}]\label{thm:pgph-real}
For the same interaction, take $A$ supported in any finite set $\Lambda_0\subset\mathbb Z$ and put
$\Lambda_k=\{j\in\mathbb Z:\dist(j,\Lambda_0)\le k\}$.
For integers $0\le\ell\le L$ and $t\in\R$,
\begin{equation}\label{eq:pgph-real}
 \no{\tau_{\Lambda_L}^t(A)-\tau_{\Lambda_\ell}^t(A)}
 \le\no{A}|\Lambda_0|e^{4|t|\Omega_0}
       \sum_{k\ge\ell+1}\Omega_k^*(4|t|).
\end{equation}
\end{theorem}

\begin{proof}[Proof of \Cref{lem:Araki}]
\proofstep{Step 1: Verify the finite-range input.}
Recall~\eqref{eq:hamiltonian}:
\[
H_n = \sum_{x=1}^{n-1} h_x,\quad h_x=h_x^\dagger,\quad
 \supp h_x\subseteq\{x,x+1\},\quad \no{h_x}\le 1.
\]
Assign $h_y$ to $S_y=\{y,y+1\}$ and extend the interaction by zero to $\mathbb Z$. Its interaction tails in the notation of \cite{perez2023locality} obey
\begin{equation}\label{eq:araki-interaction-tails}
 \begin{aligned}
 \Omega_k&:=\sup_j\sum_{j\in Z,\,\operatorname{diam}Z\ge k}\no{h_Z}
 \le\kappa_{\mathrm{loc}}\mathbf1_{\{k\le 1\}},\qquad\kappa_{\mathrm{loc}}=2,\\
 \sum_{k\ge1}e^{\zeta k}\Omega_k
 &\le\kappa_{\mathrm{loc}} e^{\zeta}=:K_\zeta<\infty.
 \end{aligned}
\end{equation}
Here, the parameter $\zeta$ is a free parameter that will be chosen later, and the constant $K_\zeta$ depends only on $\zeta$.

Evaluate \eqref{eq:pgph-generating} at $z=e^\zeta$ and argument $4|s|$, and use \eqref{eq:araki-interaction-tails}. For $\zeta>4|s|\kappa_{\mathrm{loc}}$ and $k\ge\ell+1$, the factor $e^{-(\zeta-4|s|\kappa_{\mathrm{loc}})k}$ is at most $e^{-(\zeta-4|s|\kappa_{\mathrm{loc}})(\ell+1)}$, giving
\begin{equation}\label{eq:araki-complex-series-tail}
\sum_{k=\ell+1}^L e^{4|s|\Omega_0 k}\Omega_k^*(4|s|)
\le e^{4|s|K_\zeta-(\zeta-4|s|\kappa_{\mathrm{loc}})(\ell+1)},
\qquad \zeta>4|s|\kappa_{\mathrm{loc}}.
\end{equation}
\proofstep{Step 2: Prove the norm and truncation bounds at imaginary time.}
At $t=0$, the two conclusions of \Cref{lem:Araki} that we need to establish are, uniformly for $|u|\le u_*$,
\begin{equation}\label{eq:araki-imaginary-target}
 \begin{aligned}
 \no{B_x(-iu)}&\le C\no{B_x},\\
 \no{(\id-\cE_{x,r})B_x(-iu)}
 &\le Ce^{-cr\log(r+1)}\no{B_x}.
 \end{aligned}
\end{equation}
The second inequality is the stronger $t=0$ bound in the lemma. To prove it, we first approximate $B_x(-iu)$ by an operator supported on a finite interval, then use \eqref{eq:local-transfer} to bound the error of the conditional-expectation truncation.
Take $I_0=N_d(x)$, so that $\supp B_x\subseteq I_0$ and $|I_0|\le2d+1$. For each integer $m\ge0$, put $I_m=N_{d+m}(x)$ and define the local approximation
\[
 B_x^{(m)}(-iu):=
\tau_{I_m}^{-iu}(B_x)
=e^{uH_{I_m}}B_xe^{-uH_{I_m}},
 \qquad H_I=\sum_{S_y\subseteq I}h_y.
\]
Applying \Cref{thm:pgph-complex} with $s=-iu$ and the interaction extended by zero outside $[n]$, we obtain, uniformly for $|u|\le u_*$ and for every $\zeta>4u_*\kappa_{\mathrm{loc}}$,
\begin{equation}\label{eq:araki-local-approx}
 \begin{aligned}
 \no{e^{uH_n}B_xe^{-uH_n}}&\le
 e^{2u_*\kappa_{\mathrm{loc}}|I_0|+4u_*K_\zeta}\no{B_x},\\
 \no{B_x(-iu)-B_x^{(m)}(-iu)}
 &\le e^{2u_*\kappa_{\mathrm{loc}}|I_0|+4u_*K_\zeta}
          e^{-(\zeta-4u_*\kappa_{\mathrm{loc}})m}\no{B_x}\,.
 \end{aligned}
\end{equation}
First fix $\zeta_0=4u_*\kappa_{\mathrm{loc}}+1$. The first inequality gives the first bound in~\eqref{eq:araki-imaginary-target}, where $C_0=e^{2u_*\kappa_{\mathrm{loc}}(2d+1)+4u_*K_{\zeta_0}}$. The same choice in the second inequality gives the exponential approximation error:
\begin{equation}\label{eq:araki-approx-error}
\no{B_x(-iu)-B_x^{(m)}(-iu)}\leq C_0e^{-m}\no{B_x}\,.
\end{equation}
Next, we observe that $B_x^{(m)}(-iu)$ is supported in $N_{d+m}(x)$. For sufficiently large $r$, take $m=\lfloor r-d\rfloor$, so that this support lies in $N_r(x)$, and apply \eqref{eq:local-transfer}:
\[
 \begin{aligned}
 \no{(\id-\cE_{x,r})B_x(-iu)}\le2\no{B_x(-iu)-B_x^{(m)}(-iu)}\leq 2C_0e^{d+1}e^{-r}\no{B_x}
 \end{aligned}
\]
Thus, it remains to improve~\eqref{eq:araki-approx-error} to the faster decay required in \eqref{eq:araki-imaginary-target} by adding a logarithmic factor to the exponent.

For any $m>0$, we define
\[
 \zeta_m=\frac{\log(m+1)}{2}.
\]
For sufficiently large $m$, this satisfies $\zeta_m>4u_*\kappa_{\mathrm{loc}}$ and gives
\[
 K_{\zeta_m}=\kappa_{\mathrm{loc}} e^{\zeta_m}=\kappa_{\mathrm{loc}}\sqrt{m+1}.
\]
Consequently, the part of the exponent in the approximation error that depends on $m$ obeys
\[
 \begin{aligned}
 4u_*K_{\zeta_m}-(\zeta_m-4u_*\kappa_{\mathrm{loc}})m
 \le4u_*\kappa_{\mathrm{loc}}\sqrt{m+1}+4u_*\kappa_{\mathrm{loc}} m
       -\frac{m\log(m+1)}{2}\le-\frac{m\log(m+1)}{4}
 \end{aligned}
\]
for sufficiently large $m$. This completes the proof of \eqref{eq:araki-imaginary-target}.

\proofstep{Step 3: Propagate the initial approximation only in real time.}
The first line of \eqref{eq:araki-local-approx}, with $\zeta=\zeta_0$, gives
\begin{equation}\label{eq:araki-real-norm}
 \begin{aligned}
 B_x(t-iu)&=e^{itH_n}B_x(-iu)e^{-itH_n},\\
 \no{B_x(t-iu)}&=\no{B_x(-iu)}\le C_0\no{B_x}.
 \end{aligned}
\end{equation}
This proves the first line of \eqref{eq:complex-cone}.

For an integer $r\ge0$, define the global evolution $D(t)$ and the supported approximation $Q_r(t)$ separately:
\begin{equation}\label{eq:araki-real-approximants}
 \begin{aligned}
 m&=\lfloor r/2\rfloor,\ D=\cE_{x,m}(B_x(-iu)),\  D(t)=e^{itH_n}De^{-itH_n},\\
 Q_r(t)&=e^{itH_{N_r(x)}}De^{-itH_{N_r(x)}},
 \qquad\supp Q_r(t)\subseteq N_r(x).
 \end{aligned}
\end{equation}
To bound $\no{(\id-\cE_{x,r})B_x(t-iu)}$, it suffices to bound $\no{B_x(t-iu)-Q_r(t)}$. We first notice
\[
 \no{B_x(t-iu)-Q_r(t)}\le\no{B_x(t-iu)-D(t)}+\no{D(t)-Q_r(t)}\,.
\]
We then bound \red{the} two terms separately.

Set $C_I:=2C_0e^{d+1}$. For $m\ge d$, put $k=\lfloor m-d\rfloor$, so $\supp B_x^{(k)}(-iu)\subseteq N_m(x)$ and $k\ge m-d-1$. Apply \eqref{eq:local-transfer} with $B=B_x(-iu)$, radius $m$, and $B^{(m)}=B_x^{(k)}(-iu)$, and then apply \eqref{eq:araki-approx-error} with approximation radius $k$:
\begin{equation}\label{eq:araki-real-truncation-initial}
 \begin{aligned}
 \no{B_x(-iu)-D}
\le2\no{B_x(-iu)-B_x^{(k)}(-iu)}\le2C_0e^{-k}\no{B_x}\le C_Ie^{-m}\no{B_x}.
 \end{aligned}
\end{equation}
For $0\le m<d$, \eqref{eq:local-transfer} with $B^{(m)}=0$ and \eqref{eq:araki-real-norm} give
\[
 \no{B_x(-iu)-D}\le2C_0\no{B_x}\le C_Ie^{-m}\no{B_x}.
\]
Thus the last bound in \eqref{eq:araki-real-truncation-initial} holds for all $m\ge0$. Apply the first line of \eqref{eq:local-transfer} with $B=B_x(-iu)$ and radius $m$. Together with \eqref{eq:araki-real-norm}, \eqref{eq:araki-real-truncation-initial}, and the definitions in \eqref{eq:araki-real-approximants}, this gives
\begin{equation}\label{eq:araki-real-initial-error}
 \begin{aligned}
 \no{B_x(t-iu)-D(t)}
 &=\no{B_x(-iu)-D}\le C_Ie^{-m}\no{B_x},\\
 \no D&\le\no{B_x(-iu)}\le C_0\no{B_x},
 \qquad\supp D\subseteq N_m(x).
 \end{aligned}
\end{equation}

To bound $\no{D(t)-Q_r(t)}$, first evaluate \eqref{eq:pgph-generating} at $z=e^a$ and argument $4|t|$, and use \eqref{eq:araki-interaction-tails} with $\zeta=a$. For any $a>0$ and integer $\ell\ge0$,
\begin{equation}\label{eq:araki-real-series-tail}
 \begin{aligned}
 \sum_{k\ge\ell+1}\Omega_k^*(4|t|)
 &\le e^{-a(\ell+1)}\sum_{k\ge0}e^{ak}\Omega_k^*(4|t|)=e^{-a(\ell+1)}
   \exp\!\left(4|t|\sum_{j\ge1}e^{aj}\Omega_j\right)\\
 &\le e^{4|t|K_a-a(\ell+1)},
 \qquad v_a:=4(\kappa_{\mathrm{loc}}+K_a).
 \end{aligned}
\end{equation}
In \eqref{eq:pgph-real}, choose $A=D$, $\Lambda_0=N_m(x)$, $\ell=r-m$, and $L\ge\ell$ large enough to contain the chain. The zero extension of the interaction gives
\begin{equation}\label{eq:araki-real-restriction-sets}
 \begin{aligned}
 |\Lambda_0|&=|N_m(x)|\le2m+1,\\
 \Lambda_\ell\cap[n]&=N_r(x),\qquad
 H_{\Lambda_\ell}=H_{N_r(x)},\\
 [n]&\subseteq\Lambda_L,\qquad H_{\Lambda_L}=H_n.
 \end{aligned}
\end{equation}
Substitute \eqref{eq:araki-real-restriction-sets} into \eqref{eq:pgph-real}, then apply \eqref{eq:araki-real-series-tail} with $a=1$, the norm bound in \eqref{eq:araki-real-initial-error}, and $\Omega_0\le\kappa_{\mathrm{loc}}$ from \eqref{eq:araki-interaction-tails}:
\begin{equation}\label{eq:araki-real-restriction}
 \begin{aligned}
 \no{D(t)-Q_r(t)}
 &\le\no D\,|N_m(x)|e^{4|t|\Omega_0}
       \sum_{k\ge r-m+1}\Omega_k^*(4|t|)\\
 &\le C_0\no{B_x}(2m+1)e^{v_1|t|-(r-m+1)}.
 \end{aligned}
\end{equation}
Choose $m=\lfloor r/2\rfloor$. The triangle inequality, \eqref{eq:araki-real-initial-error}, and \eqref{eq:araki-real-restriction} give
\begin{equation}\label{eq:araki-complex-approx-error}
 \begin{aligned}
 \no{B_x(t-iu)-Q_r(t)}
 &\le\no{B_x(t-iu)-D(t)}+\no{D(t)-Q_r(t)}\\
 &\le\no{B_x}\bigl(C_Ie^{-m}
       +C_0(2m+1)e^{v_1|t|-(r-m+1)}\bigr)\\
 &\le C_2e^{v_1|t|-r/4}\no{B_x},
 \qquad C_2:=eC_I+\left(1+\frac4{e}\right)C_0.
 \end{aligned}
\end{equation}
Apply the second line of \eqref{eq:local-transfer} with $B=B_x(t-iu)$ and $B^{(r)}=Q_r(t)$, whose support is given in \eqref{eq:araki-real-approximants}, and then use \eqref{eq:araki-complex-approx-error}. Applying the same line of \eqref{eq:local-transfer} with $B^{(r)}=0$ and using \eqref{eq:araki-real-norm} gives the independent bound $2C_0\no{B_x}$. Combining these two bounds yields
\begin{equation}\label{eq:araki-complex-truncation}
 \begin{aligned}
 \no{(\id-\cE_{x,r})B_x(t-iu)}
 &\le2\min\{C_0,C_2e^{v_1|t|-r/4}\}\no{B_x}\\
 &\le C\min\{1,e^{v_1|t|-r/4}\}\no{B_x},
 \qquad C:=2\max\{C_0,C_2\}.
 \end{aligned}
\end{equation}
This concludes the proof of the second inequality in \eqref{eq:complex-cone}.

\proofstep{Step 4: Prove the real-time bound for quasi-local initial matrices.}
The norm assumption in \eqref{eq:quasilocal-initial} gives
\begin{equation}\label{eq:araki-quasilocal-norm}
 \no{D_x(t)}=\no{e^{itH_n}D_xe^{-itH_n}}=\no{D_x}\le M,
\end{equation}
which proves the first line of \eqref{eq:quasilocal-real-cone}. To prove the second inequality in \eqref{eq:quasilocal-real-cone}, the idea is similar to that in Step 3.

For an integer $r\ge0$, define
\begin{equation}\label{eq:araki-quasilocal-approximants}
 \begin{aligned}
 m&=\lfloor r/2\rfloor,\qquad D_x^{(m)}=\cE_{x,m}(D_x),\\
 D_x^{(m)}(t)&=e^{itH_n}D_x^{(m)}e^{-itH_n},\\
 Q_{x,r}(t)&=e^{itH_{N_r(x)}}D_x^{(m)}e^{-itH_{N_r(x)}},
 \qquad\supp Q_{x,r}(t)\subseteq N_r(x).
 \end{aligned}
\end{equation}
Substitute radius $m$ in \eqref{eq:quasilocal-initial} and substitute $B=D_x$, $r=m$ in the first line of \eqref{eq:local-transfer}. With the definitions in \eqref{eq:araki-quasilocal-approximants}, this yields
\begin{equation}\label{eq:araki-quasilocal-initial-error}
 \begin{aligned}
 \no{D_x(t)-D_x^{(m)}(t)}
 &=\no{D_x-D_x^{(m)}}\le Me^{-\mu m},\\
 \no{D_x^{(m)}}&\le M,
 \qquad\supp D_x^{(m)}\subseteq N_m(x).
 \end{aligned}
\end{equation}
Apply \eqref{eq:pgph-real} with $A=D_x^{(m)}$ and the sets in \eqref{eq:araki-real-restriction-sets}, so $\ell=r-m$. Then use \eqref{eq:araki-real-series-tail} with $a=\mu$, \eqref{eq:araki-quasilocal-initial-error}, and $\Omega_0\le\kappa_{\mathrm{loc}}$ from \eqref{eq:araki-interaction-tails}:
\begin{equation}\label{eq:araki-quasilocal-restriction}
 \begin{aligned}
 \no{D_x^{(m)}(t)-Q_{x,r}(t)}
 &\le\no{D_x^{(m)}}|N_m(x)|e^{4|t|\Omega_0}
       \sum_{k\ge r-m+1}\Omega_k^*(4|t|)\\
 &\le M(2m+1)e^{v_\mu|t|-\mu(r-m+1)}.
 \end{aligned}
\end{equation}
Adding \eqref{eq:araki-quasilocal-initial-error} and \eqref{eq:araki-quasilocal-restriction}, and choosing $m=\lfloor r/2\rfloor$, gives
\begin{equation}\label{eq:araki-quasilocal-approx-error}
 \begin{aligned}
 \no{D_x(t)-Q_{x,r}(t)}
 &\le\no{D_x(t)-D_x^{(m)}(t)}
       +\no{D_x^{(m)}(t)-Q_{x,r}(t)}\\
 &\le M\bigl(e^{-\mu m}
       +(2m+1)e^{v_\mu|t|-\mu(r-m+1)}\bigr)\\
 &\le C_\mu^{(0)}M e^{v_\mu|t|-\mu r/4},
 \qquad C_\mu^{(0)}:=e^\mu+\left(1+\frac4{\mu e}\right).
 \end{aligned}
\end{equation}
Applying \eqref{eq:local-transfer} gives
\begin{equation}\label{eq:araki-quasilocal-truncation}
 \begin{aligned}
 \no{(\id-\cE_{x,r})D_x(t)} \le2M\min\{1,C_\mu^{(0)}e^{v_\mu|t|-\mu r/4}\}\le2C_\mu^{(0)}M\min\{1,e^{v_\mu|t|-\mu r/4}\}.
 \end{aligned}
\end{equation}
This concludes the proof of~\eqref{eq:quasilocal-real-cone}.
\red{For noninteger $r\ge0$, use $\cE_{x,r}=\cE_{x,\lfloor r\rfloor}$ and $e^{-c\lfloor r\rfloor}\le e^c e^{-cr}$ to extend both~\eqref{eq:complex-cone} and~\eqref{eq:quasilocal-real-cone} to all real radii, increasing the constants if necessary.}
\end{proof}

\subsection{The auxiliary local gap, the boundary perturbation of the Gibbs density, and clustering on separate intervals}\label{app:gap}

The estimates for spectral conditional expectations in \Cref{prop:thermal} require three inputs: a variance bound on an isolated interval, a local comparison with the Gibbs density obtained by cutting its boundary, and decay of correlations between separated multiplication maps. We establish them below, after recalling the two literature results used in their proofs.

For a full-rank density matrix $\eta$, we recall the isometry in \eqref{eq:rho-space}--\eqref{eq:Gamma}:
\begin{equation}\label{eq:static-kms-coordinates}
 \begin{aligned}
 \Gamma_{\eta}^{1/2}(O)&:=\eta^{1/4}O\eta^{1/4},\qquad
 \|\Gamma_{\eta}^{1/2}(O)\|_2=\|O\|_{\eta},\\
 \langle X,Y\rangle_2&:=\Tr(X^\dagger Y),\qquad
 \|\mathcal T\|_{2\to2}:=\sup_{X\ne0}\frac{\|\mathcal T(X)\|_2}{\|X\|_2},\\
 \Gamma_{\eta}^{1/2}(\idmat)&=\sqrt{\eta},\qquad\|\sqrt{\eta}\|_2=1.
 \end{aligned}
\end{equation}
Write $\Gamma^{1/2}=\Gamma_{\rhoB}^{1/2}$ and define the algebra of local multiplication maps
\[
 \mathfrak M(S):=\operatorname{span}\{X\mapsto BXC:\supp B,\supp C\subseteq S\}.
\]
Recall $\cB_{B,C}(O)=\rhoB^{-1/4}B\rhoB^{1/4}\,O\,
\rhoB^{1/4}C\rhoB^{-1/4}$ \red{from}~\eqref{eq:weighted-multiplication}\red{. By}~\eqref{eq:weighted-algebra}, \red{we have}
\begin{equation}\label{eq:doubled-coefficient-support}
 \begin{aligned}
 \Gamma^{1/2}\cB_{B,C}\Gamma^{-1/2}&=(X\mapsto BXC),\\
 \Gamma^{1/2}\mathfrak A(S)\Gamma^{-1/2}&=\mathfrak M(S)\red{,}
 \end{aligned}
\end{equation}
where $\mathfrak A(S)$ is the algebra of finite sums of $\cB_{B,C}$ with $B,C$ supported on $S$.

Next, we recall a result in \cite[Lemma~III.4, Eq.~(3.7), and Theorem~V.1, Eq.~(5.1)]{BergamaschiChen}.

\begin{theorem}[Auxiliary gap]\label{thm:bc-block-gap}
Let $\widetilde H=\sum_{j=1}^{m-1}k_{j,j+1}$ act on a chain of $m$ qudits of dimension $2^{q}$, where
\[
 \supp k_{j,j+1}\subseteq\{j,j+1\},\qquad
 k_{j,j+1}=k_{j,j+1}^\dagger,\qquad
 \no{k_{j,j+1}}\le1\qquad(1\le j<m).
\]
For $\beta'>0$, put $\eta=e^{-\beta'\widetilde H}/\Tr e^{-\beta'\widetilde H}$, and let $\mathcal P_j$ be the complete Pauli family on qudit $j$. Then
\begin{equation}\label{eq:bc-block-gap-input}
 \sum_j\sum_{P\in\mathcal P_j}\|[P,O]\|_\eta^2
 \ge\gamma_{\beta',q}
       \|O-\Tr(\eta O)\idmat\|_\eta^2,
 \qquad\gamma_{\beta',q}>0,
\end{equation}
uniformly in $m$.
\end{theorem}

\begin{theorem}[Equal-time clustering in matrix form; consequence of {\cite[Theorems~1--2]{NachtergaeleSims}}]\label{thm:ns-clustering}
Let $\mathcal F(Z)=\mathcal F(Z)^\dagger\in\mathfrak M(Z)$ be linear maps on a tensor product of matrix spaces, each of dimension at most $N$. Suppose, for some $\zeta>0$,
\begin{equation}\label{eq:ns-interaction-input}
 M_\zeta:=\sup_z\sum_{Z\ni z}|Z|N^{2|Z|}
       e^{\zeta\operatorname{diam}Z}\|\mathcal F(Z)\|_{2\to2}<\infty.
\end{equation}
For $\mathbb H=\sum_Z\mathcal F(Z)$, suppose there is a matrix $F$ such that
\[
 \|F\|_2=1,\qquad \mathbb H(F)=0,\qquad
 \Tr[X^\dagger\mathbb H(X)]
 \ge\gamma\bigl(\|X\|_2^2-|\Tr(F^\dagger X)|^2\bigr)
 \quad(X\in\mathcal B(\mathcal H)),\qquad\gamma>0.
\]
For $\mathcal A_S\in\mathfrak M(S)$, $\mathcal B_T\in\mathfrak M(T)$ with disjoint $S,T$, the general-support bound following Theorem~1 of the cited paper reads
\begin{equation}\label{eq:ns-general-lr-input}
 \|[e^{it\mathbb H}\mathcal A_Se^{-it\mathbb H},\mathcal B_T]\|_{2\to2}
 \le2|S||T|N^{2|S|}(e^{2M_\zeta|t|}-1)
       e^{-\zeta\dist(S,T)}
       \|\mathcal A_S\|_{2\to2}\|\mathcal B_T\|_{2\to2}.
\end{equation}
Here exponentials and products of maps denote their composition on matrices.
Using \eqref{eq:ns-general-lr-input} \red{in the proof of \cite[Theorem~2]{NachtergaeleSims}, at imaginary time zero,} gives
\begin{equation}\label{eq:ns-clustering-input}
 \begin{aligned}
 &\left|\Tr[F^\dagger\mathcal A_S(\mathcal B_T(F))]
   -\Tr[F^\dagger\mathcal A_S(F)]\Tr[F^\dagger\mathcal B_T(F)]\right|\\
 &\qquad\le C(1+|T|)e^{C|S|-\mu\dist(S,T)}
       \|\mathcal A_S\|_{2\to2}\|\mathcal B_T\|_{2\to2},
 \qquad\mu=\frac{\gamma\zeta}{4M_\zeta+\gamma}>0.
 \end{aligned}
\end{equation}
Here $C$ depends only on $N,\zeta,M_\zeta,\gamma$.
\end{theorem}

Now, we are ready to state the three lemmas and provide their proofs one by one.

\begin{lemma}[Auxiliary local gap]\label{lem:aux-gap}
For every isolated interval $I$, put $H_I=\sum_{S_y\subseteq I}h_y$ and $\rho_I=e^{-\beta H_I}/\Tr e^{-\beta H_I}$. There is $\gamma>0$, depending only on $\beta,q$, such that
\begin{equation}\label{eq:isolated-gap}
 \sum_{x\in I,a}\|[P_{x,a},O]\|_{\rho_I}^2
 \ge\gamma\|O-\Tr(\rho_I O)\idmat_I\|_{\rho_I}^2
\end{equation}
for every observable $O$ on $I$, uniformly in $I$.
\end{lemma}
Lemma~\ref{lem:aux-gap} is a direct consequence of Theorem~\ref{thm:bc-block-gap}.

\begin{proof}[Proof of Lemma~\ref{lem:aux-gap}]

The isolated Hamiltonian $H_I=\sum_{S_y\subseteq I}h_y$ is a nearest-neighbor chain on the $m=|I|$ qudits of $I$, with terms of norm at most one by~\eqref{eq:hamiltonian}. Apply \Cref{thm:bc-block-gap} with $\widetilde H=H_I$, $\beta'=\beta$, and $\eta=\rho_I$. On site $x$, the complete Pauli family consists of $\idmat$ and the $P_{x,a}$, and $[\idmat,O]=0$. This gives~\eqref{eq:isolated-gap} with $\gamma=\gamma_{\beta,q}$. For $|I|=1$, where $H_I=0$ and $\rho_I=2^{-q}\idmat_I$, the bound also follows from the identity $\sum_a\|[P_{x,a},O]\|_{\rho_I}^2=2\cdot4^q\|O-\Tr(\rho_IO)\idmat_I\|_{\rho_I}^2$, so $\gamma=\min\{\gamma_{\beta,q},2\cdot4^q\}$ covers all intervals.
\end{proof}

For an interval $I=\{a,\ldots,b\}\subseteq[n]$, define its left and right exterior intervals by
\[
 I_-:=\{1,\ldots,a-1\},\qquad
 I_+:=\{b+1,\ldots,n\},\qquad
 I^c=I_-\sqcup I_+.
\]
We define the cut Hamiltonian, the removed interaction, and the corresponding product Gibbs weight by
\begin{equation}\label{eq:boundary-cut}
 H_{\mathrm{cut}}=H_{I_-}+H_I+H_{I_+},\qquad
 V_\partial=H_n-H_{\mathrm{cut}},\qquad
 \rho_{\mathrm{cut}}=\rho_{I_-}\otimes\rho_I\otimes\rho_{I_+}.
\end{equation}
Here $\rho_J$ is the isolated Gibbs state on $J$, with $H_\varnothing=0$ and $\rho_\varnothing=1$ on the one-dimensional empty tensor factor. All tensor products below are embedded in the original site order.
For $r\ge0$, define the boundary neighborhood by
\begin{equation}\label{eq:boundary-collar}
 \mathcal C_r(I):=N_r(I)\cap N_r(I^c),\qquad
 |\mathcal C_r(I)|\le4(1+r),\qquad
 I\cup\mathcal C_r(I)\subseteq N_r(I).
\end{equation}
When $I=[n]$, set $\mathcal C_r(I)=\varnothing$.

\begin{lemma}[Boundary perturbation of the Gibbs density]\label{lem:boundary}
There exist $C\ge1$ and $c>0$, depending only on $\beta$, such that, uniformly in $n,I$, for every system matrix $O$,
\begin{equation}\label{eq:weight-equivalence}
 C^{-1}\|O\|_{\rho_{\mathrm{cut}}}\le\nr O\le C\|O\|_{\rho_{\mathrm{cut}}}.
\end{equation}
Moreover, for every $r\ge0$, the change of coordinates $W=\Gamma^{1/2}\Gamma_{\rho_{\mathrm{cut}}}^{-1/2}$ has an invertible approximant $W_r$ satisfying
\begin{equation}\label{eq:boundary-map-locality}
 \begin{aligned}
 \|W\|_{2\to2}+\|W^{-1}\|_{2\to2}
       +\|W_r\|_{2\to2}+\|W_r^{-1}\|_{2\to2}&\le C,\\
 \|W-W_r\|_{2\to2}&\le Ce^{-cr},\qquad
 W_r\in\mathfrak M(\mathcal C_r(I)).
 \end{aligned}
\end{equation}
When $I=[n]$, one can take $W_r=W=\id$.
\end{lemma}
Lemma~\ref{lem:boundary} characterizes the effect of perturbing the Gibbs density by omitting the boundary terms. The cut Hamiltonian $H_{\mathrm{cut}}$ will later be used to construct strictly local approximations to superoperators associated with the full Hamiltonian.

\begin{proof}
Cutting the boundary changes $H_n$ only by the terms crossing the two cuts. We show that the \red{quarter-power} Gibbs factors differ by a bounded, invertible matrix that can be approximated near those cuts. This will give both the norm comparison \eqref{eq:weight-equivalence} and the local change of weight \eqref{eq:boundary-map-locality}.

The terms removed from $H_n$ are those whose assigned support intervals $S_y$ meet both $I$ and $I^c$. By \eqref{eq:hamiltonian}, $S_y=\{y,y+1\}$ and $\no{h_y}\le1$, so at most one such interval crosses each boundary. Define their index set and record the resulting bounds:
\begin{equation}\label{eq:boundary-crossing-terms}
 \begin{aligned}
 \mathcal X_I&:=\{y\in[n-1]:S_y\cap I\ne\varnothing,\ S_y\cap I^c\ne\varnothing\},\\
 V_\partial&=\sum_{y\in\mathcal X_I}h_y,\qquad
 |\mathcal X_I|\le2,\qquad\no{V_\partial}\le2,\\
 S_y&\subseteq\mathcal C_{1}(I)\qquad(y\in\mathcal X_I).
 \end{aligned}
\end{equation}
To compare the two Gibbs factors, interpolate in imaginary time. For $0\le s\le\beta/4$, define
\begin{equation}\label{eq:boundary-evolution}
 D(s)=e^{-sH_n}e^{sH_{\mathrm{cut}}},\qquad
 D'(s)=-V_\partial(is)D(s),\qquad D(0)=\idmat.
\end{equation}
The differential equation in \eqref{eq:boundary-evolution} involves only the imaginary-time translates of $V_\partial$. We therefore approximate this generator first. For $r\ge1$, put $k=\lfloor r\rfloor-1$ and
\[
 V_{\partial,r}(s)=\sum_{y\in\mathcal X_I}\cE_{y,k}(h_y(is));
 \qquad V_{\partial,r}(s)=0\quad(0\le r<1).
\]
Equation~\eqref{eq:complex-cone}, with $B_y=h_y$, $d=1$, real time zero, and $u=-s$, bounds the norm and tail of each summand. The contractivity in \eqref{eq:local-transfer} bounds its truncation. Together with \eqref{eq:boundary-crossing-terms} and $N_{k}(y)\subseteq\mathcal C_r(I)$, these estimates give
\begin{equation}\label{eq:boundary-generator-locality}
 \begin{aligned}
 \sup_{0\le s\le\beta/4}(\no{V_\partial(is)}+\no{V_{\partial,r}(s)})&\le C,\\
 \sup_{0\le s\le\beta/4}\no{V_\partial(is)-V_{\partial,r}(s)}&\le Ce^{-cr},\\
 \supp V_{\partial,r}(s)&\subseteq\mathcal C_r(I).
 \end{aligned}
\end{equation}
Bounded radii are included by increasing $C$. Now evolve with this truncated generator:
\[
 D_r'(s)=-V_{\partial,r}(s)D_r(s),\qquad D_r(0)=\idmat.
\]
We will use $D_r$ to approximate $D$ and construct the desired approximation $W_r$.

Integrating the equations for $D,D_r$ and their inverses, using the first inequality in \eqref{eq:boundary-generator-locality}, yields
\begin{equation}\label{eq:boundary-propagator-bounds}
 \begin{aligned}
 \no{D(s)D(t)^{-1}}&\le e^{C(s-t)}\quad(0\le t\le s\le\beta/4),\\
 \no{D(s)}+\no{D(s)^{-1}}+\no{D_r(s)}+\no{D_r(s)^{-1}}&\le C',\\
 \supp D_r(s),\supp D_r(s)^{-1}&\subseteq\mathcal C_r(I).
 \end{aligned}
\end{equation}
The error in replacing $D$ by $D_r$ is controlled by the difference of their generators. Equations~\eqref{eq:boundary-evolution}, \eqref{eq:boundary-generator-locality}, and \eqref{eq:boundary-propagator-bounds} give
\[
 \begin{aligned}
 D(s)-D_r(s)
 &=-\int_0^sD(s)D(t)^{-1}[V_\partial(it)-V_{\partial,r}(t)]D_r(t)\,\dd t,\\
 \no{D(s)-D_r(s)}
 &\le\int_0^se^{C(s-t)}Ce^{-cr}\no{D_r(t)}\,\dd t
 \le C'e^{-cr}.
 \end{aligned}
\]
At $s=\beta/4$, write $D=D(\beta/4)$, $D_r=D_r(\beta/4)$. Thus
\begin{equation}\label{eq:boundary-expansional}
 \no D+\no{D^{-1}}+\no{D_r}+\no{D_r^{-1}}\le C,
 \qquad\no{D-D_r}\le Ce^{-cr}.
\end{equation}

We next account for the normalization of the Gibbs weights. Set $Z=\Tr e^{-\beta H_n}$, $Z_{\mathrm{cut}}=\Tr e^{-\beta H_{\mathrm{cut}}}$. The perturbation bound in \eqref{eq:boundary-crossing-terms} controls their ratio:
\begin{equation}\label{eq:boundary-partition-ratio}
 \begin{aligned}
 \left|\frac{\dd}{\dd t}\log\Tr e^{-\beta(H_{\mathrm{cut}}+tV_\partial)}\right|
 &=\beta\left|\frac{\Tr(V_\partial e^{-\beta(H_{\mathrm{cut}}+tV_\partial)})}
                         {\Tr e^{-\beta(H_{\mathrm{cut}}+tV_\partial)}}\right|
 \le\beta\no{V_\partial},\\
 |\log(Z/Z_{\mathrm{cut}})|&\le\beta\no{V_\partial}\le2\beta.
 \end{aligned}
\end{equation}
Next, $e^{-\beta H_n/4}=De^{-\beta H_{\mathrm{cut}}/4}=e^{-\beta H_{\mathrm{cut}}/4}D^\dagger$ and the isometry \eqref{eq:static-kms-coordinates} give
\begin{equation}\label{eq:boundary-factorization}
 \begin{aligned}
 \rhoB^{1/4}O\rhoB^{1/4}
 &=W(\rho_{\mathrm{cut}}^{1/4}O\rho_{\mathrm{cut}}^{1/4}),\\
 W(X)&=(Z_{\mathrm{cut}}/Z)^{1/2}DXD^\dagger,\qquad
 W_r(X)=(Z_{\mathrm{cut}}/Z)^{1/2}D_rXD_r^\dagger.
 \end{aligned}
\end{equation}
The bounds on $D,D_r$ in \eqref{eq:boundary-expansional} and on $Z/Z_{\mathrm{cut}}$ in \eqref{eq:boundary-partition-ratio} now control $W,W_r$ and their inverses. Applying the Hilbert--Schmidt multiplication bound to \eqref{eq:boundary-factorization},
\[
 \begin{gathered}
 \|W\|_{2\to2}+\|W^{-1}\|_{2\to2}
       +\|W_r\|_{2\to2}+\|W_r^{-1}\|_{2\to2}\le C,\\
 \|W-W_r\|_{2\to2}
 \le(Z_{\mathrm{cut}}/Z)^{1/2}(\no D+\no{D_r})\no{D-D_r}
 \le Ce^{-cr}.
 \end{gathered}
\]
The support statement in \eqref{eq:boundary-propagator-bounds} proves \eqref{eq:boundary-map-locality}. Finally, \eqref{eq:static-kms-coordinates} and \eqref{eq:boundary-factorization} imply
\[
 \begin{aligned}
 \nr O&=\|W(\rho_{\mathrm{cut}}^{1/4}O\rho_{\mathrm{cut}}^{1/4})\|_2
       \le\|W\|_{2\to2}\|O\|_{\rho_{\mathrm{cut}}},\\
 \|O\|_{\rho_{\mathrm{cut}}}&\le\|W^{-1}\|_{2\to2}\nr O,
 \end{aligned}
\]
which is \eqref{eq:weight-equivalence}.
\end{proof}

\begin{lemma}[Clustering on separate intervals]\label{lem:double-clustering}
For local multiplication maps $\mathcal A_S\in\mathfrak M(S)$, $\mathcal B_T\in\mathfrak M(T)$ on matrices over $I$, with disjoint $S,T\subseteq I$,
\begin{equation}\label{eq:double-clustering}
 \begin{aligned}
 &\left|\Tr[\sqrt{\rho_I}\,\mathcal A_S(\mathcal B_T(\sqrt{\rho_I}))]
   -\Tr[\sqrt{\rho_I}\,\mathcal A_S(\sqrt{\rho_I})]\,
    \Tr[\sqrt{\rho_I}\,\mathcal B_T(\sqrt{\rho_I})]\right|\\
 &\qquad\le C(1+|T|)e^{C|S|-c\dist(S,T)}
       \|\mathcal A_S\|_{2\to2}\|\mathcal B_T\|_{2\to2},
 \end{aligned}
\end{equation}
with $C,c$ independent of $I$.
\end{lemma}

Lemma~\ref{lem:double-clustering} will be used together with Lemma~\ref{lem:boundary} to trivialize a \green{Gibbs-centered} island, that is, to prove~\eqref{eq:screen} in~\Cref{prop:thermal}. It controls correlations between the terms supported inside $I$ and terms supported near the boundary.

\begin{proof}
This proof applies Theorem~\ref{thm:ns-clustering} to a suitably chosen $\mathbb{H}$.

We \red{express the commutator maps} in \eqref{eq:Pauli-energy} \red{in Hilbert--Schmidt coordinates} using \eqref{eq:static-kms-coordinates}:
\begin{equation}\label{eq:doubled-derivations}
 \begin{aligned}
 d_{x,a}&:=\Gamma_{\rho_I}^{1/2}\delta_{x,a}\Gamma_{\rho_I}^{-1/2},\qquad
 \mathbb H_I:=\sum_{x\in I,a}d_{x,a}^\dagger d_{x,a},\\
 d_{x,a}(X)&=P_{x,a}^{I}(i\beta/4)X-XP_{x,a}^{I}(-i\beta/4),\\
 P_{x,a}^{I}(z)&:=e^{izH_I}P_{x,a}e^{-izH_I}.
 \end{aligned}
\end{equation}
{\hypersetup{linkcolor=black}Here $\delta_{x,a}(O)=[P_{x,a},O]$ is the commutator in~\eqref{eq:Pauli-energy}, restricted to matrices over $I$, and $\dagger$ denotes the Hilbert-Schmidt adjoint as in~\eqref{eq:trace-adjoint}.}
For $O=\rho_I^{-1/4}X\rho_I^{-1/4}$, the quadratic form of $\mathbb H_I$ is the left side of \eqref{eq:isolated-gap}. That equation and the isometry \eqref{eq:static-kms-coordinates} therefore give
\begin{equation}\label{eq:doubled-gap}
 \begin{aligned}
 \Tr[X^\dagger\mathbb H_I(X)]
 &=\sum_{x\in I,a}\|[P_{x,a},O]\|_{\rho_I}^2
 \ge\gamma\left(\|X\|_2^2-|\Tr(\sqrt{\rho_I}X)|^2\right),\\
 \mathbb H_I(\sqrt{\rho_I})&=0,\qquad
 \Pi_I(X):=\Tr(\sqrt{\rho_I}X)\sqrt{\rho_I},\\
 \mathbb H_I&\ge\gamma(\id-\Pi_I).
 \end{aligned}
\end{equation}
Thus $\sqrt{\rho_I}$ is the unique normalized zero eigenoperator up to phase, and the gap is at least $\gamma$, uniformly in $I$.

It remains to decompose $\mathbb{H}_I$ into a sum of local terms to verify the locality hypothesis \eqref{eq:ns-interaction-input}. The coefficients in \eqref{eq:doubled-derivations} are imaginary-time translates of single-site Pauli strings. Let $\cE_{x,r}^{I}$ denote normalized partial trace outside $N_r(x)\cap I$, and truncate these coefficients:
\[
 d_{x,a}^{(r)}(X)=\cE_{x,r}^{I}(P_{x,a}^{I}(i\beta/4))X
                  -X\cE_{x,r}^{I}(P_{x,a}^{I}(-i\beta/4)).
\]
By the $t=0$ case of \Cref{lem:Araki}, applied to the chain $I$ (its constants are uniform in the chain length) with $B_x=P_{x,a}$, $d=0$, and $u_*=\beta/4$, both coefficients have uniformly bounded norms and tails of order $e^{-cr\log(r+1)}$.  This implies
\begin{equation}\label{eq:doubled-derivation-locality}
 \|d_{x,a}\|_{2\to2}+\|d_{x,a}^{(r)}\|_{2\to2}\le C,\qquad
 \|d_{x,a}-d_{x,a}^{(r)}\|_{2\to2}\le Ce^{-cr\log(r+1)}.
\end{equation}
To turn these approximants into a local interaction for $\mathbb H_I$, telescope $(d_{x,a}^{(r)})^\dagger d_{x,a}^{(r)}$ over the radius. For integer $r\ge0$, set
\begin{equation}\label{eq:doubled-interaction-shells}
 \begin{aligned}
 \mathcal F_{x,a,0}&=(d_{x,a}^{(0)})^\dagger d_{x,a}^{(0)},\\
 \mathcal F_{x,a,r}&=(d_{x,a}^{(r)})^\dagger d_{x,a}^{(r)}
               -(d_{x,a}^{(r-1)})^\dagger d_{x,a}^{(r-1)}\quad(r\ge1),\\
 \mathbb H_I&=\sum_{x\in I,a}\sum_{r\ge0}\mathcal F_{x,a,r},\\
 Z_{x,r}&:=N_r(x)\cap I,\qquad \mathcal F_{x,a,r}\in\mathfrak M(Z_{x,r}),\\
 \|\mathcal F_{x,a,r}\|_{2\to2}&\le C'e^{-c'r\log(r+1)}.
 \end{aligned}
\end{equation}
The last bound follows from \eqref{eq:doubled-derivation-locality} at \red{radii $r$ and $r-1$}. This decay is fast enough to absorb both the distance weight and the support-size weight in \eqref{eq:ns-interaction-input}. Indeed, combine terms with the same support to form $\mathcal F_I$. With $N=4^q$, $|Z_{x,r}|\le2r+1$, $\operatorname{diam}Z_{x,r}\le2r$, and at most $(4^q-1)(2r+1)$ pairs $(x,a)$ through one site, \eqref{eq:doubled-interaction-shells} gives
\begin{equation}\label{eq:doubled-interaction-norm}
 \begin{aligned}
 \sup_z\sum_{Z\ni z}|Z|4^{2q|Z|}
       e^{\zeta\operatorname{diam}Z}\|\mathcal F_I(Z)\|_{2\to2}
 &\le C\sum_{r\ge0}(1+r)^2\,4^{q(4r+2)}e^{2\zeta r}
                        e^{-c'r\log(r+1)}\\
 &<\infty\qquad(\zeta>0),
 \end{aligned}
\end{equation}
uniformly in $I$.

We have now verified both hypotheses of the clustering estimate. In \eqref{eq:ns-interaction-input}--\eqref{eq:ns-clustering-input}, substitute
\[
 (\mathbb H,\mathcal F,F,N)=(\mathbb H_I,\mathcal F_I,\sqrt{\rho_I},4^q).
\]
Equation~\eqref{eq:doubled-gap} supplies $\gamma$; \eqref{eq:doubled-interaction-norm} supplies a uniform bound on $M_\zeta$, for example at $\zeta=1$. Equation~\eqref{eq:ns-clustering-input} then proves \eqref{eq:double-clustering} with constants independent of $I$.
\end{proof}

\subsection{Proof of \texorpdfstring{\Cref{prop:thermal}}{Proposition \getrefnumber{prop:thermal}}}\label{app:projections}

In this section, we prove \Cref{prop:thermal} using the three lemmas above. The proof is organized into three parts:
\[
 \begin{aligned}
 \text{Lemmas~\ref{lem:aux-gap} and~\ref{lem:boundary}:}\quad
 \eqref{eq:isolated-gap},\ \eqref{eq:weight-equivalence}
 &\Longrightarrow\eqref{eq:conditional},\\
 \text{Lemma~\ref{lem:boundary}:}\quad
 \eqref{eq:boundary-map-locality}
 &\Longrightarrow\eqref{eq:projection-locality},\\
 \text{Lemmas~\ref{lem:boundary} and~\ref{lem:double-clustering}:}\quad
 \eqref{eq:boundary-map-locality},\ \eqref{eq:double-clustering}
 &\Longrightarrow\eqref{eq:screen}\Longrightarrow\eqref{eq:quasilocal-screen}.
 \end{aligned}
\]
\begin{proof}[Proof of \Cref{prop:thermal}]
We first work with the product weight $\rho_{\mathrm{cut}}$, defined in~\eqref{eq:boundary-cut}. The boundary comparison then transfers the variance estimate and the spectral conditional expectation to $\rhoB$.

\paragraph{Conditional variance.}
Choose an orthonormal basis $\{F_j\}$ of the KMS Hilbert space of matrices over $I^c$ with respect to the inner product $\langle X,Y\rangle_{\rho_{I^c}}=\Tr(\rho^{1/2}_{I^c}X^\dagger \rho^{1/2}_{I^c}Y)$. Then, for any matrix $O$, we have:
\begin{equation}\label{eq:product-kms-expansion}
 \rho_{\mathrm{cut}}=\rho_I\otimes\rho_{I^c},\qquad
 \rho_{I^c}:=\rho_{I_-}\otimes\rho_{I_+},\qquad
 O=\sum_jO_j\otimes F_j,\qquad
 \langle F_j,F_k\rangle_{\rho_{I^c}}=\delta_{jk}.
\end{equation}
Applying \eqref{eq:isolated-gap} to each $O_j$ in \eqref{eq:product-kms-expansion} gives
\begin{equation}\label{eq:product-conditional-variance}
 \begin{aligned}
 \inf_{B:\supp B\subseteq I^c}\|O-B\|_{\rho_{\mathrm{cut}}}^2
  &=\sum_j\|O_j-\Tr(\rho_I O_j)\idmat_I\|_{\rho_I}^2\le\gamma^{-1}\sum_j\sum_{y\in I,a}\|[P_{y,a},O_j]\|_{\rho_I}^2\\
 &=\gamma^{-1}\sum_{y\in I,a}\|[P_{y,a},O]\|_{\rho_{\mathrm{cut}}}^2.
 \end{aligned}
\end{equation}
We now transfer this estimate to the full Gibbs state $\rhoB$. Let $C_w$ be the constant in \eqref{eq:weight-equivalence}. The orthogonal projection formula \eqref{eq:projection-basic}, followed by \eqref{eq:weight-equivalence} on $O-B$, \eqref{eq:product-conditional-variance}, and \eqref{eq:weight-equivalence} on each commutator, yields
\begin{equation}\label{eq:thermal-conditional-transfer}
 \begin{aligned}
 \nr{(\id-p_I)O}^2
 &=\inf_{B:\supp B\subseteq I^c}\nr{O-B}^2\le C_w^2\inf_{B:\supp B\subseteq I^c}\|O-B\|_{\rho_{\mathrm{cut}}}^2\\
 &\le C_w^2\gamma^{-1}\sum_{y\in I,a}\|[P_{y,a},O]\|_{\rho_{\mathrm{cut}}}^2\le C_w^4\gamma^{-1}\sum_{y\in I,a}\nr{[P_{y,a},O]}^2.
 \end{aligned}
\end{equation}
This proves \eqref{eq:conditional}. For $I=[n]$, where $I^c=\varnothing$ and $\rho_{\mathrm{cut}}=\rhoB$, it is \eqref{eq:isolated-gap} itself, i.e., the global gap.

\paragraph{Local projection approximants.}
The product weight $\rho_{\mathrm{cut}}$ also gives an explicit description of $p^r_I$.

Recall $W,W_r$ from Lemma~\ref{lem:boundary} \eqref{eq:boundary-map-locality}. We first construct the orthogonal projection $p_I$ and then replace some factors in the formula to obtain the local approximants $p_I^r$. Define
\begin{equation}\label{eq:thermal-projection-embedding}
 \begin{aligned}
 V_0(Y)&:=\sqrt{\rho_I}\otimes Y,\quad V_0^\dagger(X)=\Tr_I[(\sqrt{\rho_I}\otimes \idmat_{I^c})X],\quad V_0^\dagger V_0=\id,\\
 V&:=WV_0,\qquad V_r:=W_rV_0\\
 G&:=V^\dagger V,\qquad G_r:=V_r^\dagger V_r,\\
 P&:=VG^{-1}V^\dagger,\qquad
 P_r:=V_rG_r^{-1}V_r^\dagger.
 \end{aligned}
\end{equation}
It is straightforward to check that $V_0^\dagger V_0=\id$. \red{Together with the invertibility of $W$ and $W_r$ from~\eqref{eq:boundary-map-locality}, this} implies that $V,V_r$ are injective, and that $G,G_r$ are invertible. Furthermore, recall $
\operatorname{ran}p_I=\{\idmat_I\otimes B:\ B\in\mathcal{B}(\mathcal{H}_{I^c})\}
$. We have
\[
\mathrm{ran}V=\left\{\Gamma^{1/2}\Gamma^{-1/2}_{\rho_{\mathrm{cut}}}(\sqrt{\rho_I}\otimes B):\  B\in\mathcal{B}(\mathcal{H}_{I^c})\right\}=\left\{\Gamma^{1/2}(\idmat_I\otimes B):\  B\in\mathcal{B}(\mathcal{H}_{I^c})\right\}=\Gamma^{1/2}(\operatorname{ran}p_I)\,.
\]
This implies that $P$ is the orthogonal projection onto $\operatorname{ran}V=\Gamma^{1/2}(\operatorname{ran}p_I)$, and hence
\begin{equation}\label{eqn:thermal-projection_p_appendix}
P=\Gamma^{1/2}p_I\Gamma^{-1/2}\,.
\end{equation}
We now show that $P_r$ induces a local approximation $p_I^r$ of $p_I$ with $p_I^r=\Gamma^{-1/2}P_r\Gamma^{1/2}$.

First, since $V_0$ is an isometry, the bounds on $W^{\pm1}$ and $W_r^{\pm1}$ in \eqref{eq:boundary-map-locality} make the projection formulas stable under the replacement $W\mapsto W_r$:
\begin{equation}\label{eq:thermal-projection-stability}
 \begin{gathered}
 C^{-1}\|Y\|_2\le\|V(Y)\|_2\le C\|Y\|_2,\qquad
 C^{-1}\|Y\|_2\le\|V_r(Y)\|_2\le C\|Y\|_2,\\
 \|G\|_{2\to2}+\|G^{-1}\|_{2\to2}
       +\|G_r\|_{2\to2}+\|G_r^{-1}\|_{2\to2}\le C',\\
 \|V-V_r\|_{2\to2}\le\|W-W_r\|_{2\to2}\le Ce^{-cr},\\
 \|G-G_r\|_{2\to2}+\|G^{-1}-G_r^{-1}\|_{2\to2}\le C'e^{-cr},\\
 \|P-P_r\|_{2\to2}\le C''e^{-cr}.
 \end{gathered}
\end{equation}
Combining the last inequality, $\|P_r\|_{2\to2}\le1$ ($P_r$ is the orthogonal projection onto $\operatorname{ran}V_r$), and \eqref{eq:static-kms-coordinates} gives
\[
\nm{p_I-p_I^r}\le Ce^{-cr},\qquad
 \nm{p_I^r}\le1.
\]

It remains to check that forming $p_I^r$ does not enlarge the support beyond $N_r(I)$. By \eqref{eq:boundary-map-locality}, $W_r$ acts only on the boundary neighborhood $\mathcal C_r=\mathcal C_r(I)$. The partial trace in $V_0^\dagger$ therefore leaves $G_r$ supported on the exterior part of this neighborhood, and its inverse has the same support. Consequently,
\begin{equation}\label{eq:thermal-projection-support}
 \begin{aligned}
 G_r^{\pm1}&\in\mathfrak M(\mathcal C_r\cap I^c)
              \quad\text{on the exterior matrix space},\\
 P_r&=W_r(\Pi_I\otimes G_r^{-1})W_r^\dagger,\\
 P_r&\in\mathfrak M(I\cup\mathcal C_r)\subseteq\mathfrak M(N_r(I))\,,
 \end{aligned}
\end{equation}
The first two \red{statements} use \eqref{eq:thermal-projection-embedding} and $\Pi_I(X)=\Tr(\sqrt{\rho_I}X)\sqrt{\rho_I}$ from \eqref{eq:doubled-gap}; the support inclusion is \eqref{eq:boundary-collar}. The last line implies that $p_I^r\in\mathfrak A(N_r(I))$. This proves \eqref{eq:projection-locality}.

\paragraph{Trivializing a \green{Gibbs-centered} island.} We only consider $I\mathrel{\red{\subsetneq}}[n]$; the case $I=[n]$ is trivial. Let $\mathcal A^\star=\mathcal A\in\mathfrak A(S)$, with $S\subseteq I$ and $\dist(S,I^c)\ge R$, and set
\begin{equation}\label{eq:screening-coordinates}
 \begin{aligned}
 \mathcal M&:=\Gamma^{1/2}\mathcal A\Gamma^{-1/2}
             =\mathcal M_I\otimes\id_{I^c},\\
 c_I&:=\Tr[\sqrt{\rho_I}\,\mathcal M_I(\sqrt{\rho_I})],\\
 \mathcal B_I&:=\mathcal M_I-c_I\id_I,\qquad
 \mathcal B:=\mathcal M-c_I\id=\mathcal B_I\otimes\id_{I^c},\\
 \|\mathcal B\|_{2\to2}&\le2\|\mathcal M\|_{2\to2}.
 \end{aligned}
\end{equation}

To prove \eqref{eq:screen}, we first aim to show that the compression $P\mathcal BP$ of the island $\mathcal B$ is exponentially small:
\begin{equation}\label{eq:screening-boundary}
 \nm{p_I(\mathcal A-c_I\id)p_I}=\|P\mathcal BP\|_{2\to2}
 \le C(1+|S|)e^{-cR}\|\mathcal M\|_{2\to2}.
\end{equation}
Here, the first equality uses \eqref{eqn:thermal-projection_p_appendix}. This gives a bound that is similar to \eqref{eq:screen}, but with the isolated Gibbs mean $c_I$ instead of the full Gibbs mean $c_\rho$. The interior map $\mathcal B_I$ is supported in $S$ and is \green{Gibbs-centered},
\[
 \Tr[\sqrt{\rho_I}\,\mathcal B_I(\sqrt{\rho_I})]=0.
\]
This centering allows us to use clustering to bound its correlations with maps near the boundary of $I$ when we replace $\rho_I$ by $\rhoB$. The argument proceeds in four steps:
\begin{enumerate}
 \item Use $P=VG^{-1}V^\dagger$ and the uniform bounds on $V,G^{-1}$ to reduce the target bound to an estimate of $V^\dagger\mathcal BV$, since
 \[
  P\mathcal BP=VG^{-1}(V^\dagger\mathcal BV)G^{-1}V^\dagger.
 \]
 \item Replace $W$ in $V=WV_0$ by $W_\ell$, with an error of order $e^{-c\ell}\|\mathcal M\|_{2\to2}$. Expand $W_\ell^\dagger W_\ell$ into orthogonal maps on the interior and exterior parts of the boundary neighborhood $\mathcal C_\ell(I)$ to control the correlations with $\mathcal B_I$ by \eqref{eq:double-clustering}.

 \item Choose $\ell$ properly so that both the approximation error and the sum of the correlation terms decay exponentially in $R$.

 \item Recover the bound on $P\mathcal BP$, and use $P(\sqrt{\rhoB})=\sqrt{\rhoB}$ to control the change from $c_I$ to the full Gibbs mean $c_\rho:=\ip{\idmat}{\mathcal A(\idmat)}$. Return to KMS coordinates to obtain \eqref{eq:screen}.
\end{enumerate}

For the first step, because $V$ and $G^{-1}$ are uniformly bounded by \eqref{eq:thermal-projection-stability}, it suffices to bound $V^\dagger\mathcal BV$ in $2\to2$ norm. We approximate $W$ in $V=WV_0$ by $W_\ell$ in this compression, where $W_\ell$ is defined from Lemma~\ref{lem:boundary}~\eqref{eq:boundary-map-locality} and only acts on the boundary neighborhood $\mathcal C_\ell(I)$. Take $\ell=\lfloor \theta R\rfloor$ with $0<\theta<1/2$ to be fixed below. For large $R$, the interior and exterior parts of the boundary neighborhood satisfy
\begin{equation}\label{eq:screening-collar-geometry}
 \begin{aligned}
 U&:=\mathcal C_\ell(I)\cap I,\qquad
 U_{\mathrm{ext}}:=\mathcal C_\ell(I)\cap I^c,\\
 |U|&\le4(1+\ell),\qquad
 \dist(U,S)\ge R-\ell,\qquad[\mathcal M,W_\ell]=0.
 \end{aligned}
\end{equation}
The commutation follows from \eqref{eq:boundary-map-locality} and disjoint coefficient supports. Thus replacing $W$ by $W_\ell$ incurs only the error from \eqref{eq:boundary-map-locality} at radius $\ell$:
\begin{equation}\label{eq:screening-boundary-error}
 \begin{aligned}
 &\|V^\dagger\mathcal BV-V_0^\dagger W_\ell^\dagger \mathcal BW_\ell V_0\|_{2\to2}\\
 \qquad\le &(\|W\|_{2\to2}+\|W_\ell\|_{2\to2})
          \|W-W_\ell\|_{2\to2}\|\mathcal B\|_{2\to2}\\
 \qquad\le &Ce^{-c\ell}\|\mathcal M\|_{2\to2}.
 \end{aligned}
\end{equation}

Because $[\mathcal M,W_\ell]=0$, we have $V_0^\dagger W_\ell^\dagger \mathcal BW_\ell V_0=V_0^\dagger W_\ell^\dagger W_\ell\mathcal BV_0$. We now expand $W_\ell^\dagger W_\ell$ into interior and exterior maps to control its correlations with $\mathcal B_I$. Choose a Hilbert--Schmidt orthonormal matrix-unit basis $\{e_i\}_{i=1}^{4^{q|U|}}$ on $U$, with corresponding linear maps
\[
 \mathcal E_{ij}(X_U):=e_i\Tr(e_j^\dagger X_U),\qquad
 \|\mathcal E_{ij}\|_{2\to2}=1.
\]
By the support and norm bounds in \eqref{eq:boundary-map-locality}, the map $W_\ell^\dagger W_\ell$ has the expansion
\begin{equation}\label{eq:screening-collar-expansion}
 \begin{aligned}
 W_\ell^\dagger W_\ell&=\sum_{i,j=1}^{4^{q|U|}}\mathcal E_{ij}\otimes\mathcal Z_{ij},\\
 \mathcal E_{ij}&\in\mathfrak M(U),\qquad
 \mathcal Z_{ij}\in\mathfrak M(U_{\mathrm{ext}}),\\
 \|\mathcal Z_{ij}\|_{2\to2}&\le\|W_\ell\|_{2\to2}^2\le C,\qquad
 \#\{(i,j)\}=16^{q|U|}.
 \end{aligned}
\end{equation}
For each term in the summation of~\eqref{eq:screening-collar-expansion}, we apply the clustering estimate \eqref{eq:double-clustering} with
\[
 (S,T,\mathcal A_S,\mathcal B_T)=(U,S,\mathcal E_{ij},\mathcal B_I).
\]
Here, we note that $\mathcal B_I$ is supported in $S\subseteq I$, while $\mathcal E_{ij}$ is supported in $U\subseteq I$. Because $\Tr[\sqrt{\rho_I}\,\mathcal B_I(\sqrt{\rho_I})]=0$ by \eqref{eq:screening-coordinates}, the factorized expectation vanishes. Equations~\eqref{eq:double-clustering} and \eqref{eq:screening-collar-geometry} give
\begin{equation}\label{eq:screening-collar-covariance}
 \begin{aligned}
 |\Tr[\sqrt{\rho_I}\,\mathcal E_{ij}(\mathcal B_I(\sqrt{\rho_I}))]|\le C(1+|S|)e^{C|U|-c\dist(U,S)}\|\mathcal B_I\|_{2\to2}\le C'(1+|S|)e^{C'\ell-c(R-\ell)}\|\mathcal M\|_{2\to2}.
 \end{aligned}
\end{equation}
Using the formula for $V_0^\dagger$ in \eqref{eq:thermal-projection-embedding}, the expansion \eqref{eq:screening-collar-expansion} gives the first identity below. Summing the bounds \eqref{eq:screening-collar-covariance} over its $16^{q|U|}$ terms gives the second:
\begin{equation}\label{eq:screening-product-compression}
 \begin{aligned}
 V_0^\dagger W_\ell^\dagger W_\ell\mathcal BV_0
 &=\sum_{i,j}\Tr[\sqrt{\rho_I}\,\mathcal E_{ij}(\mathcal B_I(\sqrt{\rho_I}))]\,
          \mathcal Z_{ij},\\
 \|V_0^\dagger W_\ell^\dagger W_\ell\mathcal BV_0\|_{2\to2}
 &\le C(1+|S|)e^{C_1\ell-cR}\|\mathcal M\|_{2\to2}.
 \end{aligned}
\end{equation}
Recall $\ell=\lfloor \theta R\rfloor$. Choose $\theta$ so that $C_1\theta\le c/2$. This makes both terms in \eqref{eq:screening-boundary-error} and \eqref{eq:screening-product-compression} exponentially small in $R$:
\begin{equation}\label{eq:screening-embedding-compression}
 \begin{aligned}
 \|V^\dagger\mathcal BV\|_{2\to2}
 \le C(1+|S|)(e^{-c\ell}+e^{C_1\ell-cR})\|\mathcal M\|_{2\to2}\le C'e^{-c'R}(1+|S|)\|\mathcal M\|_{2\to2}.
 \end{aligned}
\end{equation}
Returning to the projection formula \eqref{eq:thermal-projection-embedding}, its uniform bounds \eqref{eq:thermal-projection-stability} turn \eqref{eq:screening-embedding-compression} into
\begin{equation}\label{eq:screening-isolated-centering}
 \begin{aligned}
 \|P\mathcal BP\|_{2\to2}\le\|V\|_{2\to2}^2\|G^{-1}\|_{2\to2}^2\|V^\dagger\mathcal BV\|_{2\to2}\le C(1+|S|)e^{-cR}\|\mathcal M\|_{2\to2}.
 \end{aligned}
\end{equation}
Finally, $P$ fixes $\sqrt{\rhoB}$ because $p_I(\idmat)=\idmat$ in~\eqref{eq:projection-identity}. {\hypersetup{linkcolor=black}Hence \eqref{eq:screening-isolated-centering} bounds the difference between the isolated Gibbs mean $c_I$ and the full Gibbs mean $c_\rho$.} Specifically,
\begin{equation}\label{eq:screening-full-centering}
 \begin{aligned}
 P(\sqrt{\rhoB})&=\sqrt{\rhoB},\qquad\|\sqrt{\rhoB}\|_2=1,\\
 c_\rho&:=\ip{\idmat}{\mathcal A(\idmat)}
          =\Tr[\sqrt{\rhoB}\,\mathcal M(\sqrt{\rhoB})],\\
 |c_\rho-c_I|&=|\Tr[\sqrt{\rhoB}\,(P\mathcal BP)(\sqrt{\rhoB})]|
                \le\|P\mathcal BP\|_{2\to2},\\
 \|P(\mathcal M-c_\rho\id)P\|_{2\to2}
 &\le2\|P\mathcal BP\|_{2\to2}
 \le C(1+|S|)e^{-cR}\|\mathcal M\|_{2\to2}.
 \end{aligned}
\end{equation}
Apply \eqref{eq:static-kms-coordinates} to \eqref{eq:screening-full-centering}, using $P=\Gamma^{1/2}p_I\Gamma^{-1/2}$ and $\|\mathcal M\|_{2\to2}=\nm{\mathcal A}$, to obtain \eqref{eq:screen} for large $R$. The bound for small $R$ follows from $\nm{p_I}\le1$ in \eqref{eq:projection-basic}.

\paragraph{Quasi-local islands.} Let $\mathcal A_x^\star=\mathcal A_x$ satisfy \eqref{eq:coefficient-locality} and $\ip{\idmat}{\mathcal A_x(\idmat)}=0$. To apply \eqref{eq:screen}, we choose the interval $I=N_r(x)$ and truncate $\mathcal A_x$ on a smaller neighborhood. At radius $\ell=\lfloor r/4\rfloor$, symmetrize an approximant from \eqref{eq:coefficient-locality}, using the adjoint identity in \eqref{eq:weighted-algebra}:
\begin{equation}\label{eq:screening-quasilocal-approximant}
 \begin{aligned}
 \mathcal A_{x,\ell}^\star&=\mathcal A_{x,\ell}\in\mathfrak A(N_\ell(x)),\qquad
 \nm{\mathcal A_{x,\ell}}\le C,\\
 \nm{\mathcal A_x-\mathcal A_{x,\ell}}&\le Ce^{-c\ell},\\
 c_\ell&:=\ip{\idmat}{\mathcal A_{x,\ell}(\idmat)},\qquad
 |c_\ell|\le\nm{\mathcal A_{x,\ell}-\mathcal A_x}\le Ce^{-c\ell}.
 \end{aligned}
\end{equation}
The last bound uses centering and $\nr{\idmat}=1$ from \eqref{eq:normalization-identities}, so the change in the mean is no larger than the discarded tail. The supported estimate \eqref{eq:screen} applies with
\[
 I=N_r(x),\qquad S=N_\ell(x),\qquad
 R=r-\ell,\qquad\mathcal A=\mathcal A_{x,\ell},\qquad |S|\le2\ell+1.
\]
Combining this instance of \eqref{eq:screen} with \eqref{eq:screening-quasilocal-approximant} and $\nm{p_{N_r(x)}}\le1$ from \eqref{eq:projection-basic} gives
\[
 \begin{aligned}
 \nm{p_{N_r(x)}\mathcal A_xp_{N_r(x)}}
 &\le\nm{\mathcal A_x-\mathcal A_{x,\ell}}+|c_\ell|
      +\nm{p_{N_r(x)}(\mathcal A_{x,\ell}-c_\ell\id)p_{N_r(x)}}\\
 &\le Ce^{-c\ell}+C(2\ell+2)e^{-c(r-\ell)}
 \le C'e^{-c'r}.
 \end{aligned}
\]
This proves \eqref{eq:quasilocal-screen}.
\end{proof}

\bibliographystyle{alpha}
\bibliography{sbi}

@article{PoulinWocjan09,
  title={Sampling from the thermal quantum Gibbs state and evaluating partition functions with a quantum computer},
  author={Poulin, David and Wocjan, Pawel},
  journal={Physical review letters},
  volume={103},
  number={22},
  pages={220502},
  year={2009},
  publisher={APS}
}

@misc{cps26,
      title={Modified logarithmic Sobolev inequality for 1D non-commuting Hamiltonians},
      author={Capel, {\'A}ngela and P{\'e}rez-Garc{\'\i}a, David and Matteo Scandi},
      year={2026},
      eprint={2610.03683},
      archivePrefix={arXiv},
      primaryClass={quant-ph},
      url={https://arxiv.org/abs/2610.03683},
}

@misc{chen2026sizeindependentlogarithmicsobolevinequalities,
      title={Size-independent Logarithmic Sobolev Inequalities for Quantum Gibbs Samplers},
      author={Hongrui Chen and Zhiyan Ding and Jianfeng Lu},
      year={2026},
      eprint={2610.31571},
      archivePrefix={arXiv},
      primaryClass={quant-ph},
      url={https://arxiv.org/abs/2610.31571},
}

@misc{gao2026completemodifiedlogarithmicsobolev,
      title={Complete Modified Logarithmic Sobolev Inequalities for one-dimensional quantum Spin Chain},
      author={Li Gao and Jingyu Guo and Sihao Wang},
      year={2026},
      eprint={2610.28600},
      archivePrefix={arXiv},
      primaryClass={quant-ph},
      url={https://arxiv.org/abs/2610.28600},
}

@misc{rf26,
      title={Rapid mixing of Gibbs samplers via quantum Dobrushin--Shlosman conditions},
      author={Rouz{\'e}, Cambyse and Fran{\c{c}}a, Daniel Stilck},
      year={2026},
      eprint={2610.03578},
      archivePrefix={arXiv},
      primaryClass={quant-ph},
      url={https://arxiv.org/abs/2610.03578},
}

@article{Motta20,
  title={Determining eigenstates and thermal states on a quantum computer using quantum imaginary time evolution},
  author={Motta, Mario and Sun, Chong and Tan, Adrian T. K. and O'Rourke, Matthew J. and Ye, Erika and Minnich, Austin J. and Brand{\~a}o, Fernando G. S. L. and Chan, Garnet Kin-Lic},
  journal={Nature Physics},
  volume={16},
  number={2},
  pages={205--210},
  year={2020}
}

@inproceedings{BrandaoSvore17,
  title={Quantum speed-ups for solving semidefinite programs},
  author={Brand{\~a}o, Fernando G. S. L. and Svore, Krysta M.},
  booktitle={58th Annual IEEE Symposium on Foundations of Computer Science (FOCS 2017)},
  pages={415--426},
  year={2017}
}

@article{vAGGdW20,
  title={Quantum {SDP}-solvers: better upper and lower bounds},
  author={van Apeldoorn, Joran and Gily{\'e}n, Andr{\'a}s and Gribling, Sander and de Wolf, Ronald},
  journal={Quantum},
  volume={4},
  pages={230},
  year={2020}
}

@article{KieferovaWiebe17,
  title={Tomography and generative training with quantum {B}oltzmann machines},
  author={Kieferov{\'a}, M{\'a}ria and Wiebe, Nathan},
  journal={Physical Review A},
  volume={96},
  number={6},
  pages={062327},
  year={2017}
}

@article{Amin18,
  title={Quantum {B}oltzmann machine},
  author={Amin, Mohammad H. and Andriyash, Evgeny and Rolfe, Jason and Kulchytskyy, Bohdan and Melko, Roger},
  journal={Physical Review X},
  volume={8},
  number={2},
  pages={021050},
  year={2018}
}

@article{Araki69,
  title={Gibbs states of a one dimensional quantum lattice},
  author={Araki, Huzihiro},
  journal={Communications in Mathematical Physics},
  volume={14},
  number={2},
  pages={120--157},
  year={1969}
}

@article{Davies74,
  title={Markovian master equations},
  author={Davies, E. Brian},
  journal={Communications in Mathematical Physics},
  volume={39},
  number={2},
  pages={91--110},
  year={1974}
}

@article{Davies76,
  title={Markovian master equations. {II}},
  author={Davies, E. Brian},
  journal={Mathematische Annalen},
  volume={219},
  number={2},
  pages={147--158},
  year={1976}
}

@article{TD00,
  title={Problem of equilibration and the computation of correlation functions on a quantum computer},
  author={Terhal, Barbara M. and DiVincenzo, David P.},
  journal={Physical Review A},
  volume={61},
  number={2},
  pages={022301},
  year={2000}
}

@article{TOVPV11,
  title={Quantum {M}etropolis sampling},
  author={Temme, Kristan and Osborne, Tobias J. and Vollbrecht, Karl G. and Poulin, David and Verstraete, Frank},
  journal={Nature},
  volume={471},
  number={7336},
  pages={87--90},
  year={2011}
}

@article{KastoryanoTemme13,
  title={Quantum logarithmic {S}obolev inequalities and rapid mixing},
  author={Kastoryano, Michael J. and Temme, Kristan},
  journal={Journal of Mathematical Physics},
  volume={54},
  number={5},
  pages={052202},
  year={2013}
}

@article{CKBG23,
  title={Quantum thermal state preparation},
  author={Chen, Chi-Fang and Kastoryano, Michael J. and Brand{\~a}o, Fernando G. S. L. and Gily{\'e}n, Andr{\'a}s},
  journal={arXiv preprint arXiv:2303.18224},
  year={2023}
}

@article{CKBG25,
  title={Efficient quantum thermal simulation},
  author={Chen, Chi-Fang and Kastoryano, Michael J. and Brand{\~a}o, Fernando G. S. L. and Gily{\'e}n, Andr{\'a}s},
  journal={Nature},
  volume={646},
  pages={561--566},
  year={2025}
}

@article{GCDK24,
  title={Quantum generalizations of {G}lauber and {M}etropolis dynamics},
  author={Gily{\'e}n, Andr{\'a}s and Chen, Chi-Fang and Doriguello, Jo{\~a}o F. and Kastoryano, Michael J.},
  journal={arXiv preprint arXiv:2405.20322},
  year={2024}
}

@article{RFA24,
  title={Efficient thermalization and universal quantum computing with quantum {G}ibbs samplers},
  author={Rouz{\'e}, Cambyse and Stilck Fran{\c{c}}a, Daniel and Alhambra, {\'A}lvaro M.},
  journal={Proceedings of the 57th Annual ACM Symposium on Theory of Computing (STOC 2025)},
  pages={1488--1495},
  year={2025},
  note={arXiv:2403.12691}
}

@article{RFA25,
  title={Optimal quantum algorithm for {G}ibbs state preparation},
  author={Rouz{\'e}, Cambyse and Stilck Fran{\c{c}}a, Daniel and Alhambra, {\'A}lvaro M.},
  journal={Physical Review Letters},
  volume={136},
  pages={060601},
  year={2026},
  note={arXiv:2411.04885}
}

@article{TongZhan25,
  title={Fast mixing of weakly interacting fermionic systems at any temperature},
  author={Tong, Yu and Zhan, Yongtao},
  journal={PRX Quantum},
  volume={6},
  pages={030301},
  year={2025},
  note={arXiv:2501.00443}
}

@article{vsmid2025rapid,
  title={Rapid mixing of quantum Gibbs samplers for weakly-interacting quantum systems},
  author={{\v{S}}m{\'\i}d, {\v{S}}t{\v{e}}p{\'a}n and Meister, Richard and Berta, Mario and Bondesan, Roberto},
  journal={arXiv preprint arXiv:2510.04954},
  year={2025}
}

@inproceedings{bakshi2026dobrushin,
  title={A Dobrushin condition for quantum Markov chains: Rapid mixing and conditional mutual information at high temperature},
  author={Bakshi, Ainesh and Liu, Allen and Moitra, Ankur and Tang, Ewin},
  booktitle={Proceedings of the 58th Annual ACM Symposium on Theory of Computing},
  pages={1508--1516},
  year={2026}
}

@article{bakshi2026rapid,
  title={Rapid mixing for high-temperature Gibbs states with arbitrary external fields},
  author={Bakshi, Ainesh and Tan, Xinyu},
  journal={arXiv preprint arXiv:2604.08408},
  year={2026}
}

@article{bergamaschi2026fast,
  title={Fast mixing of all-to-all quantum systems at high temperatures},
  author={Bergamaschi, Thiago},
  journal={arXiv preprint arXiv:2606.26090},
  year={2026}
}

@article{SMBB25,
  title={Polynomial-time quantum Gibbs sampling for the weak and strong coupling regime of the Fermi-Hubbard model at any temperature},
  author={{\v{S}}m{\'\i}d, {\v{S}}t{\v{e}}p{\'a}n and Meister, Richard and Berta, Mario and Bondesan, Roberto},
  journal={Nature Communications},
  volume={16},
  number={1},
  pages={10736},
  year={2025},
  publisher={Nature Publishing Group UK London}
}

@article{farrell2026preparing,
  title={Preparing thermal states of frustrated quantum spin systems using 139 qubits},
  author={Farrell, Roland C and Zhan, Yongtao and Katschke, Lucas and Pollet, Lode and Rosen, Ilan T and Halimeh, Jad C},
  journal={arXiv preprint arXiv:2605.26245},
  year={2026}
}

@article{BCGLPR23,
  title={Rapid thermalization of spin chain commuting Hamiltonians},
  author={Bardet, Ivan and Capel, {\'A}ngela and Gao, Li and Lucia, Angelo and P{\'e}rez-Garc{\'\i}a, David and Rouz{\'e}, Cambyse},
  journal={Physical Review Letters},
  volume={130},
  number={6},
  pages={060401},
  year={2023},
  publisher={APS}
}

@article{Ciccarello22,
  title={Quantum collision models: Open system dynamics from repeated interactions},
  author={Ciccarello, Francesco and Lorenzo, Salvatore and Giovannetti, Vittorio and Palma, G Massimo},
  journal={Physics Reports},
  volume={954},
  pages={1--70},
  year={2022},
  publisher={Elsevier}
}

@article{chen2026overcoming,
  title={Overcoming the {Lamb} Shift in System-Bath Interaction Models via {KMS} Detailed Balance: High-Accuracy Thermalization with Time-Bounded Interactions},
  author={Chen, Hongrui and Ding, Zhiyan and Zhang, Ruizhe},
  journal={arXiv preprint arXiv:2604.15616},
  year={2026}
}

@article{ong2026rigorous,
  title={Rigorous error bounds for dissipative thermal state preparation from weak system-bath coupling},
  author={Ong, Christopher and Parameswaran, SA and Placke, Benedikt and Hahn, Dominik},
  journal={arXiv preprint arXiv:2605.03011},
  year={2026}
}

@article{scandi2026thermalization,
  title={Thermalization in open many-body systems and KMS detailed balance},
  author={Scandi, Matteo and Alhambra, {\'A}lvaro M},
  journal={Physical Review X},
  volume={16},
  number={1},
  pages={011040},
  year={2026},
  publisher={APS}
}

@misc{CB23,
      title={Fast Thermalization from the Eigenstate Thermalization Hypothesis},
      author={Chi-Fang Chen and Fernando G. S. L. Brandão},
      year={2023},
      eprint={2112.07646},
      archivePrefix={arXiv},
      primaryClass={quant-ph},
      url={https://arxiv.org/abs/2112.07646},
}

@article{shtanko2021preparing,
  title={Preparing thermal states on noiseless and noisy programmable quantum processors},
  author={Shtanko, Oles and Movassagh, Ramis},
  journal={arXiv preprint arXiv:2112.14688},
  year={2021}
}

@article{lloyd2026quantum,
  title={Quantum thermal state preparation for near-term quantum processors},
  author={Lloyd, Jerome and Abanin, Dmitry A},
  journal={Physical Review X},
  volume={16},
  number={3},
  pages={031053},
  year={2026},
  publisher={APS}
}

@article{wang2025beyond,
  title={Beyond Lindblad dynamics: Rigorous guarantees for thermal and ground state preservation under system bath interactions},
  author={Wang, Ke and Ding, Zhiyan},
  journal={arXiv preprint arXiv:2512.03457},
  year={2025}
}

@article{hahn2025towards,
  title={Towards efficient quantum thermal state preparation via local driving: Lindbladian simulation with provable guarantees},
  author={Hahn, Dominik and Parameswaran, SA and Placke, Benedikt},
  journal={arXiv preprint arXiv:2505.22816},
  year={2025}
}

@article{ding2026simple,
  title={Simple and efficient end-to-end quantum thermal and ground state preparation},
  author={Ding, Zhiyan and Zhan, Yongtao and Preskill, John and Lin, Lin},
  journal={Nature Physics},
  pages={1--5},
  year={2026},
  publisher={Nature Publishing Group UK London}
}

@article{DLL25,
  title={Efficient Quantum Gibbs Samplers with Kubo--Martin--Schwinger Detailed Balance Condition},
  author={Ding, Zhiyan and Li, Bowen and Lin, Lin},
  journal={Communications in Mathematical Physics},
  volume={406},
  number={3},
  pages={67},
  year={2025},
  publisher={Springer}
}

@article{perez2023locality,
  title={Locality Estimates for Complex Time Evolution in {1D}},
  author={P{\'e}rez-Garc{\'\i}a, David and P{\'e}rez-Hern{\'a}ndez, Antonio},
  journal={Communications in Mathematical Physics},
  volume={399},
  number={2},
  pages={929--970},
  year={2023},
  publisher={Springer}
}

@article{pilatowskycameo2026,
      title={One-dimensional quantum Gibbs states in constant circuit depth},
      author={Saúl Pilatowsky-Cameo and Georgios Styliaris and Ainesh Bakshi and Daniel Malz},
      year={2026},
      journal={arXiv preprint arXiv:2609.35973},
      archivePrefix={arXiv},
}

@article{chending2026,
      title={{Spectral analysis beyond the spectral gap: Rapid mixing of quantum Gibbs samplers in one dimension}},
      author={Hongrui Chen and Zhiyan Ding},
      year={2026},
      journal={{arXiv preprint arXiv:2610.05848}},
      archivePrefix={arXiv},
}

@article{kim2026rapidmixingquantumspin,
      title={Rapid mixing of quantum spin chains at any finite temperature},
      author={Leeseok Kim},
      year={2026},
      journal={arXiv preprint arXiv:2610.01190},
}

@article{KACR,
  title={Rapid Thermalization of Dissipative Many-Body Dynamics of Commuting Hamiltonians},
  author={Kochanowski, Jan and Alhambra, {\'A}lvaro M. and Capel, {\'A}ngela and Rouz{\'e}, Cambyse},
  journal={Communications in Mathematical Physics},
  volume={406},
  pages={176},
  year={2025},
  doi={10.1007/s00220-025-05353-y},
  note={arXiv:2404.16780}
}

@article{BergamaschiChen,
  title={Fast Mixing of Quantum Spin Chains at All Temperatures},
  author={Bergamaschi, Thiago and Chen, Chi-Fang},
  journal={arXiv preprint arXiv:2510.08533},
  year={2026},
  note={arXiv:2510.08533v2}
}

@article{SlezakEtAl,
  title={Polynomial-Time Thermalization and Gibbs Sampling from System--Bath Couplings},
  author={Slezak, Samuel and Scandi, Matteo and Alhambra, {\'A}lvaro M. and Stilck Fran\c{c}a, Daniel and Rouz{\'e}, Cambyse},
  journal={arXiv preprint arXiv:2601.16154},
  year={2026}
}

@article{CKG,
  title={An Efficient and Exact Noncommutative Quantum {G}ibbs Sampler},
  author={Chen, Chi-Fang and Kastoryano, Michael J. and Gily{\'e}n, Andr{\'a}s},
  journal={arXiv preprint arXiv:2311.09207},
  year={2023}
}

@article{KastoryanoBrandao,
  title={Quantum {G}ibbs Samplers: The Commuting Case},
  author={Kastoryano, Michael J. and Brand{\~a}o, Fernando G. S. L.},
  journal={Communications in Mathematical Physics},
  volume={344},
  pages={915--957},
  year={2016},
  doi={10.1007/s00220-016-2641-8}
}

@article{NachtergaeleSims,
  title={{Lieb--Robinson} Bounds and the Exponential Clustering Theorem},
  author={Nachtergaele, Bruno and Sims, Robert},
  journal={Communications in Mathematical Physics},
  volume={265},
  number={1},
  pages={119--130},
  year={2006},
  doi={10.1007/s00220-006-1556-1},
  note={arXiv:math-ph/0506030}
}

\end{document}